\documentclass[english]{book}
\usepackage[T1]{fontenc}
\usepackage[latin9]{inputenc}
\usepackage[a4paper]{geometry}
\usepackage{color}
\usepackage{babel}
\usepackage{verbatim}
\usepackage{longtable}
\usepackage{float}
\usepackage{rotfloat}
\usepackage{url}
\usepackage{amssymb}
\usepackage{graphicx}
\usepackage[unicode=true,
 bookmarks=true,bookmarksnumbered=false,bookmarksopen=false,
 breaklinks=false,pdfborder={0 0 1},backref=false,colorlinks=false]
 {hyperref}
\hypersetup{pdftitle={GENIE Physics and User Manual}}
\usepackage{breakurl}

\makeatletter

\providecommand{\tabularnewline}{\\}
\floatstyle{ruled}
\newfloat{algorithm}{tbp}{loa}[chapter]
\providecommand{\algorithmname}{Algorithm}
\floatname{algorithm}{\protect\algorithmname}

\makeatother

\begin{document}

\title{\textbf{\huge{}The GENIE Neutrino Monte Carlo Generator}\textbf{\Large{}}\\
\textbf{\Huge{}}\\
\textbf{PHYSICS \& USER MANUAL}\textbf{\Huge{}}\\
\textbf{\textit{\normalsize{}(version 2.10.0)}}}

\author{{\small{}Costas Andreopoulos$^{2,\:5}$, Christopher Barry$^{2}$,
Steve Dytman$^{3}$, Hugh Gallagher$^{6}$, Tomasz Golan$^{1,\:4}$,
}\\
{\small{}Robert Hatcher$^{1}$, Gabriel Perdue$^{1}$, and Julia Yarba$^{1}$}
\\
{\scriptsize{}}\\
{\scriptsize{}$^{1}$ Fermi National Accelerator Laboratory, Batavia,
Illinois 60510, USA}\\
{\scriptsize{}$^{2}$University of Liverpool, Dept. of Physics, Liverpool
L69 7ZE, UK}\\
{\scriptsize{}$^{3}$ University of Pittsburgh, Dept. of Physics and
Astronomy, Pittsburgh PA 15260, USA}\\
{\scriptsize{}$^{4}$ University of Rochester, Dept. of Physics and
Astronomy, Rochester NY 14627, USA}\\
{\scriptsize{}$^{5}$STFC Rutherford Appleton Laboratory, Particle
Physics Dept., Oxfordshire OX11 0QX, UK}\\
{\scriptsize{}$^{6}$ Tufts University, Dept. of Physics and Astronomy,
Medford MA, 02155, USA}\\
{\small{}}\\
\textsc{\normalsize{} }}

\maketitle

\chapter*{$\;\;\;$}

\author{\textcolor{red}{This is a }\textit{\textcolor{red}{draft}}\textcolor{red}{{}
document. }\\
\textcolor{red}{Please send comments and corrections to costas.andreopoulos@stfc.ac.uk}}

\chapter*{Preface}

GENIE \cite{GENIE} is a suite of products for the experimental neutrino
physics community. This suite includes i) a modern software framework
for implementing neutrino event generators, a state-of-the-art comprehensive
physics model and tools to support neutrino interaction simulation
for realistic experimental setups (aka the ``Generator''), ii) extensive
archives of neutrino, charged-lepton and hadron scattering data and
software to produce a comprehensive set of data/MC comparisons (aka
the ``Comparisons''), and iii) a generator tuning framework and
fitting applications (aka the ``Tuning''). These products come with
different licenses. The Generator is an open-source product, whereas
the Comparisons and the Tuning are distributed only under special
agreement.\\
\\
This book provides the definite guide for the GENIE Generator: It
presents the software architecture and a detailed description of its
physics model and official tunes. In addition, it provides a rich
set of data/MC comparisons that characterise the physics performance
of GENIE. Detailed step-by-step instructions on how to install and
configure the Generator, run its applications and analyze its outputs
are also included.\\

\chapter*{Key}
\begin{itemize}
\item Class names: \textit{Italic} (eg. \textit{TRandom3}, \textit{GHepRecord},
...)
\item GENIE application / library names: \textit{Italic} (eg. \textit{gT2Kevgen},
\textit{gevdump}, ...)
\item External packages: Normal (eg. ROOT, PYTHIA6, ...)
\item Object names: `Normal' (eg. `flux', ...)
\item Typed-in commands: \texttt{\textbf{\small{}Courier+Small+Bold}} (eg.
\texttt{\textbf{\small{}\$ gevdump -f /data/file.ghep.root}})
\item Fragments of typed-in commands: \texttt{\textbf{\small{}`Courier+Small+Bold'}}
(eg. \texttt{\textbf{\small{}`-n 100'}})
\item Environmnental variables: \texttt{\textbf{\small{}`Courier+Small+Bold'}}
(eg. \texttt{\textbf{\small{}\$GENIE}})
\item Filenames and paths: `\textit{Italic}' (eg. \textit{`/data/flux/atmospheric/bglrs/numu.root'})
\item URLs: \textit{Italic} (eg.\textit{ http://www.genie-mc.org})
\end{itemize}
Notes:
\begin{itemize}
\item A leading \$, \& or \% in typed-in commands represent your command
shell prompt. Don't type that in.
\end{itemize}
\tableofcontents{}

\chapter{Introduction}

\section{GENIE project overview}

Over the last few years, throughout the field of high energy physics
(HEP), we have witnessed an enormous effort committed to migrating
many popular procedural Monte Carlo Generators into their C++ equivalents
designed using object-oriented methodologies. Well-known examples
are the GEANT \cite{Agostinelli:2002hh}, HERWIG \cite{Bahr:2008pv}
and PYTHIA \cite{Sjostrand:2007gs} Monte Carlo Generators. This reflects
a radical change in our approach to scientific computing. Along with
the eternal requirement that the modeled physics be correct and extensively
validated with external data, the evolving nature of computing in
HEP has introduced new requirements. These requirements relate to
the way large HEP software systems are developed and maintained, by
wide geographically-spread collaborations over a typical time-span
of \$\textbackslash{}sim\$25 years during which they will undergo
many (initially unforeseen) extensions and modifications to accommodate
new needs. This puts a stress on software qualities such re-usability,
maintainability, robustness, modularity and extensibility. Software
engineering provides many well proven techniques to address these
requirements and thereby improve the quality and lifetime of HEP software.
In neutrino physics, the requirements for a new physics generator
are more challenging for three reasons: the lack of a `canonical'
procedural generator, theoretical and phenomenological challenges
in modeling few-GeV neutrino interactions, and the rapidly evolving
experimental and theoretical landscape. \\

The long-term goal of this project is for GENIE to become a `canonical'
neutrino event generator whose validity will extend to all nuclear
targets and neutrino flavors over a wide spectrum of energies ranging
from $\sim$1 MeV to $\sim$1 PeV. Currently, emphasis is given to
the few-GeV energy range, the challenging boundary between the non-perturbative
and perturbative regimes which is relevant for the current and near
future long-baseline precision neutrino experiments using accelerator-made
beams. The present version provides comprehensive neutrino interaction
modelling in the energy from, approximately, $\sim$100 MeV to a few
hundred GeV.

GENIE\footnote{\textbf{GENIE} stands for \textbf{G}enerates \textbf{E}vents for \textbf{N}eutrino
\textbf{I}nteraction \textbf{E}xperiments} is a ROOT-based \cite{Brun:1997pa} Neutrino MC Generator. It was
designed using object-oriented methodologies and developed entirely
in C++ over a period of more than three years, from 2004 to 2007.
Its first official physics release (v2.0.0) was made available on
August 2007. GENIE has already being adopted by the majority of neutrino
experiments, including those as the JPARC and NuMI neutrino beamlines,
and will be an important physics tool for the exploitation of the
world accelerator neutrino program.

The project is supported by a group of physicists from all major neutrino
experiments operating in this energy range, establishing GENIE as
a major HEP event generator collaboration. Many members of the GENIE
collaboration have extensive experience in developing and maintaining
the legacy Monte Carlo Generators that GENIE strives to replace, which
guarantees knowledge exchange and continuation. The default set of
physics models in GENIE have adiabatically evolved from those in the
NEUGEN \cite{Gallagher:2002sf} package, which was used as the event
generator by numerous experiments over the past decade.

\section{Neutrino Interaction Simulation: Challenges and Significance}

Neutrinos have played an important role in particle physics since
their discovery half a century ago. They have been used to elucidate
the structure of the electroweak symmetry groups, to illuminate the
quark nature of hadrons, and to confirm our models of astrophysical
phenomena. With the discovery of neutrino oscillations using atmospheric,
solar, accelerator, and reactor neutrinos, these elusive particles
now take center stage as the objects of study themselves. Precision
measurements of the lepton mixing matrix, the search for lepton charge-parity
(CP) violation, and the determination of the neutrino masses and hierarchy
will be major efforts in HEP for several decades. The cost of this
next generation of experiments will be significant, typically tens
to hundreds of millions of dollars. A comprehensive, thoroughly tested
neutrino event generator package plays an important role in the design
and execution of these experiments, since this tool is used to evaluate
the feasibility of proposed projects and estimate their physics impact,
make decisions about detector design and optimization, analyze the
collected data samples, and evaluate systematic errors. With the advent
of high-intensity neutrino beams from proton colliders, we have entered
a new era of high-statistics, precision neutrino experiments which
will require a new level of accuracy in our knowledge, and simulation,
of neutrino interaction physics.

While object-oriented physics generators in other fields of high energy
physics were evolved from well established legacy systems, in neutrino
physics no such `canonical' MC exists. Until quite recently, most
neutrino experiments developed their own neutrino event generators.
This was due partly to the wide variety of energies, nuclear targets,
detectors, and physics topics being simulated. Without doubt these
generators, the most commonly used of which have been GENEVE \cite{Cavanna:2002se},
NEUT \cite{Hayato:2002sd}, NeuGEN \cite{Gallagher:2002sf}, NUANCE
\cite{Casper:2002sd} and NUX \cite{Rubbia:2001}, played an important
role in the design and exploration of the previous and current generation
of accelerator neutrino experiments. Tuning on the basis of unpublished
data from each group's own experiment has not been unusual making
it virtually impossible to perform a global, independent evaluation
for the state-of-the-art in neutrino interaction physics simulations.
Moreover, limited manpower and the fragility of the overextended software
architectures meant that many of these legacy physics generators were
not keeping up with the latest theoretical ideas and experimental
measurements. A more recent development in the field has been the
direct involvement of theory groups in the development of neutrino
event generators, such as the NuWRO \cite{Juszczak:2005zs} and GiBUU
\cite{Leitner:2006ww} packages, and the inclusion of neutrino scattering
in the long-established FLUKA hadron scattering package \cite{Fasso:2003xz}.

Simulating neutrino interactions in the energy range of interest to
current and near-future experiments poses significant challenges.
This broad energy range bridges the perturbative and non-perturbative
pictures of the nucleon and a variety of scattering mechanisms are
important. In many areas, including elementary cross sections, hadronization
models, and nuclear physics, one is required to piece together models
with different ranges of validity in order to generate events over
all of the available phase space. This inevitably introduces challenges
in merging and tuning models, making sure that double counting and
discontinuities are avoided. In addition there are kinematic regimes
which are outside the stated range of validity of all available models,
in which case we are left with the challenge of developing our own
models or deciding which model best extrapolates into this region.
An additional fundamental problem in this energy range is a lack of
data. Most simulations have been tuned to bubble chamber data taken
in the 70's and 80's. Because of the limited size of the data samples
(important exclusive channels might only contain a handful of events),
and the limited coverage in the space of ($\nu/\overline{\nu},E_{\nu}$,
A), substantial uncertainties exist in numerous aspects of the simulations.

The use cases for GENIE are also informed by the experiences of the
developers and users of the previous generation of procedural codes.
Dealing with these substantial model uncertainties has been an important
analysis challenge for many recent experiments. The impact of these
uncertainties on physics analyses have been evaluated in detailed
systematics studies and in some cases the models have been fit directly
to experimental data to reduce systematics. These `downstream' simulation-related
studies can often be among the most challenging and time-consuming
in an analysis.

To see the difficulties facing the current generation of neutrino
experiments, one can look no further than the K2K and MiniBooNE experiments.
Both of these experiments have measured a substantially different
Q$^{2}$ distribution for their quasielastic-like events when compared
with their simulations, which involve a standard Fermi Gas model nuclear
model \cite{Ishida:2002xd,miniBoone:2007ru}. The disagreement between
nominal Monte Carlo and data is quite large - in the lowest Q$^{2}$
bin of MiniBooNE the deficit in the data is around 30\% \cite{miniBoone:2007ru}.
It seems likely that the discrepancies seen by both experiments have
a common origin. However the two experiments have been able to obtain
internal consistency with very different model changes - the K2K experiment
does this by eliminating the charged current (CC) coherent contribution
in the Monte Carlo \cite{Sanchez:2006hp} and the MiniBooNE experiment
does this by modifying certain parameters in their Fermi Gas model
\cite{miniBoone:2007ru}. Another example of the rapidly evolving
nature of this field is the recently reported excess of low energy
electron-like events by the MiniBooNE collaboration \cite{AguilarArevalo:2008rc}.
These discrepancies have generated significant new theoretical work
on these topics over the past several years \cite{Paschos:2005km,Singh:2006bm,AlvarezRuso:2007tt,Bodek:2007wb,Harvey:2007rd,Buss:2007ar,Benhar:2005dj,Amaro:2006if}.
The situation is bound to become even more interesting, and complicated,
in the coming decade, as new high-statistics experiments begin taking
data in this energy range. Designing a software system that can be
responsive to this rapidly evolving experimental and theoretical landscape
is a major challenge. 

One of the aims of this manual is to describe the ways in which the
GENIE neutrino event generator addresses these challenges. These solutions
rely heavily on the power of modern software engineering, particularly
the extensibility, modularity, and flexibility of object oriented
design, as well as the combined expertise and experience of the collaboration
with previous procedural codes.

\chapter{Neutrino Interaction Physics Modeling}

\label{ch:physics}

\section{Introduction}

The set of physics models used in GENIE incorporates the dominant
scattering mechanisms from several MeV to several hundred GeV and
are appropriate for any neutrino flavor and target type. Over this
energy range, many different physical processes are important. These
physics models can be broadly categorized into nuclear physics models,
cross section models, and hadronization models.

Substantial uncertainties exist in modeling neutrino-nucleus interactions,
particularly in the few-GeV regime which bridges the transition region
between perturbative and non-perturbative descriptions of the scattering
process. For the purposes of developing an event generator this theoretical
difficulty is compounded by the empirical limitation that previous
experiments often did not publish results in these difficult kinematic
regions since a theoretical interpretation was unavailable.

In physics model development for GENIE we have been forced to pay
particular attention to this `transition region', as for few-GeV experiments
it dominates the event rate. In particular the boundaries between
regions of validity of different models need to be treated with care
in order to avoid theoretical inconsistencies, discontinuities in
generated distributions, and double-counting. In this brief section
we will describe the models available in GENIE and the ways in which
we combine models to cover regions of phase space where clear theoretical
or empirical guidance is lacking.

\section{Nuclear Physics Model}

The importance of the nuclear model depends strongly on the kinematics
of the reaction. Nuclear physics plays a large role in every aspect
of neutrino scattering simulations at few-GeV energies and introduces
coupling between several aspects of the simulation. The relativistic
Fermi gas (RFG) nuclear model is used for all processes. GENIE uses
the version of Bodek and Ritchie which has been modified to incorporate
short range nucleon-nucleon correlations \cite{Bodek:1981wr}. This
is simple, yet applicable across a broad range of target atoms and
neutrino energies. The best tests of the RFG model come from electron
scattering experiments \cite{Moniz:1971mt}. At high energies, the
nuclear model requires broad features due to shadowing and similar
effects. At the lower end of the GENIE energy range, the impulse approximation
works very well and the RFG is often successful. The nuclear medium
gives the struck nucleon a momentum and average binding energy which
have been determined in electron scattering experiments. Mass densities
are taken from review articles \cite{DeJager:1987qc}. For $A<$20,
the modified Gaussian density parameterization is used. For heavier
nuclei, the 2-parameter Woods-Saxon density function is used. Thus,
the model can be used for {\em all} nuclei. Presently, fit parameters
for selected nuclei are used with interpolations for other nuclei.
All isotopes of a particular nucleus are assumed to have the same
density.

It is well known that scattering kinematics for nucleons in a nuclear
environment are different from those obtained in scattering from free
nucleons. For quasi-elastic and elastic scattering, Pauli blocking
is applied as described in Sec. \ref{sec:xsec}. For nuclear targets
a nuclear modification factor is included to account for observed
differences between nuclear and free nucleon structure functions which
include shadowing, anti-shadowing, and the EMC effect \cite{Bodek:2002ps}.

Nuclear reinteractions of produced hadrons is simulated using a cascade
Monte Carlo which will be described in more detail in a following
section. The struck nucleus is undoubtedly left in a highly excited
state and will typically de-excite by emitting nuclear fission fragments,
nucleons, and photons. At present de-excitation photon emission is
simulated only for oxygen \cite{Ejiri:1993rh,Kobayashi:2005ut} due
to the significance of these 3-10 MeV photons in energy reconstruction
at water Cherenkov detectors. Future versions of the generator will
handle de-excitation photon emission from additional nuclear targets.

\section{Cross section model}

\label{sec:xsec}

The cross section model provides the calculation of the differential
and total cross sections. During event generation the total cross
section is used together with the flux to determine the energies of
interacting neutrinos. The cross sections for specific processes are
then used to determine which interaction type occurs, and the differential
distributions for that interaction model are used to determine the
event kinematics. While the differential distributions must be calculated
event-by-event, the total cross sections can be pre-calculated and
stored for use by many jobs sharing the same physics models. Over
this energy range neutrinos can scatter off a variety of different
`targets' including the nucleus (via coherent scattering), individual
nucleons, quarks within the nucleons, and atomic electrons.

\textbf{Quasi-Elastic Scattering:} Quasi-elastic scattering (e.g.
$\nu_{\mu}+n\rightarrow\mu^{-}+p$) is modeled using an implementation
of the Llewellyn-Smith model \cite{LlewellynSmith:1972zm}. In this
model the hadronic weak current is expressed in terms of the most
general Lorentz-invariant form factors. Two are set to zero as they
violate G-parity. Two vector form factors can be related via CVC to
electromagnetic form factors which are measured over a broad range
of kinematics in electron elastic scattering experiments. Several
different parametrizations of these electromagnetic form factors including
Sachs \cite{Sachs:1962zzc}, BBA2003 \cite{Budd:2003wb} and BBBA2005
\cite{Bradford:2006yz} models are available with BBBA2005 being the
default. Two form factors - the pseudo-scalar and axial vector, remain.
The pseudo-scalar form factor is assumed to have the form suggested
by the partially conserved axial current (PCAC) hypothesis \cite{LlewellynSmith:1972zm},
which leaves the axial form factor F$_{A}$(Q$^{2}$) as the sole
remaining unknown quantity. F$_{A}(0)$ is well known from measurements
of neutron beta decay and the Q$^{2}$ dependence of this form factor
can only be determined in neutrino experiments and has been the focus
of a large amount of experimental work over several decades. In GENIE
a dipole form is assumed, with the axial vector mass m$_{A}$ remaining
as the sole free parameter with a default value of 0.99 GeV/c$^{2}$. 

For nuclear targets, the struck a suppression factor is included from
an analytic calculation of the rejection factor in the Fermi Gas model,
based on the simple requirement that the momentum of the outgoing
nucleon exceed the fermi momentum $k_{F}$ for the nucleus in question.
Typical values of $k_{F}$ are 0.221 GeV/c for nucleons in $^{12}$C,
0.251 GeV/c for protons in $^{56}$Fe, and 0.256 GeV/c for neutrons
in $^{56}$Fe.

\textbf{Elastic Neutral Current Scattering:} Elastic neutral current
processes are computed according to the model described by Ahrens
et al. \cite{Ahrens:1986xe}, where the axial form factor is given
by: 
\begin{equation}
G_{A}(Q^{2})=\frac{1}{2}\frac{G_{A}(0)}{(1+Q^{2}/M_{A}^{2})^{2}}(1+\eta).
\end{equation}
 The adjustable parameter $\eta$ includes possible isoscalar contributions
to the axial current, and the GENIE default value is $\eta=0.12$.
For nuclear targets the same reduction factor described above is used.

\textbf{Baryon Resonance Production:} The production of baryon resonances
in neutral and charged current channels is included with the Rein-Sehgal
model \cite{Rein:1981wg}. This model employs the Feynman-Kislinger-Ravndal
\cite{Feynman:1971wr} model of baryon resonances, which gives wavefunctions
for the resonances as excited states of a 3-quark system in a relativistic
harmonic oscillator potential with spin-flavor symmetry. In the Rein-Sehgal
paper the helicity amplitudes for the FKR model are computed and used
to construct the cross sections for neutrino-production of the baryon
resonances. From the 18 resonances of the original paper we include
the 16 that are listed as unambiguous at the latest PDG baryon tables
and all resonance parameters have been updated. In our implementation
of the Rein-Sehgal model interference between neighboring resonances
has been ignored. Lepton mass terms are not included in the calculation
of the differential cross section, but the effect of the lepton mass
on the phase space boundaries is taken into account. For tau neutrino
charged current interactions an overall correction factor to the total
cross section is applied to account for neglected elements (pseudoscalar
form factors and lepton mass terms) in the original model. The default
value for the resonance axial vector mass m$_{A}$ is 1.12 GeV/c$^{2}$,
as determined in the global fits carried out in Reference \cite{Kuzmin:2006dh}.

\textbf{Coherent Neutrino-Nucleus Scattering:} Coherent scattering
results in the production of forward going pions in both charged current
($\nu_{\mu}+A\rightarrow\mu^{-}+\pi^{+}+A$) and neutral current ($\nu_{\mu}+A\rightarrow\nu_{\mu}+\pi^{0}+A$)
channels. Coherent neutrino-nucleus interactions are modeled according
to the Rein-Sehgal model \cite{Rein:1983pf}. Since the coherence
condition requires a small momentum transfer to the target nucleus,
it is a low-Q$^{2}$ process which is related via PCAC to the pion
field. The Rein-Sehgal model begins from the PCAC form at Q$^{2}$=0,
assumes a dipole dependence for non-zero Q$^{2}$, with $m_{A}=1.00$
GeV/c$^{2}$, and then calculates the relevant pion-nucleus cross
section from measured data on total and inelastic pion scattering
from protons and deuterium \cite{Yao:2006px}. The GENIE implementation
is using the modified PCAC formula described in a recent revision
of the Rein-Sehgal model \cite{Rein:2006di} that includes lepton
mass terms.

\textbf{Non-Resonance Inelastic Scattering:} Deep (and not-so-deep)
inelastic scattering (DIS) is calculated in an effective leading order
model using the modifications suggested by Bodek and Yang \cite{Bodek:2002ps}
to describe scattering at low Q$^{2}$. In this model higher twist
and target mass corrections are accounted for through the use of a
new scaling variable and modifications to the low Q$^{2}$ parton
distributions. The cross sections are computed at a fully partonic
level (the ${\nu}q{\rightarrow}lq'$ cross sections are computed for
all relevant sea and valence quarks). The longitudinal structure function
is taken into account using the Whitlow R ($R=F_{L}/2xF_{1}$) parameterization
\cite{Whitlow:1990gk}. The default parameter values are those given
in \cite{Bodek:2002ps}, which are determined based on the GRV98 LO
parton distributions \cite{Gluck:1998xa}. An overall scale factor
of 1.032 is applied to the predictions of the Bodek-Yang model to
achieve agreement with the measured value of the neutrino cross section
at high energy (100 GeV). This factor is necessary since the Bodek-Yang
model treats axial and vector form modifications identically and would
therefore not be expected to reproduce neutrino data perfectly. This
overall DIS scale factor needs to be recalculated when elements of
the cross section model are changed.

The same model can be extended to low energies; it is the model used
for the nonresonant processes that compete with resonances in the
few-GeV region.

\textbf{Quasi-Elastic Charm Production:} QEL charm production is modeled
according to the Kovalenko local duality inspired model \cite{Kovalenko:1990zi}
tuned by the GENIE authors to recent NOMAD data \cite{Bischofberger:2005ur}.

\textbf{Deep-Inelastic Charm Production:} DIS charm production is
modeled according to the Aivazis, Olness and Tung model \cite{Aivazis:1993kh}.
Charm-production fractions for neutrino interactions are taken from
\cite{DeLellis:2002pr}, and utilize both Peterson \cite{Peterson:1982ak}
and Collins-Spiller \cite{Collins:1984ms} fragmentation functions,
with Peterson fragmentation functions being the default. The charm
mass is adjustable and is set to 1.43 GeV/c$^{2}$ by default.

\textbf{Inclusive Inverse Muon Decay:} Inclusive Inverse Muon Decay
cross section is computed using an implementation of the Bardin and
Dokuchaeva model \cite{Bardin:1986dk} taking into account all 1-loop
radiative corrections.

\textbf{Neutrino-Electron Elastic Scattering:} The cross sections
for all ${\nu}e-$ scattering channels other than Inverse Muon Decay
is computed according to \cite{Marciano:2003eq}. Inverse Tau decay
is neglected.

\subsubsection{Modeling the transition region}

\label{sec:transition}

As discussed, for example, by Kuzmin, Lyubushkin and Naumov \cite{Kuzmin:2005bm}
one typically considers the total ${\nu}N$ CC scattering cross section
as

\begin{center}
$\sigma^{tot}=\sigma^{QEL}\oplus\sigma^{1\pi}\oplus\sigma^{2\pi}\oplus...\oplus\sigma^{1K}\oplus...\oplus\sigma^{DIS}$ 
\par\end{center}

In the absence of a model for exclusive inelastic multi-particle neutrinoproduction,
the above is usually being approximated as

\begin{center}
$\sigma^{tot}=\sigma^{QEL}\oplus\sigma^{RES}\oplus\sigma^{DIS}$ 
\par\end{center}

assuming that all exclusive low multiplicity inelastic reactions proceed
primarily through resonance neutrinoproduction. For the sake of simplicity,
small contributions to the total cross section in the few GeV energy
range, such as coherent and elastic ${\nu}e^{-}$ scattering, were
omitted from the expression above. In this picture, one should be
careful in avoiding double counting the low multiplicity inelastic
reaction cross sections.

In GENIE release the total cross sections is constructed along the
same lines, adopting the procedure developed in NeuGEN \cite{Gallagher:2002sf}
to avoid double counting. The total inelastic differential cross section
is computed as

\begin{center}
$\frac{{\displaystyle d^{2}\sigma^{inel}}}{{\displaystyle dQ^{2}dW}}=\frac{{\displaystyle d^{2}\sigma^{RES}}}{{\displaystyle dQ^{2}dW}}+\frac{{\displaystyle d^{2}\sigma^{DIS}}}{{\displaystyle dQ^{2}dW}}$ 
\par\end{center}

The term $d^{2}\sigma^{RES}/dQ^{2}dW$ represents the contribution
from all low multiplicity inelastic channels proceeding via resonance
production. This term is computed as

\begin{center}
$\frac{{\displaystyle d^{2}\sigma^{RES}}}{{\displaystyle dQ^{2}dW}}={\sum\limits _{k}}\bigl(\frac{{\displaystyle d^{2}\sigma^{R/S}}}{{\displaystyle dQ^{2}dW}}\bigr)_{k}\cdot{\Theta}(Wcut-W)$ 
\par\end{center}

where the index $k$ runs over all baryon resonances taken into account,
$W_{cut}$ is a configurable parameter and $(d^{2}\sigma_{{\nu}N}^{RS}/dQ^{2}dW)_{k}$
is the Rein-Seghal model prediction for the $k^{th}$ resonance cross
section.

The DIS term of the inelastic differential cross section is expressed
in terms of the differential cross section predicted by the Bodek-Yang
model appropriately modulated in the ``resonance-dominance\textquotedbl{}
region $W<W_{cut}$ so that the RES/DIS mixture in this region agrees
with inclusive cross section data \cite{MacFarlane:1983ax,Berge:1987zw,Ciampolillo:1979wp,Colley:1979rt,Bosetti:1981ip,Mukhin:1979bd,Baranov:1979sx,Barish:1978pj,Baker:1982ty,Eichten:1973cs}
and exclusive 1-pion \cite{Lerche:1978cp,Ammosov:1988xb,Grabosch:1988gw,Bell:1978qu,Kitagaki:1986ct,Allen:1980ti,Allen:1985ti,Allasia:1990uy,Campbell:1973wg,Barish:1978pj,Radecky:1981fn}
and 2-pion \cite{Day:1984nf,Kitagaki:1986ct} cross section data:

\begin{center}
\begin{eqnarray*}
\frac{d^{2}\sigma^{DIS}}{dQ^{2}dW} & = & \frac{d^{2}\sigma^{DIS,BY}}{dQ^{2}dW}\cdot{\Theta}(W-Wcut)+\\
 & + & \frac{d^{2}\sigma^{DIS,BY}}{dQ^{2}dW}\cdot{\Theta}(Wcut-W)\cdot{\sum\limits _{m}}f_{m}
\end{eqnarray*}
 
\par\end{center}

In the above expression, $m$ refers to the multiplicity of the hadronic
system and, therefore, the factor $f_{m}$ relates the total calculated
DIS cross section to the DIS contribution to this particular multiplicity
channel. These factors are computed as $f_{m}=R_{m}{\cdot}P_{m}^{had}$
where $R_{m}$ is a tunable parameter and $P_{m}^{had}$ is the probability,
taken from the hadronization model, that the DIS final state hadronic
system multiplicity would be equal to $m$. The approach described
above couples the GENIE cross section and hadronic multiparticle production
model \cite{Yang:2007zzt}.

\subsubsection{Cross section model tuning}

\begin{figure}[htb]
\center \includegraphics[width=30pc]{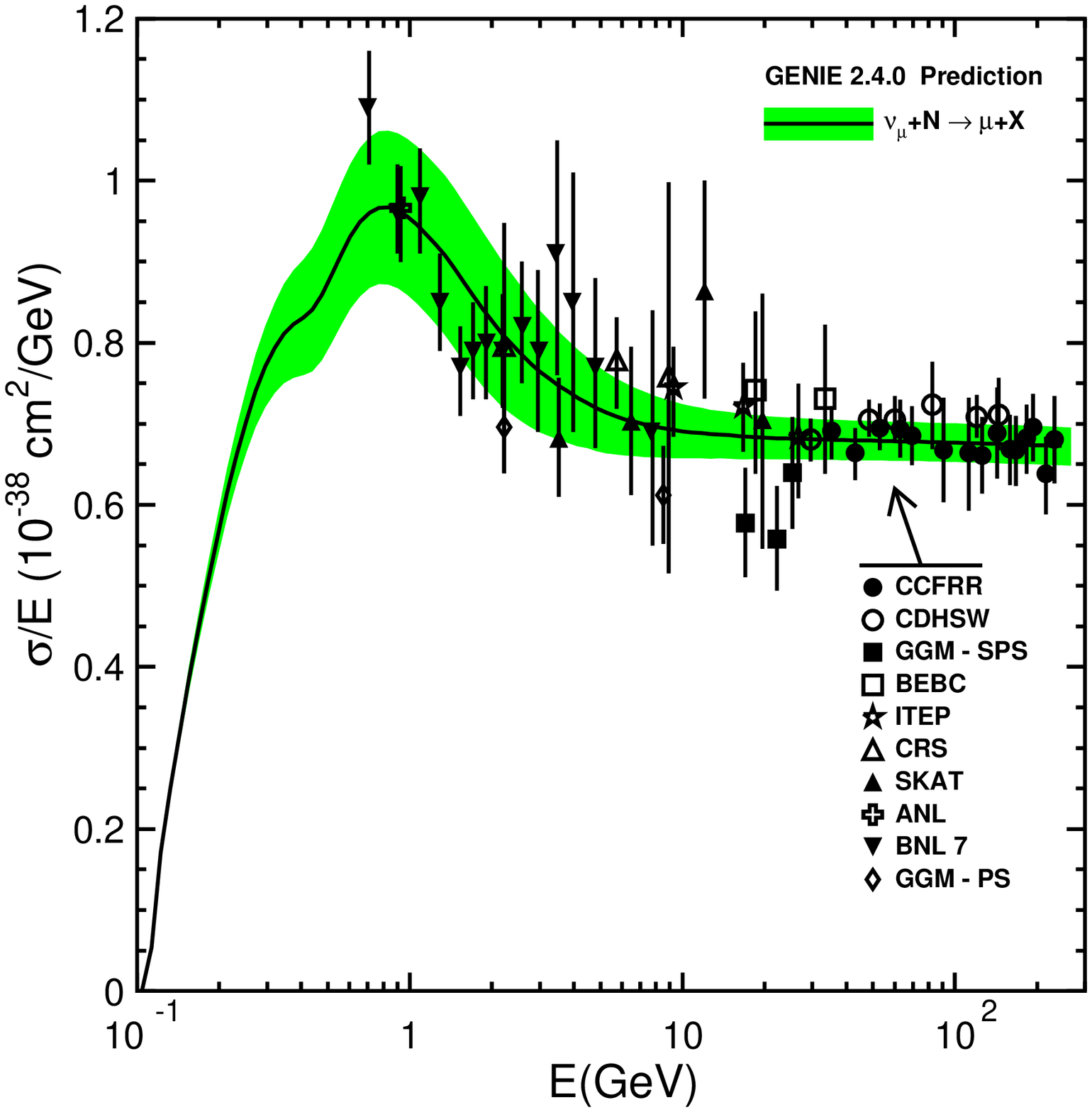}
\caption{$\nu{\mu}$ charged current scattering from an isoscalar target. The
shaded band indicates the estimated uncertainty on the free nucleon
cross section. Data are from \cite{MacFarlane:1983ax} (CCFRR), \cite{Berge:1987zw}
(CDHSW), \cite{Ciampolillo:1979wp} (GGM-SPS), \cite{Colley:1979rt,Bosetti:1981ip}
(BEBC), \cite{Mukhin:1979bd} (ITEP), \cite{Baranov:1979sx} (CRS,
SKAT), \cite{Barish:1978pj} (ANL), \cite{Baker:1982ty} (BNL) and
\cite{Eichten:1973cs} (GGM-PS).}

\label{fig:XSecErrEnvelope} 
\end{figure}

As mentioned previously, the quasi-elastic, resonance production,
and DIS models employ form factors, axial vector masses, and other
parameters which have been determined by others in their global fits
\cite{Bradford:2006yz,Kuzmin:2006dh}. In order to check the overall
consistency of our model, and to verify that we have correctly implemented
the DIS model, predictions are compared to electron scattering inclusive
data \cite{Gallagher:2004nq,e49a10} and neutrino structure function
data \cite{Bhattacharya:2009zz}. The current default values for transition
region parameters are $W_{cut}$=1.7 GeV/c$^{2}$, $R_{2}(\nu p)=R_{2}(\overline{\nu}n)$=0.1,
$R_{2}(\nu n)=R_{2}(\overline{\nu}p)$=0.3, and $R_{m}$=1.0 for all
$m>2$ reactions. These are determined from fits to inclusive and
exclusive (one and two-pion) production neutrino interaction channels.
For these comparisons we rely heavily on online compilations of neutrino
data \cite{Whalley:2004sz} and related fitting tools \cite{Andreopoulos:2005vd}
that allow one to include some correlated systematic errors (such
as arising from flux uncertainties). The GENIE default cross section
for $\nu{\mu}$ charged current scattering from an isoscalar target,
together with the estimated uncertainty on the total cross section,
as evaluated in \cite{Adamson:2007gu} are shown in Fig. \ref{fig:XSecErrEnvelope}.

\section{Neutrino-induced Hadron Production}

\subsubsection{Introduction}

\label{agkysec:0} In neutrino interaction simulations the hadronization
model (or fragmentation model) determines the final state particles
and 4-momenta given the nature of a neutrino-nucleon interaction (CC/NC,
$\nu$/$\bar{\nu}$, target neutron/proton) and the event kinematics
($W^{2}$, $Q^{2}$, $x$, $y$). 
The modeling of neutrino-produced hadronic showers is important for
a number of analyses in the current and coming generation of neutrino
oscillation experiments:

{\em Calorimetry:} Neutrino oscillation experiments like MINOS
which use calorimetry to reconstruct the shower energy, and hence
the neutrino energy, are sensitive to the modelling of hadronic showers.
These detectors are typically calibrated using single particle test
beams, which introduces a model dependence in determining the conversion
between detector activity and the energy of neutrino-produced hadronic
systems \cite{Adamson:2007gu}.

{\em NC/CC Identification:} Analyses which classify events as charged
current (CC) or neutral current (NC) based on topological features
such as track length in the few-GeV region rely on accurate simulation
of hadronic particle distributions to determine NC contamination in
CC samples.

{\em Topological Classification:} Analyses which rely on topological
classifications, for instance selecting quasi-elastic-like events
based on track or ring counting depend on the simulation of hadronic
systems to determine feeddown of multi-particle states into selected
samples. Because of the wide-band nature of most current neutrino
beams, this feeddown is non-neglible even for experiments operating
in beams with mean energy as low as 1 GeV \cite{miniBoone:2007ru,Hiraide:2008yu}.

{\em $\nu_{e}$ Appearance Backgrounds:} A new generation of $\nu_{\mu}\rightarrow\nu_{e}$
appearance experiments are being developed around the world, which
hope to find evidence of charge-parity (CP) violation in the lepton
sector. In these experiments background is dominated by neutral pions
generated in NC interaction. The evaluation of NC backgrounds in these
analysis can be quite sensitive to the details of the NC shower simulation
and specifically the $\pi^{0}$ shower content and transverse momentum
distributions of hadrons \cite{Sanchez:2008zza}.

\begin{figure*}
\centering \includegraphics[width=1\textwidth]{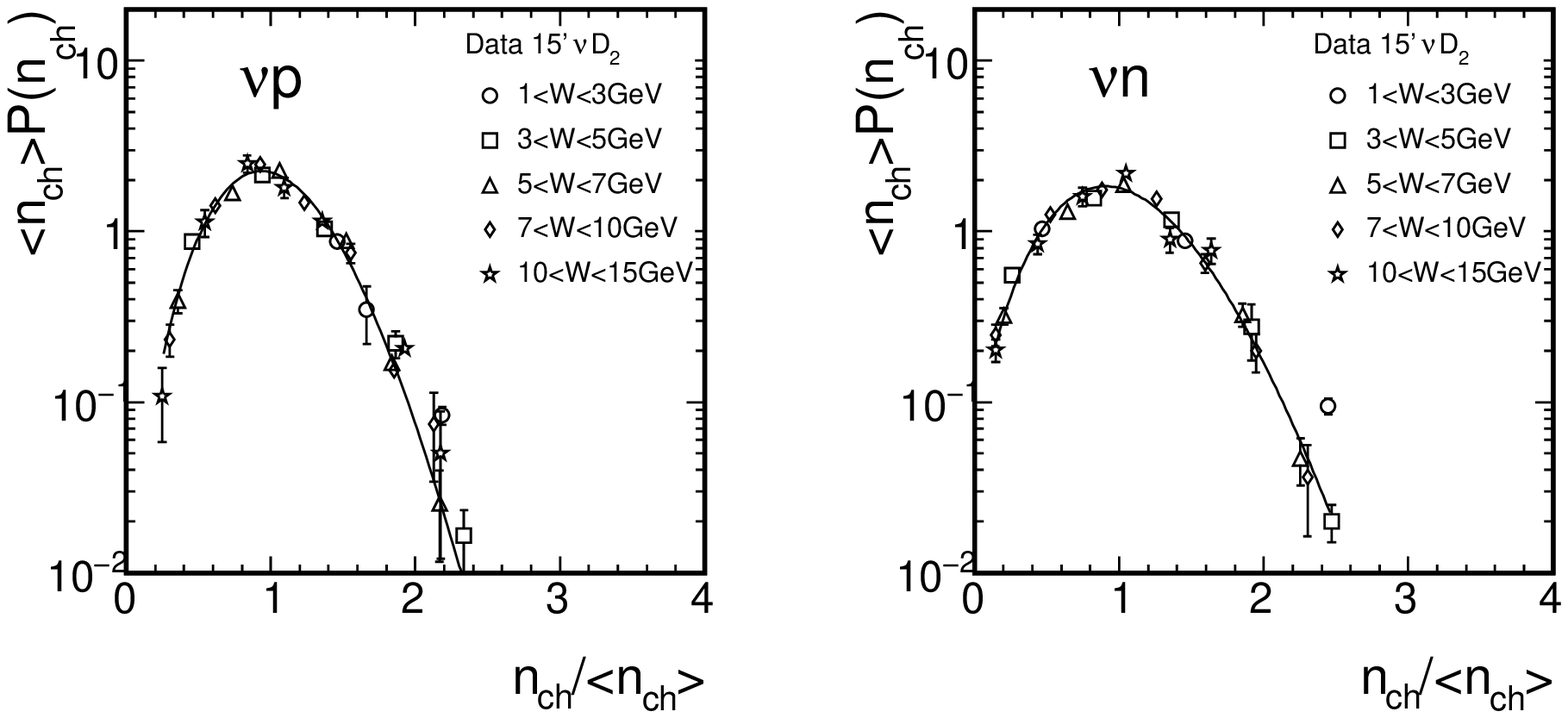}\caption{$\nu p$ (left) and $\nu n$ interactions. The curve represents a
fit to the Levy function. Data points are taken from \cite{Zieminska:1983bs}.}

\label{fig:kno_levy} 
\end{figure*}


In order to improve Monte Carlo simulations for the MINOS experiment,
a new hadronization model, referred to here as the `AGKY model', was
developed. We use the PYTHIA/JETSET \cite{Sjostrand:2006za} model
to simulate the hadronic showers at high hadronic invariant masses.
We also developed a phenomenological description of the low invariant
mass hadronization since the applicability of the \\
 PYTHIA/JETSET model, for neutrino-induced showers, is known to deteriorate
as one approaches the pion production threshold. We present here a
description of the AGKY hadronization model and the tuning and validation
of this model using bubble chamber experimental data.

\subsubsection{The AGKY Model}

\label{agkysec:1} The AGKY model, which is now the default hadronization
model in the neutrino Monte Carlo generators NEUGEN \cite{Gallagher:2002sf}
and GENIE-2.0.0 \cite{Andreopoulos:2006cz}, includes a phenomenological
description of the low invariant mass region based on Koba-Nielsen-Olesen
(KNO) scaling \cite{Koba:1972ng}, while at higher masses it gradually
switches over to the PYTHIA/JETSET model. The transition from the
KNO-based model to the \\
 PYTHIA/JETSET model takes place gradually, at an intermediate invariant
mass region, ensuring the continuity of all simulated observables
as a function of the invariant mass. This is accomplished by using
a transition window {[}$W_{min}^{tr},W_{max}^{tr}${]} over which
we linearly increase the fraction of neutrino events for which the
hadronization is performed by the PYTHIA/JETSET model from 0\% at
$W_{min}^{tr}$ to 100\% at $W_{max}^{tr}$. The default values used
in the AGKY model are: 
\begin{equation}
W_{min}^{tr}=2.3\mbox{ GeV}/\mbox{c}^{2},W_{max}^{tr}=3.0\mbox{ GeV}/\mbox{c}^{2}.
\end{equation}

The kinematic region probed by any particular experiment depends on
the neutrino flux, and for the 1-10 GeV range of importance to oscillation
experiments, the KNO-based phenomenological description plays a particularly
crucial role. The higher invariant mass region where PYTHIA/JETSET
is used is not accessed until a neutrino energy of approximately 3
GeV is reached, at which point 44.6\% of charged current interactions
are non-resonant inelastic and are hadronized using the KNO-based
part of the model. For 1 GeV neutrinos this component is 8.3\%, indicating
that this model plays a significant role even at relatively low neutrino
energies. At 9 GeV, the contributions from the KNO-based and PYTHIA/JETSET
components of the model are approximately equal, with each handling
around 40\% of generated CC interactions. The main thrust of this
work was to improve the modeling of hadronic showers in this low invariant
mass / energy regime which is of importance to oscillation experiments.

The description of AGKY's KNO model, used at low invariant masses,
can be split into two independent parts: 
\begin{itemize}
\item Generation of the hadron shower particle content 
\item Generation of hadron 4-momenta 
\end{itemize}
These two will be described in detail in the following sections.

The neutrino interactions are often described by the following kinematic
variables: 
\begin{eqnarray}
Q^{2} & = & 2E_{\nu}(E_{\mu}-p_{\mu}^{L})-m^{2}\nonumber \\
\nu & = & E_{\nu}-E_{\mu}\nonumber \\
W^{2} & = & M^{2}+2M\nu-Q^{2}\nonumber \\
x & = & Q^{2}/2M\nu\nonumber \\
y & = & \nu/E_{\nu}
\end{eqnarray}
 where $Q^{2}$ is the invariant 4-momentum transfer squared, $\nu$
is the neutrino energy transfer, $W$ is the effective mass of all
secondary hadrons (invariant hadronic mass), $x$ is the Bjorken scaling
variable, $y$ is the relative energy transfer, $E_{\nu}$ is the
incident neutrino energy, $E_{\mu}$ and $p_{\mu}^{L}$ are the energy
and longitudinal momentum of the muon, $M$ is the nucleon mass and
$m$ is the muon mass.

For each hadron in the hadronic system, we define the variables $z=E_{h}/\nu$,
$x_{F}=2p_{L}^{*}/W$ and $p_{T}$ where $E_{h}$ is the energy in
the laboratory frame, $p_{L}^{*}$ is the longitudinal momentum in
the hadronic c.m.s., and $p_{T}$ is the transverse momentum.

\paragraph{Low-$W$ model: Particle content}

\label{agkysec:3} At low invariant masses the AGKY model generates
hadronic systems that typically consist of exactly one baryon ($p$
or $n$) and any number of $\pi$ and K mesons that are kinematically
possible and consistent with charge conservation.

For a fixed hadronic invariant mass and initial state (neutrino and
struck nucleon), the method for generating the hadron shower particles
generally proceeds in four steps:

{\em Determine $\langle n_{ch}\rangle$:} Compute the average charged
hadron multiplicity using the empirical expression: 
\begin{equation}
\langle n_{ch}\rangle=a_{ch}+b_{ch}\ \ln W^{2}
\end{equation}
 The coefficients $a_{ch}$, $b_{ch}$, which depend on the initial
state, have been determined by bubble chamber experiments.

{\em Determine $\langle n\rangle$:} Compute the average hadron
multiplicity as $\langle n_{tot}\rangle=1.5\langle n_{ch}\rangle$
\cite{Wittek:1988ke}.

{\em Deterimine n:} Generate the actual hadron multiplicity taking
into account that the multiplicity dispersion is described by the
KNO scaling law \cite{Koba:1972ng}: 
\begin{equation}
\langle n\rangle\times P(n)=f(n/\langle n\rangle)
\end{equation}
 where $P(n)$ is the probability of generating $n$ hadrons and $f$
is the universal scaling function which can be parametrized by the
Levy function \footnote{The Levy function: $Levy(z;c)=2e^{-c}c^{cz+1}/\Gamma(cz+1)$}
($z=n/\langle n\rangle$) with an input parameter $c$ that depends
on the initial state. Fig.\ref{fig:kno_levy} shows the KNO scaling
distributions for $\nu p$ (left) and $\nu n$ (right) CC interactions.
We fit the data points to the Levy function and the best fit parameters
are $c_{ch}=7.93\pm0.34$ for the $\nu p$ interactions and $c_{ch}=5.22\pm0.15$
for the $\nu n$ interactions.

{\em Select particle types:} Select hadrons up to the generated
hadron multiplicity taking into account charge conservation and kinematic
constraints. The hadronic system contains any number of mesons and
exactly one baryon which is generated based on simple quark model
arguments. Protons and neutrons are produced in the ratio 2:1 for
$\nu p$ interactions, 1:1 for $\nu n$ and $\bar{\nu}p$, and 1:2
for $\bar{\nu}n$ interactions. Charged mesons are then created in
order to balance charge, and the remaining mesons are generated in
neutral pairs. The probablilities for each are 31.33$\%$ ($\pi^{0},\pi^{0}$),
62.66$\%$ ($\pi^{+},\pi^{-}$), and 6$\%$ strange meson pairs. The
probability of producing a strange baryon via associated production
is determined from a fit to $\Lambda$ production data: 
\begin{equation}
P_{hyperon}=a_{hyperon}+b_{hyperon}\ \ln W^{2}
\end{equation}

TABLE \ref{tab:par} shows the default average hadron multiplicity
and dispersion parameters used in the AGKY model.

\begin{table}
\center \caption{Default average hadron multiplicity and dispersion parameters used
in the AGKY model.}

\label{tab:par} %
\begin{tabular}{|l|r|r|r|r|}
\hline 
 & $\nu p$ & $\nu n$ & $\bar{\nu}p$ & $\bar{\nu}n$\tabularnewline
\hline 
$a_{ch}$  & 0.40 \cite{Zieminska:1983bs}  & -0.20 \cite{Zieminska:1983bs} & 0.02 \cite{Barlag:1981wu}  & 0.80 \cite{Barlag:1981wu}\tabularnewline
$b_{ch}$  & 1.42 \cite{Zieminska:1983bs}  & 1.42 \cite{Zieminska:1983bs} & 1.28 \cite{Barlag:1981wu}  & 0.95 \cite{Barlag:1981wu}\tabularnewline
$c_{ch}$  & 7.93 \cite{Zieminska:1983bs}  & 5.22 \cite{Zieminska:1983bs} & 5.22  & 7.93\tabularnewline
$a_{hyperon}$  & 0.022  & 0.022  & 0.022  & 0.022\tabularnewline
$b_{hyperon}$  & 0.042  & 0.042  & 0.042  & 0.042\tabularnewline
\hline 
\end{tabular}
\end{table}

\paragraph{Low-$W$ model: Hadron system decay}

\label{agkysec:4}

Once an acceptable particle content has been generated, the available
invariant mass needs to be partitioned amongst the generated hadrons.
The most pronounced kinematic features in the low-$W$ region result
from the fact that the produced baryon is much heavier than the mesons
and exhibits a strong directional anticorrelation with the current
direction.

Our strategy is to first attempt to reproduce the experimentally measured
final state nucleon momentum distributions. We then perform a phase
space decay on the remnant system employing, in addition, a $p_{T}$-based
rejection scheme designed to reproduce the expected meson transverse
momentum distribution. The hadronization model performs its calculation
in the hadronic c.m.s., where the z-axis is in the direction of the
momentum transfer. Once the hadronization is completed, the hadronic
system will be boosted and rotated to the LAB frame. The boost and
rotation maintains the $p_{T}$ generated in the hadronic c.m.s.

\begin{figure}
\centering \includegraphics[width=1\columnwidth]{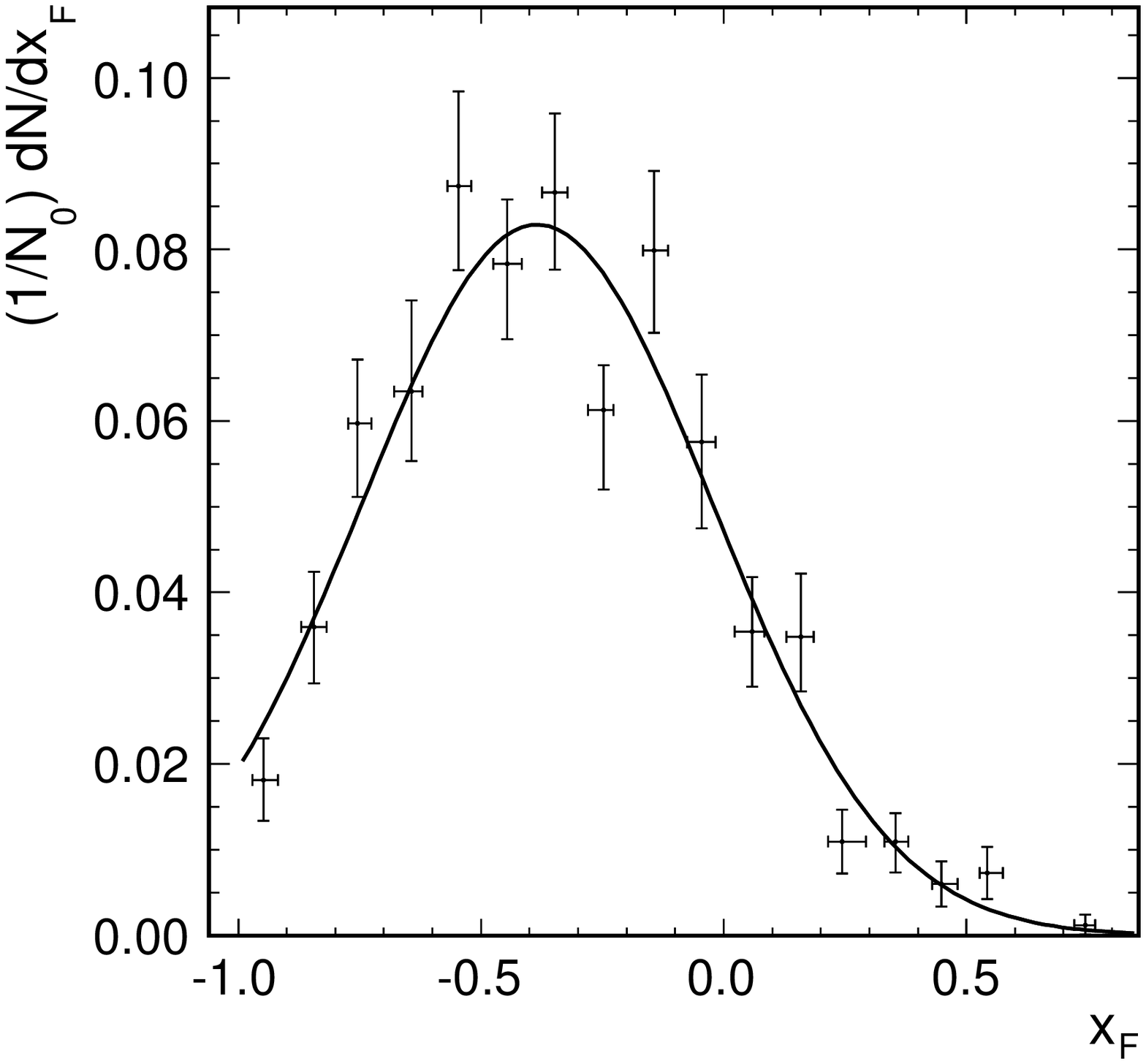}
\caption{\label{fig:xf} Nucleon $x_{F}$ distribution data from Cooper $et\ al.$
\cite{CooperSarkar:1982ay} and the AGKY parametrization (solid line).
}
\end{figure}

In more detail, the algorithm for decaying a system of $N$ hadrons
is the following:

{\em Generate baryon:} Generate the baryon 4-momentum $P_{N}^{*}=(E_{N}^{*},{\bf p}_{N}^{*})$
using the nucleon $p_{T}^{2}$ and $x_{F}$ PDFs which are parametrized
based on experimental data \cite{Derrick:1977zi,CooperSarkar:1982ay}.
The $x_{F}$ distribution used is shown in Fig.\ref{fig:xf}. We do
not take into account the correlation between $p_{T}$ and $x_{F}$
in our selection.

{\em Remnant System:} Once an accepted $P_{N}^{*}$ has been generated,
calculate the 4-momentum of the remaining N-1 hadrons, (the ``remnant''
hadronic system) as $P_{R}^{*}=P_{X}^{*}-P_{N}^{*}$ where $P_{X}^{*}=(W,0)$
is the initial hadron shower 4-momentum in the hadronic c.m.s.

{\em Decay Remnant System:} Generate an unweighted phase space
decay of the remnant hadronic system. The decay takes place at the
remnant system c.m.s. after which the particles are boosted back to
the hadronic c.m.s. The phase space decay employs a rejection method
suggested in \cite{Clegg:1981js}, with a rejection factor $e^{-A*p_{T}}$
for each meson. This causes the transverse momentum distribution of
the generated mesons to fall exponentially with increasing $p_{T}^{2}$.
Here $p_{T}$ is the momentum component perpendicular to the current
direction.

Two-body hadronic systems are treated as a special case. Their decay
is performed isotropically in the hadronic c.m.s. and no $p_{T}$-based
suppression factor is applied.

\paragraph{High-$W$ model: PYTHIA}

\label{agkysec:5} The high invariant mass hadronization is performed
by the PYTHIA model \cite{Sjostrand:2006za}. The PYTHIA program is
a standard tool for the generation of high-energy collisions, comprising
a coherent set of physics models for the evolution from a few-body
hard process to a complex multihadronic final state. It contains a
library of hard processes and models for initial- and final-state
parton showers, multiple parton-parton interactions, beam remnants,
string fragmentation and particle decays. The hadronization model
in PYTHIA is based on the Lund string fragmentation framework \cite{Andersson:1983ia}.
In the AGKY model, all but four of the PYTHIA configuration parameters
are set to be the default values. Those four parameters take the non-default
values tuned by NUX \cite{Rubbia:2001}, a high energy neutrino MC
generator used by the NOMAD experiment: 
\begin{itemize}
\item $P_{s\bar{s}}$ controlling the ${s\bar{s}}$ production suppression:
\\
 (PARJ(2))=0.21. 
\item $P_{\langle p_{T}^{2}\rangle}$ determining the average hadron $\langle p_{T}^{2}\rangle$:
\\
 (PARJ(21))=0.44. 
\item $P_{ngt}$ parameterizing the non-gaussian $p_{T}$ tails: \\
 (PARJ(23))=0.01. 
\item $P_{Ec}$ an energy cutoff for the fragmentation process: \\
 (PARJ(33))=0.20. 
\end{itemize}

\subsubsection{Data/MC Comparisons}

\label{agkysec:6}

The characteristics of neutrino-produced hadronic systems have been
extensively studied by several bubble chamber experiments. The bubble
chamber technique is well suited for studying details of charged hadron
production in neutrino interactions since the detector can provide
precise information for each track. However, the bubble chamber has
disadvantages for measurements of hadronic system characteristics
as well. The detection of neutral particles, in particular of photons
from $\pi^{0}$ decay, was difficult for the low density hydrogen
and deuterium experiments. Experiments that measured neutral pions
typically used heavily liquids such as neon-hydrogen mixtures and
Freon. While these exposures had the advantage of higher statistics
and improved neutral particle identification, they had the disadvantage
of introducing intranuclear rescattering which complicates the extraction
of information related to the hadronization process itself.

We tried to distill the vast literature and focus on the following
aspects of $\nu$/$\bar{\nu}$ measurements made in three bubble chambers
- the Big European Bubble Chamber (BEBC) at CERN, the 15-foot bubble
chamber at Fermilab, and the SKAT bubble chamber in Russia. Measurements
from the experiments of particular interest for tuning purposes can
be broadly categorized as multiplicity measurements and hadronic system
measurements. Multiplicity measurements include averaged charged and
neutral particle ($\pi^{0}$) multiplicities, forward and backward
hemisphere average multiplicities and correlations, topological cross
sections of charged particles, and neutral - charged pion multiplicity
correlations. Hadronic system measurements include fragmentation functions
($z$ distributions), $x_{F}$ distributions, $p_{T}^{2}$ (transverse
momentum squared) distributions, and $x_{F}-\langle p_{T}^{2}\rangle$
correlations (``seagull'' plots).

The systematic errors in many of these measurements are substantial
and various corrections had to be made to correct for muon selection
efficiency, neutrino energy smearing, \textit{etc}. The direction
of the incident $\nu$/$\bar{\nu}$ is well known from the geometry
of the beam and the position of the interaction point. Its energy
is unknown and is usually estimated using a method based on transverse
momentum imbalance. The muon is usually identified through the kinematic
information or by using an external muon identifier (EMI). The resolution
in neutrino energy is typically 10\% in the bubble chamber experiments
and the invariant hadronic mass $W$ is less well determined.

The differential cross section for semi-inclusive pion production
in neutrino interactions 
\begin{equation}
\nu+N\rightarrow\mu^{-}+\pi+X\label{nuN}
\end{equation}
 may in general be written as: 
\begin{equation}
\frac{d\sigma(x,Q^{2},z)}{dxdQ^{2}dz}=\frac{d\sigma(x,Q^{2})}{dxdQ^{2}}D^{\pi}(x,Q^{2},z),
\end{equation}
 where $D^{\pi}(x,Q^{2},z)$ is the pion fragmentation function. Experimentally
$D^{\pi}$ is determined as: 
\begin{equation}
D^{\pi}(x,Q^{2},z)=[N_{ev}(x,Q^{2})]^{-1}dN/dz.
\end{equation}

In the framework of the Quark Parton Model (QPM) the dominant mechanism
for reactions (\ref{nuN}) is the interaction of the exchanged $W$
boson with a d-quark to give a u-quark which fragments into hadrons
in neutrino interactions, leaving a di-quark spectator system which
produces target fragments. In this picture the fragmentation function
is independent of $x$ and the scaling hypothesis excludes a $Q^{2}$
dependence; therefore the fragmentation function should depend only
on $z$. There is no reliable way to separate the current fragmentation
region from the target fragmentation region if the effective mass
of the hadronic system ($W$) is not sufficiently high. Most experiments
required $W>W_{0}$ where $W_{0}$ is between 3 GeV/$c^{2}$ and 4
GeV/$c^{2}$ when studying the fragmentation characteristics. The
caused difficulties in the tuning of our model because we are mostly
interested in the interactions at low hadronic invariant masses.

\begin{figure*}
\centering \includegraphics[width=1\textwidth]{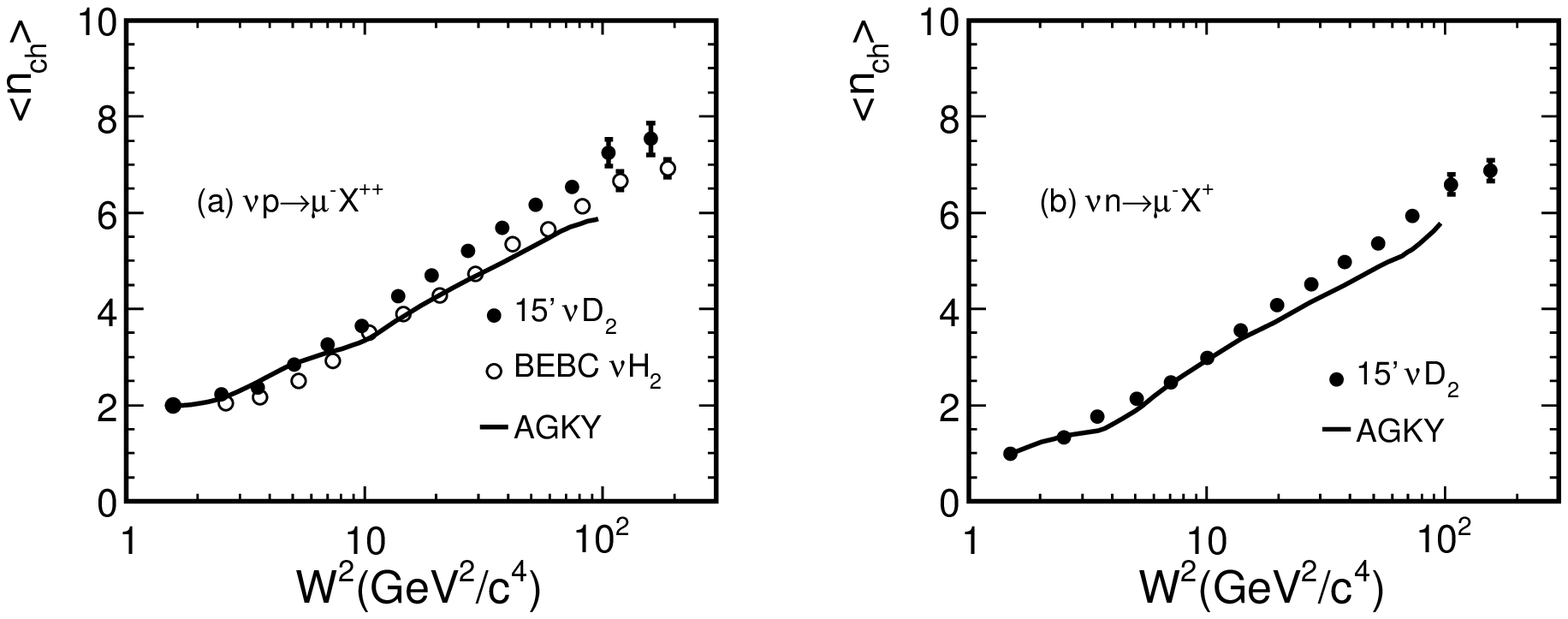}
\caption{$\langle n_{ch}\rangle$ as a function of $W^{2}$. (a) $\nu p$ events.
(b) $\nu n$ events. Data points are taken from \cite{Zieminska:1983bs,Allen:1981vh}.}

\label{fig:cMulCh} 
\end{figure*}

\begin{figure*}
\centering \includegraphics[width=1\textwidth]{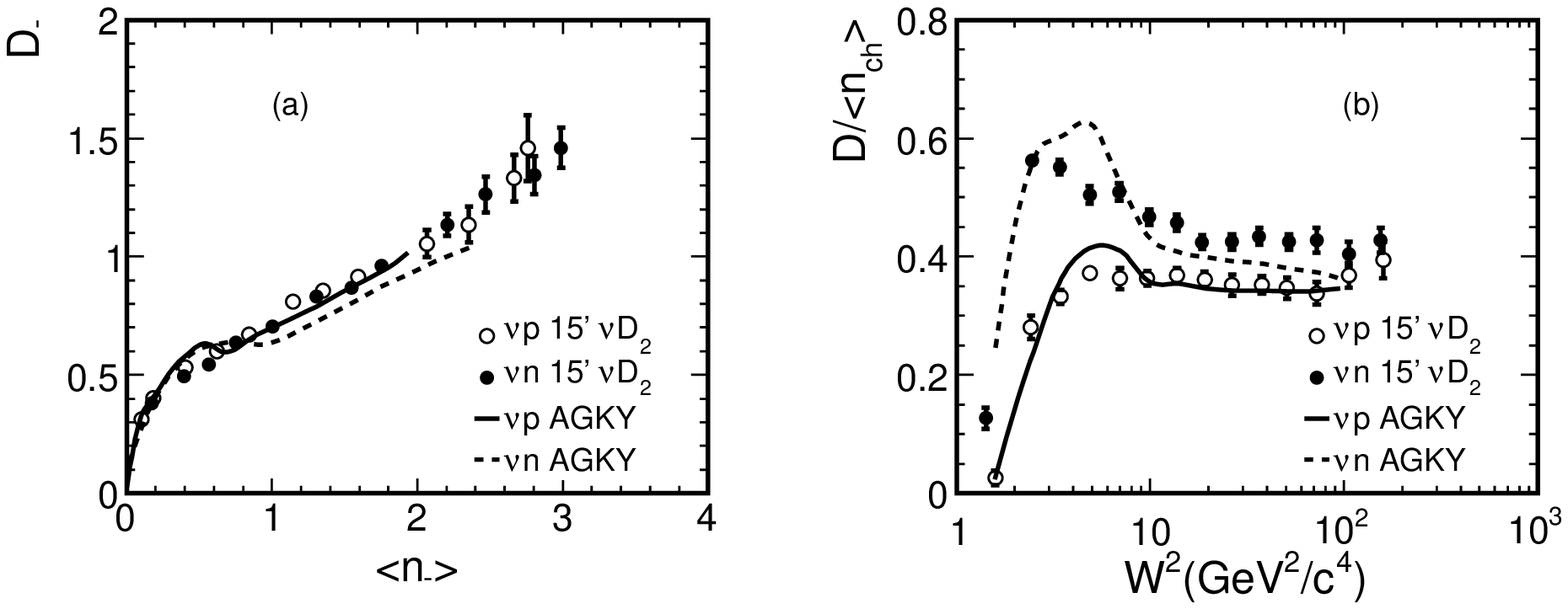}
\caption{$D_{-}=(\langle n_{-}^{2}\rangle-\langle n_{-}\rangle^{2})^{1/2}$
as a function of $\langle n_{-}\rangle$. (b) $D/\langle n_{ch}\rangle$
as a function of $W^{2}$. Data points are taken from \cite{Zieminska:1983bs}.}

\label{fig:cDispCh} 
\end{figure*}

We determined the parameters in our model by fitting experimental
data with simulated CC neutrino free nucleon interactions uniformly
distributed in the energy range from 1 to 61 GeV. The events were
analyzed to determine the hadronic system characteristics and compared
with published experimental data from the BEBC, Fermilab 15-foot,
and SKAT bubble chamber experiments. We reweight our MC to the energy
spectrum measured by the experiment if that information is available.
This step is not strictly necessary for the following two reasons:
many observables (mean multiplicity, dispersion, \textit{etc.}) are
measured as a function of the hadronic invariant mass $W$, in which
case the energy dependency is removed; secondly the scaling variables
($x_{F}$, $z$, \textit{etc.}) are rather independent of energy according
to the scaling hypothesis.

Some experiments required $Q^{2}>1\mbox{GeV}^{2}$ to reduce the quasi-elastic
contribution, $y<0.9$ to reduce the neutral currents, and $x>0.1$
to reduce the sea-quark contribution. They often applied a cut on
the muon momentum to select clean CC events. We apply the same kinematic
cuts as explicitly stated in the papers to our simulated events. The
hadronization model described here is used only for continuum production
of hadrons, resonance-mediated production is described as part of
the resonance model \cite{Rein:1981wg}. Combining resonance and non-resonant
inelastic contributions to the inclusive cross section requires care
to avoide double-counting \cite{Bodek:2003wd}, and the underlying
model used here includes a resonant contribution which dominates the
cross section at threshold, but whose contribution gradually diminishes
up to a cutoff value of W=1.7 GeV/c$^{2}$, above which only non-resonant
processes contribute. All of the comparisons shown in this paper between
models and data include the resonant contribution to the models unless
it is explicitly excluded by experimental cuts.

Fig.\ref{fig:cMulCh} shows the average charged hadron multiplicity
$\langle n_{ch}\rangle$ (the number of charged hadrons in the final
state, \textit{i.e.} excluding the muon) as a function of $W^{2}$.
$\langle n_{ch}\rangle$ rises linearly with $\ln(W^{2})$ for $W>2\mbox{GeV}/c^{2}$.
At the lowest $W$ values the dominant interaction channels are single
pion production from baryon resonances: 
\begin{eqnarray}
\nu+p & \rightarrow & \mu^{-}+p+\pi^{+}\label{res1}\\
\nu+n & \rightarrow & \mu^{-}+p+\pi^{0}\label{res2}\\
\nu+n & \rightarrow & \mu^{-}+n+\pi^{+}\label{res3}
\end{eqnarray}
 Therefore $\langle n_{ch}\rangle$ becomes 2 (1) for $\nu p$ ($\nu n$)
interactions as $W$ approaches the pion production threshold. For
$\nu p$ interactions there is a disagreement between the two measurements
especially at high invariant masses, which is probably due to differences
in scattering from hydrogen and deuterium targets. 
Our parameterization of low-$W$ model was based on the Fermilab 15-foot
chamber data. Historically the PYTHIA/JETSET program was tuned on
the BEBC data. The AGKY model uses the KNO-based empirical model at
low invariant masses and it uses the PYTHIA/JETSET program to simulation
high invariance mass interactions. Therefore the MC prediction agrees
better with the Fermilab data at low invariant masses and it agrees
better with the BEBC data at high invariant masses.

\begin{figure*}
\centering \includegraphics[width=1\textwidth]{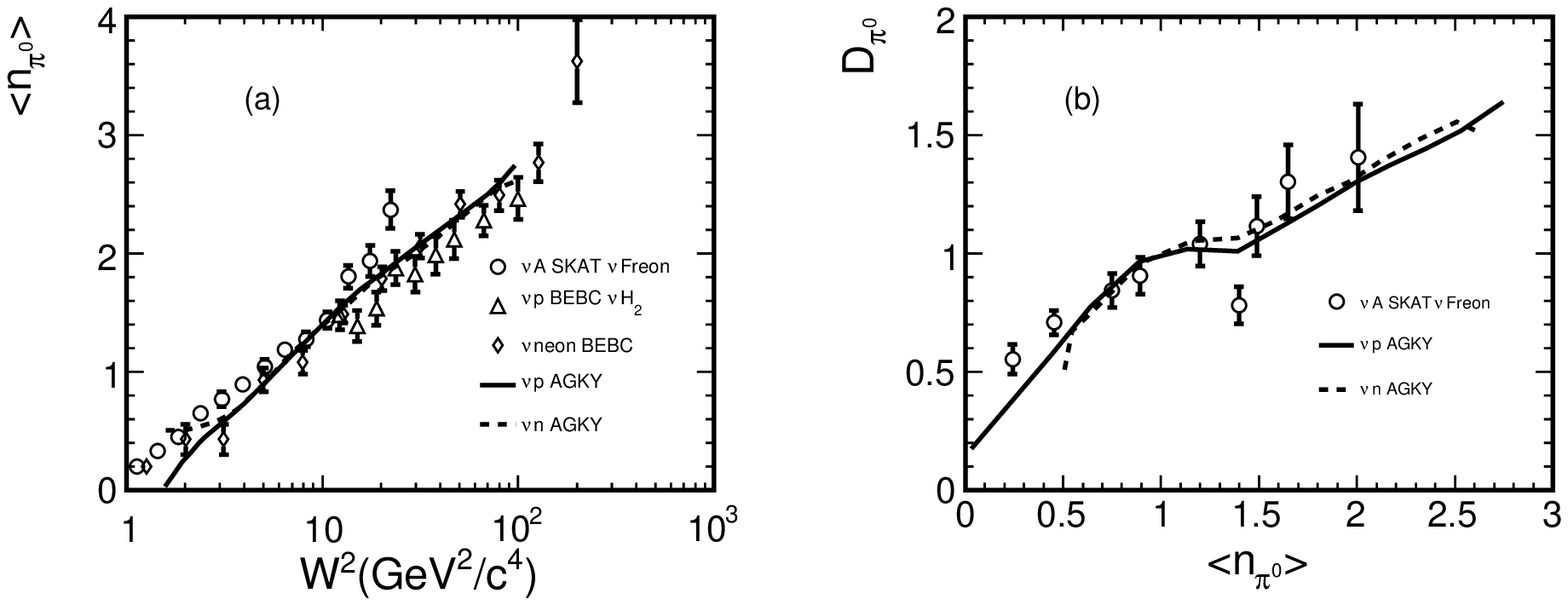}
\caption{$\pi^{0}$ mesons as a function of $W^{2}$. (b) Dispersion of the
distributions in multiplicity as a function of the average multiplicity
of $\pi^{0}$ mesons. Data points are taken from \cite{Wittek:1988ke,Ivanilov:1984gh,Grassler:1983ks}}

\label{fig:cMulPi0} 
\end{figure*}

\begin{figure*}
\centering \includegraphics[width=1\textwidth]{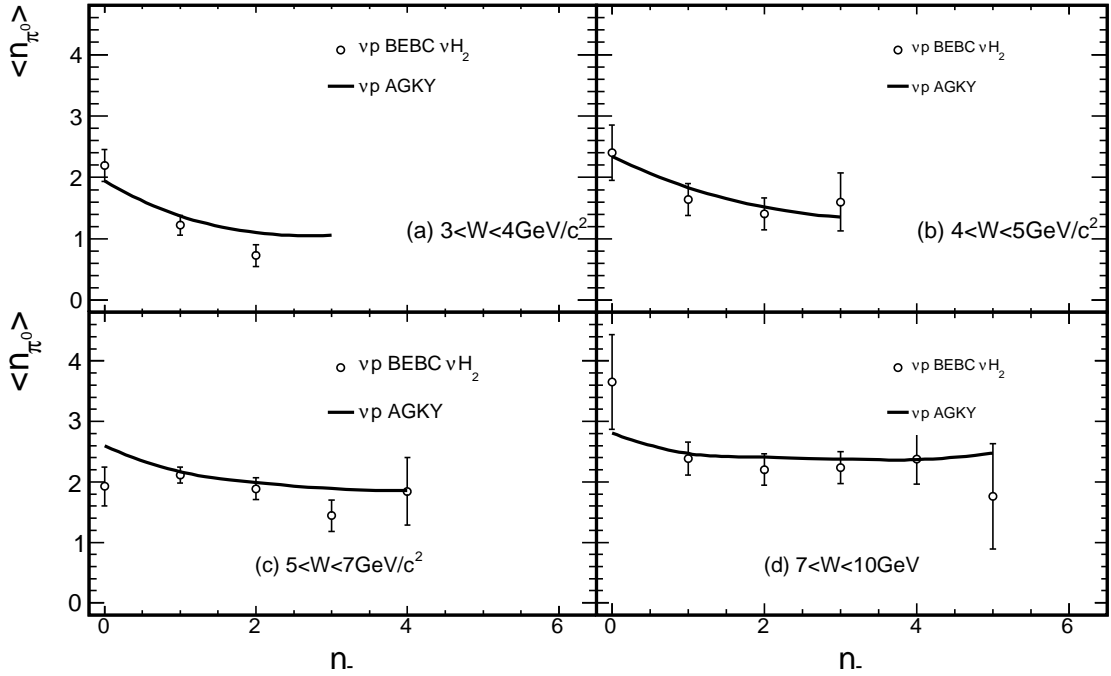}
\caption{$\pi^{0}$ multiplicity $\langle n_{\pi^{0}}\rangle$ as a function
of the number of negative hadrons $n_{-}$ for different intervals
of $W$. Data points are taken from \cite{Grassler:1983ks}.}

\label{fig:cCorr_Pi0_Ch} 
\end{figure*}

\begin{figure*}
\centering \includegraphics[width=1\textwidth]{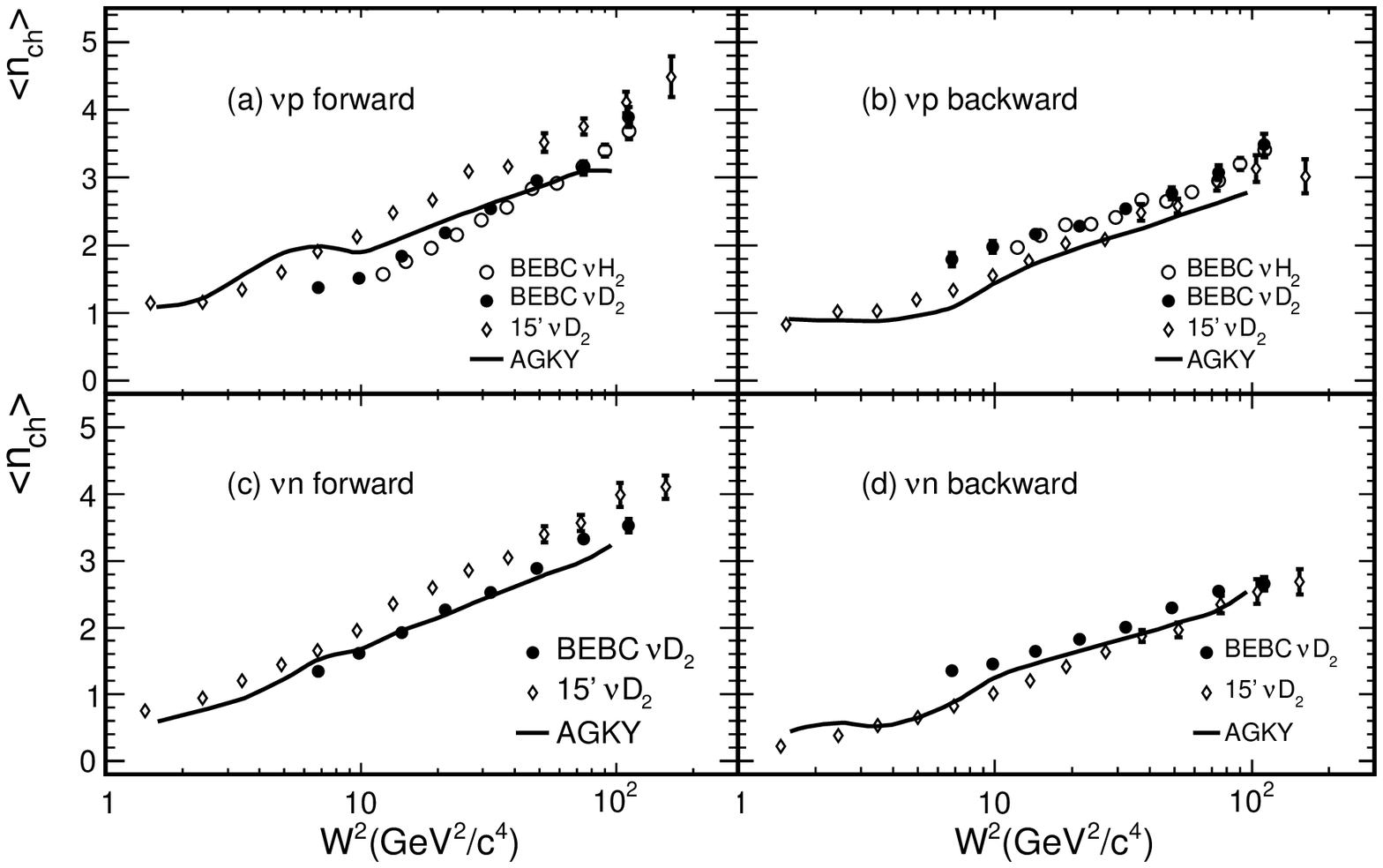}
\caption{$W^{2}$: (a) $\nu p$, forward, (b) $\nu p$, backward, (c) $\nu n$,
forward, (d) $\nu n$, backward. Data points are taken from \cite{Zieminska:1983bs,Grassler:1983ks,Allasia:1984ua}.}

\label{fig:cMulFB} 
\end{figure*}

Fig.\ref{fig:cDispCh}(a) shows the dispersion $D_{-}=(\langle n_{-}^{2}\rangle-\langle n_{-}\rangle^{2})^{1/2}$
of the negative hadron multiplicity as a function of $\langle n_{-}\rangle$.
Fig.\ref{fig:cDispCh}(b) shows the ratio $D/\langle n_{ch}\rangle$
as a function of $W^{2}$. The dispersion is solely determined by
the KNO scaling distributions shown in Fig.\ref{fig:kno_levy}. The
agreement between data and MC predictions is satisfactory. 

Fig.\ref{fig:cMulPi0}(a) shows the average $\pi^{0}$ multiplicity
$\langle n_{\pi^{0}}\rangle$ as a function of $W^{2}$. Fig.\ref{fig:cMulPi0}(b)
shows the dispersion of the distributions in multiplicity as a function
of the average multiplicity of $\pi^{0}$ mesons. As we mentioned
it is difficult to detect $\pi^{0}$'s inside a hydrogen bubble chamber.
Also shown in the plot are some measurements using heavy liquids such
as neon and Freon. In principle, rescattering of the primary hadrons
can occur in the nucleus. Some studies of inclusive negative hadron
production in the hydrogen-neon mixture and comparison with data obtained
by using hydrogen targets indicate that these effects are negligible
\cite{Berge:1978fr}. The model is in good agreement with the data.
$\langle n_{\pi^{0}}\rangle$ is 0(1/2) for $\nu p$($\nu n$) interactions
when the hadronic invariant mass approaches the pion production threshold,
which is consistent with the expectation from the reactions (\ref{res1}-\ref{res3}).
The model predicts the same average $\pi^{0}$ multiplicity for $\nu p$
and $\nu n$ interactions for $W>2\mbox{GeV}/c^{2}$.

Fig.\ref{fig:cCorr_Pi0_Ch} shows the average $\pi^{0}$ multiplicities
$\langle n_{\pi^{0}}\rangle$ as a function of the number of negative
hadrons $n_{-}$ for various $W$ ranges. At lower $W$, $\langle n_{\pi^{0}}\rangle$
tends to decrease with $n_{-}$, probably because of limited phase
space, while at higher $W$ $\langle n_{\pi^{0}}\rangle$ is rather
independent of $n_{-}$ where there is enough phase space. Our model
reproduces the correlation at lower $W$ suggested by the data. However,
another experiment measured the same correlation using neon-hydrogen
mixture and their results indicate that $\langle n_{\pi^{0}}\rangle$
is rather independent of $n_{-}$ for both $W>4\mbox{GeV}/c^{2}$
and $W<4\mbox{GeV}/c^{2}$ \cite{Ammosov:1978vt}. Since events with
$\pi^{0}$ but with 0 or very few charged pions are dominant background
events in the $\nu_{e}$ appearance analysis, it is very important
to understand the correlation between the neutral pions and charged
pions; this should be a goal of future experiments \cite{Drakoulakos:2004gn}.

Fig.\ref{fig:cMulFB} shows the average charged-hadron multiplicity
in the forward and backward hemispheres as functions of $W^{2}$.
The forward hemisphere is defined by the direction of the current
in the total hadronic c.m.s. There is a bump in the MC prediction
in the forward hemisphere for $\nu p$ interactions at $W\sim2\mbox{GeV}/c^{2}$
and there is a slight dip in the backward hemisphere in the same region.
This indicates that the MC may overestimate the hadrons going forward
in the hadronic c.m.s. at $W\sim2\mbox{GeV}/c^{2}$ and underestimate
the hadrons going backward. One consequence could be that the MC overestimates
the energetic hadrons since the hadrons in the forward hemisphere
of hadronic c.m.s. get more Lorentz boost than those in the backward
hemisphere when boosted to the LAB frame. This may be caused by the
way we determine the baryon 4-momentum and preferably select events
with low $p_{T}$ in the phase space decay. These effects will be
investigated further for improvement in future versions of the model.

\begin{figure}
\centering \includegraphics[width=1\columnwidth]{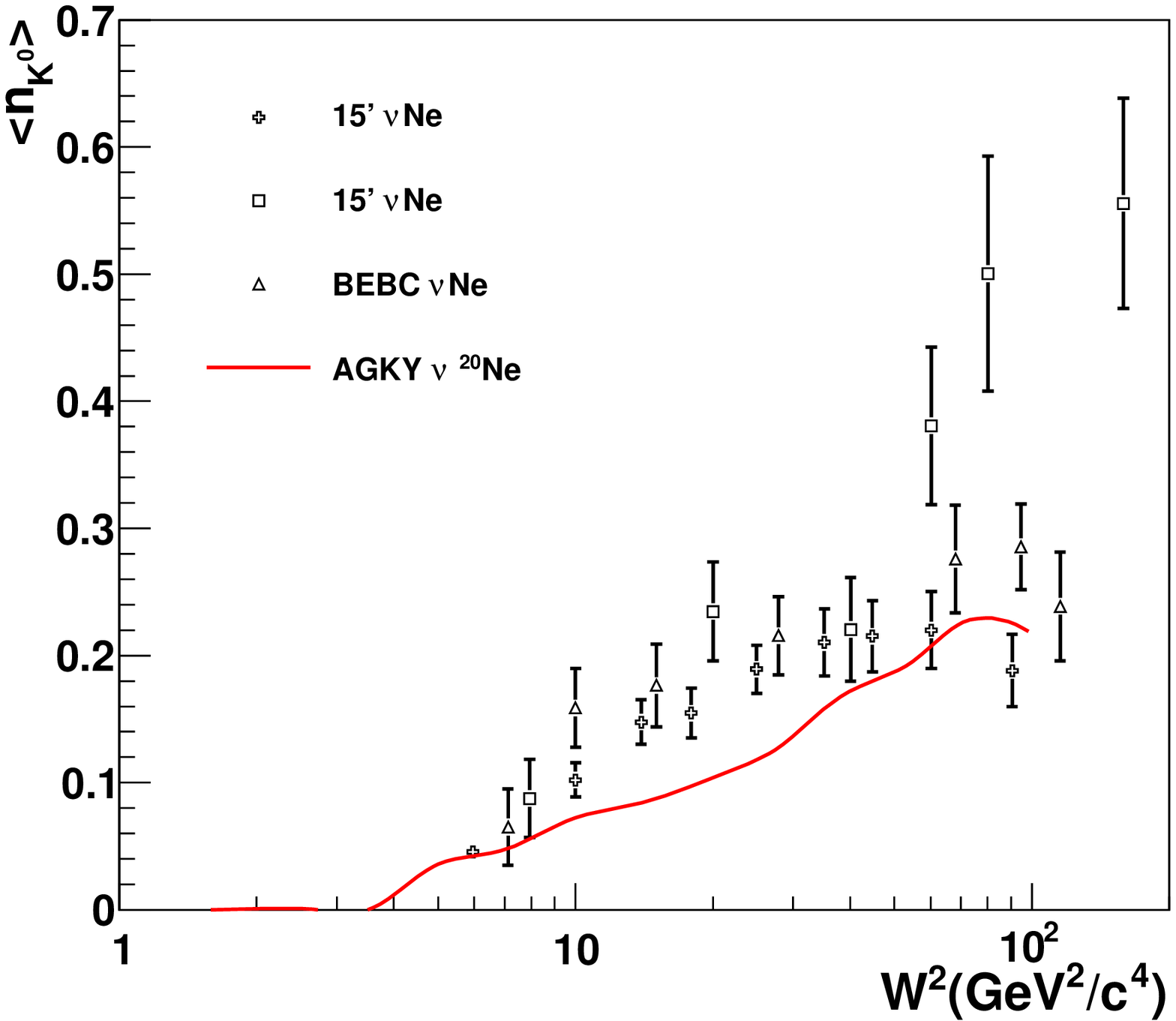}
\caption{\cite{Bosetti:1982vk,Baker:1986xx,DeProspo:1994ac}. \label{fig:kaons}}
\end{figure}

\begin{figure}
\centering \includegraphics[width=1\columnwidth]{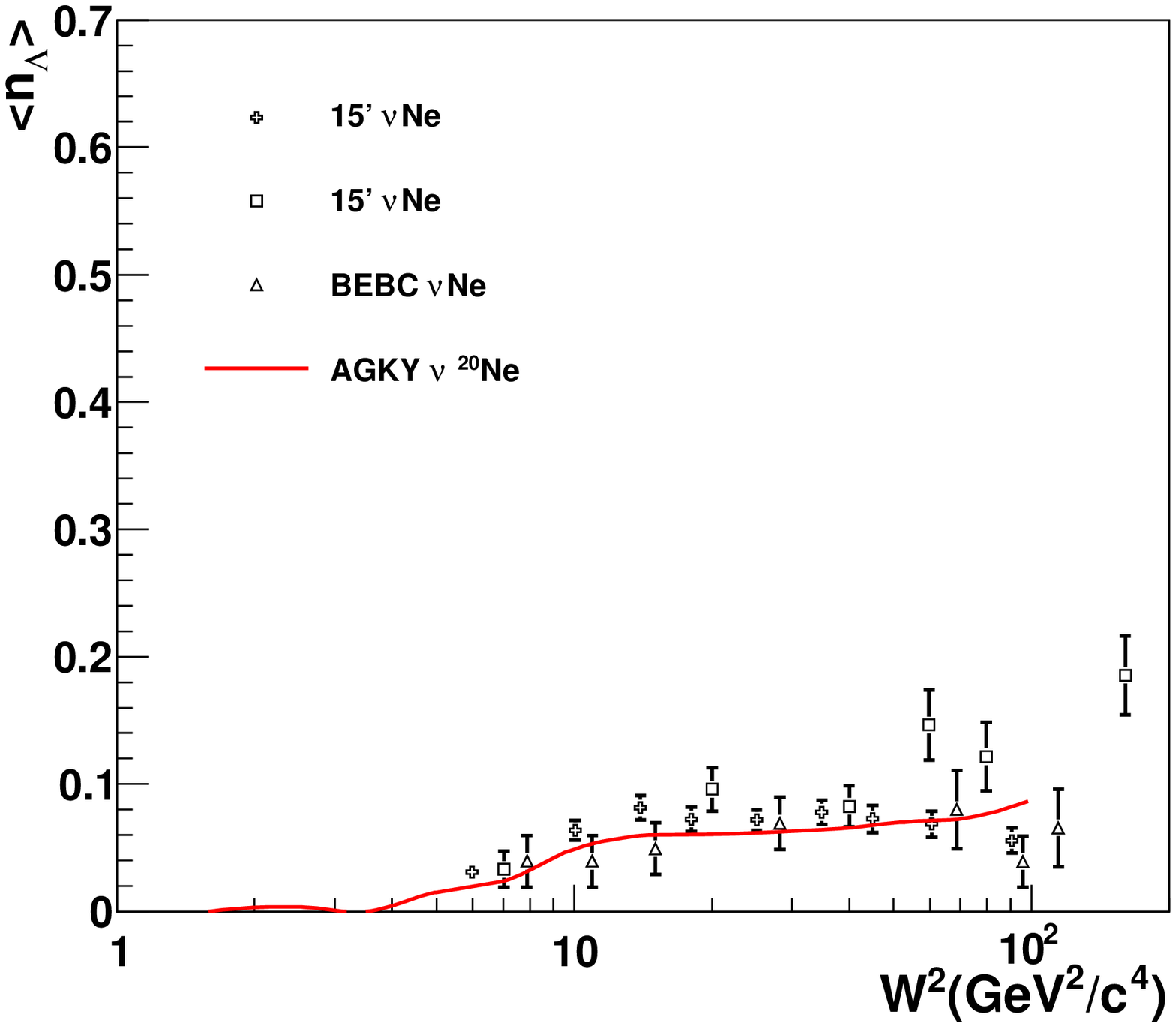}
\caption{\cite{Bosetti:1982vk,Baker:1986xx,DeProspo:1994ac}. \label{fig:lambdas}}
\end{figure}

The production of strange particles via associated production is shown
in Figures \ref{fig:kaons} and \ref{fig:lambdas}. The production
of kaons and lambdas for the KNO-based model are in reasonable agreement
with the data, while the rate of strange meson production from JETSET
is clearly low. We have investigated adjusting JETSET parameters to
produce better agreement with data. While it is possible to improve
the agreement with strange particle production data, doing so yields
reduced agreement with other important distributions, such as the
normalized charged particle distributions.

\begin{figure*}
\centering \includegraphics[width=1\textwidth]{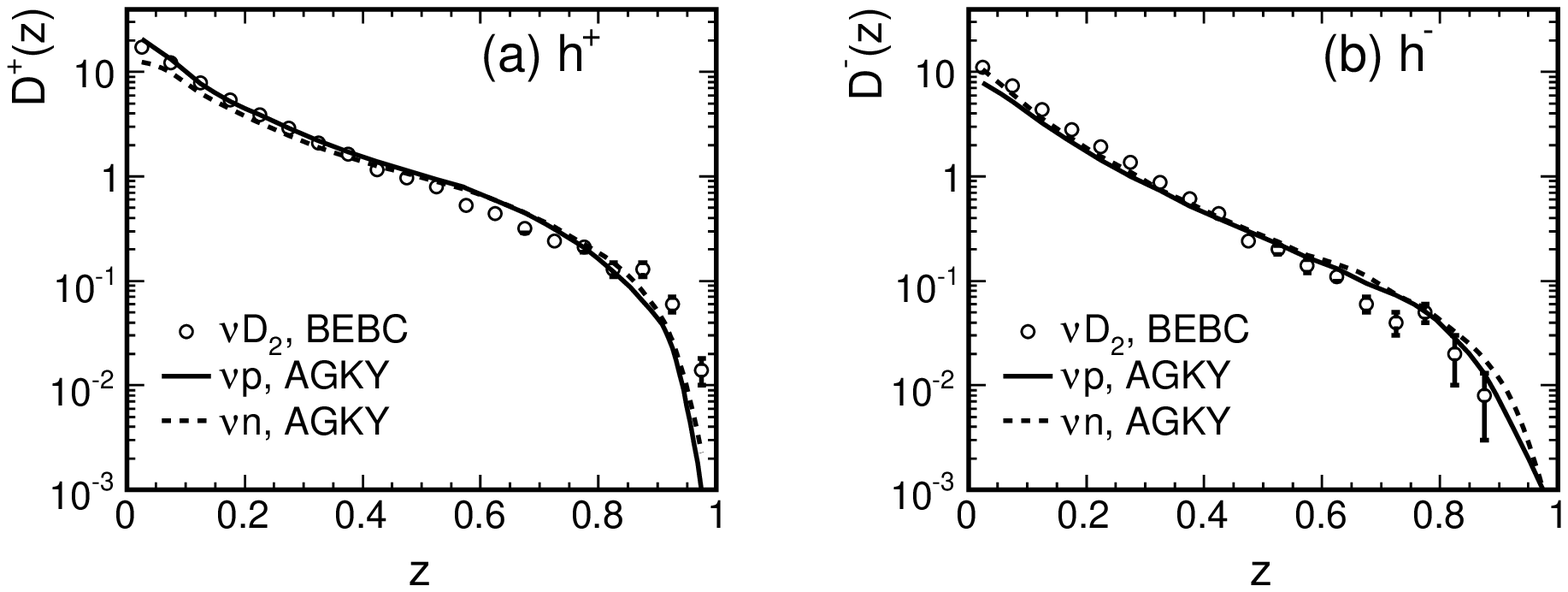}
\caption{$W^{2}>5(GeV/c^{2})^{2}$, $Q^{2}>1(GeV/c)^{2}$. Data points are
taken from \cite{Allasia:1984ua}.}

\label{fig:cZ} 
\end{figure*}

Fig.\ref{fig:cZ} shows the fragmentation functions for positive and
negative hadrons. The fragmentation function is defined as: $D(z)=\frac{1}{N_{ev}}\cdot\frac{dN}{dz}$,
where $N_{ev}$ is the total number of interactions (events) and $z=E/\nu$
is the fraction of the total energy transfer carried by each final
hadron in the laboratory frame. The AGKY predictions are in excellent
agreement with the data.

\begin{figure*}
\centering \includegraphics[width=1\textwidth]{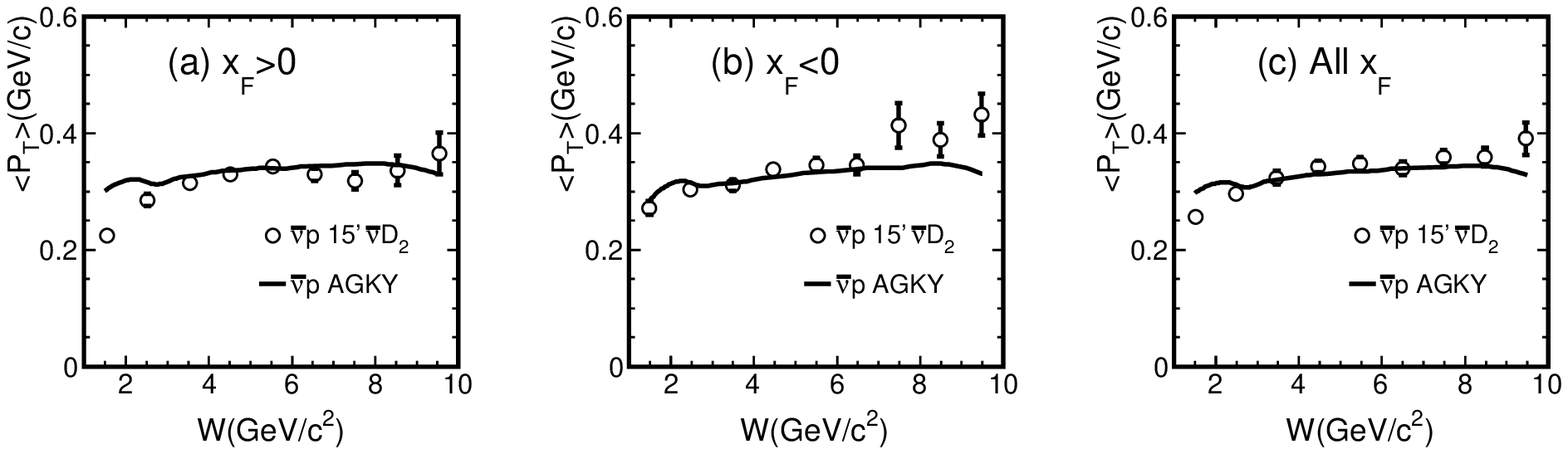}
\caption{$W$ for the selections (a) $x_{F}>0$, (b) $x_{F}<0$, and (c) all
$x_{F}$. Data points are taken from \cite{Derrick:1981br}.}

\label{fig:cPt_W} 
\end{figure*}

Fig.\ref{fig:cPt_W} shows the mean value of the transverse momentum
with respect to the current direction of charged hadrons as a function
of $W$. The MC predictions match the data reasonably well. In the
naive QPM, the quarks have no transverse momentum within the struck
nucleon, and the fragments acquire a $P_{T}^{frag}$ with respect
to the struck quark from the hadronization process. The average transverse
momentum $\langle P_{T}^{2}\rangle$ of the hadrons will then be independent
of variables such as $x_{BJ}$, $y$, $Q^{2}$, $W$, \textit{etc.},
apart from trivial kinematic constraints and any instrumental effects.
Both MC and data reflect this feature. However, in a perturbative
QCD picture, the quark acquires an additional transverse component,
$\langle P_{T}^{2}\rangle^{QCD}$, as a result of gluon radiation.
The quark itself may also have a primordial $\langle P_{T}^{2}\rangle^{prim}$
inside the nucleon. These QCD effects can introduce dependencies of
$\langle P_{T}^{2}\rangle$ on the variables $x_{BJ}$, $y$, $Q^{2}$,
$W$, $z$, \textit{etc.}

\begin{figure*}
\centering \includegraphics[width=1\textwidth]{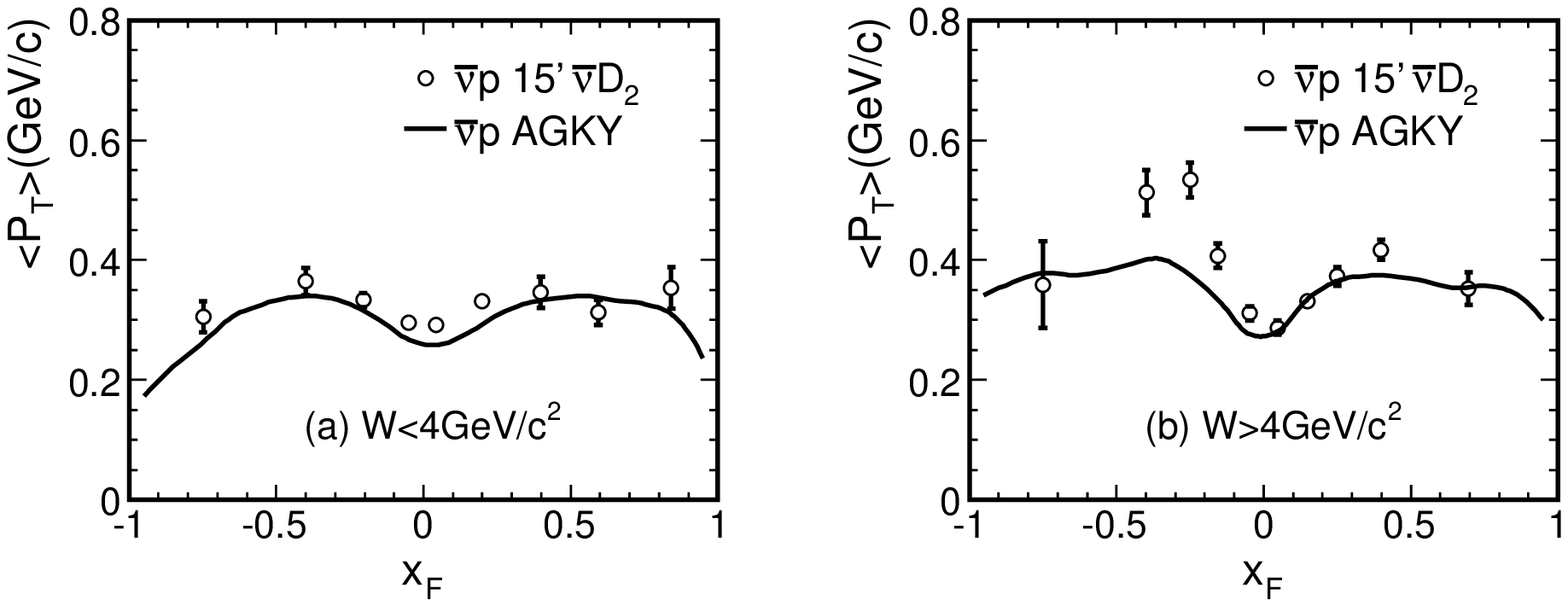}
\caption{$x_{F}$ for $\bar{\nu}p$. (a) $W<4\mbox{GeV}/c^{2}$, (b) $W>4\mbox{GeV}/c^{2}$.
Data points are taken from \cite{Derrick:1981br}.}

\label{fig:cPt_xf} 
\end{figure*}

Fig.\ref{fig:cPt_xf} shows the mean value of the transverse momentum
of charged hadrons as a function of $x_{F}$, where $x_{F}=\frac{p_{L}^{*}}{p_{Lmax}^{*}}$
is the Feynman-x. As is well known, $\langle p_{T}\rangle$ increases
with increasing $|x_{F}|$ with a shape called the seagull effect.
This effect is reasonably well modeled by the AGKY model.

\subsubsection{Conclusions}

\label{agkysec:7}

In this section we have described the GENIE hadronic mutiparticle
production model tuned for experiments in the few-GeV energy regime.
The model exhibits satisfactory agreement with wide variety of data
for charged, neutral pions as well as strange particles. Several upcoming
expriments will have high-statistics data sets in detectors with excellent
energy resolution, neutral particle containment, and particle identification.
These experiments are in some cases considering possible running with
cryogenic hydrogen and deuterium targets. These experiments will be
operating in this few-GeV regime and have the potential to fill in
several gaps in our understanding that will help improve hadronic
shower modeling for oscillation experiments.

The upcoming generation of experiments have all the necessary prerequisites
to significantly address the existing experimental uncertainties in
hadronization at low invariant mass. These result from the fact that
these detectors have good containment for both charged and neutral
particles, high event rates, good tracking resolution, excellent particle
identification and energy resolution, and the possibility of collecting
data on free nucleons with cryogenic targets. The latter offers the
possibility of addressing the challenge of disentangling hadronization
modeling from intranuclear rescattering effects. Charged current measurements
of particular interest will include clarifying the experimental discrepancy
at low invariant mass between the existing published results as shown
in Fig.\ref{fig:cMulFB}, the origin of which probably relates to
particle misidentification corrections \cite{Grassler:1983ks}. This
discrepancy has a large effect on forward/backward measurements, and
a succesful resolution of this question will reduce systematic differences
between datasets in a large class of existing measurements. In addition,
measurements of transverse momentum at low invariant masses will be
helpful in model tuning. Measurements of neutral particles, in particular
multiplicity and particle dispersion from free targets at low invariant
mass, will be tremendously helpful. The correlation between neutral
and charged particle multiplicities at low invariant mass is particularly
important for oscillation simulations, as it determines the likelihood
that a low invariant mass shower will be dominated by neutral pions.

\clearpage

\section{Intranuclear Hadron Transport}

The hadronization model describes particle production from free targets
and is tuned primarily to bubble chamber data on hydrogen and deuterium
targets \cite{Zieminska:1983bs,Derrick:1977zi,Allen:1981vh,Ivanilov:1984gh,Grassler:1983ks,Allasia:1984ua,Berge:1978fr,Ammosov:1978vt}.
Hadrons produced in the nuclear environment may rescatter on their
way out of the nucleus, and these reinteractions significantly modify
the observable distributions in most detectors.

It is also well established that hadrons produced in the nuclear environment
do not immediately reinteract with their full cross section. The basic
picture is that during the time it takes for quarks to materialize
as hadrons, they propagate through the nucleus with a dramatically
reduced interaction probability. This was implemented in GENIE as
a simple `free step' at the start of the intranuclear cascade during
which no interactions can occur. The `free step' comes from a formation
time of 0.342 fm/c according to the SKAT model \cite{Baranov:1984rv}.

\begin{figure}
\begin{minipage}[t]{0.45\columnwidth}%
\includegraphics[width=0.95\columnwidth]{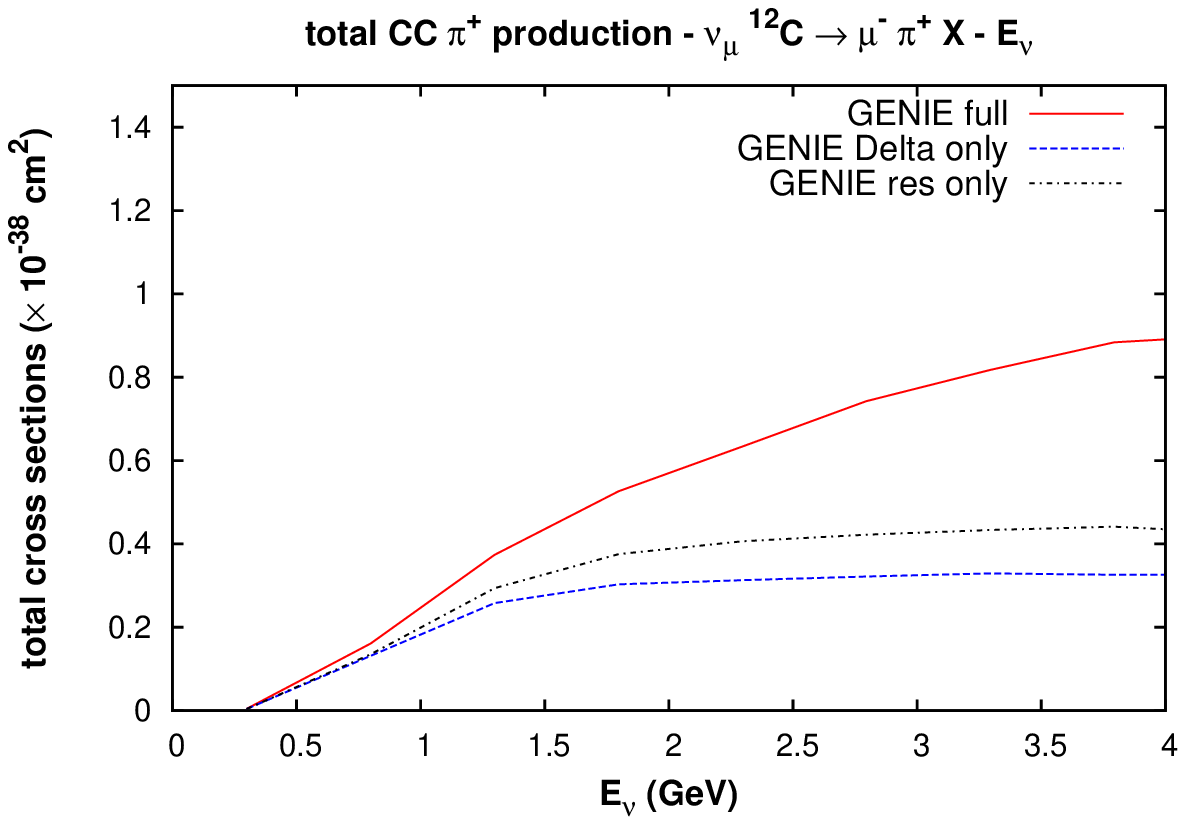}

\caption{$\pi^{+}$ total cross section resulting from $\nu_{\mu}^{12}C$ interactions.
Different lines show results including all sources, all resonances,
and the $\Delta$ resonance alone. The nonresonant processes are significant
in GENIE.}
\label{fig:inuke:numuC12_1pip}%
\end{minipage}\ \ \ \ \ \ \ \  %
\begin{minipage}[t]{0.45\columnwidth}%
\includegraphics[width=0.95\columnwidth]{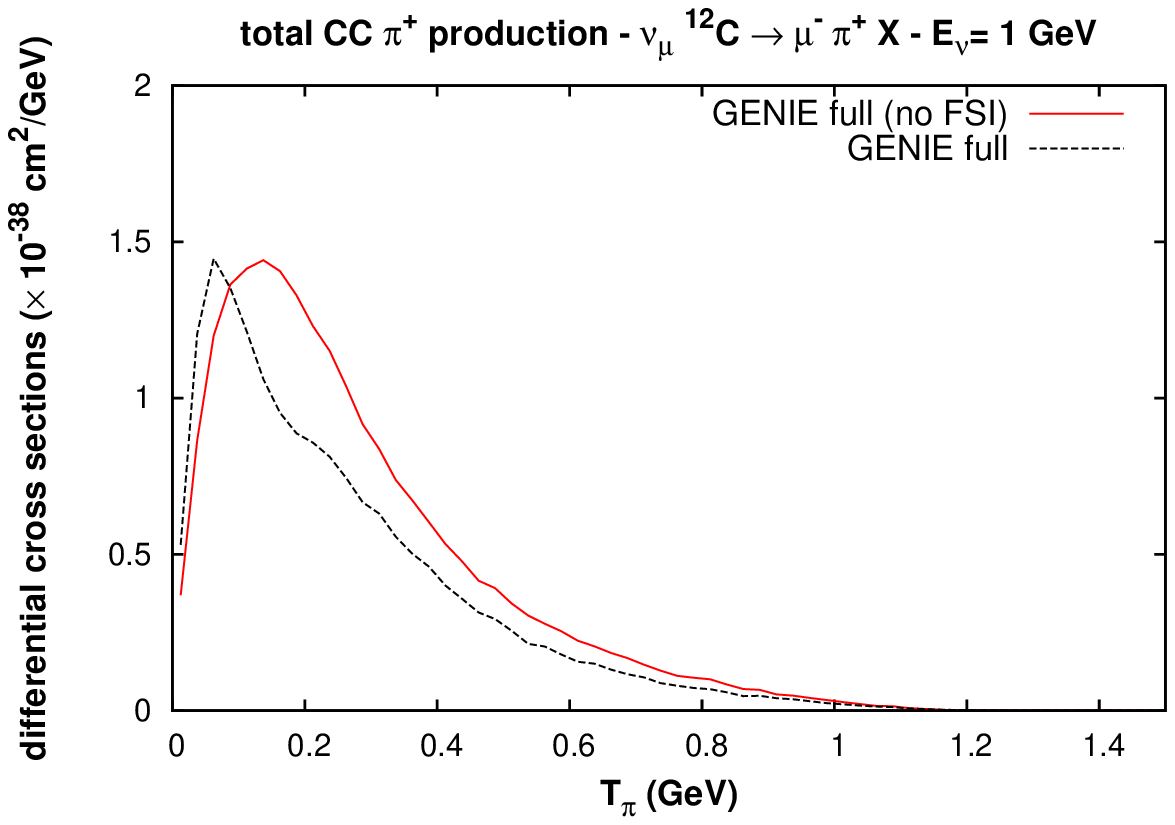}

\caption{Comparison of the $\pi^{+}$ momentum distribution due to the bare
resonance interaction and what is seen in the final state. }
\label{fig:inuke:pi-spectrum-fsi-effect}%
\end{minipage}
\end{figure}

Intranuclear hadron transport in GENIE is handled by a subpackage
called INTRANUKE. INTRANUKE is an intranuclear cascade simulation
and has gone through numerous revisions since the original version
was developed for use by the Soudan 2 Collaboration \cite{Merenyi:1992gf}.
The sensitivity of a particular experiment to intranuclear rescattering
depends strongly on the detector technology, the energy range of the
neutrinos, and the physics measurement being made. INTRANUKE simulates
rescattering of pions and nucleons in the nucleus. When produced inside
a nucleus, hadrons have a typical mean free path (MFP) of a few femtometers.
Detectors in a neutrino experiment are almost always composed of nuclei
today. Therefore, the hadrons produced in the primary interaction
(what the neutrino does directly) often (e.g. $\sim$30\% in iron
for few GeV neutrinos) have a FSI. There are many possibilities from
benign to dangerous. For example, a quasielastic (QE) interaction
that emits a proton can end up with a final state of 3 protons, 2
neutrons, and a few photons with finite probability. For a 1 GeV muon
neutrino QE interaction in carbon, the probability of a final state
different than 1 proton is 35\% (GENIE). A possibility even worse
is a pion production primary interaction where the pion is absorbed.
Such an event occurs for 20\% (GENIE) of pion production events and
can look like a QE event. At minimum, the wrong beam energy will be
measured for these events as the topology is often mistaken. A high
quality Monte Carlo code is the only way to understand the role of
these events. Fig. \ref{fig:inuke:numuC12_1pip} shows the pion energies
that are relevant to a $\nu_{\mu}C$ experiment at 1 GeV; we must
understand the interactions of pions of up to about 0.8 GeV kinetic
energy. We see that the $\Delta$ resonance dominates the response
for pion production, but provides only about half of all pions. Fig.
\ref{fig:inuke:pi-spectrum-fsi-effect} shows that the spectrum of
pions is significantly altered by FSI. 

The best way to understand the effects of FSI is to measure the \textit{cross
sections} for as many final states as possible with neutrino beams.
At this time, the storehouse for this kind of data is very bare. Dedicated
cross section experiments such as SciBooNE and MINERvA will bridge
this gap, but we will always be dependent on hadron-nucleus and photon-nucleus
experiments for some information. These experiments measure very useful
properties of hadrons propagating in nuclei. Although hadron beams
are always composed free particles, neutrino experiments need the
properties of hadrons produced off-shell in the nucleus. (Pion photoproduction
experiments provide useful bridge reactions. Pion FSI are always an
important part of all theory calculations for these experiments; the
models always come from pion-nucleus data.) The correct attitude is
to validate FSI models for neutrino-nucleus with hadron-nucleus data,
then use these models to make first predictions of the upcoming dedicated
cross section experiments.

Various models are available. Quantum mechanical models for hadron-nucleus
experiments would be the most correct, but difficulties in tracking
multiple particles make such a calculation impossible. Semi-classical
models have some success in describing pion-nucleus interaction data
and are now being applied to neutrino interactions \cite{Leitner:2006ww}.
However, intranuclear cascade (INC) models \cite{Harp:1974zz,Mashnik:LAUR057321,Mashnik:2005ay,Mashnik:PrivCom}
provided the most important means to understand pion-nucleus data
where the final state was highly inelastic, i.e. the kinds of data
most important for neutrinos. In the semi-classical or INC models,
the hadron sees a nucleus of (largely) isolated nucleons (neutrons
and protons). The probability of interaction is governed by the \textit{free}
cross section and the density of nucleons, 
\begin{equation}
\lambda(E,r)=\frac{1}{\sigma_{hN,tot}*\rho(r)}\label{eq:mfp}
\end{equation}
 The actual class of interaction is then chosen according to the cross
sections for various reactions for free nucleons, sometimes modified
for nuclear medium effects.

\subsubsection{Survey of models}

Semi-classical models have advanced significantly due to the work
of the Giessen group in building a new program called GiBUU \cite{Leitner:2006ww}.
The strong interaction section \cite{Engel:1993jh,Buss:2007ar} is
the most complete part of the package. The dominant interaction of
pions is through resonance formation and they are handled with care.
Nucleons in the nucleus are corrected for binding with a local potential
well and for Fermi motion with a local Fermi momentum. Resonance production
is corrected for the nearby nucleons in a local density approximation.
Nonresonant reactions are added by hand. Allowing for the nonlocality
of the interaction is an important recent advance. The classical part
of the model comes from the use of free cross sections with corrections
rather than quantum mechanical amplitudes for interactions. Thus,
GiBUU could be called a very sophisticated INC model. The passage
of a hadron through the nuclear medium is then handled by a set of
coupled integro-differential equations. Thus, required computer resources
are significant.

GENIE, NEUT, and FLUKA have more standard INC models. They use free
cross sections for interactions but also apply medium corrections
of various kinds. These corrections are less complete and more empirical
than what is found in GiBUU. These models are most applicable for
higher energy hadrons (roughly pions with kinetic energy larger than
300 MeV and nucleons above 200 MeV), where the mean free path is long
compared to the inter-nuclear spacing of roughly 1.8fm and the pion
Compton wavelength.

Peanut (FLUKA) \cite{Battistoni:2007zzb} received a major effort
in 1995-9 and is very well adapted to describe processes from 10 MeV
to 100 GeV. They include effects such as coherence time, refraction,
and pre-equilibrium/compound nucleus processes which simulate known
quantum mechanical features. NEUT FSI is based on the work of Salcedo,
Oset, Vicente-Vacas, and Garcia-Recio \cite{Salcedo:1987md}. This
is a ``$\Delta$ dominance'' model such as were common in the 1980's
when pion-nucleus physics was important in nuclear physics. It has
the advantage of doing a careful job simulating the pion-nucleus interaction
through $\Delta$(1232) intermediate states.

\subsubsection{Systematics of hadron-nucleus data}

Each nucleus has $A$ nucleons ($Z$ protons +$N$ neutrons). All
nuclei of interest to neutrino physics are either bound or slightly
unbound. Nuclear densities show saturation because of short range
repulsion. Therefore, the typical nucleus is approximately a sphere
of radius proportional to $A^{1/3}$. The charge density of light
nuclei ($A$<20) is found to be Gaussian or modified Gaussian. Heavier
nuclei are described by the Woods-Saxon shape, 
\begin{equation}
\rho(r)=N_{0}\frac{1}{1+e^{(r-c)/z}}
\end{equation}
 where $c$ describes the size and $a$ describes the width of the
surface of a nucleus. For example, $c$=4.1 fm and $z$=0.55 fm for
$^{56}$Fe. To good accuracy, $c$ is the radius where the density
falls to half the central value with $c\sim1.2fm*A^{1/3}$ and $z\sim0.55fm$.

Hadrons interact with nuclei in a variety of ways. We use historical
definitions of final states that come from interpretation of experiments.
In \textit{elastic scattering}, the final state nucleus is in its
ground state and the hadron has same charge as the beam particle.
If the hadron scatters inelastically, the residual nucleus can be
in the ground state or the nucleus can break apart. At low excitation
energies ($<\sim$ 10 MeV), the residual nucleus decays to a photon
and the ground state. (This is important in analysis of SuperKamiokande
data.) At higher excitation energies, one or more nucleons are emitted.
Final state interactions increase this number. If there is a hadron
of the same type but different charge in the final state, we call
it \textit{charge exchange}. For example, the reaction $\pi^{-}p\rightarrow\pi^{0}n$
is very common inside nuclei. As a boson, the pion can disappear inside
the nucleus. Pion initiated reactions with no pions in the final state
are called \textit{absorption}. (This provides an important background
process to neutrino quasielastic scattering.) For incident nucleons,
most of these labels apply exactly. Since they can't be absorbed,
final states with 2 or more nucleons are called \textit{spallation}.
If the hadron has enough energy, a pion (a second pion if the initial
hadron is a pion) can be produced in the nucleus. We call those events
\textit{pion production}.

For low energy incident particles, these definitions are clean. At
higher energies, the states mix and confusion can result. For example,
a reaction $\pi^{+}{}^{12}C\rightarrow\pi^{+}\pi^{0}{}^{12}C$ can
be inelastic, charge exchange, or pion production. Definitions we
use call it pion production. A way to avoid difficulties is to measure
inclusive cross sections; there, the energy and angular distribution
of a particular particle are determined. In each case, various reactions
are possible but models can be tested without ambiguity.

Because the MFP is so short for hadron interactions, elastic scattering
cross sections look very diffractive. In fact, the angular distribution
can be calculated with a quantum mechanical model using a black disk
for the nucleus. This wave property is very difficult to simulate
in a semi-classical or INC model.

Another consequence of the short MFP is seen in the total reaction
cross section ($\sigma_{reac}=\sigma_{tot}-\sigma_{elas}$). For hadrons,
this is close to the nuclear size, $\pi R^{2}$. For example, $\sigma_{reac}$
for protons and neutrons of 0.4-1GeV is flat at a value of about 300$mb$=30$fm^{2}$
for carbon and about 80$fm^{2}$ for iron. These corresponds to a
radius, $R$, of about 3 and 5 fm. These values are close to the radius
where the nuclear density is about half of the central value. If we
divide these values by $A^{1/3}$, the result is close to the commonly
used value of 1.2 fm. The pion-nucleus reaction cross section at kinetic
energies of about 85-315 MeV is dominated by the effect of the $\Delta$(1232)
resonance. Thus, the effective size of the nucleus here is at a radius
where the density is about 1\% of the central density. For total cross
sections, the A dependence is often a power relation, $\sigma\propto A^{\alpha}$,
but $\alpha$ will vary from the expected value of 2/3 due to more
complicated dynamics. The total cross section for pion-nucleus has
a power of about 0.8 for a wide range in energy. The $A$ dependence
of $\alpha$\cite{Ashery:1981tq,Navon:1983xj} varies between 0.55
and 0.8 for the components of the total reaction cross section as
a function of energy and process.

Nevertheless, many inelastic cross sections have prominent contributions
from \textit{quasifree} interactions. Here, the hadrons in the final
state have the kinematics as though they came from a single interaction
between the incident particle and a nucleon in the nuclear medium.
The name comes from the fact that nucleons in the nuclear medium are
in a bound state and therefore not free. If the nucleon were free,
the scattered particles would have a single energy at each angle.
The struck nucleons have momentum (called Fermi motion), giving particles
a range of momentum at a given angle. The largest momentum a nucleon
can have is well-defined in the Fermi Gas model, is approximate in
real nuclei. It is called the \textit{Fermi momentum} and its value
is approximately 250 MeV/c. In heavy nuclei, the average binding energy
is about 25 MeV. Thus, the peak due to quasifree scattering from a
bound nucleon is shifted by about 40 MeV from the free case and the
width is roughly 100 MeV.

This process has been widely studied for electron and pion probes.
If it could be studied with neutrinos, the same structure would be
seen. The so-called quasielastic peak is prominent in the inclusive
scattering cross section. At high excitation energies (lower kinetic
energy for the scattered particle), a second peak is found for quasifree
pion production from a bound nucleon. Final state interactions are
more important in the details in this case. Consider the case of $\pi^{+}$
interactions in carbon at 245 MeV. Evidence for quasifree pion scattering
is strong. A scattered $\pi^{+}$ is tagged on one side of the beam
and the spectrum of protons is measured on the other side. A prominent
peak is seen close to the angle where protons would be if the target
was a free proton. The same correlation is seen between 2 protons
where the $\pi^{+}$ is absorbed on a quasideuteron in the nuclear
medium. Strong evidence for quasifree pion scattering and absorption
is seen. Calculations with an INC model are in excellent agreement
with these data.

\begin{figure}
\begin{minipage}[t]{0.48\columnwidth}%
\includegraphics[width=0.95\columnwidth]{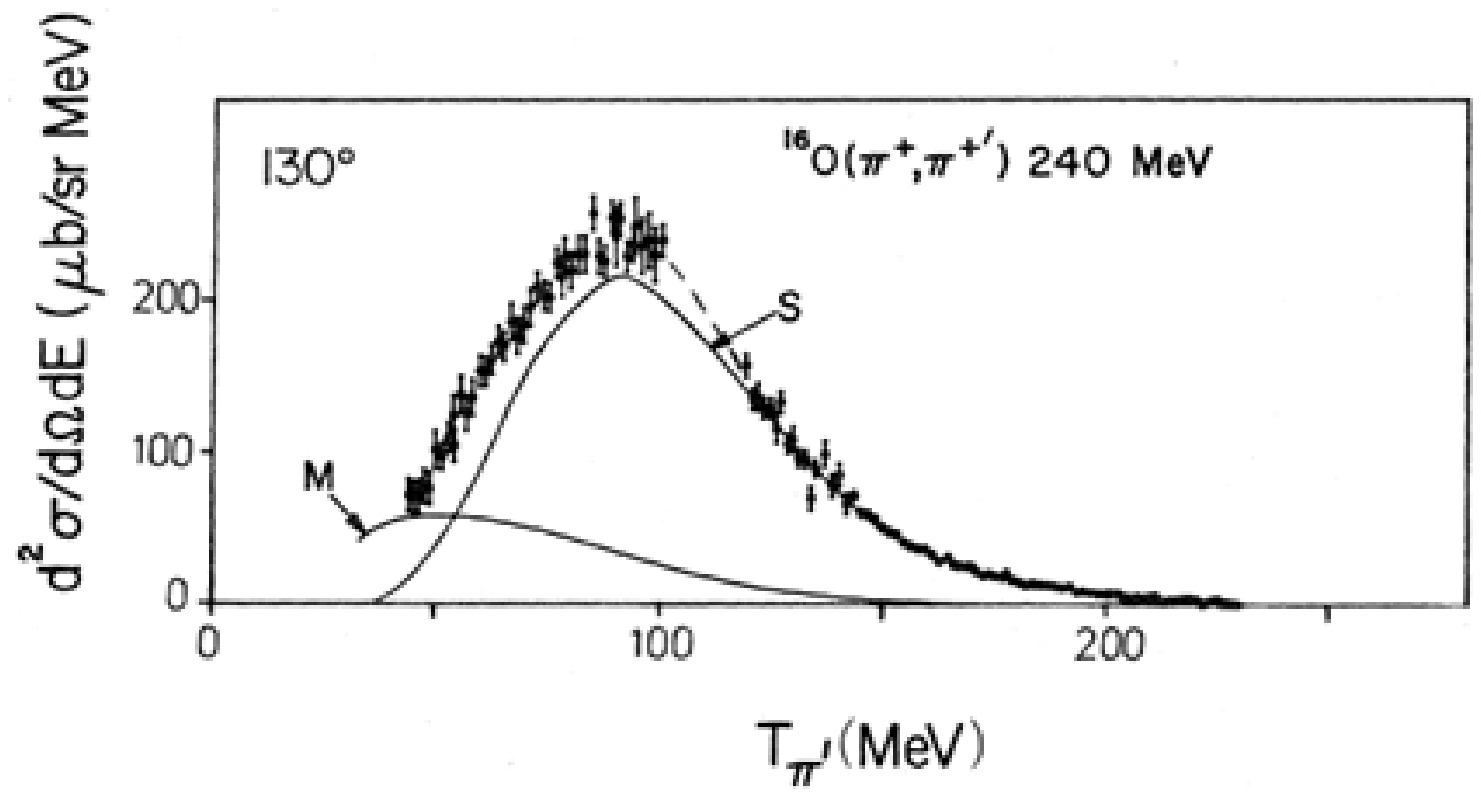}

\caption{Inclusive $\pi^{+}$scattering data from Ingram, et al. compared with
separate curves for single and multiple scattering contributions.}

\label{fig:inuke:ingram-pip-o-240-multscat}%
\end{minipage}\ \ \ \ \ \ \ \  %
\begin{minipage}[t]{0.48\columnwidth}%
\includegraphics[width=0.95\columnwidth]{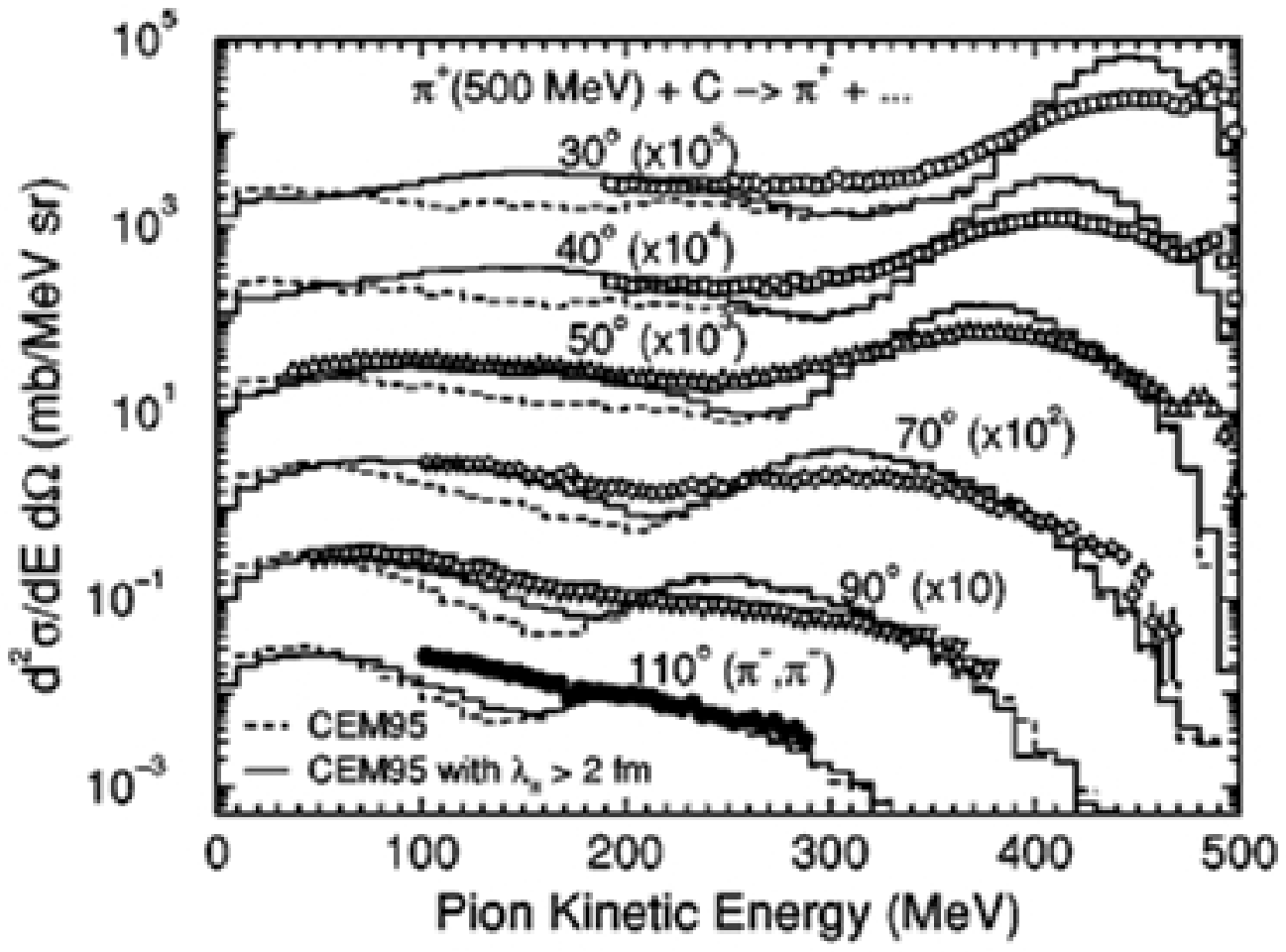}

\caption{Inclusive $\pi^{-}$ scattering data from Zumbro, et al. compared
with INC calculations of Mashnik, et al. }

\label{fig:inuke:mashnik-zumbro-pim-c-500}%
\end{minipage}
\end{figure}

The energy distribution of $\pi^{+}$ detected at 130$^{\circ}$ \cite{Ingram:1982bn}
shows a peak close to where scattering from a free proton would be
seen. Since Fig. \ref{fig:inuke:ingram-pip-o-240-multscat} is for
a H$_{2}$O target, scattering from H is seen as a gap at about 130
MeV (cross section is too large to show). Pions interacting with oxygen
nuclei produce a peak at about 100 MeV. Calculations show it is dominated
by events with a single scattering (S). At low energies, the distribution
is modified by events with more than one scattering (M). At forward
angles, the contributions from multiple scattering aremore important.

If incident particles have a higher energy, complications can be found.
With light targets, FSI effects are small and quasifree scattering
and pion production peaks are seen. However, INC calculations have
trouble getting the shape right, particularly in the region between
the peaks. Fig. \ref{fig:inuke:mashnik-zumbro-pim-c-500} is for $\pi^{-}$
scattering from $^{12}C$ at 500 MeV \cite{Zumbro:1993hr}. For $\pi^{+}$
absorption, the quasifree process would be $\pi^{+}d\rightarrow pp$
since pions are highly unlikely to be absorbed on a single nucleon.
LADS data \cite{Kotlinski:2000hp} for $\pi^{+}$ absorption in $Ar$
(A=40) shows the largest strength for the $pp$ final state but this
is less than half of the total cross section.

\subsubsection{INC models}

\begin{figure}
\begin{minipage}[t]{0.48\columnwidth}%
\includegraphics[width=1\columnwidth]{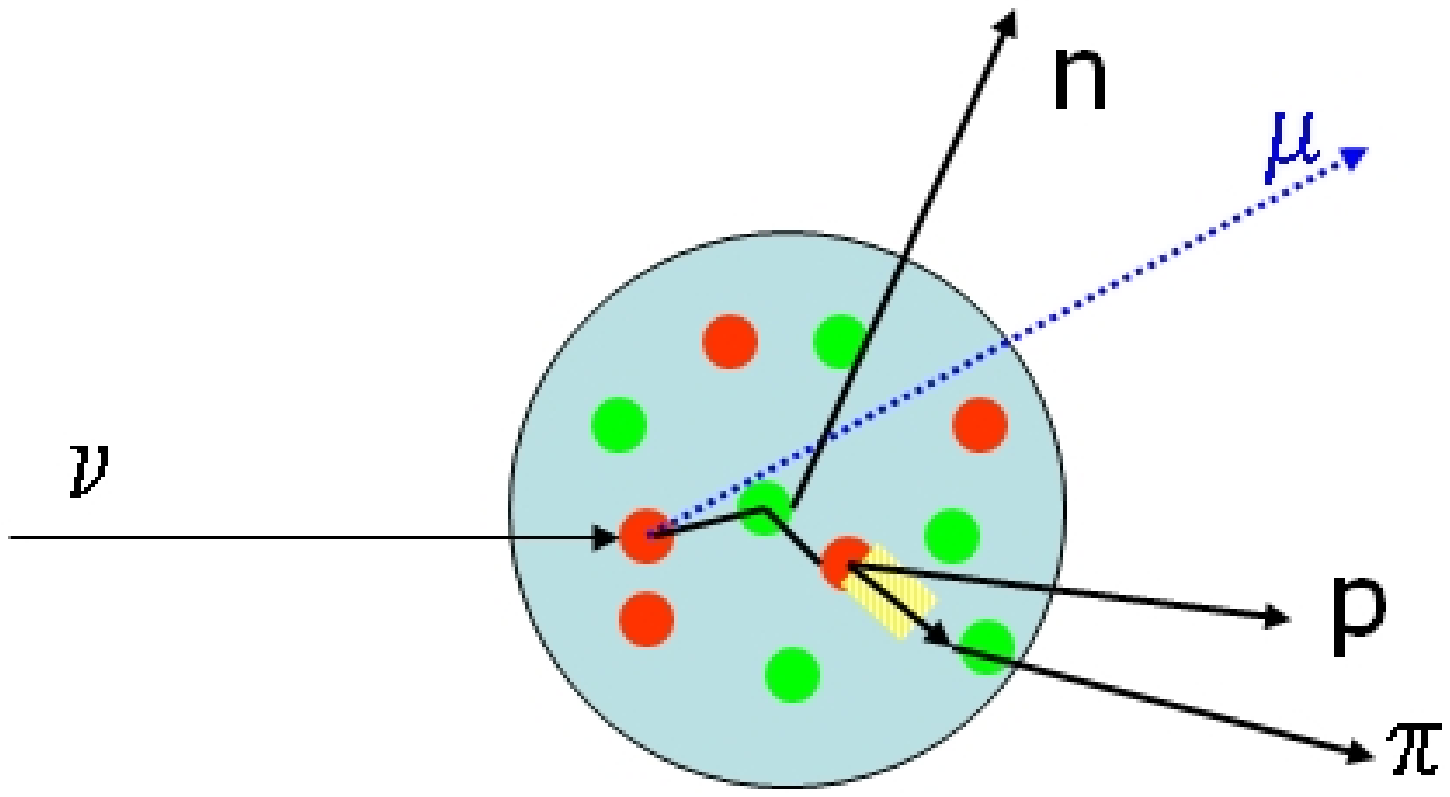}

\caption{Schematic diagram for reaction involving typical FSI process.}

\label{fig:inuke:genie-diag-fsi}%
\end{minipage}\ \ \ \ \ \ \ \ %
\begin{minipage}[t]{0.48\columnwidth}%
\includegraphics[width=1\columnwidth]{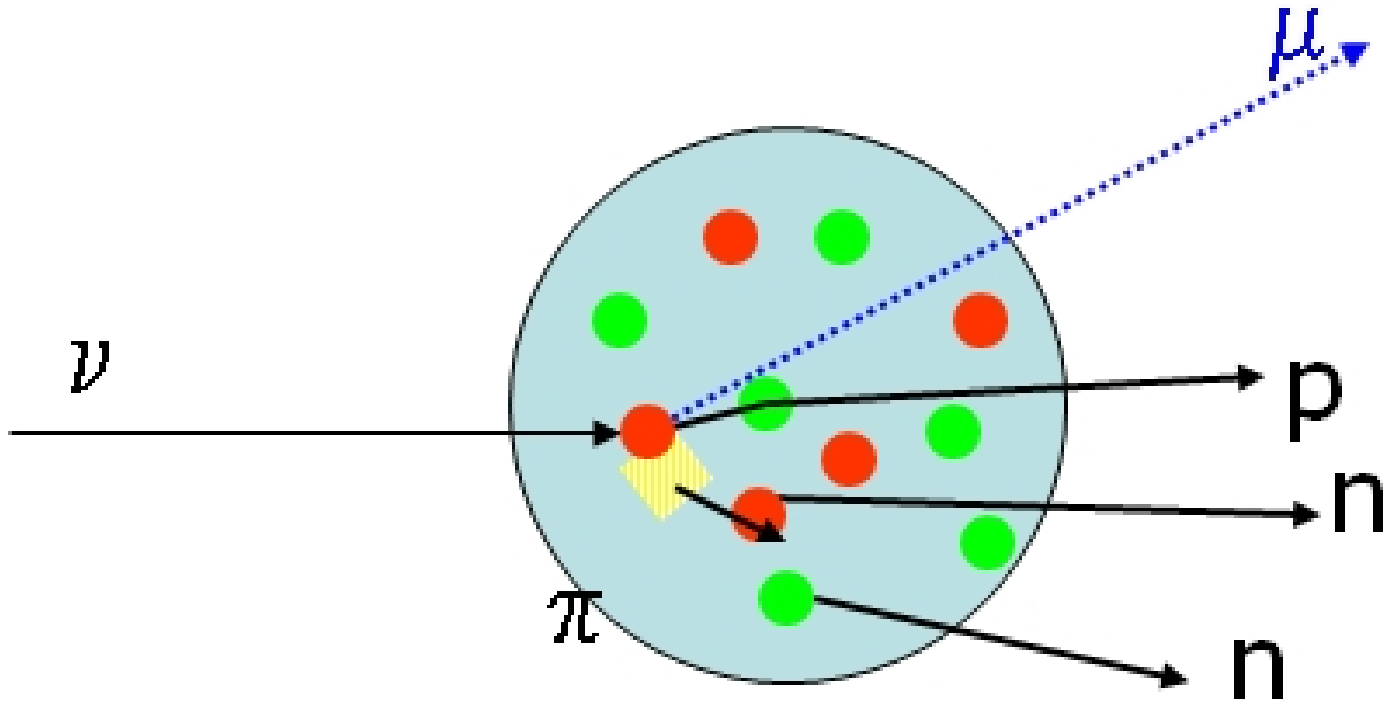}

\caption{Schematic diagram for reaction where pion is produced then absorbed
in the same nucleus.}

\label{fig:inuke:genie-diag-piabs}%
\end{minipage}
\end{figure}

Prominence of the quasifree reaction mechanism shows why INC models
are valuable. These models assume the nucleus is an ensemble of nucleons
which have Fermi motion and binding energy. The incident particle
interacts in a series of encounters with single nucleons called a
cascade (see Figs. \ref{fig:inuke:genie-diag-fsi}, \ref{fig:inuke:genie-diag-piabs}).

All interactions are governed by the cross section for the free process,
e.g. $\pi^{+}n\rightarrow\pi^{+}n$ or $pp\rightarrow pp$. Probability
of interaction is governed by a mean free path according to Eqn.~\ref{eq:mfp}.
Cross sections for pions, kaons, protons, and photons interacting
with free nucleons are fit with a partial wave analysis with results
provided by the GWU group \cite{Arndt:2006bf,GWU:Web}. Nucleon densities
come from compilations; note that neutron and protons have very similar
densities even for nuclei such as lead.

The problems with INC models must be considered. Since interactions
are governed by cross sections rather than quantum mechanical amplitudes,
the nuclear model is often very simple. The simplest and most general
nuclear model is the Fermi gas which is the basis for all neutrino-nucleus
event generator models. Effects of nucleon correlations must be included
empirically. Both the struck nucleon and the scattered hadron are
likely to be off-shell. Although this effect has been shown to be
`moderate', it is difficult to simulate in a semi-classical model.
Thus, there is no definite prescription for an INC model; many versions
exist with a wide range of applicability.

\begin{figure}
\begin{minipage}[t]{0.48\columnwidth}%
\includegraphics[width=0.95\columnwidth]{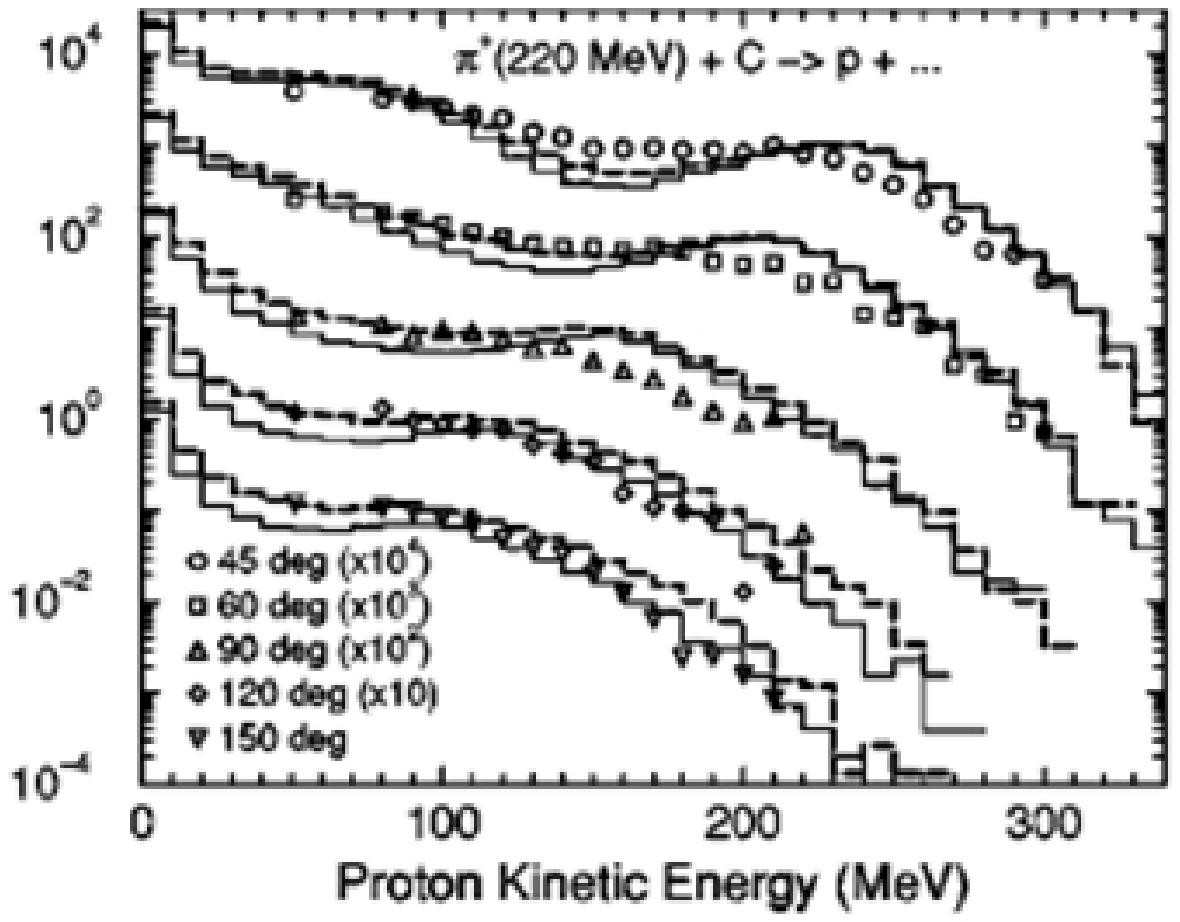}

\caption{Mashnik, et al. INC calculations compared with McKeown, et al. data.}

\label{fig:inuke:mashnik-pip-220-carbon}%
\end{minipage}\ \ \ \ \ \ \ \ %
\begin{minipage}[t]{0.48\columnwidth}%
\includegraphics[width=0.95\columnwidth]{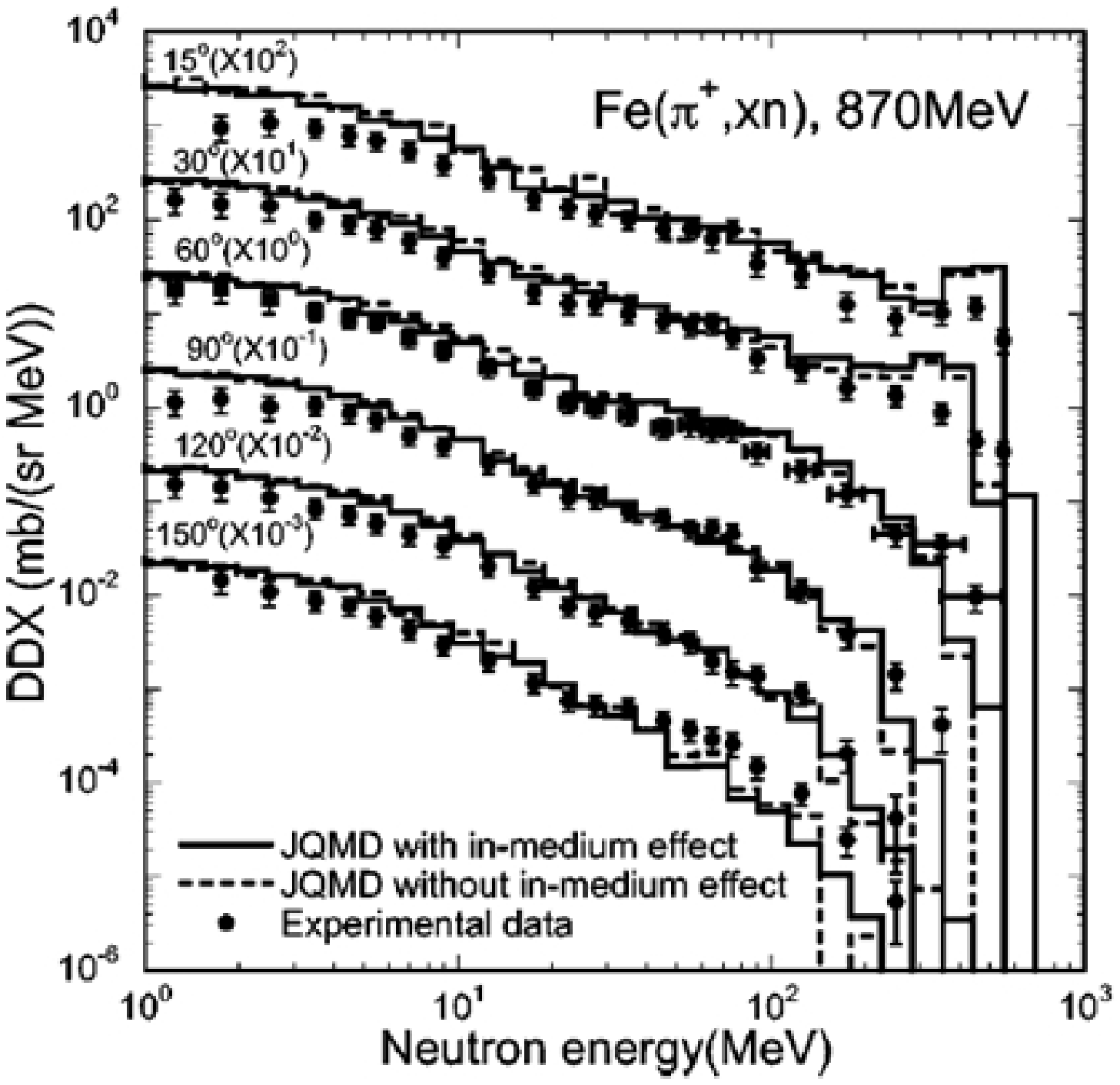}

\caption{Mashnik, et al. INC calculations compared with Iwamoto, et al. data. }

\label{fig:inuke:mashnik-iwamoto-pip-fe-870-n}%
\end{minipage}
\end{figure}

The successes of INC models are large. For many reactions, they are
the only models available for comparison. They were first used for
pion production in proton-nucleus interactions by Metropolis and Harp
\cite{Harp:1974zz}. A general INC model (CEM03) developed by Mashnik
and collaborators \cite{Mashnik:LAUR057321,Mashnik:2005ay,Mashnik:PrivCom}
has been applied with success to a wide range of pion- and proton-nucleus
data \cite{Mashnik:2000up}. Examples are shown in Figs. \ref{fig:inuke:mashnik-zumbro-pim-c-500},
\ref{fig:inuke:mashnik-pip-220-carbon} and \ref{fig:inuke:mashnik-iwamoto-pip-fe-870-n};
we will show similar comparisons for the GENIE FSI model. 

The FSI model in FLUKA is PEANUT. This uses a more sophisticated INC
model than CEM03. Various nuclear and quantum mechanical corrections
are applied. The result is impressive agreement with a wide variety
of data.

Treatment of pion absorption is somewhat different in the INC models
than the $\Delta$ dominance models. In the latter, pions first rescatter
off a nucleon (off-shell) and then absorbed on another. There are
other mechanisms which should be included. Salcedo, Oset, Vicente-Vacas,
and Garcia-Recio \cite{Salcedo:1987md} include both S-wave absorption
and 3-body absorption. In INC, the fundamental process for pion absorption
is $\pi^{+}d\rightarrow pp$ and this is often the only process included.
Since the density of nucleons is much smaller in deuterium as compared
with real nuclei, an empirical factor (with a value often about 3)
must be included.

\subsubsection{The INTRANUKE / hA FSI model}

The first FSI model is in the spirit of the other models in GENIE.
It is simple and empirical, data-driven. Rather than calculate a cascade
of hadronic interactions as is done in a complete INC model, we use
the total cross section for each possible nuclear process for pions
and nucleons as a function of energy up to 1.2 GeV. Thus, it is called
\textit{hA}. The emphasis is on iron because the first application
was to MINOS where production of high energy pions is important. At
low energies (50-300 MeV), there is sufficient data \cite{Ashery:1981tq,Navon:1983xj,Carroll:1976hj,Clough:1974qt,bauhoff}
for a good description. At high energies, only a few data points are
available. Here, we use results obtained for the CEM03 model. Although
the calculations are complete, they are not in good agreement with
the existing total cross section data. Therefore, the calculations
are normalized to the data at low energies. Elastic data at high energy
are used to extrapolate the model to 1.2 GeV.

The \textit{hA} model also handles proton and neutron rescattering.
The same reactions are possible except that neither can be absorbed.
Still, multinucleon knockout is highly probable. Although much less
data is available for nucleons than pions, CEM03 was tuned primarily
for them.

\begin{figure}
\begin{minipage}[t]{0.48\columnwidth}%
\includegraphics[width=0.95\columnwidth]{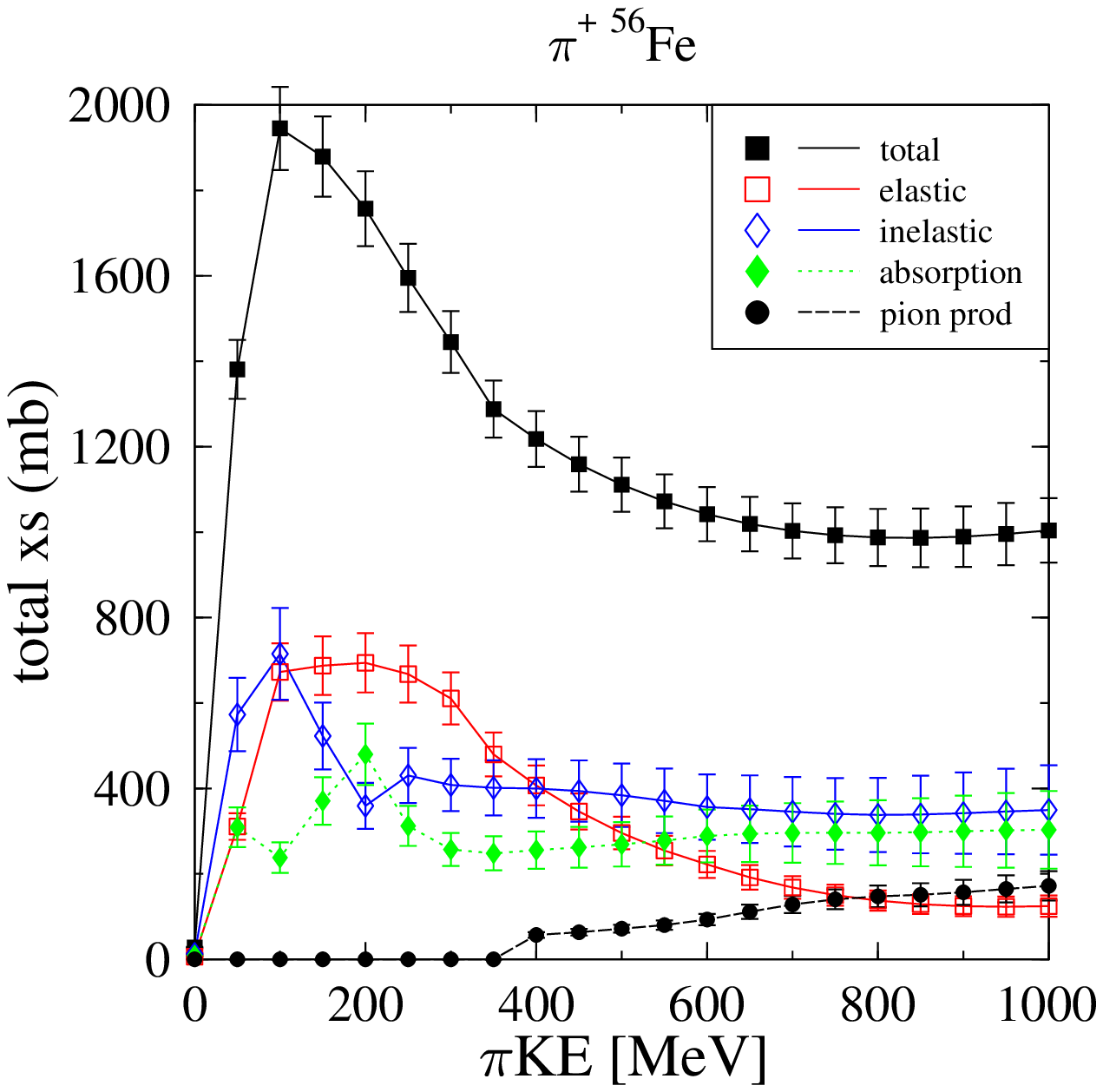}

\caption{$\pi^{+}Fe$ reactions used in GENIE hA model. Final states are chosen
according to these values.}

\label{fig:inuke:pife-xs}%
\end{minipage}\ \ \ \ \ \ \ \ %
\begin{minipage}[t]{0.48\columnwidth}%
\includegraphics[width=0.95\columnwidth]{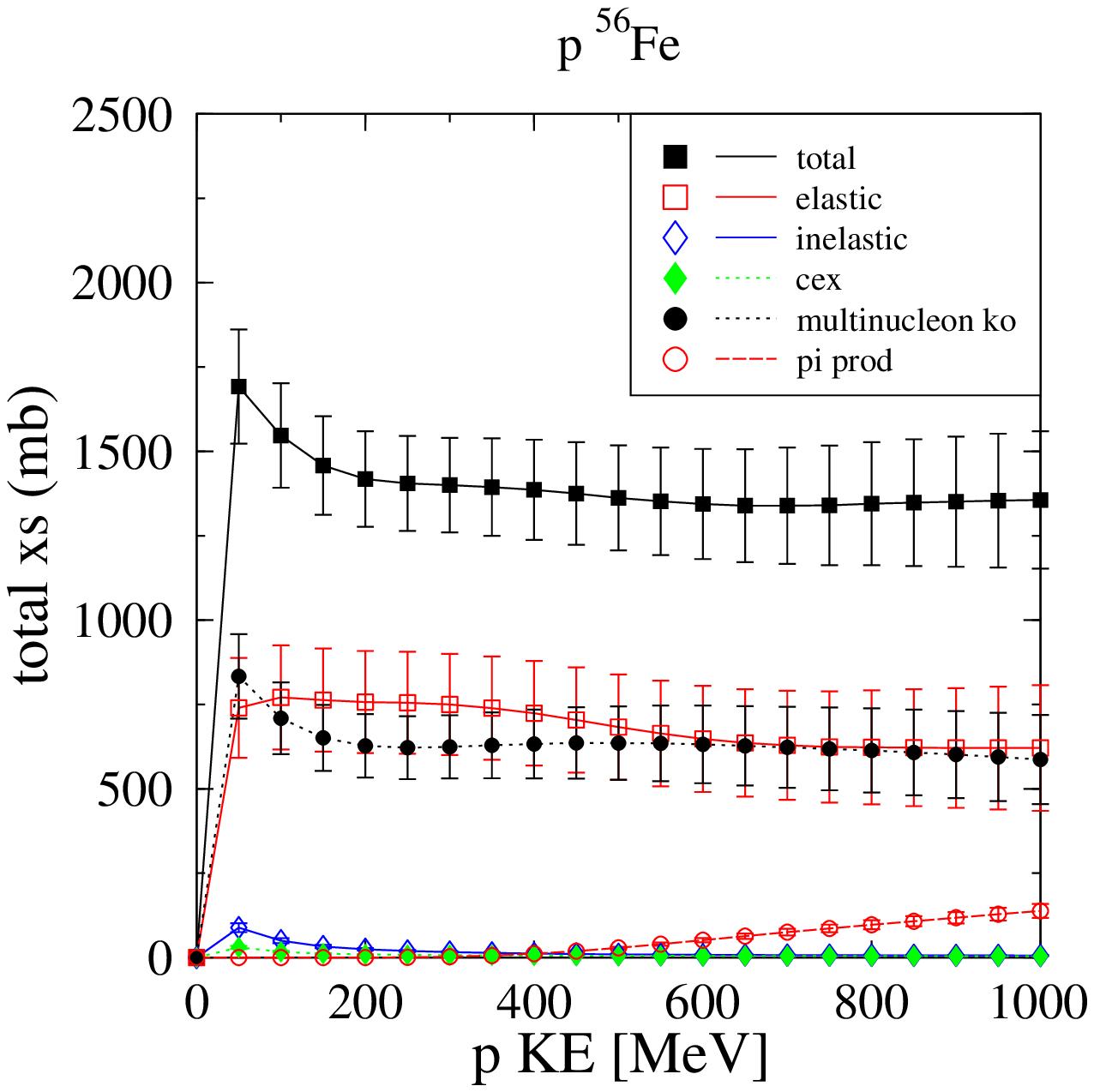}

\caption{Same for $pFe$ reactions.}

\label{fig:inuke:pfe-xs}%
\end{minipage}
\end{figure}

The values used for $\pi^{+}$ and $p$ are shown in Figs. \ref{fig:inuke:pife-xs}
and \ref{fig:inuke:pfe-xs}. Data values are used for energies below
315 MeV for all cross sections. Total cross section data is available
across the entire range. Data for total and total reaction cross sections
are used across the entire range in energy. Cross sections for targets
other than iron are obtained by scaling by $A^{2/3}$. As discussed
above, this is a reasonable approximation. Because such a large range
is covered, processes such as pion production must be included. Here,
we use the CEM03 calculations. 

\begin{figure}
\begin{minipage}[t]{0.48\columnwidth}%
\includegraphics[width=0.95\columnwidth]{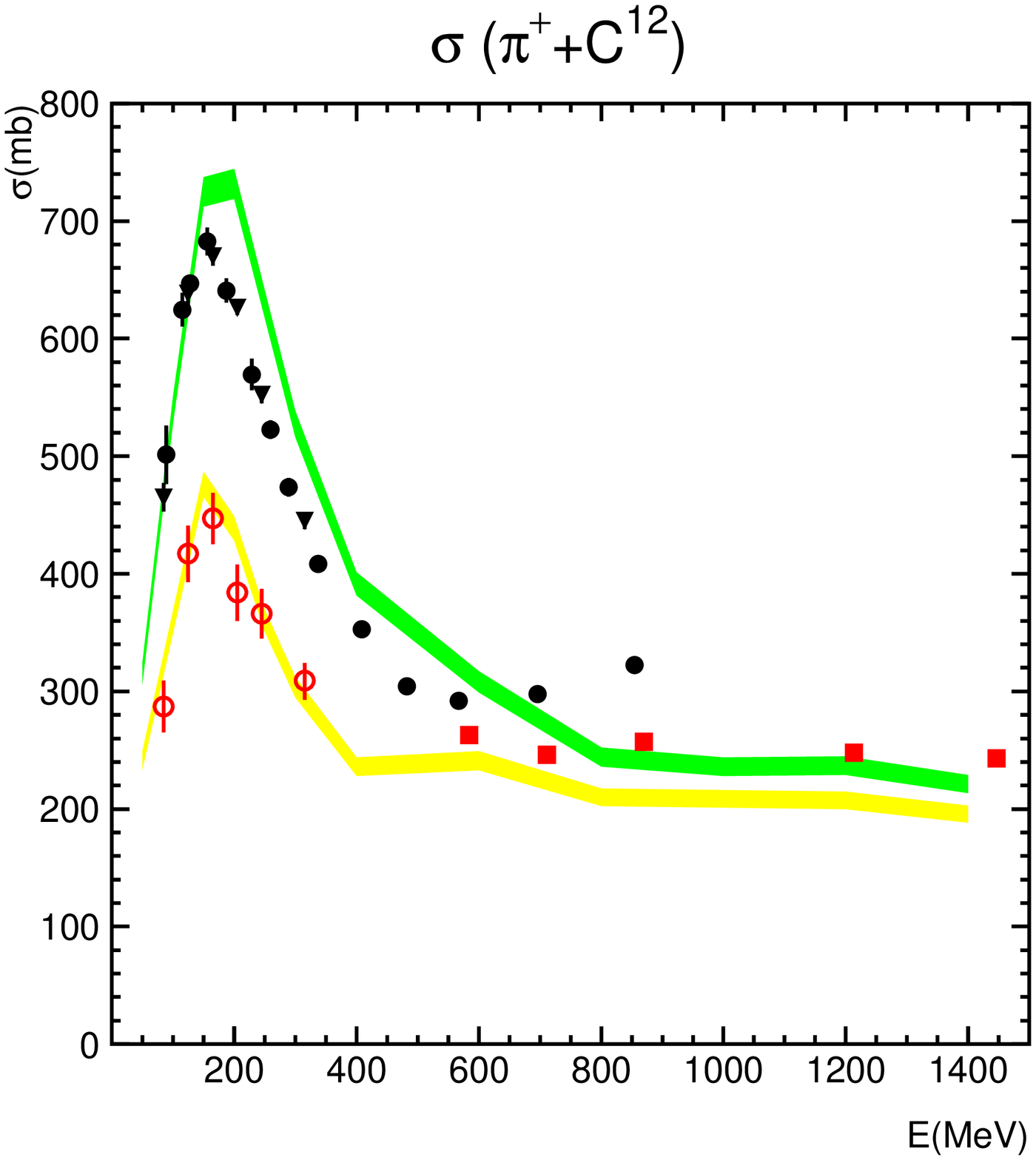}

\caption{$\pi^{+}C$ reactions. GENIE hA model is used. Total cross section
is determined with proper mean free path in a carbon nucleus.}

\label{fig:inuke:c121}%
\end{minipage}\ \ \ \ \ \ \ \ %
\begin{minipage}[t]{0.48\columnwidth}%
\includegraphics[width=0.95\columnwidth]{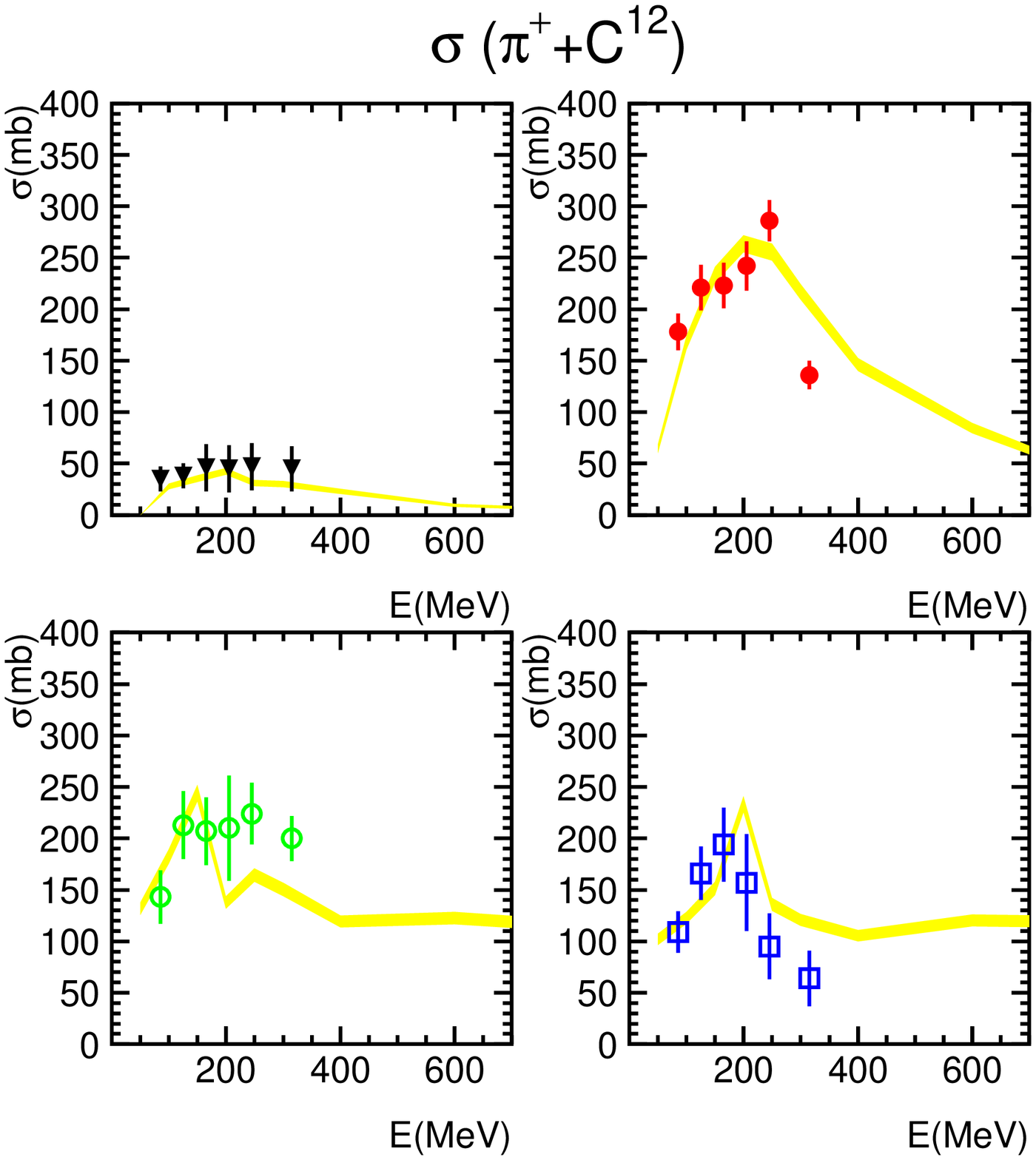}

\caption{Component cross sections come from the corresponding iron cross sections
scaled by $A^{2/3}$. }

\label{fig:inuke:c12breakdown}%
\end{minipage}
\end{figure}

The total cross section is calculated from the mean free path and
can be checked against data. In addition the accuracy of the $A^{2/3}$
scaling can be checked with data from another target. We show the
total and component cross sections for the model compared with carbon
data in Figs. \ref{fig:inuke:c121} and \ref{fig:inuke:c12breakdown}.
(Agreement for iron has less information and is equal in quality.)

All the data points in Figs. \ref{fig:inuke:c121} and \ref{fig:inuke:c12breakdown}
have error bars. These are either taken from the data or estimated.
These provide the range of values sampled during reweighting exercises.
This is an excellent way to estimate model dependent errors in a neutrino
oscillation experiment (see `Event Reweighting' chapter). The ability
to reweight is an important feature of this model.

This is the default FSI model in GENIE v2.4.0, the public version
as of now. It uses identical cross section for $\pi^{+}$ and $\pi^{-}$
and for $p$ and $n$. For isoscalar targets (e.g. $^{12}C$ and $^{16}O$,
this is no issue for the pions because of isospin symmetry. For targets
such as lead, this is a 10\% effect. The charges of particles in the
final state tend to reflect the charge of the probe. For example,
final states for $\pi^{+}$ have more protons than neutrons while
the opposite holds for $\pi^{-}$. Cross sections for $\pi^{0}$ beams
can't be measured. This code uses isospin symmetry to calculate them
from the charged pions. The total reaction cross sections for $p$
and $n$ are very similar, plots are shown in the next section. Charges
of final state particles tend to be more positive for incident protons.

Pion absorption and nucleon spallation reactions can knock out large
numbers of nucleons. This is seen strongly in data. More detailed
calculations (see below) show an average of 10 nucleons ejected from
iron in pion absorption. To simplify the code, the \textit{hA} model
limits this to 5. For MINOS, this is never an issue.

Angular and energy distributions of particles are estimated. For elastic
scattering, template angular distributions from relevant data are
used. These distributions are very forward peaked, so it's not an
important simplification. For final states with more than 1 hadron,
particles are distributed by phase space. This gives the correct limits,
but the energy distribution changes somewhat when the $\Delta$ resonance
dominates. The effect of these approximations have not yet been simulated,
but they are unlikely to be an important effect in the MINOS experiment.
One of the most significant errors is in the treatment of the quasielastic
scattering. Only the incident particle is put in the final state and
it's energy and angle distribution are both flat.

Since the elastic cross section can't be generated in an INC model,
it has to be added on. For the \textit{hA} model, we chose an empirical
method. The size of the nucleus is increased by $\Delta$R which is
proportional to the de Broglie wavelength. This nicely matches the
data for all energies.

Almost all of the problems in the last paragraphs will be fixed in
GENIE v2.6.0. Changes due to isospin in either hadron or nucleus will
be greatly improved. The number of final states sampled will be increased.
Inelastic final states will be assumed to be dominated by quasielastic
events. (This approximation can be checked against data and will be
discussed in the next section.)

\subsubsection{The INTRANUKE / hN FSI model}

The second FSI model in GENIE (\textit{hN}) is a full INC model. It
includes interactions of pions, nucleons, kaons, and photons in all
nuclei. The basis is the angular distributions as a function of energy
for about 14 reactions from threshold to 1.2 GeV. All this information
comes from the GWU group \cite{Arndt:2006bf,GWU:Web}. A preliminary
version of the hN model is scheduled to be in GENIE v2.6.0, but the
\textit{hA} model will still be the default.

As a full INC model, all reactions on all nuclei can be calculated.
None of the restrictions that apply to \textit{hA} model are relevant.
Although the choice of interaction points through the MFP is identical
in the 2 models, the cascade is fully modeled in the \textit{hN} model.
For example, there is a small but finite probability of knocking out
every nucleon in an event.

One new feature of this code is the inclusion of nucleon pre-equilibrium
and compund nuclear processes. The present model is simple, but effective.
This is important to give an improved description of the vertex energy
deposition.

The code was designed to minimize the number of parameters. One parameter
scales the absorption MFP and is fit to the pion total absorption
cross section. Separate values for the $\Delta R$ values for pions
and nucleons are fit to the total reaction cross sections. All particles
get a free step when they are produced; this simulates the effect
of $\Delta$ resonance propagation in a simple way. It is used to
adjust the normalization of certain inclusive scattering distributions.
A shift in the energy of nucleons in the nucleus is used to put the
quasielastic peak (see Fig. 2L) (similar to what is used in electron
scattering).

The validation of this new code comes in 2 parts- the total cross
sections for various processes (e.g. Fig. \ref{fig:inuke:c121}) which
test the overall propagation of particles and the inclusive cross
sections (e.g. Figs. \ref{fig:inuke:mashnik-zumbro-pim-c-500}, \ref{fig:inuke:mashnik-pip-220-carbon}
and \ref{fig:inuke:mashnik-iwamoto-pip-fe-870-n}). Each is important.
Previous validations emphasize the total cross sections because this
sets the overall flow of particles into each topology. Previous neutrino
experiments emphasize topology. Future experiments are expected to
put emphasis into the distribution of particles in energy and angle
as beam and detector technology improve.

\begin{figure}
\begin{minipage}[t]{0.48\columnwidth}%
\includegraphics[width=0.95\columnwidth]{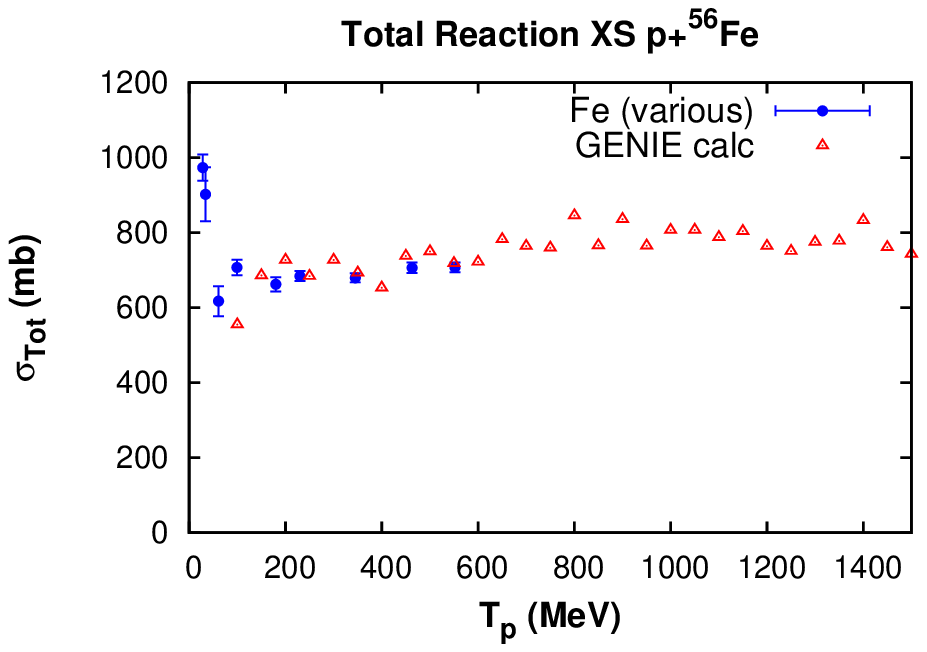}

\caption{$pFe$ reactions from the GENIE \textit{hN} model. }

\label{fig:inuke:sigreac-p-fe}%
\end{minipage}\ \ \ \ \ \ \ \ %
\begin{minipage}[t]{0.48\columnwidth}%
\includegraphics[width=0.95\columnwidth]{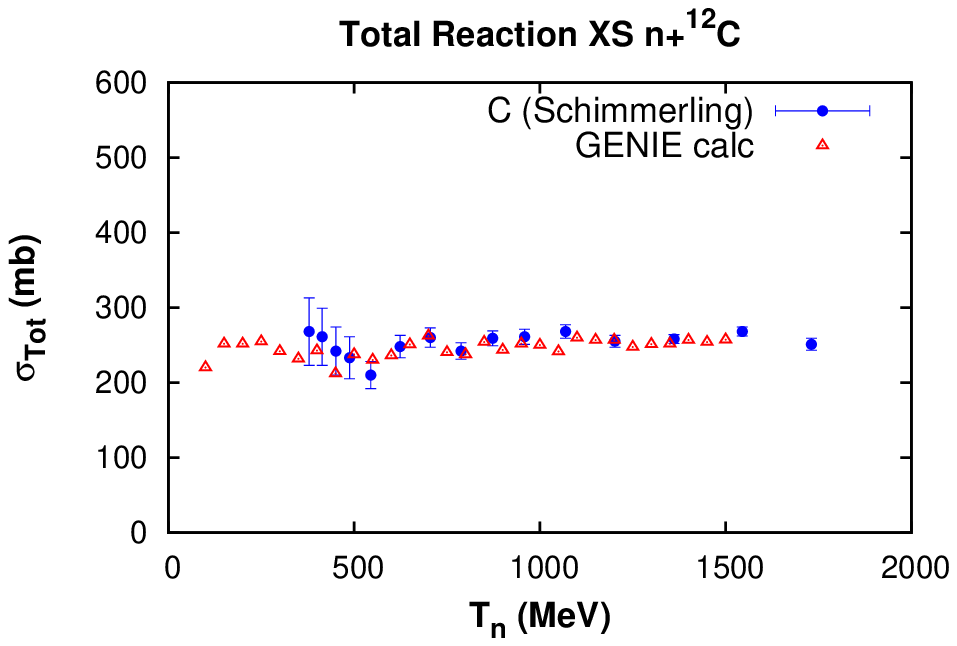}

\caption{Total reaction cross sections for $nC$ reactions from the GENIE\textit{
hN} model. }

\label{fig:inuke:sigreac-n-c}%
\end{minipage}
\end{figure}

The component total cross section data is limited to hadron energy
of less than $\sim$350 MeV. The exception is the total reaction cross
section which has been measured for $\pi^{+}$, $\pi^{-}$, $p$,
and $n$ up to roughly 1 GeV. Figs. \ref{fig:inuke:sigreac-p-fe}
and \ref{fig:inuke:sigreac-n-c} show $\sigma_{reac}$ for protons
in iron and neutrons in carbon respectivelly. The energy dependence
is flat and we see the cross section approximately equal to the nuclear
area as discussed in the introduction. Agreement of the model is excellent.

\begin{figure}
\begin{minipage}[t]{0.48\columnwidth}%
\includegraphics[width=0.95\columnwidth]{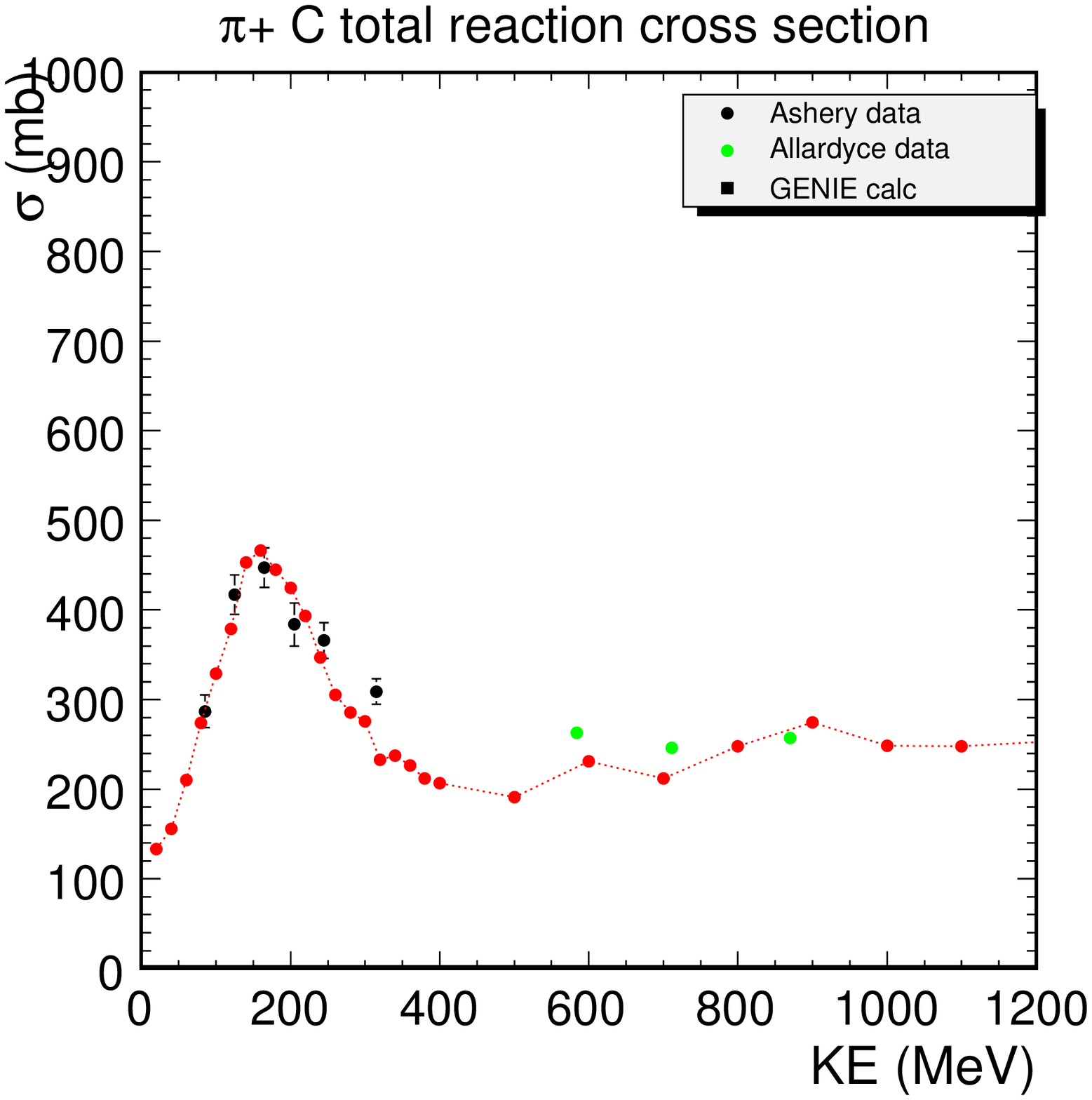}

\caption{$\pi^{+}C$ using the GENIE\textit{ hN} model compared to data.}

\label{fig:inuke:pip-c-reactot-19}%
\end{minipage}\ \ \ \ \ \ \ \ %
\begin{minipage}[t]{0.48\columnwidth}%
\includegraphics[width=0.95\columnwidth]{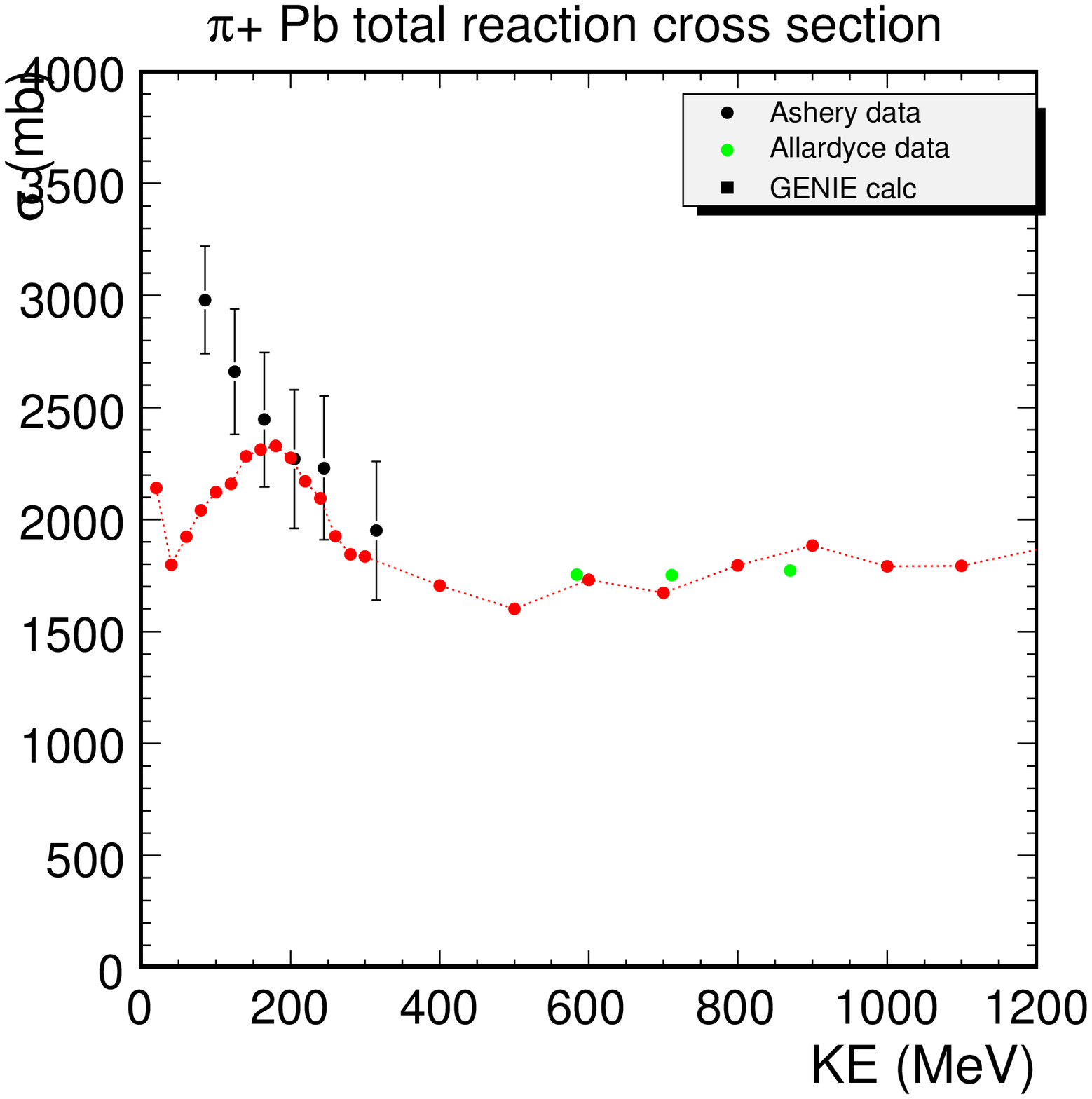}

\caption{Total reaction cross sections for $\pi^{+}Pb$ using the GENIE \textit{hN}
model. }

\label{fig:inuke:pip-pb-reactot-19}%
\end{minipage}
\end{figure}

In Figs. \ref{fig:inuke:pip-c-reactot-19} and \ref{fig:inuke:pip-pb-reactot-19},
we show $\sigma_{reac}$ for pions. The agreement is excellent except
at low energies for heavier targets; this is still under study.

\begin{figure}
\begin{minipage}[t]{0.48\columnwidth}%
\includegraphics[width=0.95\columnwidth]{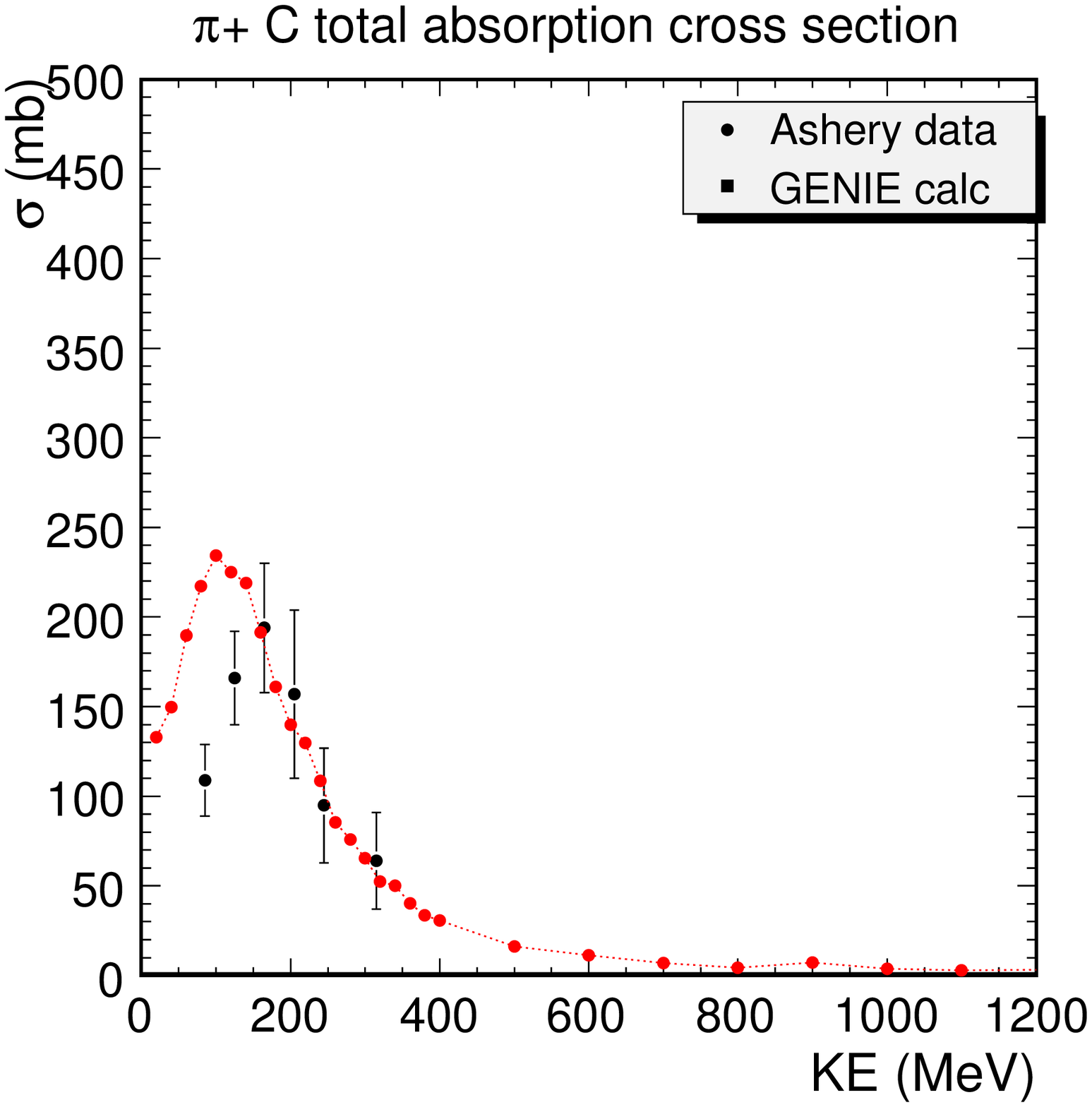}

\caption{$\pi^{+}C$ using the GENIE\textit{ hN} model compared to data. }

\label{fig:inuke:pip-c-abstot-19}%
\end{minipage}\ \ \ \ \ \ \ \ %
\begin{minipage}[t]{0.48\columnwidth}%
\includegraphics[width=0.95\columnwidth]{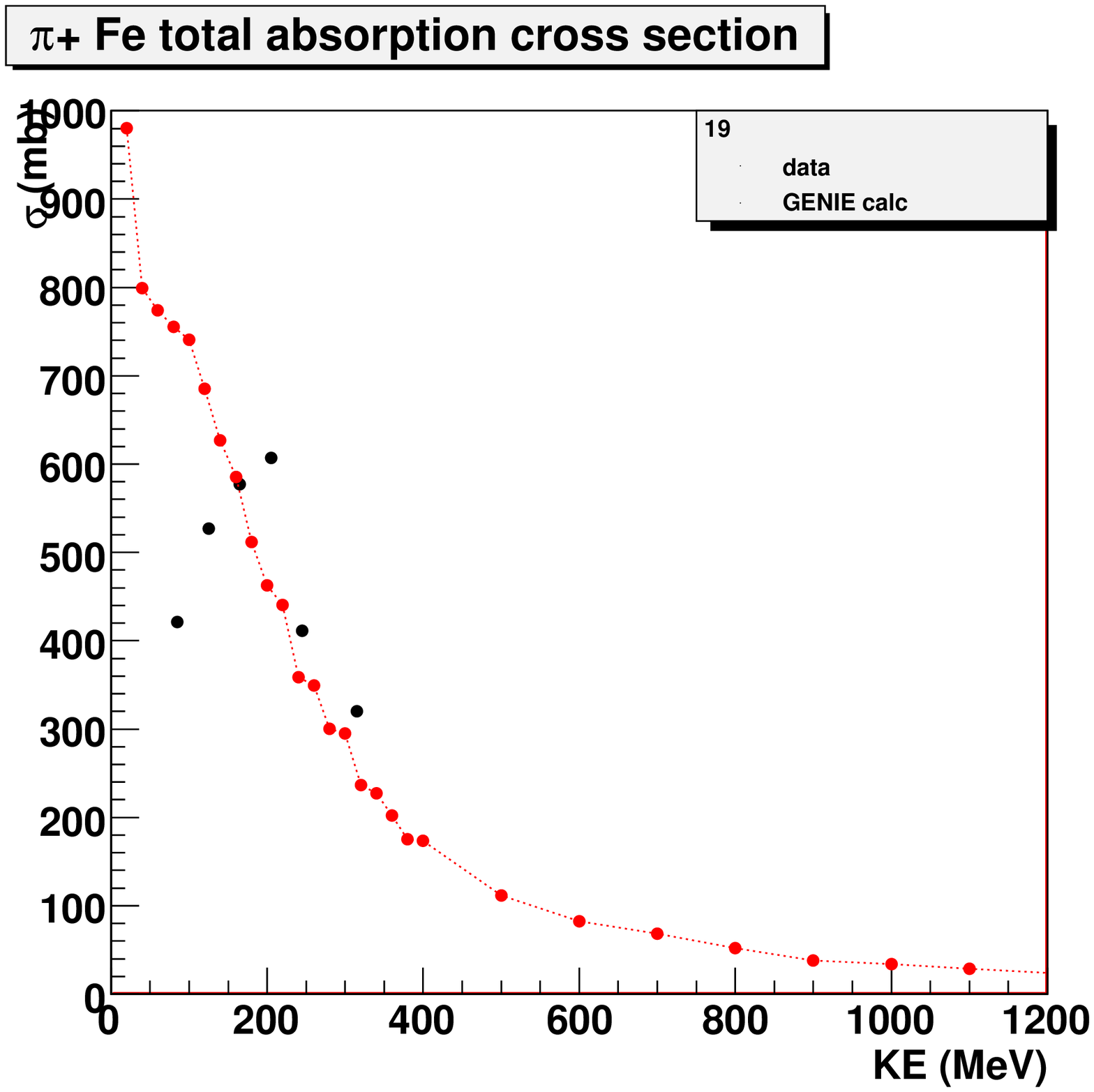}

\caption{Total reaction cross sections for $\pi^{+}Pb$ using the GENIE\textit{
hN} model.}

\label{fig:inuke:pip-fe-abstot-19}%
\end{minipage}
\end{figure}

With the significant interest in absorption, we show 2 examples of
that total cross section in Figs. \ref{fig:inuke:pip-c-abstot-19}
and \ref{fig:inuke:pip-fe-abstot-19}. Overall agreement is very good,
but the problem in $\sigma_{reac}$ at low energies for heavy targets
is shown to be in the absorption channel.

\begin{figure}
\begin{minipage}[t]{0.48\columnwidth}%
\includegraphics[width=0.95\columnwidth]{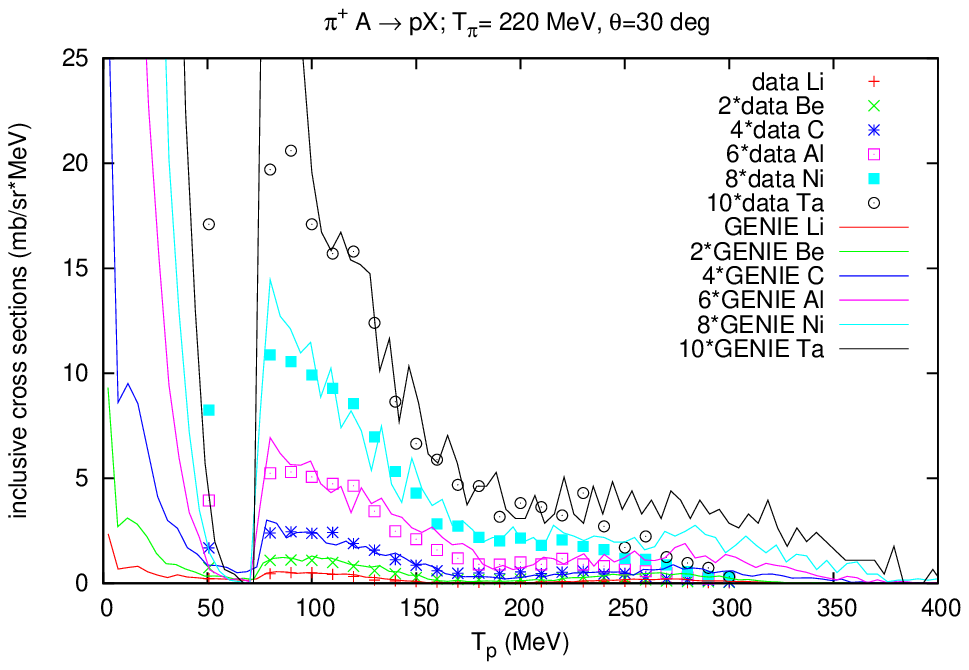}

\caption{$\pi^{+}$ interacting in various nuclei. In each distribution, protons
are detected at 30$^{\circ}$. Data is from McKeown, et al. These
protons come from both absorption and scattering processes.}

\label{fig:inuke:mck-pip-A-220-30-comp}%
\end{minipage}\ \ \ \ \ \ \ \ %
\begin{minipage}[t]{0.48\columnwidth}%
\includegraphics[width=0.95\columnwidth]{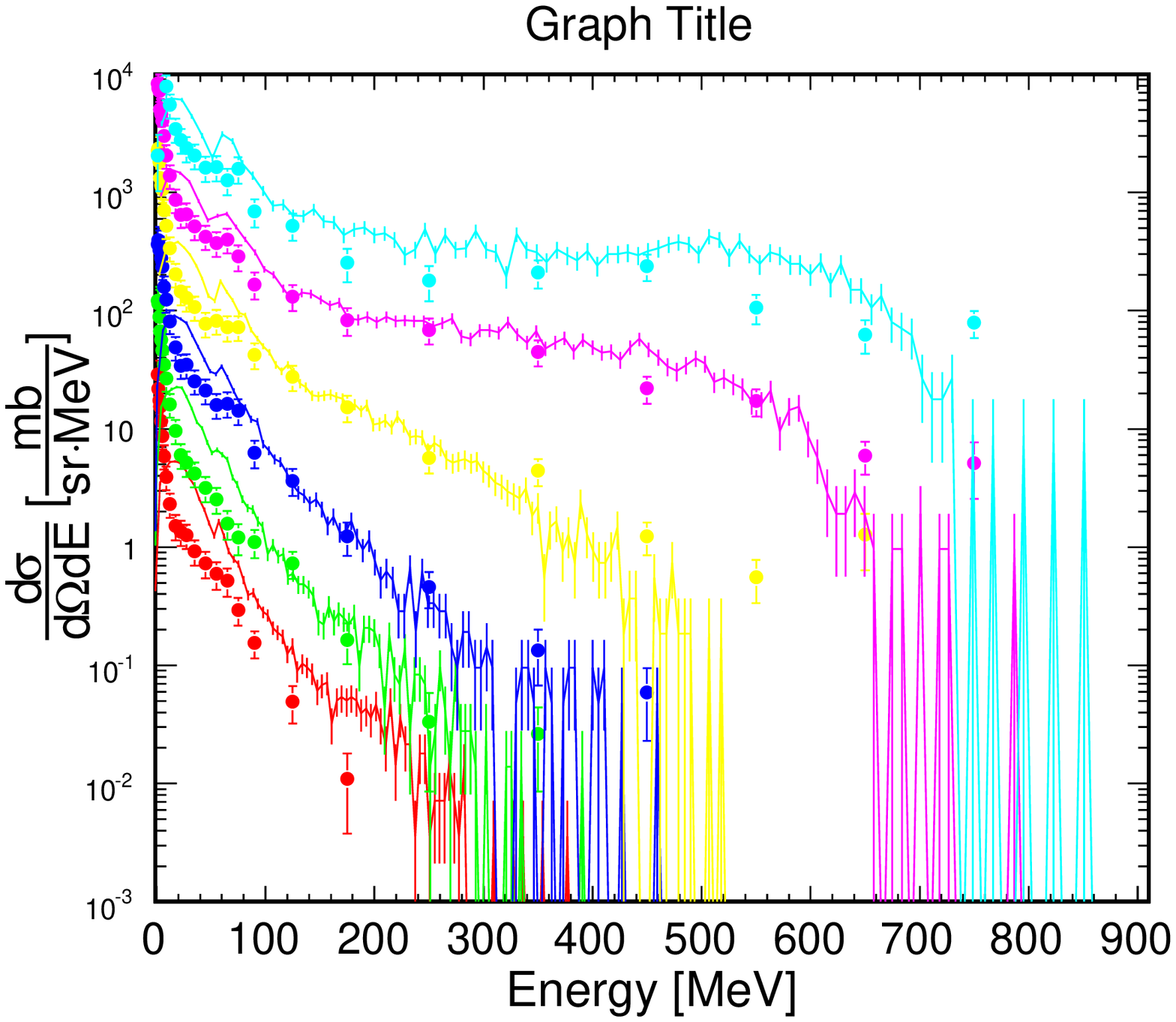}

\caption{Inclusive cross sections for neutrons emitted from 870 MeV $\pi^{+}$
interacting in iron. In each case, neutrons are detected at different
angles. Data is from Iwamoto, et al. These neutrons come from predominantly
the absorption process. }

\label{fig:inuke:iwa-pip-870-n-55}%
\end{minipage}
\end{figure}

Continuing with absorption, we show 2 examples similar to Figs. \ref{fig:inuke:mashnik-pip-220-carbon}
and \ref{fig:inuke:mashnik-iwamoto-pip-fe-870-n} in Figs. \ref{fig:inuke:mck-pip-A-220-30-comp}
and \ref{fig:inuke:iwa-pip-870-n-55}. The agreement shown here is
excellent as the details of pion reactions are explored across a wide
kinematic range.

\begin{figure}
\begin{minipage}[t]{0.48\columnwidth}%
\includegraphics[width=0.95\columnwidth]{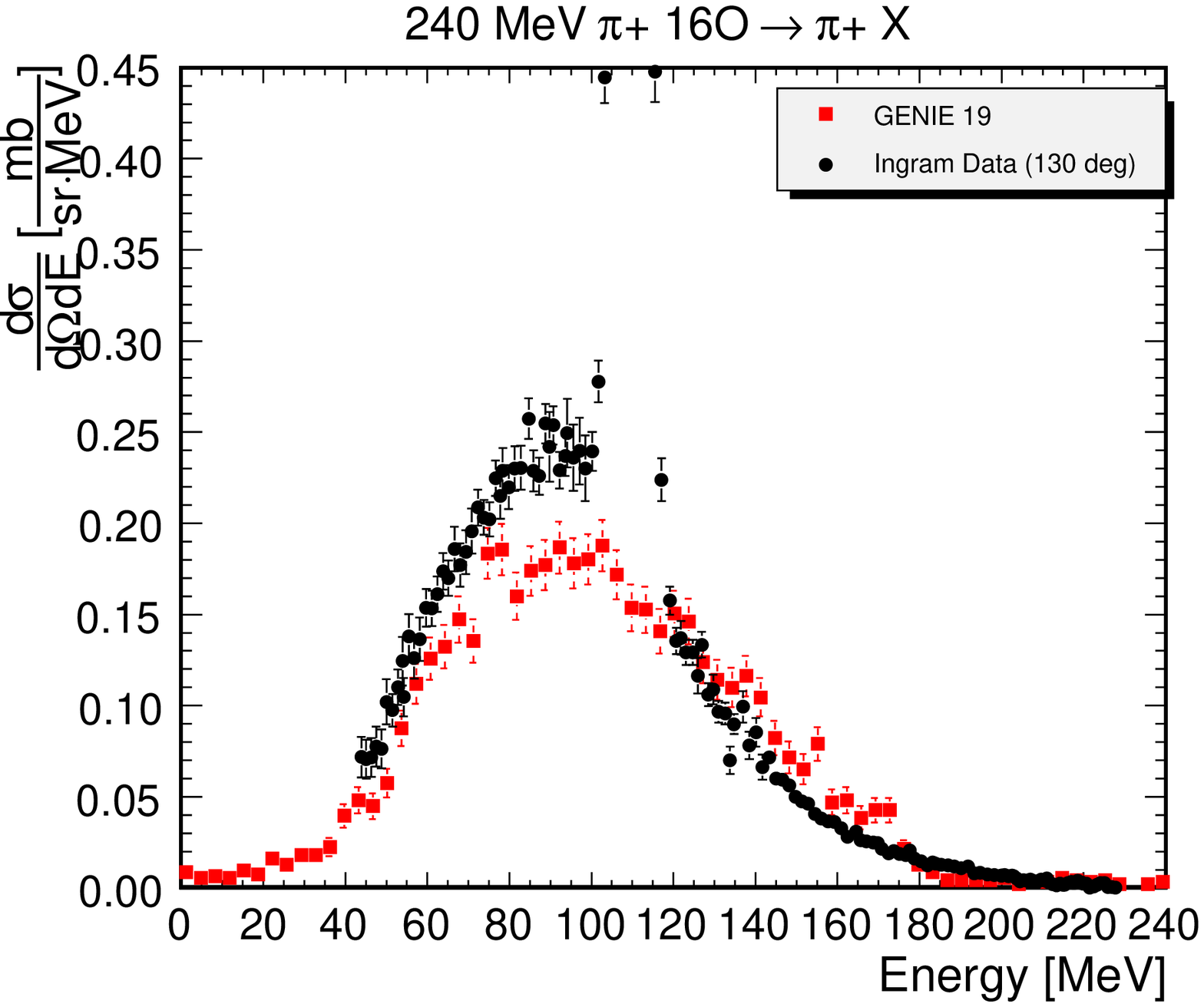}

\caption{$\pi^{+}$ scattered at 130$^{\circ}$ from 240 MeV $\pi^{+}$ interacting
with oxygen. At this back angle, the spectrum of $\pi^{+}$ is dominated
by the quasifree mechanism. Data is from Ingram, et al.}

\label{fig:inuke:ing-pip-o-240-130-55}%
\end{minipage}\ \ \ \ \ \ \ \ %
\begin{minipage}[t]{0.48\columnwidth}%
\includegraphics[width=0.95\columnwidth]{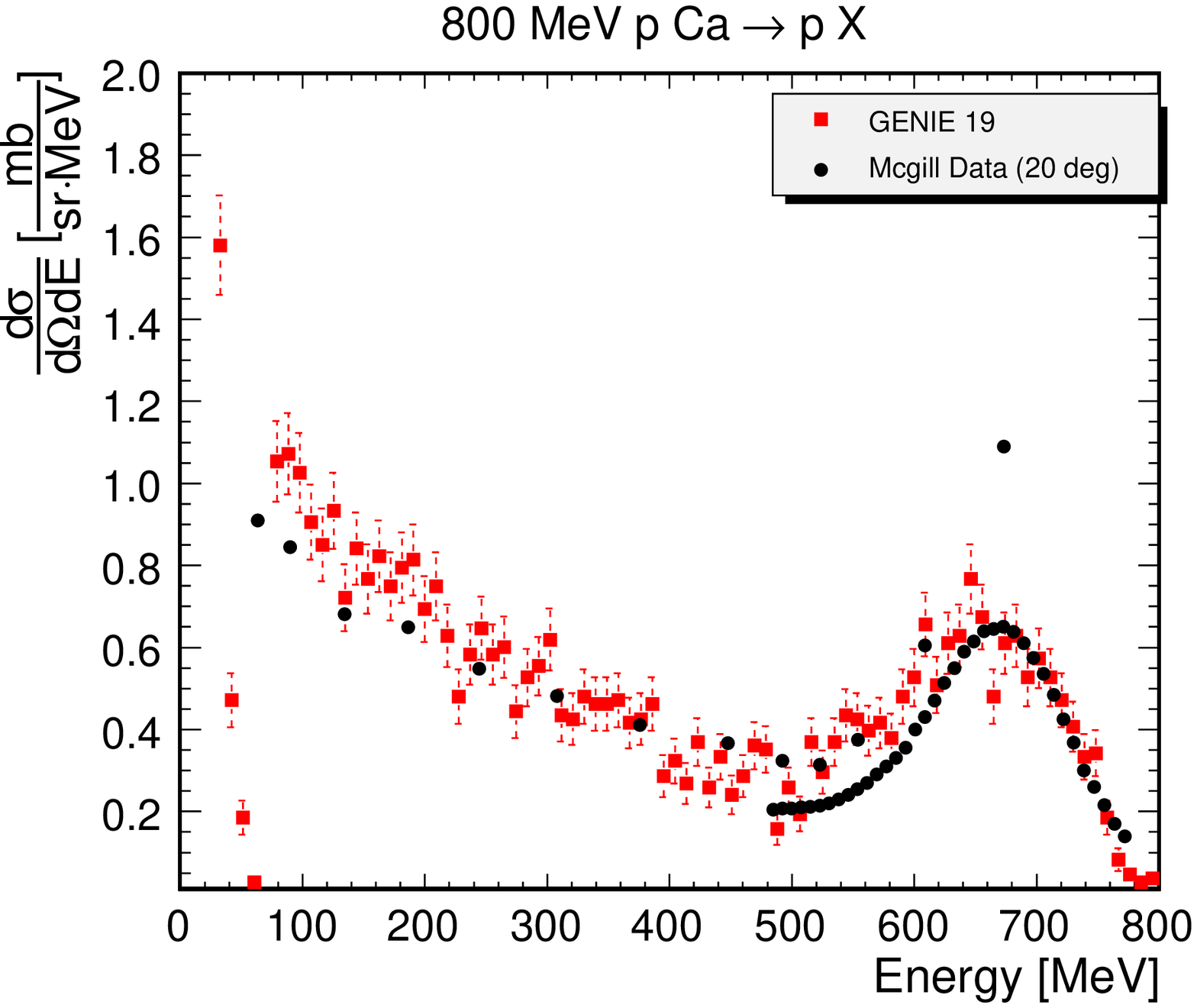}

\caption{Inclusive cross sections for protons scattered at 20$^{\circ}$ from
800 MeV protons interacting with calcium. Data is from McGill, et
al and Chrien, et al. There is a known absolute normalization difference
between the 2 experiments but it is not available to us. }

\label{fig:inuke:mcgill-pp-ca-20-new}%
\end{minipage}
\end{figure}

The last example of this new code is for scattering processes. When
hadrons interact in the nuclear medium, the quasifree scattering process
is important; that has been seen in numerous data sets. In Figs. \ref{fig:inuke:ing-pip-o-240-130-55}
and \ref{fig:inuke:mcgill-pp-ca-20-new} we show examples for pion
and proton scattering. For the pion case, a back angle is shown; here,
the quasielastic mechanism dominates. For protons, the beam energy
is large enough that the multiple scattering process is sampled over
a wide range in energy. The agreement is excellent.

\subsubsection{Conclusions}

We have reviewed strong interactions as they will be applied to neutrino
experiments of the near future. The basic premise is that hadron-nucleus
experiments are the best way (definitely now, likely also in the future)
to validate FSI models. General properties have been identified from
data, the general blackness of nuclei to hadrons along with the importance
of quasifree mechanisms.

Various models were discussed with a focus on INC models. Although,
they are not the most theoretically viable, the role of INC models
is significant because they can 'easily' describe a wide range of
data.

The 2 FSI models in GENIE are described in some detail. The \textit{hA}
model is simpler and more empirical. Although it isn't the most accurate,
it is very fast and straightforward to reweight. The \textit{hN} model
is a full INC calculation which is much more accurate. In v2.4, the
\textit{hA} model is the only FSI model. For version 2.6, both will
be included but \textit{hA} will still be the default. The \textit{hA}
model will be applicable to all nuclei from helium to lead for kinetic
energies up to 1.2 GeV for pions and nucleons. Its main value will
be for high energy neutrinos and in reweighting. The \textit{hN} model
is nearly complete for this round. It will be valid for energies above
50 MeV and will provide a very complete description of many final
states.

Thus, each GENIE FSI model has independent validity. An important
component of any simulation is the estimation of systematic errors.
A comparison of the results using each model can show model dependence.
Varying parameters inside the \textit{hA} model is the best way to
assess systematic errors due to FSI.

\chapter{The GENIE Software Framework}

\label{ch:fmwk}

{[}to be added in future revision{]}

\chapter{Downloading \& Installing GENIE}

\section{Understanding the versioning scheme}

\subsubsection*{SVN Tags}

In the GENIE version numbering scheme, releases are tagged in the
SVN source-code repository as \textbf{\textit{R-major\_minor\_revision}}\footnote{For example, tag R-1\_99\_1 corresponds to GENIE vrs 1.99.1, tag R-2\_0\_2
corresponds to GENIE vrs 2.0.2 etc.}. When a number of significant functionality improvements or additions
have been made, the major index is incremented. The minor index is
incremented in case of significant fixes and/or minor feature additions.
The revision number is incremented for minor bug fixes and updates.

\subsubsection*{Version number semantics}
\begin{itemize}
\item Versions with even minor number (eg 2.0.{*}, 2.4.{*}) correspond to
stable, fully validated physics production releases\footnote{To the dismay of mathematicians, our versioning scheme uses 0 as am
even number.}.
\item Versions with odd minor number (eg. 2.3.{*}, 2.5.{*}) correspond to
release candidates tagged during the validation stage preceding the
release of a production version.
\item Production versions, and candidate releases, always have an even revision
number. 
\item The SVN head has a nominal version number of 999.999.999.
\end{itemize}

\subsubsection*{Release codenames}

The major production-quality releases are code-named after modern
extinct or endangered species (series of production releases: \textbf{\textit{Auk}},
\textbf{\textit{Blueback}}, \textbf{\textit{Cheetah}}, \textbf{\textit{Dodo}},
\textbf{\textit{Elk}}, \textbf{\textit{Fox}}, \textbf{\textit{Gazelle}},
\textbf{\textit{Hippo}}, \textbf{\textit{Ibex}},... ).

\subsubsection*{Release qualifiers}

The GENIE releases are marked as:
\begin{itemize}
\item \textbf{\textit{pro}} : Validated production-quality versions recommended
for physics studies.
\item \textbf{\textit{old}}: Older `pro' versions that have been greatly
superseded by newer versions. Versions marked as `old' become unsupported.
We appreciate that experiments get highly attached on specific versions
due to the enormous amount of work invested in generating high statistics
samples and calculating MC-dependent corrections and systematics.
We strive to support `pro' versions for a minimum of two years.
\item \textbf{\textit{rc}}: Release candidates. You may not use for physics
studies.
\item \textbf{\textit{special}}: Special releases prepared for a particular
study or event such as a) the evaluation of an experiment systematic
with an appropriately modified version of GENIE, or b) a GENIE tutorial
or a summer / winter school. You may not use these releases outside
the intended context. 
\end{itemize}

\section{Obtaining the source code}

The official GENIE source code is maintained at a SubVersion repository
hosted at HepForge\footnote{http://www.hepforge.org}. The development
version and a host of frozen physics releases are available from the
repository. Alternatively, you can download compressed archives stored
at the HepForge archive area, or you can create and download such
archives using the web interface to the GENIE SubVersion repository.
Details are given below. Further general information can be found
at
\begin{itemize}
\item The HepForge documentation page: \textit{http://www.hepforge.org/docs/}
\item The SVN book: \textit{http://svnbook.red-bean.com/}
\end{itemize}
The code repository can be accessed anonymously via HTTP, without
a HepForge account. You need to have a SubVersion client installed
and you probably already do. If not, binaries are readily available
for most platforms (see \textit{http://subversion.apache.org/}).\\
\\
\\
You can check-out the generator SVN trunk by typing: \\
\\
\texttt{\textbf{\small{}\$ svn co http://svn.hepforge.org/genie/generator/trunk}}\texttt{\textbf{\textit{\small{}
<local\_dir>}}}\\
\\
You can check-out frozen releases by typing: \\
\\
\texttt{\textbf{\small{}\$ svn co http://svn.hepforge.org/genie/generator/branches/}}\texttt{\textbf{\textit{\small{}<genie\_tag>
<local\_dir>}}}\\
\\
Make the appropriate substitutions for \texttt{\textbf{\textit{\small{}<genie\_tag>}}}
and \texttt{\textbf{\textit{\small{}<local\_dir>}}}. To view the available\texttt{\textbf{\textit{\small{}
<genie\_tag>}}}s see the GENIE release table on the web, or just type:
\\
\\
\texttt{\textbf{\small{}\$ svn list http://svn.hepforge.org/genie/generator/branches/
}}~\\
{\small \par}

Alternatively, compressed archives for recent stable version releases
are posted at the HepForge archive area:\\
\textit{ }\url{http://www.hepforge.org/downloads/genie}\\
\\
You can also download a compressed archive of the latest development
version (created automatically upon your request) using the repository's
web interface. Visit the repository trunk at:\\
\textit{\url{http://projects.hepforge.org/genie/trac/browser/trunk}}\\
Then click on `Download in other formats: Zip Archive' towards the
end of the page. \\

Write access to the Generator repository, as well as to other GENIE
products including the Comparisons and the Tuning require a HepForge
account and it is permitted only for GENIE collaborators. Special
limited accounts may be setup for regular GEBINE contributors.\\

\section{3rd Party Sofwtare}

A typical GENIE installation\footnote{A minimal installation that can be used for event generation / physics
studies.} requires the following external packages\footnote{The implicit assumption here is that you start with a `working system'
where some basic tools, such as the gcc compiler suite, make, autoconf,
PERL, CVS and SVN clients etc, are already installed. Instructions
are given assuming that you are using the bash shell but it is trivial
to adapt these instructions for your own shell.}:
\begin{itemize}
\item \textbf{\textit{ROOT }}
\item \textbf{\textit{GSL}}\\
The GNU Scientific Library
\item \textbf{\textit{LHAPDF}} \\
The Les Houches Accord PDF interface, a PDFLIB successor.
\item \textbf{\textit{PYTHIA6}}\\
The well known LUND Monte Carlo package used by GENIE for particle
decays and string fragmentation (for neutrino interactions of high
invariant mass).
\item \textbf{\textit{log4cpp}} \\
A C++ library for message logging.
\item \textbf{\textit{libxml2}}\\
The C XML library for the GNOME project.
\end{itemize}
The installation of external packages is described in detail in their
corresponding web pages. Additional detailed instructions can also
be found at Appendix \ref{app:ThirdPartySoftw}  of this manual.

\section{Preparing your environment}

A number of environmental variables need to bee set or updated before
using GENIE.
\begin{itemize}
\item Set the \textbf{`}\texttt{GENIE}\textbf{'} environmental variable
to point at the top level GENIE directory
\item Set the \textbf{`}\texttt{ROOTSYS}\textbf{'} environmental variable
to point at the top level ROOT directory
\item Set the \textbf{`}\texttt{LHAPATH}\textbf{'} environmental variable
to point to LHAPDF's PDF data files
\item Append `\$\texttt{ROOTSYS}/\textit{bin}' and `\$\texttt{GENIE}/\textit{bin}'
to your \textbf{`}\texttt{PATH}\textbf{'}
\item Append `\$\texttt{ROOTSYS}/\textit{lib}', `\$\texttt{GENIE}/\textit{lib}'
and the paths to the log4cpp, libxml2, LHPADF and PYTHIA6 libraries
to your \textbf{`}\texttt{LD\_LIBRARY\_PATH}\textbf{'} environmental
variable (or to your \textbf{`}\texttt{DYLD\_LIBRARY\_PATH}\textbf{'}
environmental variable if you are using GENIE on MAC OS X).
\end{itemize}
It is more convenient to create a GENIE setup script and execute it
before using GENIE.\\
A setup script should look like the following:\\

\begin{verbatim}
#!/bin/bash

export GENIE=/path/to/genie/top/directory

export ROOTSYS=/path/to/root/top/directory
export LHAPATH=/path/to/lhapdf/PDFSets/

export PATH=$PATH:\
$ROOTSYS/bin:\
$GENIE/bin

export LD_LIBRARY_PATH=$LD_LIBRARY_PATH:\
/path/to/log4cpp/library:\
/path/to/libxml2/library:\
/path/to/lhapdf/libraries:\
/path/to/pythia6/library:\
$ROOTSYS/lib:\
$GENIE/lib

\end{verbatim}

Assuming that the above script is named `\textit{genie\_setup}', you
can execute it by typing:\\
\\
\texttt{\textbf{\small{}\$ source genie\_setup}}~\\
{\small \par}

\section{Configuring GENIE}

A configuration script is provided with the GENIE source code to help
you configure your GENIE installation (enable / disable features and
specify paths to external packages). To see what configuration options
are available, type:\\
\\
\texttt{\textbf{\small{}\$ cd \$GENIE}}\\
\texttt{\textbf{\small{}\$ ./configure -{}-help}}\\
\\
This will generate a screen output that looks like the following: 

\begin{verbatim}

FLAG              DESCRIPTION      
                   
--prefix          Install path (for make install)      

enable/disable options 
prefix with either --enable or --disable 
(eg. --enable-lhapdf --disable-flux-drivers)

profiler           GENIE code profiling using Google perftools.
doxygen-doc        Generate doxygen documentation at build time.
dylibversion       Adds version number in dynamic library names.
lowlevel-mesg      Disable (rather than filter out) some prolific
                   debug level messages known to slow GENIE down.
debug              Adds -g compiler option to request debug info.
lhapdf             Use the LHAPDF library.
cernlib            Use the CERN libraries.
flux-drivers       Enable built-in flux drivers.
geom-drivers       Enable built-in detector geometry drivers.
mueloss            Muon energy loss modeling. 
validation-tools   GENIE physics model validation tools.
test               Build test programs.
t2k                Enable T2K-specific event generation application.
fnal               Enable FNAL experiment-specific event generation application.
atmo               Enable atmospheric neutrino event generation application.
nucleon-decay      Enable nucleon decay event generation application.
rwght              Enable event reweighting tools.
masterclass        Enable GENIE neutrino masterclass application.     

with options for 3rd party software
prefix with --with (eg. --with-lhapdf-lib=/some/path)

optimiz-level      Compiler optimiz. level (O,O2,O3,OO,Os)
profiler-lib       Path to profiler library
doxygen-path       Path to doxygen binary
pythia6-lib        Path to PYTHIA6 library
cern-lib           Path to CERN libraries
lhapdf-inc         Path to LHAPDF includes
lhapdf-lib         Path to LHAPDF libraries
libxml2-inc        Path to libxml2 includes
libxml2-lib        Path to libxlm2 library
log4cpp-inc        Path to log4cpp includes
log4cpp-lib        Path to log4cpp library
\end{verbatim}

By default all options required for a minimal installation that can
be used for physics event generation are enabled and non-essential
features are disabled. Typically, the folowing should be sufficient
for most users:\\
\\
\texttt{\textbf{\small{}\$ cd \$GENIE}}~\\
\texttt{\textbf{\small{}\$ ./configure}}\\
\\
Not specifying any configuration option (like above) is equivalent
to specifying:\\
\texttt{\textbf{\small{}-{}-disable-profiler }}~\\
\texttt{\textbf{\small{}-{}-disable-doxygen-doc }}~\\
\texttt{\textbf{\small{}-{}-enable-dylibversion }}~\\
\texttt{\textbf{\small{}-{}-disable-lowlevel-mesg }}~\\
\texttt{\textbf{\small{}-{}-disable-debug }}~\\
\texttt{\textbf{\small{}-{}-enable-lhapdf }}~\\
\texttt{\textbf{\small{}-{}-disable-cernlib }}~\\
\texttt{\textbf{\small{}-{}-enable-flux-drivers }}~\\
\texttt{\textbf{\small{}-{}-enable-geom-drivers }}~\\
\texttt{\textbf{\small{}-{}-enable-mueloss }}~\\
\texttt{\textbf{\small{}-{}-disable-validation-tools }}~\\
\texttt{\textbf{\small{}-{}-disable-test}}~\\
\texttt{\textbf{\small{}-{}-disable-t2k }}~\\
\texttt{\textbf{\small{}-{}-disable-fnal}}~\\
\texttt{\textbf{\small{}-{}-disable-atmo}}~\\
\texttt{\textbf{\small{}-{}-disable-nucleon-decay}}~\\
\texttt{\textbf{\small{}-{}-disable-rwght}}~\\
\texttt{\textbf{\small{}-{}-disable-masterclass}}\\
\\
The default optimization level is set to O2 and \texttt{\textbf{\small{}-{}-prefix}}
is set to /usr/local.\\
\\
The configuration script can, in principle, \textit{auto-detect the
paths} to required external packages installed at your system if no
path is given explicitly. On some occasions, before scanning your
system for external products, the configuration script will check
whether some rather standard environmental variables have been set
(from example, before searching for the PYTHIA6 / JETSET library,
the configure script will check whether a `\texttt{\small{}PYTHIA6}'
environmental variable has been set. See \texttt{\textbf{\small{}`./configure
-{}-help}}' for more information).\\
\\
Obviously, if you want greater control over the configuration options
(so that you do not depend on pre-set defaults that may one day change),
if you want to modify some other default options or if the script
fails to discover some external product path, then do set the configure
script options explicitly.

\section{Building GENIE}

Once GENIE has been properly configured, you are ready to build it.
Just type:\\
\\
\texttt{\textbf{\small{}\$ cd \$GENIE}}\\
\texttt{\textbf{\small{}\$ gmake}}\\
\\
On successful completion you should be able to find many libraries
located in \textit{\$GENIE/lib} and some applications and scripts
in \textit{\$GENIE/bin. }\\
\textit{}\\
You may stop the building procedure here and start using GENIE now!
However, some users may prefer to take their installation one step
further and type:\\
\\
\texttt{\textbf{\small{}\$ gmake install}}~\\
\texttt{\textbf{\small{}}}~\\
If \textit{/some/path} was the location specified via the \texttt{\textbf{\small{}-{}-prefix}}
configuration flag, then \texttt{\textbf{\small{}`gmake install'}}
will:
\begin{itemize}
\item move all executables and scripts to \textit{/some/path/bin,}
\item move all libraries to \textit{/some/path/lib,} and 
\item move all headers to \textit{/some/path/include/GENIE}. 
\end{itemize}
If you do run \texttt{\textbf{\small{}`gmake install'}}, before running
GENIE you need to update your `\texttt{\small{}LD\_LIBRARY\_PATH}'
(or `\texttt{\small{}DYLD\_LIBRARY\_PATH}' on MAC OS X) and `\texttt{\small{}PATH}'
environmental variables accordingly.\\
\\
Whether you stop the installation procedure after the `\texttt{\textbf{\small{}gmake}}'
or `\texttt{\textbf{\small{}gmake install}}' step is probably more
a matter of personal taste \footnote{I find it easier to manage multiple GENIE installations if I stop
after the `\texttt{\textbf{\small{}gmake}}' step.}. Whatever you choose should work given that your system's paths have
been properly set. \\
\\
Assuming now that the GENIE installation has been completed without
apparent errors, we are going to provide instructions for a couple
of simple post-installation tests to verify that GENIE has been properly
built.

\section{Performing simple post-installation tests}

Here are few simple things you can do in order to try out your installation:\\

\begin{enumerate}
\item Generate a $\nu_{\mu}+O^{16}$ ($\nu_{\mu}$ PDG code: 14, $O^{16}$
PDG code: 1000080160) event sample (10k events) between 0 and 10 GeV,
using a simple histogram-based description of the T2K $\nu_{\mu}$
flux (ROOT \textit{TH1D} object `h30000' stored in \textit{`\$}\texttt{\textbf{\small{}GENIE}}\textit{/data/flux/t2kflux.root}').
Use pre-calculated cross-sections (later, you will learn how to calculate
these on your own) which can be downloaded from \url{http://www.hepforge.org/archive/genie/data/}.
\\
\\
The commands used here will be explained in the next section:\\
\\
\texttt{\textbf{\small{}\$ gevgen -n 10000 -p 14 -t 1000080160 -e
0,10 -{}-run 100 }}~\\
\texttt{\textbf{\small{}$\qquad$-f \$GENIE/data/flux/t2kflux.root,h30000
}}~\\
\texttt{\textbf{\small{}$\qquad$-{}-seed 2989819 -{}-cross-sections
/some/path/xsec.xml }}~\\
\texttt{\textbf{\small{}}}~\\
A `\textit{genie-mcjob-<run number>.status}' status file is created.
It is updated periodicaly with job statistics and the most recent
event dump. When the job is completed a `\textit{gntp.<run number>.ghep.root}'
file, containing the generated event tree, is written-out. To print-out
the first 200 events from the event file you just generated, type:\texttt{\textbf{\small{}}}~\\
\texttt{\textbf{\small{}\$ gevdump -f gntp.100.ghep.root -n 200}}{\small \par}
\item Generate a 10,000 event sample of $\pi^{+}+O^{16}$interactions for
$\pi^{+}$'s of 200 MeV kinetic energy. \\
($\pi^{+}$ PDG code: 211, $O^{16}$ PDG code: 1000080160):\\
\\
\texttt{\textbf{\small{}\$ gevgen\_hadron -n 10000 -p 211 -t 1000080160
-k 0.2 -{}-seed 9839389}}~\\
\texttt{\textbf{\small{}}}~\\
{\small \par}
\end{enumerate}
If everything seems to work then the GENIE is really `out of the bottle'.
Continue reading the Physics and User Manual to find out more about
running the GENIE applications bundled in your installation.

\chapter{Generating Neutrino Event Samples}

\section{Introduction}

{[}to be added in future revision{]}

\section{Preparing event generation inputs: Cross-section splines}

When generating neutrino interaction events, most CPU-cycles are spent
on calculating neutrino interaction cross sections. In order to select
an interaction channel for a neutrino scattered off a target at a
particular energy, the differential cross section for each possible
channel is integrated over the kinematic phase space available at
this energy. With $\sim10^{2}$ possible interaction modes per initial
state and with $\sim10^{5}$ differential cross section evaluations
per cross section integration then $\sim10^{7}$ differential cross
section evaluations are required just in order to select an interaction
channel for a given initial state. Had you been simulating events
in a realistic detector geometry ($\sim10^{2}$ different isotopes)
then the number of differential cross section evaluations, before
even starting simulating the event kinematics, would rise to $\sim10^{9}$.
It is therefore advantageous to pre-calculate the cross section data.
The event generation drivers can be instructed to to load the pre-computed
data and estimate the cross section by numerical interpolation, rather
than by performing numerous CPU-intensive differential cross section
integrations. The cross section data are written out in XML format
and, when loaded into GENIE, they are used for instantiating \textit{Spline}
objects.

\subsection{The XML cross section splines file format}

The XML file format is particularly wekk-suited for moving data between
different GENIE applications. This is the only intended usage of these
files. If you wish to use GENIE's cross section splines in another
context, eg. within your analysis code, then we recommend converting
them from XML to ROOT format using utilities provided by GENIE (See
Section \ref{sub:UsingSplinesInUserPrograms}). Although you should
never have to read the XML cross section file, it is generally usefull
that you do have an understanding of how it is structured so as to
be able to diagnose problems. 

All XML splines are stored within `\texttt{\textbf{\small{}<genie\_xsec\_spline\_list>}}'
tags:\\

\begin{verbatim}
<?xml version="1.0" encoding="ISO-8859-1"?>
<!-- generated by genie::XSecSplineList::SaveSplineList() -->
<genie_xsec_spline_list version="2.00" uselog="1"> 
... ... ...
... ... ...
</genie_xsec_spline_list> 
\end{verbatim}

The `uselog=''1''' flag indicates that the spline knots are spaced
`logarithmically' in energy (This is the default GENIE option so that
there is higher knot density where the cross section changes more
rapidly). The data for each spline are stored within `\texttt{\textbf{\small{}<spline>}}'
tags\footnote{In the description below, the curly braces within tags are to be `viewed'
as a single value of the specified type with the specified semantics. }:

\begin{verbatim}
<spline 
    name   = "{algorithm/reaction; string}" 
    nknots = "{number of knots; int}"> 
<knot> 
  <E>    {energy; double}        </E> 
  <xsec> {cross section; double} </xsec> 
</knot> 
<knot> 
  <E>    {energy; double}        </E> 
  <xsec> {cross section; double} </xsec> 
</knot> 
... ...
</spline>
\end{verbatim}

Each spline is named by combining the names of the cross section algorithm
and its configuration with a string interaction code. These rather
long names are built automatically by GENIE and used for retrieving
the correct spline\footnote{GENIE takes the safest route and checks both the `reaction mode' and
`cross section algorithm'. It will not use cross section spline data
calculated by a cross section algorithm A, if an alternative cross
section algorithm B is currently in use.} from the spline pool. For example, a spline named `\textit{\small{}genie::DISPartonModelPXSec/CC-Default/nu:-12;tgt:1000260560;N:2112;q:-1(s);proc:Weak{[}CC{]},DIS}'
indicates that it was computed using the cross section algorithm `\textit{\small{}genie::DISPartonModelPXSec}'
run in the `\textit{\small{}CC-Default}' configuration for an interaction
channel with the following string code: `\textit{\small{}nu:-12;tgt:1000260560;N:2112;q:-1(s);proc:Weak{[}CC{]},DIS}'
(indicating a DIS CC $\nu_{\mu}$$Fe^{56}$ scattering process of
a sea $\bar{d}$ quark in a bound neutron). The spline knots are listed
in increasing energy, going up to a maximum value specified during
the spline construction. One of the knots falls exactly on the energy
threshold for the given process so as to improve the accuracy of numerical
interpolation around threshold. The energy and cross section values
are given in the natural system of units ($\hbar=c=1$) used internally
within GENIE (Note that the more widespread cross section units, $10^{-38}$$cm^{2}$,
are used when the cross section data are exported to a ROOT format
for inclusion in user analysis code. See Section\ref{sub:UsingSplinesInUserPrograms}).

\subsection{Downloading pre-computed cross section splines \label{sub:spline-download}}

Cross section spline XML files are kept in:\textit{ \url{http://www.hepforge.org/archive/genie/data/}}\\
\\
You need to select the file corresponding to the version of GENIE
you are using.\\
\\
Typically I post cross section spline files for all modeled processes
for $\nu_{e},$ $\bar{\nu_{e}}$, $\nu_{\mu},$ $\bar{\nu_{\mu}}$,
$\nu_{\tau},$ $\bar{\nu_{\tau}}$ scattered off free-nucleons (p,
n) and off a large set of nuclear targets (the $\sim40$ isotopes
that can be found in the T2K detector geometries\footnote{$N^{14}$, $N^{15}$, $O^{16}$, $O^{17}$, $O^{18}$, $Al^{27}$,
$C^{12}$, $C^{13}$, $H^{2}$, $Cl^{35}$, $Cl^{37}$, $Pb^{204}$,
$Pb^{206}$, $Pb^{207}$, $Pb^{208}$, $Cu^{63}$, $Cu^{65}$, $Zn^{64}$,
$Zn^{66}$, $Zn^{67}$, $Zn^{68}$, $Zn^{70}$, $Ar^{36}$, $Ar^{38}$,
$Ar^{40}$, $Si^{28}$, $Si^{29}$, $Si^{30}$, $B^{10}$, $B^{11}$,
$Na^{23},$ $Fe^{54},$ $Fe^{56},$ $Fe57,$ $Fe^{58}$, $Co^{59}$.}). Using the posted free-nucleon cross section data is easy / fast
to calculate cross section splines for any set of nuclear targets.
\\
\\
Any reasonable request for providing additional cross section splines
will be satisfied.

\subsection{Generating cross section splines \label{sub:spline-generate}}

Cross section spline calculation is very CPU-intensive. It is recommended
that, for the default GENIE configuration, you use the officially
distributed files. However, the information provided in this section
will allow you to generate your own cross section spline files, should
you need to.

\subsubsection{The \textit{gmkspl} spline generation utility}

\subsubsection*{Name}

\textit{gmkspl} -- A GENIE utility for generating the cross section
splines for a specified set of modeled processes for a specified list
of initial states. The cross section splines are written out in an
XML file in the format expected by all other GENIE programs.

\subsubsection*{Source}

The source code for this utility may be found in \textit{`\$}\texttt{\textbf{\small{}GENIE}}\textit{/src/stdapp/gMakeSplines.cxx}'.

\subsubsection*{Synopsis}

\texttt{\textbf{\small{}\$ gmkspl -p }}\texttt{\small{}neutrino\_code}\texttt{\textbf{\small{}
<-t }}\texttt{\small{}target\_codes}\texttt{\textbf{\small{}, -f }}\texttt{\small{}geometry}\texttt{\textbf{\small{}>
{[}-n }}\texttt{\small{}nknots}\texttt{\textbf{\small{}{]} {[}-e max\_energy{]}}}\\
\texttt{\textbf{\small{}$\qquad${[}<-{}-output-cross-sections | -o>
}}\texttt{\small{}xml\_file}\texttt{\textbf{\small{}{]} {[}-{}-input-cross-sections
}}\texttt{\small{}xml\_file}\texttt{\textbf{\small{}{]}}}\\
\texttt{\textbf{\small{}$\qquad${[}-{}-seed }}\texttt{\small{}rnd\_seed\_num}\texttt{\textbf{\small{}{]}
{[}-{}-event-generator-list }}\texttt{\small{}list\_name}\texttt{\textbf{\small{}{]}
{[}-{}-message-thresholds }}\texttt{\small{}xml\_file}{]}\\
\\
where {[}{]} marks optional arguments, and <> marks a list of arguments
out of which only one can be selected at any given time.

\subsubsection*{Description }

The following options are available:\textbf{}\\
\textbf{}\\
\textbf{-p} Specifies the neutrino PDG codes.\\

Multiple neutrino codes can be specified as a comma separated list.\textbf{}\\
\textbf{}\\
\textbf{-t} Specifies the target PDG codes.\\

Multiple target PDG codes can be specified as a comma separated list.\textbf{
}The PDG2006 conventions is used (10LZZZAAAI). So, for example, $O^{16}$
code = 1000080160, $Fe^{56}$ code = 1000260560. For more details
see Appendix \ref{cha:AppendixStatAndPdgCodes}. \textbf{}\\
\textbf{}\\
\textbf{-f} Specifies a ROOT file containing a ROOT/GEANT detector
geometry description.\textbf{}\\
\textbf{}\\
\textbf{-n} Specifies the number of knots per spline.\\

By default GENIE is using 15 knots per decade of the spline energy
range and at least 30 knots overall.\textbf{}\\
\textbf{}\\
\textbf{-e} Specifies the maximum neutrino energy in the range of
each spline. \\

By default the maximum energy is set to be the declared upper end
of the validity range of the event generation thread responsible for
generating the cross section spline.\\
\\
\textbf{--output-cross-sections, -o} Specifies the name (incl. full
path) of an output cross-section XML file.\\

By default GENIE writes-out the calculated cross section splines in
an XML file named `\textit{xsec\_splines.xml}' created at the current
directory.\textbf{}\\
\textbf{}\\
\textbf{--input-cross-sections} Specifies the name (incl. full path)
of the output XML file. \\

An input cross-section file could be specified when it is possible
to recycle previous calculations. It is, sometimes, possible to recycle
cross-section calculations for scattering off free nucleons when calculating
nuclear cross-sections.\textbf{ }\\
\textbf{}\\
\textbf{--seed }Specifies the random number seed for the current job.\\

This setting will only ne relevant if MC intergation methods are employed
for cross-section calculation.\\
\\
\textbf{--event-generator-list }List of event generators to load.
\textbf{}\\

The list of event generators to load affects the list of processes
that can be simulated and, for which, cross-section calculations need
to be calculated by this application. By default, GENIE is loading
a list of of tuned and fully-validated generators which allow comprehensive
neutrino interaction modelling the medium-energy range.\textbf{ }Valid
settings are the XML block names appearing in \textit{\$}\texttt{\textbf{\small{}GENIE}}\textit{/config/EventGeneratorListAssembler.xml}'.
Please, make sure you read Sec. \ref{sec:ObtainingSpecialSamples}
explaining why, almost invariantly, for physics studies you should
be using a comprehensive collection of event generators.\texttt{\textbf{\small{} }}\textbf{}\\
\textbf{}\\
\textbf{--message-thresholds }Specifies the GENIE verbosity level.\\

The verbosity level is controlled with an XML file allowing users
to customize the threshold of each message stream. See \textit{`\$}\texttt{\textbf{\small{}GENIE}}\textit{/config/Messenger.xml}'
for the XML schema. The \textit{`Messenger.xml' }file contains the
default thresholds used by GENIE. The \textit{`Messenger\_laconic.xml'
}and \textit{`Messenger\_rambling.xml' }files define, correspondingly,
less and more verbose configurations.

\subsubsection*{Examples}
\begin{figure}
\center

\includegraphics[scale=0.6]{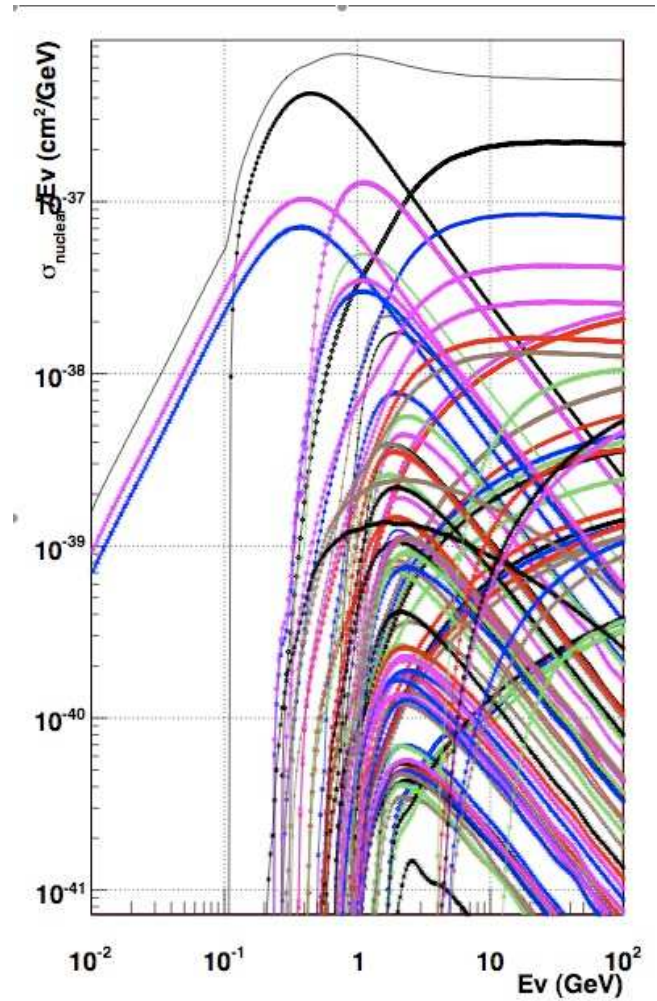}

\caption{Cross section splines just for $\nu_{\mu}Fe^{56}$ processes modeled
in GENIE. The large number of splines and the fine numerical integration
stepping makes spline calculation a very CPU-intensive process. }
\label{fig:NuMuFeSplines}
\end{figure}

\begin{enumerate}
\item To calculate cross-sections for $\nu_{\mu}$ (PDG code: 14) and $\bar{\nu_{\mu}}$
(PDG code: -14) scattering off $Fe^{56}$ (PDG code: 1000260560),
and build splines with 150 knots in the energy range up to 20 GeV,
type \\
\\
\texttt{\textbf{\small{}\$ gmkspl -p 14,-14 -t 1000260560 -n 150 -e
20}} \\
\\
The cross section splines will be saved in an output XML file named
\textit{`xsec\_splines.xml} ' (default name).\\

\item To calculate the CCQE cross-section for $\nu_{\mu}$ (PDG code: 14)
and $\bar{\nu_{\mu}}$ (PDG code: -14) scattering off all the targets
in the input ROOT geometry file `\textit{/data/mygeometry.root}' and
write out the splines in a file named \textit{`mysplines.xml}', type\\
\\
\texttt{\textbf{\small{}\$}} \texttt{\textbf{\small{}gmkspl -p 14,-14
-f /data/mygeometry.root -o mysplines.xml -{}-event-generator-list
CCQE}}~\\
\texttt{\textbf{\small{}}}~\\
{\small \par}
\end{enumerate}
Generating cross-section splines is a CPU-intensive task as a large
number of processes (see Fig. \ref{fig:NuMuFeSplines}) and numerical
integration of steeply peaked differential cross-sections over extended,
multi-dimensional kinematical phase spaces. When cross-section calculations
are needed for multiple targets, it is often impractical to generate
all splines in a single job. The task is typically split into smaller
jobs which can be run on parallel in a batch farm. Batch submission
scripts used by GENIE developers can be found in `\textit{\$}\texttt{\textbf{\small{}GENIE}}\textit{/src/scripts/production/batch/}'
and easily adapted to match user needs. Detailed documentation is
available within the scripts. The multiple XML outputs of all the
\textit{gmkspl }jobs can be merged into a single XML file using GENIE's
\textit{gspladd} utility. (See Section \ref{sub:gspladd}.) It is
worth highlighting that, for faster results, it is preferable if one
organizes the jobs as `single neutrino + multiple nuclear targets'
rather than `multiple neutrinos + single nuclear target': In the former
case intermediate, CPU-intensive free-nucleon cross-section calculations,
for the given neutrino species, will be recycled in the nuclear target
cross-section calculations. For even faster results one can calculate
the free-nucleon cross-section splines first, then feed the output
into a nuclear cross-section spline calculation. Because of the way
nuclear effects are currently handled, nuclear cross-section calculations
can recycle CPU-intensive free-nucleon calculations resulting in a
dramatic speed improvement. To feed-in free-nucleon cross-sections
in a nuclear cross-section calculation job, use the \textit{gmkspl}
--input-cross-sections option. Note that, if you feed-in cross-sections,
the calculated cross-sections can not extend higher in energy than
the input cross-sections.

\subsubsection{The \textit{gspladd} spline merging utility \label{sub:gspladd}}

\subsubsection*{Name}

\textit{gspladd} -- A GENIE utility for merging many separate XML
cross section files into a single XML file.

\subsubsection*{Source}

The source code for this utility may be found in \textit{`\$}\texttt{\textbf{\small{}GENIE}}\textit{/src/stdapp/gSplineAdd.cxx}'.

\subsubsection*{Synopsis}

\texttt{\textbf{\small{}\$ gspladd -f }}\texttt{file\_list}\texttt{\textbf{\small{}
-d }}\texttt{directory\_list}\texttt{\textbf{\small{} -o }}\texttt{output\_file}

\subsubsection*{Description }

The following options are available:\textbf{}\\
\textbf{}\\
\textbf{\textcolor{magenta}{-f }}Specifies input XML files. Multiple
input files can be specified as a comma separated list.\textbf{}\\
\textbf{}\\
\textbf{\textcolor{magenta}{-d }}Specifies input directories. Multiple
input files can be specified as a comma separated list. All XML files
found in each directory will be included.\textbf{}\\
\textbf{}\\
\textbf{\textcolor{magenta}{-o}} Specifies the name of the output
XML file\textbf{.} \\

\subsubsection*{Notes}
\begin{itemize}
\item At least 2 XML files must be specified as inputs for the \textit{gspladd
}application to work.
\end{itemize}

\subsubsection*{Examples}
\begin{enumerate}
\item To merge `\textit{/data/iron/xsec.xml}' and `\textit{/data/oxygen/xsec.xml}'
into `\textit{./xsec\_all.xml}', type:\\
\\
\texttt{\textbf{\small{}\$ gspladd -f /data/iron/xsec.xml,/data/oxygen/xsec.xml
-o xsec\_all.xml}}\\

\item To merge `\textit{./xsec\_Fe56.xml}' and all the cross section spline
files found in `\textit{/scratch/job1}' and `\textit{/scratch/job2}'
into `\textit{./xsec\_all.xml}', type\\
\\
\texttt{\textbf{\small{}\$}} \texttt{\textbf{\small{}gspladd -f xsec\_Fe56.xml
-d /scratch/job1/,/scratch/job2 -o xsec\_all.xml}}{\small \par}
\end{enumerate}

\subsection{Re-using splines for modified GENIE configurations}

You should \textit{never} be doing that (unless you are absolutely
sure about what you are doing). The safest assumption is that changes
in GENIE, either a change of default model parameter or a change of
a default model, \textit{invalidates} previously generated cross section
splines as the cross section models (used for generating these splines)
may be affected.

\subsection{Using cross section splines in your analysis program \label{sub:UsingSplinesInUserPrograms}}

As seen before, GENIE's \textit{gmkspl }utility writes-out cross section
values in XML format. While this format is particularly well-suited
for moving data between GENIE components, it is not the most usefull
format from the perspective of a user who wishes to read and interpolate
these cross section data in different contexts within his/her analysis
code. 

GENIE provides the \textit{gspl2root} utility to convert XML cross
section splines into a ROOT formats. The XML cross section data for
each process and initial state are converted into a single ROOT \textit{TGraph}
objects. All ROOT \textit{TGraph} objects corresponding to the same
initial state are written-out in the same ROOT \textit{TDirectory}
which is named after the given initial state. Multiple \textit{TDirectory}
objects can be saved in a single output ROOT file. ROOT \textit{TGraph}
objects support numerical interpolation via the `\textit{TGraph::Eval(double)}'
method, so, essentially, \textbf{one can write-out all GENIE cross
section `functions' one needs into a single ROOT file}. More details
on this particularly useful feature are given next.

\subsubsection{The \textit{gspl2root }spline file conversion utility}

\subsubsection*{Name}

\textit{gspl2root} - A GENIE utility to convert XML cross section
files into a ROOT format.

\subsubsection*{Source}

The source code for this utility may be found in \textit{`\$}\texttt{\textbf{\small{}GENIE}}\textit{/src/stdapp/gSplineXnml2Root.cxx}'.

\subsubsection*{Synopsis}

\texttt{\textbf{\small{}\$ gspl2root }}{\small \par}

\texttt{\textbf{\small{}-f }}\texttt{input\_xml\_file}\texttt{\textbf{\small{} }}{\small \par}

\texttt{\textbf{\small{}-p }}\texttt{neutrino\_pdg\_code}\texttt{\textbf{\small{}
-t }}\texttt{target\_pdg\_code }

\texttt{\textbf{\small{}{[}-e }}\texttt{maximum\_energy}\texttt{\textbf{\small{}{]}
{[}-o }}\texttt{output\_root\_file}\texttt{\textbf{\small{}{]} {[}-w{]}}}\\
\\
where {[}{]} denotes an optional argument.\\

\subsubsection*{Description}

The following options are available:\\

\textbf{\textcolor{magenta}{-f }}Specifies the input XML cross section
spline file. \\

\textbf{\textcolor{magenta}{-p}} Specifies the neutrino PDG code.\\

\textbf{\textcolor{magenta}{-t }}Specifies the target PDG code (format:
10LZZZAAAI).\\

\textbf{\textcolor{magenta}{-e}} Specifies the maximum energy for
the generated graphs.\\

\textbf{\textcolor{magenta}{-o}} Specifies the output ROOT file name.\\

\textbf{\textcolor{magenta}{-w}} Instructs \textit{gspl2root} to write-out
plots in a postscipt file.\\

\subsubsection*{Notes}
\begin{itemize}
\item The spline data written-out have the energies given in $GeV$and the
cross sections given in in $10^{-38}$$cm^{2}$.
\end{itemize}

\subsubsection*{Examples}
\begin{enumerate}
\item In order to extract all $\nu_{\mu}$+$n$, $\nu_{\mu}$+$p$ and $\nu_{\mu}$+$O^{16}$
cross section splines from the input XML file `\textit{mysplines.xml}',
convert splines into a ROOT format and save them into a single ROOT
file `\textit{xsec.root}', type:\\
\\
\texttt{\textbf{\small{}\$ gspl2root -f mysplines.xml -p 14 -t 1000000010
-o xsec.root}}\\
\texttt{\textbf{\small{}\$ gspl2root -f mysplines.xml -p 14 -t 1000010010
-o xsec.root}}\\
\texttt{\textbf{\small{}\$ gspl2root -f mysplines.xml -p 14 -t 1000080160
-o xsec.root}}\\
\\
A large number of graphs (one per simulated process and appropriate
totals) will be generated in each case. Each set of plots is saved
into its own ROOT \textit{TDirectory} named after the specified initial
state. \\
\\
The stored graphs can be used for cross section interpolation. For
instance, the `\textit{xsec.root}' file generated in this example
will contain a `nu\_mu\_O16' \textit{TDirectory} (generated by the
last command) which will include cross section graphs for all $\nu_{\mu}$+$O^{16}$
processes. To extract the $\nu_{\mu}$+$O^{16}$ DIS CC cross section
graph for hit u valence quarks in a bound proton and evaluate the
cross section at energy E, type:\\
\\
\texttt{\textbf{\small{}root{[}0{]} TFile file(``xsec.root'',''read'');}}~\\
\texttt{\textbf{\small{}root{[}1{]} TDirectory {*} dir = (TDirectory{*})
file->Get(\textquotedbl{}nu\_mu\_O16\textquotedbl{});}}~\\
\texttt{\textbf{\small{}root{[}2{]} TGraph {*} graph = (TGraph{*})
dir->Get(\textquotedbl{}dis\_cc\_p\_uval\textquotedbl{});}}~\\
\texttt{\textbf{\small{}root{[}3{]} cout <\textcompwordmark{}< graph->Eval(E)
<\textcompwordmark{}< endl;}}{\small \par}
\end{enumerate}

\section{Simple event generation cases}

This section will introduce \textit{gevgen,} a generic GENIE event
generation application. This particular application has access to
the full suite of GENIE physics models but will only handle relatively
simple flux and geometry setups. It doesn't use any of the atmospheric,
JPARC, NuMI or other specialized flux drivers included in GENIE and
doesn't use ROOT/Geant-4 based detector geometries. A reader interested
in the more specialized event generation applications included in
GENIE can jump to Chapter \ref{cha:UsingFluxAndGeo}.

\subsection{The \textit{\normalsize{}gevgen} generic event generation application
\label{sub:gevgen}}

\subsubsection*{Name}

\textit{gevgen} - A generic GENIE event generation application for
simple event generation cases. The application handles event generation
for neutrinos scattered off a given target (or `target mix'). It doesn't
support event generation over ROOT/Geant4-based detector geometries.
It handles mono-energetic flux neutrinos or neutrino fluxes described
in simple terms (either via a functional form, a vector file or a
ROOT \textit{TH1D} histogram).

\subsubsection*{Source}

The source code for this utility may be found in \textit{`\$}\texttt{\textbf{\small{}GENIE}}\textit{/src/stdapp/gEvGen.cxx}'.

\subsubsection*{\texttt{\small{}Synopsis}}

\texttt{\textbf{\small{}\$ gevgen {[}-h{]} {[}-r }}\texttt{\small{}run\#}\texttt{\textbf{\small{}{]}
-n }}\texttt{\small{}nev}\texttt{\textbf{\small{} -p }}\texttt{\small{}neutrino\_pdg}\texttt{\textbf{\small{}
-t }}\texttt{\small{}target\_pdg}\texttt{\textbf{\small{} -e }}\texttt{\small{}energy}\texttt{\textbf{\small{}
{[}-f }}\texttt{\small{}flux}\texttt{\textbf{\small{}{]} }}{\small \par}

\texttt{\textbf{\small{}{[}-w{]} {[}-seed }}\texttt{\small{}random\_number\_seed}\texttt{\textbf{\small{}{]}
{[}-{}-cross-section }}\texttt{\small{}xml\_file}\texttt{\textbf{\small{}{]}
{[}-{}-event-generator-list }}\texttt{\small{}list\_name}\texttt{\textbf{\small{}{]}}}{\small \par}

\texttt{\textbf{\small{}{[}-{}-message-thresholds }}\texttt{\small{}xml\_file}\texttt{\textbf{\small{}{]}
{[}-{}-unphysical-event-mask }}\texttt{\small{}mask}\texttt{\textbf{\small{}{]}
{[}-{}-event-record-print-level }}\texttt{\small{}level}\texttt{\textbf{\small{}{]}}}{\small \par}

\texttt{\textbf{\small{}{[}-{}-mc-job-status-refresh-rate }}\texttt{\small{}rate}\texttt{\textbf{\small{}{]}
{[}-{}-cache-file }}\texttt{\small{}root\_file}\texttt{\textbf{\small{}{]}}}
\\
\\
where {[}{]} denotes an optional argument.

\subsubsection*{Description}

The following options are available:
\begin{itemize}
\item \textbf{\textcolor{magenta}{-h}} Prints-out help on \textit{gevgen}
syntax and exits.
\item \textbf{\textcolor{magenta}{-r}} Specifies the MC run number. 
\item \textbf{\textcolor{magenta}{-n}} Specifies the number of events to
generate. 
\item \textbf{\textcolor{magenta}{p }}Specifies the neutrino PDG code.
\item \textbf{\textcolor{magenta}{-t}} Specifies the target PDG code(s).
\\
\\
The PDG2006 convention is used (10LZZZAAAI). So, for example, $O^{16}$
code = 1000080160, $Fe^{56}$ code = 1000260560. For more details
see Appendix \ref{cha:AppendixStatAndPdgCodes}. \\
\\
Multiple targets (a `target mix') can be specified as a comma-separated
list of PDG codes, each followed by its corresponding weight fraction
in brackets as in:\\
`\texttt{\textbf{\small{}code1{[}fraction1{]},code2{[}fraction2{]},...}}'.\\
For example, to use a target mix of 95\% O16 and 5\% H type: \\
`\texttt{\textbf{\small{}-t 1000080160{[}0.95{]},1000010010{[}0.05{]}}}'.
\\

\item \textbf{\textcolor{magenta}{-e }}Specifies the neutrino energy or
energy range.\\
\\
For example, specifying `\texttt{\textbf{\small{}-e 1.5}}' will instruct
\textit{gevgen} to generate events at 1.5 GeV. \\
\\
If what follows `\texttt{\textbf{\small{}-e}}' is a comma separated
pair of values then gevgen will interpret that as an `energy range'.
For example, specifying `\texttt{\textbf{\small{}-e 0.5,2.3}}' will
be interpreted as the {[}0.5 GeV, 2.3 GeV{]} range. If an energy range
is specified then \textit{gevgen} expects the `\texttt{\textbf{\small{}-f}}'
option to be set as well so as to describe the energy spectrum of
flux neutrinos over that range (see below).\\

\item \textbf{\textcolor{magenta}{-f}} Specifies the neutrino flux spectrum.\\
\\
This generic event generation driver allows to specify the flux in
any one of three simple ways:

\begin{itemize}
\item As a `function'.\\
For example, in order to specify a flux that has the $x^{2}+4e^{-x}$
functional form, type:\\
 `\texttt{\textbf{\small{}-f `x{*}x+4{*}exp(-x)}}''\\
 
\item As a `vector file'.\\
The file should contain 2 columns corresponding to energy (in GeV),
flux (in arbitrary units).\\
For example, in order to specify that the flux is described by the
vector file `\textit{/data/fluxvec.data}', type:\\
`\texttt{\textbf{\small{}-f /data/fluxvec.data}}'\\
 
\item As a `1-D histogram (\textit{TH1D}) in a ROOT file'.\\
The general syntax is: `\texttt{\textbf{\small{}-f /full/path/file.root,object\_name}}'.\\
For example, in order to specify that the flux is described by the
`nue' \textit{TH1D} object in `\textit{/data/flux.root}', type:\\
`\texttt{\textbf{\small{}-f /data/flux.root,nue}}'
\end{itemize}
\item \textbf{-w} Forces generation of weighted events. \\
\\
This option is relevant only if a neutrino flux is specified via the
`\texttt{\textbf{\small{}-f}}' option. In this context `weighted'
refers to an event generation biasing in selecting an initial state
(a flux neutrino and target pair at a given neutrino energy). Internal
weighting schemes for generating event kinematics can still be enabled
independently even if `\texttt{\textbf{\small{}-}}w' is not set. Don't
use this option unless you understand what the internal biasing does
and how to analyze the generated sample. The default option is to
generated unweighted events.
\item \textbf{\textcolor{magenta}{--seed }}Specifies the random number seed
for the current job.
\item \textbf{\textcolor{magenta}{--cross-sections}} Specifies the name
(incl. full path) of an input XML file with pre-computed neutrino
cross-sections
\item \textbf{\textcolor{magenta}{--event-generator-list}}\textbf{ }Specifies
the list of event generators to use in the MC job. \\
\\
By default, GENIE is loading a list of of tuned and fully-validated
generators which allow comprehensive neutrino interaction modelling
the medium-energy range.\textbf{ }Valid settings are the XML block
names appearing in \textit{\$}\texttt{\textbf{\small{}GENIE}}\textit{/config/EventGeneratorListAssembler.xml}'.
Please, make sure you read Sec. \ref{sec:ObtainingSpecialSamples}
explaining why, almost invariantly, for physics studies you should
be using a comprehensive collection of event generators.\texttt{\textbf{\small{} }}{\small \par}
\item \textbf{\textcolor{magenta}{--message-thresholds}}\textbf{ }Specifies
the GENIE verbosity level. \\
\\
The verbosity level is controlled with an XML file allowing users
to customize the threshold of each message stream. See \textit{`\$}\texttt{\textbf{\small{}GENIE}}\textit{/config/Messenger.xml}'
for the XML schema. The \textit{`Messenger.xml' }file contains the
default thresholds used by GENIE. The \textit{`Messenger\_laconic.xml'
}and \textit{`Messenger\_rambling.xml' }files define, correspondingly,
less and more verbose configurations.
\item \texttt{\textbf{\textcolor{magenta}{-{}-unphysical-event-mask}}} Specify
a 16-bit mask to allow certain types of unphysical events to be written
in the output event file. \\
\\
By default, all unphysical events are rejected.
\item \texttt{\textbf{\textcolor{magenta}{-{}-event-record-print-level}}}\texttt{\textbf{\small{}
}}Allows users to set the level of information shown when the event
94 record is printed in the screen. \\
\\
See GHepRecord::Print() for allowed settings.
\item \texttt{\textbf{\textcolor{magenta}{-{}-mc-job-status-refresh-rate}}}\texttt{\textbf{\small{}
}}Allows users to customize the refresh rate of the status file.
\item \texttt{\textbf{\textcolor{magenta}{-{}-cache-file}}} Allows users
to specify a ROOT file so that results of calculation cached throughout
a MC job can be re-used in subsequent MC jobs.
\end{itemize}

\subsubsection*{Examples}
\begin{enumerate}
\item To generate 20,000 $\nu_{\mu}$ (PDG code: 14) scattered off $Fe^{56}$
(PDG code: 1000260560) at an energy of 6.5 GeV, reading pre-computed
cross-sections from \textit{`/data/gxsec.xml}', and using a random
number seed of 171872, type: \\
\\
\texttt{\textbf{\small{}\$ gevgen -n 20000 -e 6.5 -p 14 -t 1000260560
-cross-sections /data/gxsec.xml -{}-seed 171872}} \\

\item To generate a similar sample as above, but with the $\nu_{\mu}$ energies,
between 1 and 4 GeV, selected from a spectrum that has the $x^{2}e^{(-x^{2}+3)/4}$
functional form, type:\textbf{\large{}}\\
\textbf{\large{}}\\
\texttt{\textbf{\small{}\$ gevgen -n 20000 -e 1,4 -p 14 -t 1000260560
-cross-sections /data/gxsec.xml -{}-seed 171872}} \texttt{\textbf{\small{}}}~\\
\texttt{\textbf{\small{}-f `x{*}x{*}exp((-x{*}x+3)/4)'}} \\

\item To generate a similar sample as above, but with the neutrino flux
described via the `\textit{/path/flux.data}' input vector file, type:\\
\\
\texttt{\textbf{\small{}\$ gevgen -n 20000 -e 1,4 -p 14 -t 1000260560
-cross-sections /data/gxsec.xml -{}-seed 171872}} \texttt{\textbf{\small{}}}~\\
\texttt{\textbf{\small{}-f /path/flux.data}}~\\
{\small \par}
\item To generate a similar sample as above, but with the neutrino flux
described a ROOT \textit{TH1D} histogram called `nu\_flux' stored
in `\textit{/path/file.root}', type:\\
\\
\texttt{\textbf{\small{}\$ gevgen -n 20000 -e 1,4 -p 14 -t 1000260560
-cross-sections /data/gxsec.xml -{}-seed 171872}} \texttt{\textbf{\small{}}}~\\
\texttt{\textbf{\small{}-f /path/file.root,nu\_flux}}\\
\\
Note that the event generation driver will use only the input histogram
bins that fall within the specified (via the `\texttt{\textbf{\small{}-e}}'
option) energy range. In the example shown above, all the neutrino
flux bins that do not fall in the 1 to 4 GeV energy range will be
neglected. The bins including 1 GeV and 4 GeV will be taken into account.
So the actual energy range used is: from the lower edge of the bin
containing 1 GeV to the upper edge of the bin containing 4 GeV. \\

\item To generate a similar sample as above, but, this time, on a target
mix that is made of 95\% O16 (PDG code: 1000080160) and 5\% H (1000010010),
type:\\
\\
\texttt{\textbf{\small{}\$ gevgen -n 30000 -e 1,4 -p 14 -cross-sections
/data/gxsec.xml -{}-seed 171872}} \texttt{\textbf{\small{}}}~\\
\texttt{\textbf{\small{}-t 1000080160{[}0.95{]},1000010010{[}0.05{]}
-f /path/file.root,nu\_flux}} 
\end{enumerate}

\paragraph*{Output files}

Typically, event generation jobs produce two files: 
\begin{itemize}
\item During job an ascii status file which contains MC job statistics and
the most recent event dump is being updated periodically. The status
file is typically named \textit{`genie-mcjob-}\texttt{\textit{\small{}<run\_number>}}\textit{.status}'
and is located in the current directory. Use --mc-job-status-refresh-rate
to adjust the refteshrate of this file. 
\item The generated events are stored in an output ROOT file, in GENIE's
native GHEP format. The event file is typically named `\texttt{\textit{\small{}<prefix>}}\textit{.}\texttt{\textit{\small{}<run\_number>}}\textit{.ghep.root}'
and is located in the current directory. In addition to the generated
event tree, the output file contains a couple of ROOT folders, `gconfig'
and `genv', containing, respectivelly, snapshots of your GENIE configuration
and running environment. Chapter \ref{cha:AnalyzingOutputs} describes
how to set-up an `event loop' and analyze the generated event sample.
\end{itemize}

\section{Obtaining special samples \label{sec:ObtainingSpecialSamples}}

\subsection{Switching reaction modes on/off}

The default behaviour of GENIE is to generate `comprehensive unweighted'
event samples. All modelled processes are included and the frequency
of process $P$ as well as the occupancy of different parts of the
kinematical phase space $\{K^{n}\}$\footnote{Such as, for example, \{$W$, $Q^{2}$\} or \{$x$, $y$\}}
reflects the value of the differetial cross section $d^{n}\sigma_{P}/d\{K^{n}\}$).

An easy way to obtain special samples is by setting the \textbf{--event-generator-list
}option available in most GENIE applications. The option controls
the list of event generators loaded into a particular GENIE MC job.
Valid settings for this option can be found in `\texttt{\textbf{\small{}\$GENIE}}\textit{/config/EventGeneratorListAssembler.xml}'
(the name of each \texttt{\small{}<param\_set> ... </param\_set>}
XML block). New parameter sets can be trivially added by the user. 

Please note that this is primarily a GENIE developer option which
users should handle with care. In the overwhelming majority of cases,
it is only poor understanding of neutrino interaction physics that
may lead one thinking that a particular setting is appropriate for
generating the special sample one requires. In general, we do not
recommend switching-off generator-level reaction modes. These modes
should be treated by the user as internal, generator-specific ``labels''.
No detector measures generator-level reaction modes like $CCQE$ or
$NC$ resonance production. Detectors measure final states / topologies
like, for example, \{1$\mu^{-}$, 0$\pi$\}, \{1$\mu^{-}$, 1$\pi^{+}$\},
\{0$\mu^{-}$, 1$\pi^{0}$\}, \{1 track, 1 shower\}, \{1 $\mu$-like
ring\} etc depending on granularity, thresholds and PID capabilities.
No final state / topology is a proxy for any particular reaction mode
(and vice versa). Intranuclear re-scattering in particular causes
significant migration between states (see Table \ref{tab:INukeTopo}).
\\
Examples: 
\begin{enumerate}
\item \{1$\mu^{-}$, 0$\pi$\} is mostly $\nu_{\mu}$ $CCQE$ but this particular
final state can also come about, for example, by $\nu_{\mu}$ resonance
production followed by intranuclear pion absorption. 
\item $\nu_{\mu}$ $CCQE$ yields mostly \{1$\mu^{-}$, 0$\pi$\} final
states but, occasionaly, can yield \{1$\mu^{-}$, 1$\pi$\} if the
recoil nucleon re-interacts. 
\item $NC$1$\pi^{0}$ final states can be caused by all 

\begin{enumerate}
\item NC elastic followed by nucleon rescattering,
\item NC resonance neutrino-production, 
\item NC non-resonance background, 
\item low-W NC DIS,
\item NC coherent scattering. 
\end{enumerate}

Each such $NC$1$\pi^{0}$ source contributes differently to the observed
pion momentum distribution.

\end{enumerate}

\subsection{Event cherry-picking}

\subsubsection{The \textit{gevpick} cherry-picking utility}

\subsubsection*{Name}

\textit{gevpick} - Reads a list of GENIE event files (GHEP format),
`cherry-picks' events with a given topology and writes them out in
a separate file. The output tree contains two additional branches
to aid book-keeping by maintaining a `link' to the source location
of each cherry-picked event. For each such event we store a) the name
of the original file and b) its original event number.

\subsubsection*{Source}

The source code for this application is in \texttt{\textbf{\small{}`\$GENIE}}\textit{/src/stapp/gEvPick.cxx}'

\subsubsection*{Synopsis}

\texttt{\textbf{\small{}gevpick }}{\small \par}

\texttt{\textbf{\small{}-i input\_file\_list}}{\small \par}

\texttt{\textbf{\small{}-t cherry\_picked\_topology}}{\small \par}

\texttt{\textbf{\small{}{[}-o output\_file\_name{]}}}~\\
\texttt{\textbf{\small{}}}~\\
where {[}{]} denotes an optional argument.\texttt{\textbf{\small{} }}{\small \par}

\subsubsection*{Description}

The following options are available: \\
\\
\textbf{\textcolor{magenta}{-i}}\textbf{ }\textbf{\small{}Specifies
the input file(s).} 

Wildcards accepted, eg \texttt{\textbf{\small{}`-i ``/data/genie/pro/gntp.{*}.ghep.root'''}}.\\
\textbf{}\\
\textbf{\textcolor{magenta}{-t }}\textbf{\small{}Specifies the event
topology to cherry-pick. }The event topology to cherry-pick can be
any of the following strings:
\begin{itemize}
\item \texttt{\textbf{\small{}`all'}}: Selet all events (basically merges
all files into one)
\item \texttt{\textbf{\small{}`numu\_cc\_1pip'}}: Selects $\nu_{\mu}$ $CC$
events with 1 $\pi^{+}$ (and no other pion) in final state.
\item \texttt{\textbf{\small{}`numu\_cc\_1pi0'}}: Selects $\nu_{\mu}$ $CC$
events with 1 $\pi^{0}$ (and no other pion) in final state.
\item \texttt{\textbf{\small{}`numu\_cc\_1pim'}}: Selects $\nu_{\mu}$ $CC$
events with 1 $\pi^{-}$ (and no other pion) in final state.
\item \texttt{\textbf{\small{}`numu\_nc\_1pip'}}: Selects $\nu_{\mu}$ $NC$
events with 1 $\pi^{+}$ (and no other pion) in final state.
\item \texttt{\textbf{\small{}`numu\_nc\_1pi0'}}: Selects $\nu_{\mu}$ $NC$
events with 1 $\pi^{0}$ (and no other pion) in final state.
\item \texttt{\textbf{\small{}`numu\_nc\_1pim'}}: Selects $\nu_{\mu}$ $NC$
events with 1 $\pi^{-}$ (and no other pion) in final state.
\item \texttt{\textbf{\small{}`numu\_cc\_hyperon'}}: Selects $\nu_{\mu}$
$CC$ events with at least 1 hyperon ( $\Sigma^{+}$, $\Sigma^{0}$,
$\Sigma^{-}$, $\Lambda^{0}$, $\Xi^{0}$, $\Xi^{-}$, $\Omega^{-}$)
in the final state.
\item \texttt{\textbf{\small{}`numubar\_cc\_hyperon'}}: Selects $\bar{\nu_{\mu}}$
$CC$ events with at least 1 hyperon in the final state.
\item \texttt{\textbf{\small{}`cc\_hyperon'}}: Selects $CC$ events with
at least 1 hyperon in the final state.
\end{itemize}
\textbf{\textcolor{magenta}{-o}}\textbf{ }\textbf{\small{}Specifies
the output file name.}{\small{} This in an optional argument. If unset,
the output file name will be constructed as: }\textit{\small{}`gntp.<topology>.ghep.root'}{\small{}.}{\small \par}

\subsubsection*{Examples}
\begin{enumerate}
\item Read all events in all `/data/pro2010a/{*}.ghep.root' files and cherry-pick
$\nu_{\mu}$ $NC$1$\pi^{0}$ events:\\
\\
\texttt{\textbf{\small{}\$ gevpick -i ``/data/pro2010a/{*}ghep.root''
-t numu\_nc\_1pi0}}~\\
\\
The cherry-picked event sample gets saved in the `\textit{gntp.numu\_nc\_1pi0.ghep.root}'
file output (default name)
\end{enumerate}

\subsubsection{Cherry-picking a new topology}

More topologies can be trivially added. Please send your request to
the GENIE authors.

\chapter{Using a Realistic Flux and Detector Geometry \label{cha:UsingFluxAndGeo}}

\section{Introduction}

The main task of GENIE is to simulate the complex physics processes
taking place when a neutrino is scattered off a nuclear target. The
generator employs advanced, heavily validated models to describe the
primary scattering process, the neutrino-induced hadronic multiparticle
production and the intra-nuclear hadron transport and re-scattering.

Event generation for realistic experimental setups presents neutrino
generators with additional computational challenges. The physics generator
is required to handle a large number of nuclear targets (ranging from
as light as $H^{1}$ to as heavy as $Pb^{208}$). Moreover, when simulating
neutrino interactions in detectors (such as the JPARC and NuMI near
detectors) exposed to a non-uniform neutrino flux changing rapidly
across the detector volume, it is particularly important to take into
account both the detailed detector geometry and the spatial dependencies
of the flux. This ensures the proper simulation of backgrounds and
avoids introducing highly non-trivial MC artifacts.

The GENIE framework provides many off-the-shelf components for simulating
neutrino interactions in realistic experimental setups. New components,
encapsulating new neutrino fluxes or detector geometry descriptions,
can be trivially added and seamlessly integrated with the GENIE neutrino
interaction physics descriptions.

\section{Components for building customized event generation applications}

GENIE provides off-the-shelf components for generating neutrino interactions
under the most realistic assumptions integrating the state-of-the-art
GENIE neutrino interaction modeling with detailed flux and detector
geometry descriptions. GENIE provides an event generation driver class,
\textit{GMCJDriver}, that can be used to setup complicated Monte Carlo
jobs involving arbitrarily complex, realistic beam flux simulations
and detector geometry descriptions. These flux descriptions are typically
derived from experiment-specific beam-line simulations while the detector
geometry descriptions are typically derived from CAD engineering drawings
mapped into the Geant4, ROOT or GDML geometry description languages.
Obviously, flux and detector geometry descriptions can take many forms,
driven by experiment-specific choices. GENIE standardizes the geometry
navigation and flux driver interfaces. These interfaces define a)
the operations that GENIE needs to perform on the geometry and flux
descriptions and b) the information GENIE needs to extract from these
in order to generate events. 

Concrete implementations of these interfaces are loaded into the GENIE
event generation drivers, extending GENIE event generation capabilities
and allow it to seamlessly integrate new geometry descriptions and
beam fluxes.

\subsection{{\normalsize{}The flux driver interface}}

In GENIE every concrete flux driver implements the \textit{GFluxI}
interface. The interface defines what neutrino flux information is
needed by the event generation drivers and how that information is
to be obtained. Each concrete flux driver implements the following
methods.
\begin{itemize}
\item \textit{const PDGCodeList \& GFluxI::FluxParticles (void) }\\
Declare the list of flux neutrinos that can be generated. This information
is used for initialization purposes, in order to construct a list
of all possible initial states in a given event generation run. 
\item \textit{double GFluxI::MaxEnergy (void) }\\
Declare the maximum energy. Again this information is used for initialization
purposes, in order to calculate the maximum possible interaction probability
in a given event generation run. Since neutrino interaction probabilities
are tiny and in order to boost the MC performance, GENIE scales all
interaction probabilities in a particular event generation run so
that the maximum possible interaction probability is 1. That maximum
interaction probability corresponds to the total interaction probability
(summed over nuclear targets and process types) for a maximum energy
neutrino following a trajectory that maximizes the density-weighted
path-lengths for each nuclear target in the geometry. GENIE adjusts
the MC run normalization accordingly to account for that internal
weighting. 
\item \textit{bool GFluxI::GenerateNext (void)} \\
Generate a flux neutrino and specify its pdg code, its weight (if
any), its 4-momentum and 4-position. The 4-position is given in the
detector coordinate system (as specified by the input geometry). Each
such flux neutrino is propagated towards the detector geometry but
is not required to cross any detector volume. GENIE will take that
neutrino through the geometry, calculate density-weighted path-lengths
for all nuclear targets in the geometry, calculate the corresponding
interactions probability off each nuclear target and decide whether
that flux neutrino should interact. If it interacts, an appropriate
\textit{GEVGDriver} will be invoked to generate the event kinematics. 
\item \textit{int GFluxI::PdgCode (void)} \\
Returns the PDG code of the flux neutrino generated by the most recent
\textit{GFluxI::GenerateNext (void)} call.
\item \textit{double GFluxI::Weight (void) }\\
Returns the weight of the flux neutrino generated by the most recent
\textit{GFluxI::GenerateNext (void)} call.
\item \textit{const TLorentzVector \& GFluxI::Momentum (void) }\\
Returns the 4-momentum of the flux neutrino generated by the most
recent \textit{GFluxI::GenerateNext (void)} call.
\item \textit{const TLorentzVector \& GFluxI::Position (void)} \\
Returns the position 4-vector of the flux neutrino generated by the
most recent \textit{GFluxI::GenerateNext (void)} call.
\item \textit{bool GFluxI::End(void)} \\
Notify that no more flux neutrinos can be thrown. This flag is typically
raised by flux drivers that simply read-in beam-line simulation outputs
(as opposed to run the beam simulation code on the fly) so as to notify
GENIE that the end of the neutrino flux file has been reached (after,
probably, having been recycled N times). The flag allows GENIE to
properly terminate the event generation run at the end-of-flux-file
irrespective of the accumulated number of events, protons on target,
or other metric of exposure.
\end{itemize}
The above correspond the the common set of operations /information
that GENIE expects to be able to perform / extract from all concrete
flux drivers. Specialized drivers may define additional information
that can be utilized in the experiment-specific event generation drivers.
One typical example of this is the flux-specific pass-through information,
that is information about the flux neutrino parents such as the parent
meson PDG code, its 4-momentum its 4-position at the production and
decay points that GENIE simply attaches to each generated event and
passes-through so as to be used in later analysis stages.

\subsection{{\normalsize{}The geometry navigation driver interface}}

In GENIE every concrete geometry driver implements the \textit{GeomAnalyzerI}
interface. The interface specifies what information about the input
geometry is relevant to the event generation and how that information
is to be obtained. Each concrete geometry driver implements methods
to
\begin{itemize}
\item \textit{const PDGCodeList \& GeomAnalyzerI::ListOfTargetNuclei (void)}\\
Declare the list of target nuclei that can be found in the geometry.
This information is used for initialization purposes, in order to
construct a list of all possible initial states in a given event generation
run. 
\item \textit{const PathLengthList \& GeomAnalyzerI::ComputeMaxPathLengths
(void)}\\
Compute the maximum density-weighted path-lengths for each nuclear
target in the geometry. Again, this is information used for initialization
purposes. The computed `worst-case' trajectory is used to calculate
the maximum possible interaction probability in a particular event
generation run which is being used internally to normalize all computed
interaction probabilities. 
\item \textit{const PathLengthList \& GeomAnalyzerI::ComputePathLengths
(const TLorentzVector \& x, const TLorentzVector \& p)}\\
Compute density-weighted path-lengths for all nuclear targets, for
a `ray' of a given 4-momentum and starting 4-position. This allows
GENIE to calculate probabilities for each flux neutrino to be scattered
off every nuclear target along its path through the detector geometry. 
\item \textit{const TVector3 \& GeomAnalyzerI::GenerateVertex (const TLorentzVector
\& x, const TLorentzVector \& p, int tgtpdg)}\\
Generate a vertex along a `ray' of a given 4-momentum and starting
4-position on a volume containing a given nuclear target. This allows
GENIE to place a neutrino interaction vertex within the detector geometry
once an interaction of a flux neutrino off a selected nuclear target
has been generated. 
\end{itemize}

\subsection{Setting-up GENIE MC jobs using fluxes and geometries}

\begin{verbatim}
{
 ...

 // get flux driver 
 GFluxI * flux_driver = new ... ; 
 
 // get geometry driver 
 GeomAnalyzerI * geom_driver = new ... ;
 
 // create the GENIE monte carlo job driver 
 GMCJDriver* mcjob_driver = new GMCJDriver; 
 mcjob_driver->UseFluxDriver(flux_driver); 
 mcjob_driver->UseGeomAnalyzer(geom_driver); 
 mcjob_driver->Configure(); 

...
}
\end{verbatim}

\section{Built-in flux drivers}

GENIE currently contains a host of concrete flux drivers that allow
GENIE to be used in many realistic, experiment-specific situations:
\begin{itemize}
\item \textit{GJPARCNuFlux}: An interface to the JPARC neutrino beam simulation
\cite{JNUBEAM:PrivCom} used at SK, nd280, and INGRID. 
\item \textit{GNuMIFlux}: An interface to the NuMI beam simulations \cite{GNuMi:PrivCom}
used at MINOS, NOvA, MINERvA and ArgoNEUT. 
\item \textit{GBartolAtmoFlux}: A driver for the BGLRS atmospheric flux
by G. Barr, T.K. Gaisser, P. Lipari, S. Robbins and T. Stanev \cite{Barr:2004br}.
\item \textit{GFlukaAtmo3DFlux}: A driver for the FLUKA 3-D atmospheric
neutrino flux by A. Ferrari, P. Sala, G. Battistoni and T. Montaruli
\cite{Battistoni:2001sw}.
\item \textit{GAstroFlux}: A driver for astrophysical neutrino fluxes. Handles
both diffuse fluxes and point sources. (Under development.)
\item \textit{GCylindTH1Flux}: A generic flux driver, describing a cylindrical
neutrino flux of arbitrary 3-D direction and radius. The radial dependence
of the neutrino flux is configurable (default: uniform per unit area).
The flux driver may be used for describing a number of different neutrino
species whose (relatively normalised) energy spectra are specified
as ROOT 1-D histograms. This driver is being used whenever an energy
spectrum is an adequate description of the neutrino flux. 
\item \textit{GSimpleNtpFlux}: An interface for a simple ntuple-based flux
that can preserve energy-position correlations without the format
being tied to any particular experimental setup (though individual
files are very much so).
\item \textit{GMonoEnergeticFlux}: A trivial flux driver throwing mono-energetic
flux neutrinos along the +z direction. More that one neutrino species
can be included, each with its own weight. The driver is being used
in simulating a single initial state at a fixed energy mainly for
probing, comparing and validating neutrino interaction models. 
\end{itemize}
New concrete flux drivers (describing the neutrino flux from other
beam-lines) can be easily developed and they can be effortlessly and
seamlessly integrated with the GENIE event generation framework.

\subsection{JPARC neutrino flux driver specifics}

\textit{GJPARCNuFlux} provides an interface to the JPARC neutrino
beam simulations (JNUBEAM \cite{JNUBEAM:PrivCom}) used at SK, nd280,
and INGRID.

{[}expand{]}

\subsection{NuMI neutrino flux driver specific}

\textit{GNuMIFlux} provides an interface to the NuMI beam simulations
used at MINOS, NOvA, MINERvA and ArgoNeut. This interface can handle
all three of the formats used so far in simulating the NuMI beamline:
Geant3-based gnumi, g4numi and flugg. It can also handle the FNAL
booster flux when that is formatted into one of the standard ntuple
layouts. These beam simulation files record hadron decays and sufficient
information to calculate new weights and energies for different positions
relative to the beam orgin. 

The driver generates a flux to cover a user specified detector \textquotedbl{}window\textquotedbl{}
after undergoing a coordinate transformation from the beam system
to that of a particular detector. The detector specific windows and
transformations are encapsulated in the `\texttt{\textbf{\small{}\$GENIE}}/\textit{src/FluxDriver/GNuMINtuple/GNuMIFlux.xml}'
file. Users can extend what is available by modifying this file and
putting a copy in a location specified by \texttt{\textbf{\small{}GXMLPATH}}=\textquotedbl{}\textit{/path/to/location}\textquotedbl{}.
Additional \textquotedbl{}\textit{param\_set}\textquotedbl{} sections
allow new configurations and these can be based on modifications of
select parameters of an existing \textquotedbl{}param\_set\textquotedbl{}
entry. Extensive documentation of the setable parameters can be found
in the XML file itself.

When the \textit{GNuMIFlux} is invoked it must be configured by passing
the method \textit{GNuMIFlux::LoadBeamSimData()} an input filename
string and a config name. The input file name may include wildcards
on the file name but not the directory path. The config name selects
a \textquotedbl{}param\_set\textquotedbl{} from the XML file. The
\textit{GNuMIFlux} object will by default declare the list of flux
neutrinos that it finds in the input files; this can be overridden
to have it ignore entries for flavors the user is not interested in.

\subsection{FLUKA and BGLRS atmospheric flux driver specifics \label{sub:AtmoFluxDrivers}}

\textit{GFlukaAtmo3DFlux} and \textit{GBartolAtmoFlux} provide, respectivelly,
an interface to the FLUKA-3D (A. Ferrari, P. Sala, G. Battistoni and
T. Montaruli \cite{Battistoni:2001sw}) and BGLRS (G. Barr, T.K. Gaisser,
P. Lipari, S. Robbins and T. Stanev \cite{Barr:2004br}) atmospheric
neutrino flux simulations. 

Both classes inherit all their functionallity from the \textit{GAtmoFlux
}base class from which they derive. \textit{GFlukaAtmo3DFlux} and
\textit{GBartolAtmoFlux} merely define the appropriate binning for
each flux simulation: The FLUKA flux is given in 40 bins of $cos\theta$,
where $\theta$ is the zenith angle, from -1 to 1 (bin width = 0.05)
and 61 equally log-spaced energy bins (20 bins per decade) with a
minimum energy of 100 MeV. The BGLRS flux is given in 20 bins of $cos\theta$
from -1 to 1 (bin width = 0.1) and and 30 equally log-spaced energy
bins (10 bins per decade) with a minimum energy of 10 GeV. For more
details please visit the FLUKA\footnote{\textit{http://pcbat1.mi.infn.it/\textasciitilde{}battist/neutrino.html}}
and BGLRS\footnote{\textit{http://www-pnp.physics.ox.ac.uk/\textasciitilde{}barr/fluxfiles/}}
flux web sites. 

Both the FLUKA and BGLRS flux simulations are distributed as ascii
data files for various locations and solar activity levels. There
is one data file per atmospheric neutrino flavor. You can specify
the input files for each neutrino flavor using the `\textit{void GAtmoFlux::SetFluxFile(int
neutrino\_code, string filename)' }method. The expected input code
is the PDG one and the input filename should include the full path
to the file. You can specify flux files for an arbitrary set of flux
neutrino flavors. Neutrino flavors for which you have not specified
a flux file will be omitted from the atmospheric neutrino event generation
job. Once you have specified flux files for all neutrino flavors you
wish to include you need to call the\textit{ `void GAtmoFlux::LoadFluxData()'
}method\textit{.}

\begin{figure}
\includegraphics[width=0.9\textwidth]{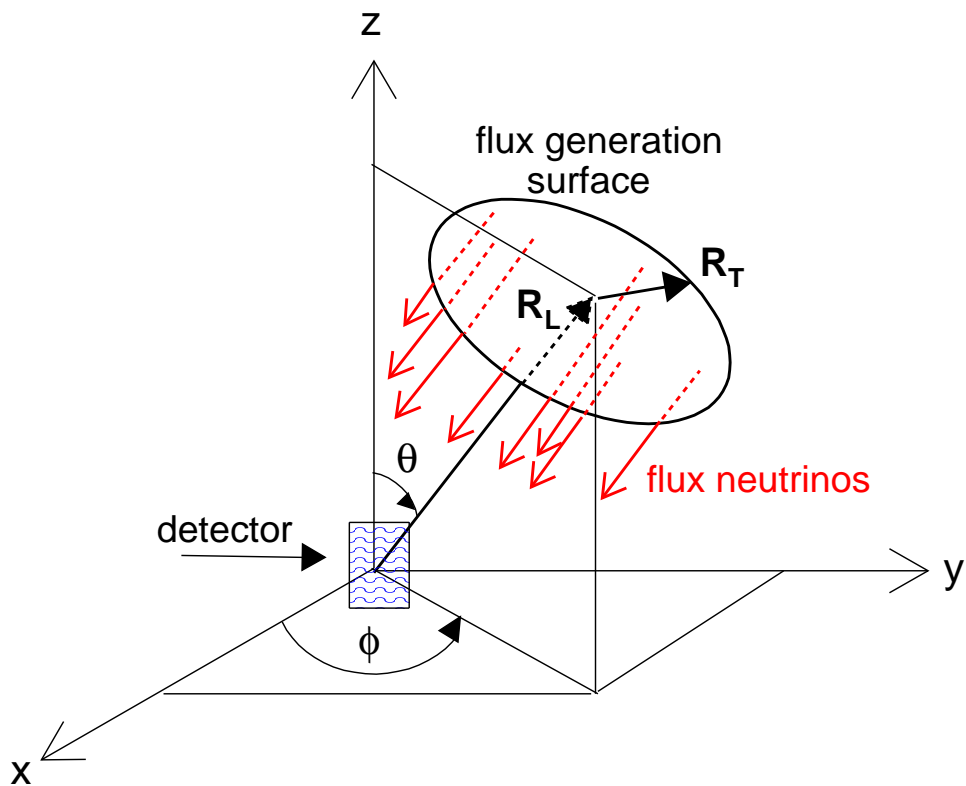} 

\caption{Construction of flux generation surface for the atmospheric neutrino
flux drivers. For a given direction, determined by the zenith angle
$\theta$ and azimuth angle $\phi$, the flux generation surface is
a circular area, with radius $R_{T}$, which is tangent to a sphere
of radius $R_{L}$ centered at the coordinate system origin. $R_{T}$
and $R_{L}$ must be appropriately chosen so that the flux generation
surface is always outside the input geometry volumes and so that,
for every given direction, the `shadow' of the generation surface
covers the entire geometry. See text for more details. \label{fig:AtmoFluxGenerationSurface}}
\end{figure}

By default, the flux neutrino position and momentum 4-vectors are
generated in the Topocentric Horizontal Coordinate System (+z: Points
towards the local zenith / +x: On same plane as local meridian, pointing
south . +y: As needed to make a right-handed coordinate system / Origin:
Input geometry centre). A rotation to a user-defined topocentric coordinate
system can be enabled by invoking the `\textit{void GAtmoFlux::SetUserCoordSystem
(TRotation \&)}' method. For a given direction, determined by the
zenith angle $\theta$ and azimuth angle $\phi$, the flux generation
surface is a circular area, with radius $R_{T}$, which is tangent
to a sphere of radius $R_{L}$ centered at the coordinate system origin.
These two radii can be set using the `\textit{void GAtmoFlux::SetRadii
(double RL, double RT)}' method. Obviously, $R_{T}$ and $R_{L}$
must be appropriately chosen so that the flux generation surface is
always outside the input geometry volume and so that, for every given
direction, the `shadow' of the generation surface covers the entire
geometry (see Fig. \ref{fig:AtmoFluxGenerationSurface}).

Energy cuts can be specified using the `\textit{void GAtmoFlux::ForceMinEnergy(double
Emin)' }and `\textit{void GAtmoFlux::ForceMaxEnergy(double Emax)'
}methods. Finally, the atmospheric neutrino flux drivers can generate
both weighted and unweighted flux neutrinos (with the unweighted-mode
used as default). In the weighted-mode the energy is generated logarithmically
and the zenith angle cosine is generated uniformly and, after a neutrino
species has been selected, the event weight is set to be the flux
histogram bin content for the given neutrino species and for the given
energy and zenith angle cosine. Using a weighted-mode may be . The
user choice can be registered using the `\textit{void GAtmoFlux::GenerateWeighted(bool
option)' }method.

\subsection{Generic histogram-based flux specifics }

\begin{figure}
\includegraphics[width=0.9\textwidth]{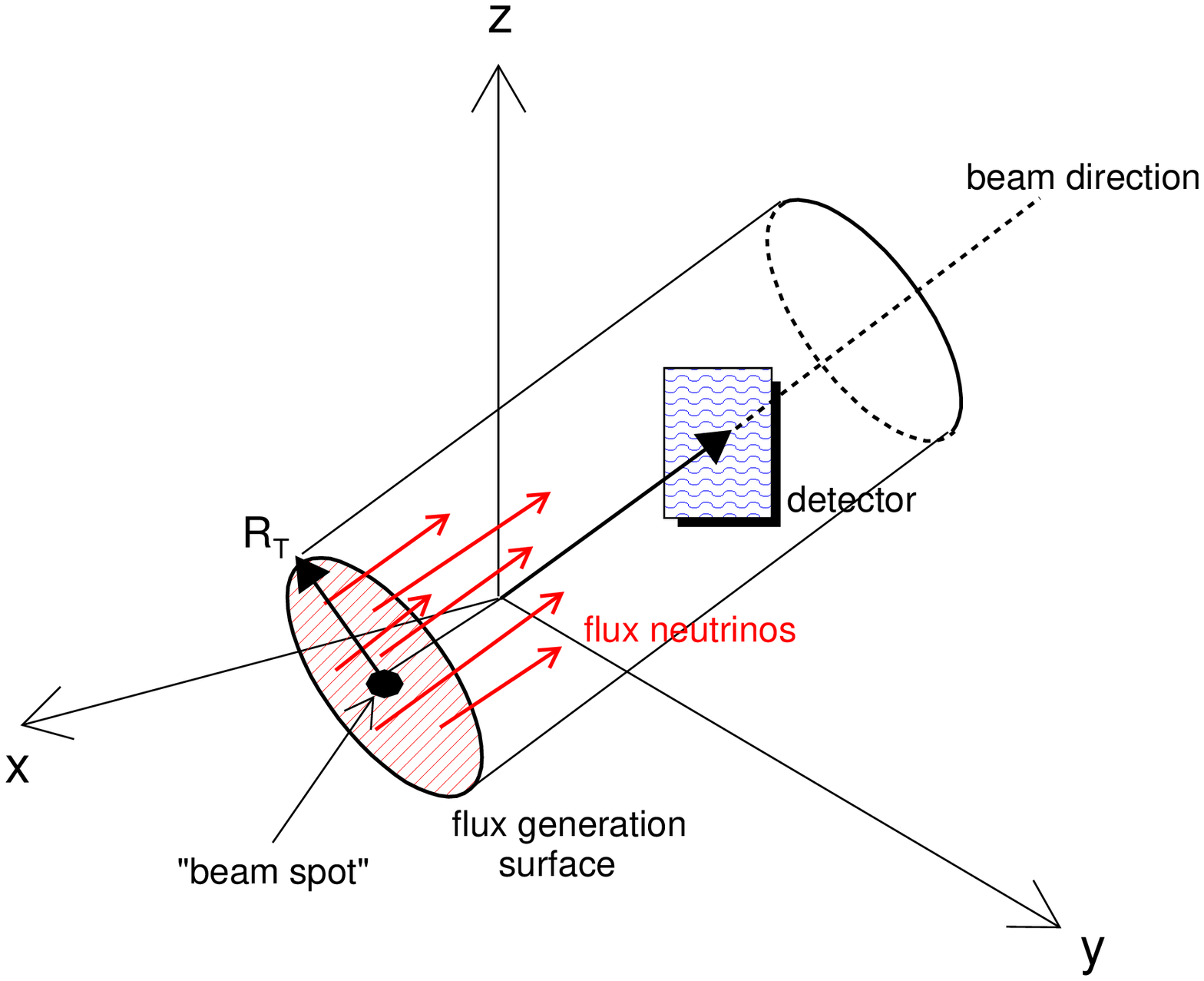}

\caption{Geometrical setup for the \textit{GCylindTH1Flux} flux drivers. The
diriver allows you to set the beam direction (in 3-D), and the radius
$R_{T}$ of the flux generation surface. To fully specify the position
of the flux generation surface in 3-D the driver allows you to set
the `beam spot' 3-vector. Additionally the $R_{T}$-dependence can
be configured. Multiple neutrino species can be generated in the flux
surface, each one with its own energy distribution and relative normalization.
See text for more details. \label{fig:GCylFluxGeom}}
\end{figure}

The \textit{GCylindTH1Flux} is generic flux driver, describing a cylindrical
neutrino flux of arbitrary 3-D direction and radius. The direction
of the flux rays (in 3-D) can be specified using the \textit{`void
GCylindTH1Flux::SetNuDirection(const TVector3 \&)'} method while the
radius of the cylinder is specified using \textit{`void GCylindTH1Flux::SetTransverseRadius(double)}'.
The flux generation surface is a circular area defined by the intersection
of the flux cylinder with a plane which is perpendicular to the flux
ray direction. To fully specify the flux neutrino generation surface
the user needs to specify the centre of that circular area (see `beam
spot' in Fig. \ref{fig:GCylFluxGeom}) using the \textit{`void GCylindTH1Flux::SetBeamSpot(const
TVector3 \& spot)}' method. Obviously the `beam spot' should be placed
upstream of the detector volume. 

The radial dependence of the neutrino flux can be configured using
the \textit{`void GCylindTH1Flux::SetRadialDependence(string rdep)'}
method. The expected input is the functional form of the $R_{T}$-dependence
(with $R_{T}$ denoted as $x$). By default, the driver is initialized
with \textit{SetRadialDependence}(``x''), so flux neutrinos are
generated uniformly per unit area.

The flux driver may be used for describing a number of different neutrino
species whose (relatively normalised) energy spectra are specified
as ROOT 1-D histograms (\textit{TH1D}). To input the energy distribution
of each neutrino species use \textit{GCylindTH1Flux::AddEnergySpectrum
(int nu\_pdgc, TH1D {*} spectrum)'}. 

Obviously, when using \textit{GCylindTH1Flux}, no energy-position
correlation is present. This may or may-not be a good approximation
depending on the specifics of your experimental setup and analysis.
If energy-position correlation is important (and known) then consider
using the \textit{GSimpleNtpFlux} flux driver. This correlation is
also built-in in the specialized JPARC, NuMI and atmospheric flux
drivers, described in this chapter, which you should be utilizing
if relevant to your application. 

\subsection{Generic ntuple-based flux specifics}

The \textit{GSimpleNtpFlux} flux driver provides an interface for
a simple ntuple-based flux that can preserve energy-position correlations
without the format being tied to any particular experimental setup
(though individual files are very much so). The basic entry consists
a TTree branch with the elements: 
\begin{itemize}
\item \textbf{px}, \textbf{py}, \textbf{pz}, \textbf{E}: 4-momentum components.
\item \textbf{vtxx}, \textbf{vtxy}, \textbf{vtxz}: Neutrino ray origin info
(detector coordinates).
\item \textbf{dist}: Distance from hadron decay to ray origin. 
\item \textbf{wgt}: Neutrino weight (generally 1.0). 
\item \textbf{metakey}: Reference back to meta-data. The \textquotedbl{}metadata\textquotedbl{}
branch has an entry per file recording general info such as the list
of neutrino flavors found in the entries, the number of protons-on-target
represented by the file (in the case of accelerator based fluxes),
the maximum energy, the minimum and maximum weights, the flux window
and a vector of strings for a record of the list of files used to
generate the GSimpleNtp file.
\end{itemize}
Additional information can be stored in conjunction with the individual
entries either by supplemental classes for branches (ala the optional
\textquotedbl{}numi\textquotedbl{} branch), or via the flexible \textquotedbl{}aux\textquotedbl{}
branch which allows arbitrary vectors of integers and doubles (name
info in the metadata allows for keeping track of what elements represent
under the assumption that all entries have identical additions).

When the \textit{GSimpleNtpFlux} is invoked it needs to be configured
by passing the method \textit{GSimpleNtpFlux::LoadBeamSimData()} an
input filename string (and a config name that is ignored). The input
file name may include wildcards on the file name but not the directory
path. Multiple gsimple flux files can also be combined into a larger
file with the use of the ROOT \textit{hadd} utility.

The \textit{GSimpleNtpFlux} is in use by some NuMI experiments as
a means of factorizing the computation necessary for the evaluation
of the \textit{GNuMIFlux} from the actual event generation. Unlike
the \textit{GNuMIFlux} files, entries can not be positioned for new
locations (which would change the entry's weight and energy) but they
also don't require the computational burden of doing so. They are
meant to be simple and fast.

\section{Built-in geometry navigation drivers}

GENIE currently contains two concrete geometry drivers which are sufficient
for all event generation cases encountered so far:
\begin{itemize}
\item \textit{ROOTGeomAnalyzer}: A geometry driver handling detector geometries
specified using ROOT. As detector geometries specified using Geant4
or \textit{GDML} can be converted into ROOT geometries, this driver
is being used in all cases where a detailed detector geometry is being
passed on to GENIE. 
\item \textit{PointGeomAnalyzer}: A trivial geometry corresponding to a
single nuclear target or a target mix (a set of nuclear targets each
with its corresponding weight fraction) at a fixed position. This
driver is being used to simulate only given initial states as a means
for probing the neutrino interaction physics modeling or in experimental
situations where the detector is being illuminated by a spatially
uniform neutrino beam and where the generated interaction vertices
do not have any spatial dependence and can be generated uniformly
within volumes of given nuclear targets. 
\end{itemize}

\subsection{ROOT geometry navigation driver specifics}

The \textit{ROOTGeomAnalyzer} works based on a probing a detailed
ROOT geometry to evaluate the mass distribution seen along individual
neutrino `rays' (a starting position in space relative to the detector
geometry and a direction). Each ray is stepped through the geometry
from one volume boundary to the next; each transition to a new volume
instantiates a new \textit{PathSegment}, which are collected into
a \textit{PathSegmentList} for the ray and which also includes information
about the ray itself.

A \textit{PathSegment} object records the information about the distance
from the ray origin to the entrance of the volume, the step length
in the volume, information about the volume (e.g. medium, material),
positions at the boundaries, and (optionally) the ROOT volume path
string (the volume hierarchy in the geometry). A neutrino ray from
the flux is passed through the geometry only once. From the information
recorded in the \textit{PathSegmentList} the \textit{GMCJDriver} can
be given the density-weighted path-lengths for all the nuclear targets.
If the \textit{GMCJDriver} decides that an interaction occurred this
\textit{PathSegmentList} is then used to properly select a vertex
position based on the chosen nuclear target.

\subsubsection{Defining units}

The \textit{ROOTGeomAnalyzer} can be configured to account for differences
in length and density units between the GENIE defaults and what is
assumed in the ROOT geometry.

\subsubsection{Defining a fiducial volume}

For ROOT geometries that include representations of material that
isn't of interest, such as the rock surrounding a cavern hall, the
\textit{ROOTGeomAnalyzer::SetTopVolName()} method allows one to consider
only the material within that volume. In more sophisticated circumstances
there might not be a volume in the ROOT geometry representing the
region in which one wants to restrict vertices. More refined limits
can be placed by configuring the \textit{ROOTGeomAnalyzer} with a
concrete implementation of the \textit{GeomVolSelectorI} interface.

A concrete implementation of the \textit{GeomVolSelectorI} interface
must provide a method for \textquotedbl{}trimming\textquotedbl{} individual
PathSegment items based on information in the segment. Trimming futher
restricts the region of the step within the volume; ranges delineate
sub-steps and by this means segments within a volume can be reduced,
split or eliminated. The implementation must also provide methods
that gets called at the start of \textit{PathSegmentList} trimming
and upon completion (these can be dummies). If the implementation
needs to know the ROOT geometry volume path hierarchy then it must
signal that.

Two useful examples of \textit{GeomVolSelectorI} are provided: \textit{GeomVolSelectorBasic}
and \textit{GeomVolSelectorFiducial}. The basic class is configurable
to select or reject whole segments based on the volume name, medium,
material and (optionally) volume path string. The fiducial class builds
on that base and add the potential for defining a elementary shape
(sphere, cylinder, box, convex polyhedron) in space that is used to
trim segments. This shape does not have to correspond to anything
represented in the ROOT geometry. The cut can be to require considering
only material within the shape or only that outside of the shape.

\section{Built-in specialized event generation applications}

This section discusses specialized GENIE-based event generation applications
included in GENIE distributions. These applications integrate the
GENIE event generation modules with very specific neutrino flux and
detector geometry descriptions. 
\begin{itemize}
\item \textit{gevgen\_t2k}: A GENIE-based event generation application for
T2K. It integrates GENIE with the JPARC neutrino beam-line simulation
(JNUBEAM) and the geometry descriptions of nd280, 2km, INGRID and
Super-K detectors. (See subsection \ref{sub:gT2Kevgen}.)
\item \textit{gevgen\_fnal}: A GENIE-based event generation application
for the Fermilab experiments (including DUNE, and the experiments
in the NuMI and Booster beam-lines). It integrates GENIE with the
Fermilab neutrino beam-line simulations and the geometry descriptions
of MINOS, NOvA, MINERvA, SBND, MicroBooNE, DUNE and other detectors.
(See subsection \ref{sub:gNuMIevgen}.)
\item \textit{gevgen\_atmo}: A GENIE-based atmospherc neutrino event generation
application. It integrates the GENIE with any of the FLUKA 3-D \cite{Battistoni:2001sw}
or BGLRS \cite{Barr:2004br} atmospheric neutrino flux simulations.
Events can be generated for either a simple target mix or a detailed
ROOT-based detector geometry (See subsection \ref{sub:gevgen_atmo}.)
\end{itemize}
Although the above applications have common options, each of the following
subsections is entirely self-contained. Please go directly to the
subsection describing the application you are interested at.

\subsection{Event generation application for the T2K experiment \label{sub:gT2Kevgen}}

\subsubsection*{Name}

\textit{gevgen\_t2k} -- A GENIE-based event generation application
for T2K. It integrates GENIE with the JPARC neutrino beam-line simulation
(JNUBEAM) and the detector geometry descriptions of nd280, 2km, INGRID
and Super-K.

\subsubsection*{Source and build options}

The source code for this application is in \texttt{\textbf{\small{}`\$GENIE}}\textit{/src/support/t2k/EvGen/gT2KEvGen.cxx}'.\\
To enable it add `\texttt{\textbf{\small{}-{}-enable-t2k}}' during
the GENIE build configuration step.

\subsubsection*{Synopsis}

\texttt{\textbf{\small{}\$ gevgen\_t2k }}{\small \par}

\texttt{\textbf{\small{}-f }}\texttt{\small{}flux}\texttt{\textbf{\small{}
{[}-p }}\texttt{\small{}POT\_normalization\_of\_flux\_file}\texttt{\textbf{\small{}{]}
{[}-R{]}}}{\small \par}

\texttt{\textbf{\small{}-g }}\texttt{\small{}geometry}\texttt{\textbf{\small{}
{[}-t }}\texttt{\small{}geometry\_top\_volume\_name}\texttt{\textbf{\small{}{]} }}{\small \par}

\texttt{\textbf{\small{}{[}-m }}\texttt{\small{}max\_path\_lengths\_xml\_file}\texttt{\textbf{\small{}{]} }}{\small \par}

\texttt{\textbf{\small{}{[}-P{]} }}\texttt{\small{}{[}pre\_gen\_flux\_prob\_name{]}}{\small \par}

\texttt{\textbf{\small{}{[}-S{]} }}\texttt{\small{}{[}output\_pre\_gen\_flux\_prob\_name{]}}{\small \par}

\texttt{\textbf{\small{}{[}-L }}\texttt{\small{}geometry\_length\_units}\texttt{\textbf{\small{}{]}
{[}-D }}\texttt{\small{}geometry\_density\_units}\texttt{\textbf{\small{}{]} }}{\small \par}

\texttt{\textbf{\small{}<-n }}\texttt{\small{}num\_of\_events}\texttt{\textbf{\small{},
-c }}\texttt{\small{}num\_of\_flux\_ntuple\_cycles}\texttt{\textbf{\small{},
-e, -E }}\texttt{\small{}exposure\_in\_POTs}\texttt{\textbf{\small{}> }}{\small \par}

\texttt{\textbf{\small{}{[}-o }}\texttt{\small{}output\_event\_file\_prefix}\texttt{\textbf{\small{}{]}
{[}-r}}\texttt{\small{} run\#}\texttt{\textbf{\small{}{]} }}{\small \par}

\texttt{\textbf{\small{}{[}-seed }}\texttt{\small{}random\_number\_seed}\texttt{\textbf{\small{}{]}
{[}-{}-cross-section }}\texttt{\small{}xml\_file}\texttt{\textbf{\small{}{]}
{[}-{}-event-generator-list }}\texttt{\small{}list\_name}\texttt{\textbf{\small{}{]}}}{\small \par}

\texttt{\textbf{\small{}{[}-{}-message-thresholds }}\texttt{\small{}xml\_file}\texttt{\textbf{\small{}{]}
{[}-{}-unphysical-event-mask }}\texttt{\small{}mask}\texttt{\textbf{\small{}{]}
{[}-{}-event-record-print-level }}\texttt{\small{}level}\texttt{\textbf{\small{}{]}}}{\small \par}

\texttt{\textbf{\small{}{[}-{}-mc-job-status-refresh-rate }}\texttt{\small{}rate}\texttt{\textbf{\small{}{]}
{[}-{}-cache-file }}\texttt{\small{}root\_file}\texttt{\textbf{\small{}{]}}} 

\texttt{\textbf{\small{}{[}-h{]}}} \\
\\
where {[}{]} denotes an optional argument and <> denotes a group of
arguments out of which only one can be set.

\subsubsection*{Description}

The following options are available: \\
\\
\textbf{\textcolor{magenta}{-f }}\textbf{Specifies the input neutrino
flux}. This option can be used to specify any of: 
\begin{itemize}
\item A JNUBEAM beam simulation output file and the detector location. The
general sytax is: \\
 \texttt{\textbf{\small{}`-f /path/flux\_file.root,detector\_loc(,neutrino\_list)}}'\\

For more information on the flux ntuples see the JNUBEAM documentation.
The ntuple has to be in ROOT format and can be generated from the
distributed HBOOK ntuples using ROOT's {\em h2root} utility. The
detector location can be any of `sk' or the near detector positions
`nd1',...,`nd6' simulated by JNUBEAM. The optional {\em neutrino\_list}
is a comma separated list neutrino PDG codes. It specifies which neutrino
flux species to to considered in the event generation job. If no such
neutrino list is specified then, by default, GENIE will consider all
neutrino species in the input flux ntuple. When a JNUBEAM ntuple is
used for describing the neutrino flux, GENIE is able to calculate
the POT exposure for the generated event sample and any one of the
exposure setting methods (`\texttt{\textbf{\small{}-e}}', `\texttt{\textbf{\small{}-E}}',
`\texttt{\textbf{\small{}-c}}', `\texttt{\textbf{\small{}-n}}', see
below) can be used.

All JNUBEAM information on the flux neutrino parent (parent PDG code,
parent 4-position and 4-momentum at the production and decay points
etc) is stored in a `flux' branch of the output event tree and is
associated with the corresponding generated neutrino event.

Example 1: \\
 To use the Super-K JNUBEAM flux ntuple from the `\textit{/t2k/flux/jnubeam001.root}'
file, type: \\
 \texttt{\textbf{\small{}`-f /t2k/flux/jnubeam001.root,sk'}} \\
 \\
 Example 2: \\
 To use the 2km flux ntuple {[}near detector position `nd1' in the
jnubeam flux simulation{]} from the `\textit{/t2k/flux/jnubeam001.root}'
file, type: \\
`\texttt{\textbf{\small{}-f /t2k/flux/jnubeam001.root,nd1'}} \\
 \\
 Example 3: \\
 To use the nd280 flux ntuple {[}near detector position `nd5' in the
jnubeam flux simulation{]} from the `\textit{/t2k/flux/jnubeam001.root}'
file, type: \\
 \texttt{\textbf{\small{}`-f /t2k/flux/jnubeam001.root,nd5'}}~\\
 \\
Example 4: \\
 To the same as above but using only the $\nu_{e}$ and $\bar{\nu_{e}}$
flux ntuple entries, type: \\
 \texttt{\textbf{\small{}`-f /t2k/flux/jnubeam001.root,nd5,12,-12'}}
\\

\item A set of flux histograms stored in a ROOT file. The general syntax
is: \\
 \texttt{\textbf{\small{}`-f /path/file.root,neutrino\_code{[}histo{]},...'}}
\\

where {\em neutrino\_code} is a standard neutrino PDG code\footnote{ $\nu_{e}$: 12, $\nu_{\mu}$: 14, $\nu_{\tau}$: 16, ${\bar{\nu}_{e}}$:
-12, ${\bar{\nu}_{\mu}}$: -14 and ${\bar{\nu}_{\tau}}$: -16} and {\em histo} is the corresponding ROOT histogram name.\\

Multiple flux histograms can be specified for different flux neutrino
species (see the example given below). The relative flux normalization
for all neutrino species should be represented correctly at the input
histogram normalization. The absolute flux normalization is not relevant:
Unlike when using JNUBEAM ntuples to describe the flux, no POT calculations
are performed when plain histogram-based flux descriptions are employed.
One can only control the MC run exposure via the number of generated
events (`\texttt{\textbf{\small{}-n}}', see below). In this case the
POT normalization of the generated sample is calculated externally.

Since there is no directional information in histogram-based descriptions
of the flux, the generated neutrino vertex is always set to (0,0,0,0).
Then it is the detector MC responsibility to rotate the interaction
vectors and plant the vertex \footnote{ This option is used only for the Super-K simulation where vertices
are distributed uniformly in volume by the detector MC (SKDETSIM).
For event generation at the more complex near detectors a JNUBEAM
ntuple-based flux description should be used so as the interaction
vertex is properly planted within the input geometry by GENIE. } Obviously no flux pass-through branch is written out in the neutrino
event tree since no such information is associated with flux neutrinos
selected from plain histograms.

Example: \\
 To use the histogram `h1' (representing the $\nu_{\mu}$ flux) and
the histogram `h2' (representing the $\nu_{e}$ flux) from the \textit{`/data/flux.root'}
file, type: \\
 \texttt{\textbf{\small{}`-f /data/flux.root,14{[}h1{]},12{[}h2{]}}}'

\end{itemize}
$ $\\
\textbf{\textcolor{magenta}{-p}}\textbf{\textcolor{magenta}{\Large{}
}}\textbf{Specifies to POT normalization of the input flux file.}
This is an optional argument. By default, it is set to the standard
JNUBEAM flux ntuple normalization of 1E+21 POT/detector (for the near
detectors) or 1E+21 POT/cm2 (for the far detector). The input normalization
factor will be used to interpret the flux weights and calculate the
POT normalization for the generated neutrino event sample. The option
is irrelevant if a simple, histogram-based description of the neutrino
flux is used (see -f option)\\
\textbf{}\\
\textbf{\textcolor{magenta}{-R}}\textbf{\small{} }\textbf{Instructs
the flux driver to start looping over the flux ntuples with a random
offset.} This is an optional argument. It may be necessary on some
occassions to avoid biases when using very large input flux files.\\
\textbf{}\\
\textbf{\textcolor{magenta}{-g}}\textbf{ Specifies the input detector
geometry.} This option can be used to specify any of: 
\begin{itemize}
\item A ROOT file containing a ROOT/Geant4-based geometry description (\textit{TGeoManager}).
\\
This is the standard option for generating events in the nd280, 2km
and INGRID detectors. \\
\\
Example: \\
To use the ROOT detector geometry description stored in the `\textit{/data/geo/nd280.root}'
file, type:\\
 \texttt{\textbf{\small{}`-g /data/geo/nd280.root}}'\\
\\
By default the entire input geometry will be used. Use the `\texttt{\textbf{\small{}-t}}'
option to allow event generation only on specific geometry volumes.
\item A mix of target materials, each with its corresponding weight. \\
This is the standard option for generating events in the Super-K detector
where the beam profile is uniform and distributing the event vertices
uniformly in the detector volume is sufficient. The target mix is
specified as a comma-separated list of nuclear PDG codes (in the PDG2006
convention: 10LZZZAAAI) followed by their corresponding weight fractions
in brackets, as in:\\
`\texttt{\textbf{\small{}-t code1{[}fraction1{]},code2{[}fraction2{]},...}}'
\\
 \\
 Example 1: \\
 To use a target mix of $88.79\%$ (weight fraction) $O^{16}$ and
$11.21\%$ $H$ (i.e. `water') type: \\
 \texttt{\textbf{\small{}`-g 1000080160{[}0.8879{]},1000010010{[}0.1121{]}'}}
\\
 \\
 Example 2: \\
 To use a target which is $100\%$ $C^{12}$, type: \\
\texttt{\textbf{\small{} `-g 1000060120}}'
\end{itemize}
\textbf{\textcolor{magenta}{-t}}\textbf{ Specifies the input top volume
for event generation.} This is an optional argument. By default, it
is set to be the `master volume' of the input geometry resulting in
neutrino events being generated over the entire geometry volume. If
the `\texttt{\textbf{\small{}-t}}' option is set, event generation
will be confined in the specified detector volume. The option can
be used to simulate events at specific sub-detectors.\\
Example:\\
To generate events in the P0D only, type:\\
\texttt{\textbf{\small{}`-t P0D}}'\\
\\
You can use the `\texttt{\textbf{\small{}-t}}' option to switch generation
on/off at multiple volumes\\
Example: \\
\texttt{\textbf{\small{}`-t +Vol1-Vol2+Vol3-Vol4'}}, or\texttt{\textbf{\small{}
}}~\\
\texttt{\textbf{\small{}`-t ``+Vol1 -Vol2 +Vol3 -Vol4'''}}~\\
This instructs the GENIE geometry navigation code to switch on volumes
`Vol1' and `Vo3' and switch off volumes `Vol2' and `Vol4'. If the
very first character is a '+', GENIE will neglect all volumes except
the ones explicitly turned on. Vice versa, if the very first character
is a `-', GENIE will keep all volumes except the ones explicitly turned
off.\\
\\
\textbf{\textcolor{magenta}{-m}}\textbf{\textcolor{magenta}{\large{}
}}\textbf{Specifies an XML file with the maximum density-weighted
path-lengths for each nuclear target in the input geometry.} This
is an optional argument. If the option is not set (and also if the
options \textbf{-P} and \textbf{-S} are not set) GENIE will scan the
input geometry to determine the maximum density-weighted path-lengths
for all nuclear targets.then, at the MC job initialization, GENIE
will scan the input geometry to determine the maximum density-weighted
path-lengths for all nuclear targets. The computed information is
used for calculating the neutrino interaction probability scale to
be used in the MC job (the tiny neutrino interaction probabilities
get normalized to a probability scale which is defined as the maximum
possible total interaction probability, corresponding to a maximum
energy neutrino in a worst-case trajectory maximizing its density-weighted
path-length, summed up over all possible nuclear targets). That probability
scale is also used to calculate the absolute, POT normalization of
a generated event sample from the POT normalization of the input JNUBEAM
flux ntuple.

Feeding-in pre-computed maximum density-weighted path-lengths results
in faster MC job initialization and ensures that the same interaction
probability scale is used across all MC jobs in a physics production
job (the geometry is scanned by a MC ray-tracing method and the calculated
safe maximum density-weighted path-lengths may differ between MC jobs).

The maximum density-weighted path-lengths for a Geant4/ROOT-based
detector geometry can be pre-computed using GENIE's \textit{gmxpl}
utility.\\
\\
\textbf{\textcolor{magenta}{-P}}\textbf{ Specifies a ROOT file with
the pre-calculated interaction probability} for each flux neutrino
in the input flux file, for the top volume and the input geometry.
This is an optional argument. This option is intended to replace the
maximum density weighted path-lengths option \textbf{-m}. This option
is new in v2.6.2. The pre-calculated interaction probability method
is specific to the flux input (JNUBEAM flux ntuples), and so has been
optimised much more than the maximum density weighted path-lengths
method. The interaction probability for each flux neutrino is pre-calcuated
before any events are generated. The maximum interaction probability
is now exact (maximally efficient) and means that the interaction
probability does not need to be recalculated, until we have decided
there has been an interaction. It is especially fast for complicated
geometries. This means that this method is up to 300 times faster
than the \textbf{-m} option. The \textbf{-P} option can be used in
one of two ways. The first is to pre-calculate the interaction probabilites
in a separate job (using the \textbf{-S} option of \emph{gevgen\_t2k},
see below). This is especially good for larger flux files with $>\mathcal{O}(100000)$
entries, as the time to pre-calculate interaction probabilities becomes
comparable to the event generation time. For small flux files, the
amount of bookkeeping when using pre-calculated interaction probabities
means that generating interaction probabilites at the start of each
job is faster (you should run \textbf{-P} with no arguments). Note
that if none of \textbf{-P}, \textbf{-S} and \textbf{-m} are set,
then GENIE will scan the input geometry to determine the maximum density-weighted
path-lengths for all nuclear targets, during initalization of the
MC job.\\
\\
\textbf{\textcolor{magenta}{-S}}\textbf{ Specifies a location to save
a ROOT file with the calculated interaction probability} for each
flux neutrino in the input flux file, for the top volume and the input
geometry. This is an optional argument. It is used to create pre-calculated
interaction probabilities for input into the \textbf{-P} option. You
should make sure to use exactly the input flux file, input geometry,
top volume name, neutrino flavours, etc... arguments in your \emph{gevgen\_t2k}
submission line for your pre-calculation of interaction probabilities
(\textbf{-S}), and your use of the pre-calculated interaction probabilites
(\textbf{-P}). The default output name is {[}flux\_file\_name{]}.{[}top\_volume\_name{]}.flxprobs.root.
This can be overridden by providing an argument to the \textbf{-S}
option. Note that running \emph{gevgen\_t2k} with this option will
not generate any events; the only output will be a ROOT file contaning
the pre-calculated interaction probabilites.\\
\\
\textbf{\textcolor{magenta}{-L}}\textbf{ Specifies the input geometry
length units.} This is an optional argument. By default, that option
is set to `mm', the length units used for the nd280 detector geometry
description. Possible options include: `m', `cm', `mm', ...\\
 \\
\textbf{\textcolor{magenta}{-D}}\textbf{ Specifies the input geometry
density units.} This is an optional argument. By default, that option
is set to `clhep\_def\_density\_unit', the density unit used for the
nd280 detector geometry description (= $\sim$1.6E-19 x g/cm3 !).
Possible options include: `kg\_m3', `g\_cm3', `clhep\_def\_density\_unit',...
\\
 \\
\textbf{\textcolor{magenta}{-c}}\textbf{ Specifies how many times
to cycle a JNUBEAM flux ntuple. }This option provides a way to set
the MC job exposure in terms of complete JNUBEAM flux ntuple cycles.
On each cycle, every flux neutrino in the ntuple will be thrown towards
the detector geometry.\\
 \\
\textbf{\textcolor{magenta}{-e}}\textbf{ Specifies how many POTs to
generate.} If this option is set, \textit{gevgen\_t2k} will work out
how many times it has to cycle through the input flux ntuple in order
to accumulate the requested statistics. The program will stop at the
earliest complete flux ntuple cycle after accumulating the required
statistics. The generated statistics will slightly overshoot the requested
number but the calculated exposure (which is also stored at the output
file) will be exact. This option is only available with JNUBEAM ntuple-based
flux descriptions.\\
 \\
\textbf{\textcolor{magenta}{-E}}\textbf{ Specifies how many POTs to
generate.} This option is similar to `\texttt{\textbf{\small{}-e}}'
but the program will stop immediately after the requested POT has
been accumulated, without waiting for the current loop over the flux
ntuple entries to be completed. The generated POT overshoot (with
respect to the requested POT) will be negligible, but the POT calculation
within a flux ntuple cycle is only approximate. This reflects the
details of the JNUBEAM beam-line simulation. This option is only available
with JNUBEAM ntuple-based flux descriptions.\\
\\
\textbf{\textcolor{magenta}{-n}}\textbf{ Specifies how many events
to generate}. Note that out of the 4 possible ways of setting the
exposure (`\texttt{\textbf{\small{}-c}}', `\texttt{\textbf{\small{}-e}}',
`\texttt{\textbf{\small{}-E}}', `\texttt{\textbf{\small{}-n}}') this
is the only available one if a plain histogram-based flux description
is used.\\
\textbf{}\\
\textbf{\textcolor{magenta}{-o}}\textbf{ Sets the prefix of the output
event file.} This is an optional argument. It allows you to override
the output event file prefix. In GENIE, the output filename is built
as:\texttt{\small{} }{\small \par}

\texttt{\small{}prefix.run\_number.event\_tree\_format.file\_format}
where, in \textit{gevgen\_t2k}, by default, \texttt{\small{}prefix}:
`gntp' and \texttt{\small{}event\_tree\_format}: `ghep' and \texttt{\small{}file\_format}:
`root'.\\
 \\
\textbf{\textcolor{magenta}{-r}}\textbf{ Specifies the MC run number.}
This is an optional argument. By default a run number of `1000' is
used.\\
\\
\textbf{\textcolor{magenta}{--seed}}\textbf{ Specifies the random
number seed} for the current job.\textbf{}\\
\textbf{}\\
\textbf{\textcolor{magenta}{--cross-sections}} Specifies the name
(incl. full path) of an input XML file with pre-computed neutrino
cross-sections\textbf{}\\
\textbf{}\\
\textbf{\textcolor{magenta}{--event-generator-list}}\textbf{ Specifies
the list of event generators to use in the MC job.} By default, GENIE
is loading a list of of tuned and fully-validated generators which
allow comprehensive neutrino interaction modelling the medium-energy
range.\textbf{ }Valid settings are the XML block names appearing in
\textit{\$}\texttt{\textbf{\small{}GENIE}}\textit{/config/EventGeneratorListAssembler.xml}'.
Please, make sure you read Sec. \ref{sec:ObtainingSpecialSamples}
explaining why, almost invariantly, for physics studies you should
be using a comprehensive collection of event generators.\texttt{\textbf{\small{} }}\textbf{}\\
\textbf{}\\
\textbf{\textcolor{magenta}{--message-thresholds}}\textbf{ Specifies
the GENIE verbosity level}. The verbosity level is controlled with
an XML file allowing users to customize the threshold of each message
stream. The XML schema can be seen in \textit{`\$}\texttt{\textbf{\small{}GENIE}}\textit{/config/Messenger.xml}'.
The \textit{`Messenger.xml' }file contains the default thresholds
used by GENIE. The \textit{`Messenger\_laconic.xml' }and \textit{`Messenger\_rambling.xml'
}files define, correspondingly, less and more verbose configurations.\texttt{\textbf{\small{}}}~\\
\texttt{\textbf{\small{}}}~\\
\textbf{\textcolor{magenta}{--unphysical-event-mask }}Specify a 16-bit
mask to allow certain types of unphysical events to be written in
the output event file. By default, all unphysical events are rejected.\\
\texttt{\textbf{\small{}}}~\\
\textbf{\textcolor{magenta}{--event-record-print-level}}\texttt{\textbf{\small{}
}}Allows users to set the level of information shown when the event
94 record is printed in the screen. See GHepRecord::Print() for allowed
settings.\texttt{\textbf{\small{}}}~\\
\texttt{\textbf{\small{}}}~\\
\textbf{\textcolor{magenta}{--mc-job-status-refresh-rate}}\texttt{\textbf{\small{}
}}Allows users to customize the refresh rate of the status file.\texttt{\textbf{\small{}}}~\\
\texttt{\textbf{\small{}}}~\\
\textbf{\textcolor{magenta}{--cache-file}} Allows users to specify
a ROOT file so that results of calculation cached throughout a MC
job can be re-used in subsequent MC jobs.\textbf{}\\
\textbf{}\\
\textbf{\textcolor{magenta}{-h}}\textbf{ }Prints out the \textit{gevgen\_t2k}
syntax and exits.{\small{}}\\
{\small \par}

\subsubsection*{Examples}
\begin{enumerate}
\item Generate events (run `1001') using the jnubeam flux ntuple in `\textit{/data/t2k/flux/07a/jnb001.root}'
and picking up the flux entries for the detector location `nd5' (which
corresponds to the `nd280m' location). The job will load the nd280
geometry from `\textit{/data/t2k/geom/nd280.root}' and interpret it
assuming the length unit is `mm' and the density unit is the default
CLHEP one. The job will stop on the first complete flux ntuple cycle
after generating 5E+17 POT. Read pre-computed cross-section splines
from `\textit{/data/t2k/xsec/xsec.xml}'. Use seed number 1982199 and,
also, use the default GENIE verbosity level.\texttt{\textbf{\small{} }}~\\
\texttt{\textbf{\small{}}}~\\
\texttt{\textbf{\small{}\$ gevgen\_t2k -r 1001 -f /data/t2k/flux/07a/jnb001.root,nd5
}}~\\
\texttt{\textbf{\small{}-g /data/t2k/geom/nd280.root -L mm -D clhep\_def\_density\_unit
}}~\\
\texttt{\textbf{\small{}-{}-cross-sections /data/t2k/xsec/xsec.xml
-e 5E+17 -{}-seed 1982199}}~\\
\texttt{\textbf{\small{} }}{\small \par}
\item As before, but now the job will stop after 100 flux ntuple cycles,
whatever POT and number of events that may correspond to.\\
\\
\texttt{\textbf{\small{}\$ gevgen\_t2k -r 1001 -f /data/t2k/flux/07a/jnb001.root,nd5
}}~\\
\texttt{\textbf{\small{}-g /data/t2k/geom/nd280.root -L mm -D clhep\_def\_density\_unit}}~\\
\texttt{\textbf{\small{}-{}-cross-sections /data/t2k/xsec/xsec.xml
-c 100 -{}-seed 1982199}}~\\
 
\item As before, but now the job will stop after generating 100000 events,
whatever POT and number of flux ntuple cycles that may correspond
to.\\
\\
\texttt{\textbf{\small{}\$ gevgen\_t2k -r 1001 -f /data/t2k/flux/07a/jnb001.root,nd5
}}~\\
\texttt{\textbf{\small{}-g /data/t2k/geom/nd280.root -L mm -D clhep\_def\_density\_unit
}}~\\
\texttt{\textbf{\small{}-{}-cross-sections /data/t2k/xsec/xsec.xml
-n 100000 -{}-seed 1982199}}~\\
{\small \par}
\item As before, but first pre-calculate interaction probilites, and then
use them to generate events.\\
\\
\texttt{\textbf{\small{}\$ gevgen\_t2k -r 1001 -f /data/t2k/flux/07a/jnb001.root,nd5
}}~\\
\texttt{\textbf{\small{}-g /data/t2k/geom/nd280.root -L mm -D clhep\_def\_density\_unit
}}~\\
\texttt{\textbf{\small{}-{}-cross-sections /data/t2k/xsec/xsec.xml
-n 100000 -{}-seed 1982199}}~\\
\texttt{\textbf{\small{}-S jnb001.nd280.global.flxprobs.root}}~\\
\texttt{\textbf{\small{}}}~\\
\texttt{\textbf{\small{}\$ gevgen\_t2k -r 1001 -f /data/t2k/flux/07a/jnb001.root,nd5
}}~\\
\texttt{\textbf{\small{}-g /data/t2k/geom/nd280.root -L mm -D clhep\_def\_density\_unit
}}~\\
\texttt{\textbf{\small{}-{}-cross-sections /data/t2k/xsec/xsec.xml
-n 100000 -{}-seed 1982199}}~\\
\texttt{\textbf{\small{}-P jnb001.nd280.global.flxprobs.root}}~\\
{\small \par}
\item Generate events (run `1001') using the jnubeam flux ntuple in `\textit{/data/t2k/flux/07a/jnb001.root}'
and picking up the flux entries for the Super-K detector location.
This time, the job will not use any detailed detector geometry description
but just ($95\%$ $O^{16}$ + $5\%$ H) target-mix. The job will stop
after generating 50000 events. As before, read pre-computed cross-section
splines from `\textit{/data/t2k/xsec/xsec.xml}'.\texttt{\textbf{\small{} }}This
time use production-mode verbosity level (set all message thresholds
to `warning').\\
\\
\texttt{\textbf{\small{}\$ gevgen\_t2k -r 1001 -f /data/t2k/flux/07a/jnb001.root,sk
}}~\\
\texttt{\textbf{\small{}-g 1000080160{[}0.95{]},1000010010{[}0.05{]}
-n 50000 -{}-seed 1982199}}~\\
\texttt{\textbf{\small{}-{}-cross-sections /data/t2k/xsec/xsec.xml
-{}-message-thresholds Messenger\_laconic.xml}}~\\
\texttt{\textbf{\small{} }}{\small \par}
\item As before, but now the flux is not described using a JNUBEAM ntuple
but a set of 1-D histograms from the `\textit{/data/flx.root}' file:
The histogram named `h1' will be used for the $\nu_{e}$ flux, `h2'
will will be used for the $\bar{\nu_{e}}$ flux, and `h3' for the
$\nu_{\mu}$ flux.\\
\\
\texttt{\textbf{\small{}\$ gevgen\_t2k -r 1001 -f /data/flx.root,12{[}h1{]},-12{[}h2{]},14{[}h3{]}
}}~\\
\texttt{\textbf{\small{}-g 1000080160{[}0.95{]},1000010010{[}0.05{]}
-n 50000 -{}-seed 1982199}}~\\
\texttt{\textbf{\small{}-{}-cross-sections /data/t2k/xsec/xsec.xml
-{}-message-thresholds Messenger\_laconic.xml }}~\\
{\small \par}
\end{enumerate}

\subsection{Event generation application for Fermilab neutrino experiments \label{sub:gNuMIevgen}}

\subsubsection*{Name}

\textit{gevgen\_fnal} -- A GENIE-based event generation application
for Fermilab neutrino experiments. It integrates the GENIE with the
Fermilab neutrino beam-line simulations and the geometry descriptions
of DUNE, MINOS, NOvA, MINERvA, ArgoNEUT, MicroBooNE, SBND and other
experiments.

\subsubsection*{Source and build options}

The source code for this application is in `\texttt{\textbf{\small{}\$GENIE}}\textit{/src/support/fnal/EvGen/gFNALExptEvGen.cxx}'.\\
To enable it add `\texttt{\textbf{\small{}-{}-enable-numi}}' during
the GENIE build configuration step.

\subsubsection*{Synopsis}

\texttt{\textbf{\small{}\$ gevgen\_fnal }}{\small \par}

\texttt{\textbf{\small{}-f }}\texttt{\small{}flux}\texttt{\textbf{\small{} }}{\small \par}

\texttt{\textbf{\small{}-g }}\texttt{\small{}geometry}\texttt{\textbf{\small{}
{[}-t }}\texttt{\small{}top\_volume\_name\_at\_geom}\texttt{\textbf{\small{}{]} }}{\small \par}

\texttt{\textbf{\small{}{[}-F }}\texttt{\small{}fiducial\_cut\_string}\texttt{\textbf{\small{}{]}
{[}-m }}\texttt{\small{}max\_path\_lengths\_xml\_file}\texttt{\textbf{\small{}{]} }}{\small \par}

\texttt{\textbf{\small{}{[}-L }}\texttt{\small{}geometry\_length\_units}\texttt{\textbf{\small{}{]}
{[}-D }}\texttt{\small{}geometry\_density\_units}\texttt{\textbf{\small{}{]}
{[}-z }}\texttt{\small{}z\_min}\texttt{\textbf{\small{}{]}}}{\small \par}

\texttt{\textbf{\small{}<-n }}\texttt{\small{}number\_of\_events}\texttt{\textbf{\small{},
-e }}\texttt{\small{}exposure\_in\_POTs}\texttt{\textbf{\small{}> }}{\small \par}

\texttt{\textbf{\small{}{[}-o }}\texttt{\small{}output\_event\_file\_prefix}\texttt{\textbf{\small{}{]}
{[}-r }}\texttt{\small{}run\#}\texttt{\textbf{\small{}{]} {[}-d }}\texttt{\small{}debug\_flags}\texttt{\textbf{\small{}{]}}}{\small \par}

\texttt{\textbf{\small{}{[}-seed }}\texttt{\small{}random\_number\_seed}\texttt{\textbf{\small{}{]}
{[}-{}-cross-section }}\texttt{\small{}xml\_file}\texttt{\textbf{\small{}{]}
{[}-{}-event-generator-list }}\texttt{\small{}list\_name}\texttt{\textbf{\small{}{]}}}{\small \par}

\texttt{\textbf{\small{}{[}-{}-message-thresholds }}\texttt{\small{}xml\_file}\texttt{\textbf{\small{}{]}
{[}-{}-unphysical-event-mask }}\texttt{\small{}mask}\texttt{\textbf{\small{}{]}
{[}-{}-event-record-print-level }}\texttt{\small{}level}\texttt{\textbf{\small{}{]}}}{\small \par}

\texttt{\textbf{\small{}{[}-{}-mc-job-status-refresh-rate }}\texttt{\small{}rate}\texttt{\textbf{\small{}{]}
{[}-{}-cache-file }}\texttt{\small{}root\_file}\texttt{\textbf{\small{}{]}}} 

\texttt{\textbf{\small{}{[}-h{]}}} \\
\\
where {[}{]} denotes an optional argument and <> denotes a group of
arguments out of which only one can be set.

\subsubsection*{Description}

The following options are available:\\
 \\
\textbf{\textcolor{magenta}{-f }}\textbf{\small{}Specifies the input
neutrino flux.} This option can be used to specify any of: 
\begin{itemize}
\item A gNuMI beam simulation output file and the detector location. The
general sytax is: \\
 \texttt{\textbf{\small{}`-f /path/flux\_file.root,detector\_loc(,neutrino\_list)}}'\\

For more information of the flux ntuples see the gNuMI documentation.
The ntuple has to be in ROOT format and can be generated from the
distributed HBOOK ntuples using ROOT's {\em h2root} utility. See
GNuMIFlux.xml for all supported detector locations. The optional {\em
neutrino\_list} is a comma separated list neutrino PDG codes. It
specifies which neutrino flux species to to considered in the event
generation job. If no such neutrino list is specified then, by default,
GENIE will consider all neutrino species in the input flux ntuple.
When a gNuMI ntuple is used for describing the neutrino flux, GENIE
is able to calculate the POT exposure for the generated event sample
and any one of the exposure setting methods (`\texttt{\textbf{\small{}-e}}',
`\texttt{\textbf{\small{}-n}}', see below) can be used. All gNuMI
information on the flux neutrino parent (parent PDG code, parent 4-position
and 4-momentum at the production and decay points etc) is stored in
a `flux' branch of the output event tree and is associated with the
corresponding generated neutrino event.

Example: \\
 To use the gNuMI flux ntuple flux.root at MINOS near detector location
`\textit{/data/flux.root}' file, type: \\
 \texttt{\textbf{\small{}`-f /data/flux.root,MINOS-NearDet'}} \\

\item A set of flux histograms stored in a ROOT file. The general syntax
is: \\
 \texttt{\textbf{\small{}`-f /path/file.root,neutrino\_code{[}histo{]},...'}}
\\

where {\em neutrino\_code} is a standard neutrino PDG code\footnote{ $\nu_{e}$: 12, $\nu_{\mu}$: 14, $\nu_{\tau}$: 16, ${\bar{\nu}_{e}}$:
-12, ${\bar{\nu}_{\mu}}$: -14 and ${\bar{\nu}_{\tau}}$: -16} and {\em histo} is the corresponding ROOT histogram name. Multiple
flux histograms can be specified for different flux neutrino species
(see the example given below). The relative flux normalization for
all neutrino species should be represented correctly at the input
histogram normalization. The absolute flux normalization is not relevan
heret: Unlike when using gNuMI ntuples to describe the flux, no POT
calculations are performed when histogram-based flux descriptions
are employed. One can only control the MC run exposure via the number
of generated events (`\texttt{\textbf{\small{}-n}}', see below). In
this case the POT normalization of the generated sample is calculated
externally.

Since there is no directional information in plain histogram-based
descriptions of the flux, the generated neutrino vertex is always
set to (0,0,0,0). Then it is the detector MC responsibility to rotate
the interaction vectors and plant the vertex \footnote{ This option is used only for the Super-K simulation where vertices
are distributed uniformly in volume by the detector MC (SKDETSIM).
For event generation at the more complex near detectors a JNUBEAM
ntuple-based flux description should be used so as the interaction
vertex is properly planted within the input geometry by GENIE. } Obviously no flux pass-through branch is written out in the neutrino
event tree since no such information is associated with flux neutrinos
selected from plain histograms.

Example: \\
 To use the histogram `h1' (representing the $\nu_{\mu}$ flux) and
the histogram `h2' (representing the $\nu_{e}$ flux) from the \textit{`/data/flux.root'}
file, type: \\
 \texttt{\textbf{\small{}`-f /data/flux.root,14{[}h1{]},12{[}h2{]}}}'

\end{itemize}
$ $\\
\textbf{\textcolor{magenta}{-g }}\textbf{\small{}Specifies the input
detector geometry.} This option can be used to specify any of: 
\begin{itemize}
\item A ROOT file containing a ROOT/Geant4-based geometry description (\textit{TGeoManager}).
\\
Example: \\
To use the ROOT detector geometry description stored in the `\textit{/data/geo/nova.root}'
file, type:\\
 \texttt{\textbf{\small{}`-g /data/geo/nova.root}}'\\
\\
By default the entire input geometry will be used. Use the `\texttt{\textbf{\small{}-t}}'
option to allow event generation only on specific geometry volumes.
\item A mix of target materials, each with its corresponding weight. \\
This is the standard option for generating events in the Super-K detector
where the beam profile is uniform and distributing the event vertices
uniformly in the detector volume is sufficient. The target mix is
specified as a comma-separated list of nuclear PDG codes (in the PDG2006
convention: 10LZZZAAAI) followed by their corresponding weight fractions
in brackets, as in:\\
`\texttt{\textbf{\small{}-t code1{[}fraction1{]},code2{[}fraction2{]},...}}'
\\
 \\
 Example 1: \\
 To use a target mix of $88.79\%$ (weight fraction) $O^{16}$ and
$11.21\%$ $H$ (i.e. `water') type: \\
 \texttt{\textbf{\small{}`-g 1000080160{[}0.8879{]},1000010010{[}0.1121{]}'}}
\\
 \\
 Example 2: \\
 To use a target which is $100\%$ $C^{12}$, type: \\
\texttt{\textbf{\small{} `-g 1000060120}}'
\end{itemize}
\textbf{\textcolor{magenta}{-t}}\textbf{ }\textbf{\small{}Specifies
the input top volume for event generation.} This is an optional argument.
By default, it is set to be the `master volume' of the input geometry
resulting in neutrino events being generated over the entire geometry
volume. If the `\texttt{\textbf{\small{}-t}}' option is set, event
generation will be confined in the specified detector volume. The
option can be used to simulate events at specific sub-detectors.\\
Example:\\
To generate events in the P0D only, type:\\
\texttt{\textbf{\small{}`-t P0D}}'\\
\\
You can use the `\texttt{\textbf{\small{}-t}}' option to switch generation
on/off at multiple volumes\\
Example: \\
\texttt{\textbf{\small{}`-t +Vol1-Vol2+Vol3-Vol4'}}, or\texttt{\textbf{\small{}
}}~\\
\texttt{\textbf{\small{}`-t ``+Vol1 -Vol2 +Vol3 -Vol4'''}}~\\
This instructs the GENIE geometry navigation code to switch on volumes
`Vol1' and `Vo3' and switch off volumes `Vol2' and `Vol4'. If the
very first character is a '+', GENIE will neglect all volumes except
the ones explicitly turned on. Vice versa, if the very first character
is a `-', GENIE will keep all volumes except the ones explicitly turned
off.\\
\\
\textbf{\textcolor{magenta}{-m}}\textbf{ }\textbf{\small{}Specifies
an XML file with the maximum density-weighted path-lengths for each
nuclear target in the input geometry.} This is an optional argument.
If the option is not set then, at the MC job initialization, GENIE
will scan the input geometry to determine the maximum density-weighted
path-lengths for all nuclear targets. The computed information is
used for calculating the neutrino interaction probability scale to
be used in the MC job (the tiny neutrino interaction probabilities
get normalized to a probability scale which is defined as the maximum
possible total interaction probability, corresponding to a maximum
energy neutrino in a worst-case trajectory maximizing its density-weighted
path-length, summed up over all possible nuclear targets). That probability
scale is also used to calculate the absolute, POT normalization of
a generated event sample from the POT normalization of the input flux
ntuple.

Feeding-in pre-computed maximum density-weighted path-lengths results
in faster MC job initialization and ensures that the same interaction
probability scale is used across all MC jobs in a physics production
job (the geometry is scanned by a MC ray-tracing method and the calculated
safe maximum density-weighted path-lengths may differ between MC jobs).

The maximum density-weighted path-lengths for a Geant4/ROOT-based
detector geometry can be pre-computed using GENIE's \textit{gmxpl}
utility.\\
\\
\textbf{\textcolor{magenta}{-L}}\textbf{ }\textbf{\small{}Specifies
the input geometry length units.} This is an optional argument. By
default it is set to `mm'. Possible options include: `m', `cm', `mm',
...\\
\\
\textbf{\textcolor{magenta}{-D}}\textbf{ }\textbf{\small{}Specifies
the input geometry density units.} This is an optional argument. By
default it is set to `g\_cm3'. Possible options include: `kg\_m3',
`g\_cm3', `clhep\_def\_density\_unit' (= $\sim$1.6E-19 x g/cm3 !),...
\\
\textbf{}\\
\textbf{\textcolor{magenta}{-F}}\textbf{ }\textbf{\small{}Applies
a fiducial cut.} This is an optional argument. Applies a fiducial
cut (for now hard-coded). Only used with ROOT-based detector geometry
descriptions. If the input string starts with \textquotedbl{}-\textquotedbl{}
then reverses sense (ie. anti-fiducial).\\
\\
\textbf{\textcolor{magenta}{-S}}\textbf{\small{} Number of rays to
use to scan geometry for max path length.} This is an optional argument.
Number of rays to use to scan geometry for max path length. Only used
with ROOT-based detector geometry descriptions (and the gNuMI ntuple-based
flux description). If `+N' : Scan the geometry using N rays generated
using flux neutrino directions pulled from the input gNuMI flux ntuple.
If `-N' : Scan the geometry using N rays x N points on each face of
a bounding box. Each ray has a uniformly distributed random inward
direction.\textbf{ }\\
\textbf{}\\
\textbf{\textcolor{magenta}{-z}}\textbf{ }\textbf{\small{}Z from which
to start flux ray in user-world coordinates.} This is an optional
argument. If left unset then flux originates on the flux window {[}No
longer attempts to determine z from geometry, generally got this wrong{]}.\\
\textbf{}\\
\textbf{\textcolor{magenta}{-o}}\textbf{ }\textbf{\small{}Sets the
prefix of the output event file.} This is an optional argument. It
allows you to override the output event file prefix. In GENIE, the
output filename is built as:

\texttt{\small{}prefix.run\_number.event\_tree\_format.file\_format}
where, in \textit{gevgen\_numi}, by default, \texttt{\small{}prefix}:
`gntp' and \texttt{\small{}event\_tree\_format}: `ghep' and \texttt{\small{}file\_format}:
`root'.\\
 \\
\textbf{\textcolor{magenta}{-r}}\textbf{ }\textbf{\small{}Specifies
the MC run number.} This is an optional argument. By default a run
number of `0' is used.\\
 \\
\textbf{\textcolor{magenta}{--seed}}\textbf{ Specifies the random
number seed} for the current job.\textbf{}\\
\textbf{}\\
\textbf{\textcolor{magenta}{--cross-sections}} Specifies the name
(incl. full path) of an input XML file with pre-computed neutrino
cross-sections\textbf{}\\
\textbf{}\\
\textbf{\textcolor{magenta}{--event-generator-list}}\textbf{ Specifies
the list of event generators to use in the MC job.} By default, GENIE
is loading a list of of tuned and fully-validated generators which
allow comprehensive neutrino interaction modelling the medium-energy
range.\textbf{ }Valid settings are the XML block names appearing in
\textit{\$}\texttt{\textbf{\small{}GENIE}}\textit{/config/EventGeneratorListAssembler.xml}'.
Please, make sure you read Sec. \ref{sec:ObtainingSpecialSamples}
explaining why, almost invariantly, for physics studies you should
be using a comprehensive collection of event generators.\texttt{\textbf{\small{} }}\textbf{}\\
\textbf{}\\
\textbf{\textcolor{magenta}{--message-thresholds}}\textbf{ Specifies
the GENIE verbosity level}. The verbosity level is controlled with
an XML file allowing users to customize the threshold of each message
stream. The XML schema can be seen in \textit{`\$}\texttt{\textbf{\small{}GENIE}}\textit{/config/Messenger.xml}'.
The \textit{`Messenger.xml' }file contains the default thresholds
used by GENIE. The \textit{`Messenger\_laconic.xml' }and \textit{`Messenger\_rambling.xml'
}files define, correspondingly, less and more verbose configurations.\texttt{\textbf{\small{}}}~\\
\texttt{\textbf{\small{}}}~\\
\textbf{\textcolor{magenta}{--unphysical-event-mask }}Specify a 16-bit
mask to allow certain types of unphysical events to be written in
the output event file. By default, all unphysical events are rejected.\\
\texttt{\textbf{\small{}}}~\\
\textbf{\textcolor{magenta}{--event-record-print-level}}\texttt{\textbf{\small{}
}}Allows users to set the level of information shown when the event
94 record is printed in the screen. See GHepRecord::Print() for allowed
settings.\texttt{\textbf{\small{}}}~\\
\texttt{\textbf{\small{}}}~\\
\textbf{\textcolor{magenta}{--mc-job-status-refresh-rate}}\texttt{\textbf{\small{}
}}Allows users to customize the refresh rate of the status file.\texttt{\textbf{\small{}}}~\\
\texttt{\textbf{\small{}}}~\\
\textbf{\textcolor{magenta}{--cache-file}} Allows users to specify
a ROOT file so that results of calculation cached throughout a MC
job can be re-used in subsequent MC jobs.\textbf{}\\
\textbf{}\\
\textbf{\textcolor{magenta}{-h}}\textbf{ }Prints out the \textit{gevgen\_fnal}
syntax and exits.\\

\subsubsection*{Examples}

\subsection{Event generation application for atmospheric neutrinos \label{sub:gevgen_atmo}}

\subsubsection*{Name}

\textit{gevgen\_atmo} -- A GENIE-based atmospheric neutrino event
generation application. It integrates GENIE with any of the FLUKA
3-D \cite{Battistoni:2001sw} or BGLRS \cite{Barr:2004br} atmospheric
neutrino flux simulations. Events can be generated for either a simple
target mix or a detailed ROOT-based detector geometry.

\subsubsection*{Source and build options}

The source code for this application is in `\texttt{\textbf{\small{}\$GENIE}}\textit{/src/support/atmo/EvGen/gAtmoEvGen.cxx}'.\\
To enable it add `\texttt{\textbf{\small{}-{}-enable-atmo}}' during
the GENIE build configuration step.

\subsubsection*{Synopsis}

\texttt{\textbf{\small{}\$ gevgen\_atmo}}{\small \par}

\texttt{\textbf{\small{}-f }}\texttt{\small{}flux}\texttt{\textbf{\small{}
-g }}\texttt{\small{}geometry}\texttt{\textbf{\small{} }}{\small \par}

\texttt{\textbf{\small{}{[}-R }}\texttt{\small{}rotation\_from\_topocentric\_hz\_frame}\texttt{\textbf{\small{}{]} }}{\small \par}

\texttt{\textbf{\small{}{[}-t }}\texttt{\small{}geometry\_top\_volume\_name}\texttt{\textbf{\small{}{]}
{[}-m }}\texttt{\small{}max\_path\_lengths\_xml\_file}\texttt{\textbf{\small{}{]} }}{\small \par}

\texttt{\textbf{\small{}{[}-L }}\texttt{\small{}geometry\_length\_units}\texttt{\textbf{\small{}{]}
{[}-D }}\texttt{\small{}geometry\_density\_units}\texttt{\textbf{\small{}{]} }}{\small \par}

\texttt{\textbf{\small{}<-n }}\texttt{\small{}number\_of\_events}\texttt{\textbf{\small{},
-e }}\texttt{\small{}exposure\_in\_terms\_of\_kton\_x\_yrs}\texttt{\textbf{\small{}>}}{\small \par}

\texttt{\textbf{\small{}{[}-E }}\texttt{\small{}energy\_range}\texttt{\textbf{\small{}{]}
{[}-o }}\texttt{\small{}output\_event\_file\_prefix}\texttt{\textbf{\small{}{]}
{[}-r }}\texttt{\small{}run\#}\texttt{\textbf{\small{}{]} }}{\small \par}

\texttt{\textbf{\small{}{[}-seed }}\texttt{\small{}random\_number\_seed}\texttt{\textbf{\small{}{]}
{[}-{}-cross-section }}\texttt{\small{}xml\_file}\texttt{\textbf{\small{}{]}
{[}-{}-event-generator-list }}\texttt{\small{}list\_name}\texttt{\textbf{\small{}{]}}}{\small \par}

\texttt{\textbf{\small{}{[}-{}-message-thresholds }}\texttt{\small{}xml\_file}\texttt{\textbf{\small{}{]}
{[}-{}-unphysical-event-mask }}\texttt{\small{}mask}\texttt{\textbf{\small{}{]}
{[}-{}-event-record-print-level }}\texttt{\small{}level}\texttt{\textbf{\small{}{]}}}{\small \par}

\texttt{\textbf{\small{}{[}-{}-mc-job-status-refresh-rate }}\texttt{\small{}rate}\texttt{\textbf{\small{}{]}
{[}-{}-cache-file }}\texttt{\small{}root\_file}\texttt{\textbf{\small{}{]}}} 

\texttt{\textbf{\small{}{[}-h{]}}} \\
\\
\\
where {[}{]} denotes an optional argument and <> denotes a group of
arguments out of which only one can be set.

\subsubsection*{Description}

The following options are available: \\
\\
\textbf{\textcolor{magenta}{-f }}\textbf{\small{}Specifies the input
neutrino flux.} This option can be used to specify the input flux
simulation data files. The general syntax is: `\texttt{\textbf{\small{}-f
simulation:/path/file{[}neutrino\_code{]},...}}'. The \texttt{\textbf{\small{}`simulation'}}
part of the option can be either `FLUKA' or `BGLRS', depending on
the origin of your input data files. GENIE will use the input tag
to use the appropriate input file format and to bin the input data
according to the choices of the FLUKA and BGLRS flux simulation authors.
See Section \ref{sub:AtmoFluxDrivers} for more details. \\
The\texttt{\textbf{ }}\texttt{\textbf{\small{}`/path/file.data{[}neutrino\_code{]}'}}
part of the option can be repeated multiple times (separated by commas),
once for each flux neutrino species you wish to consider. \\
Example 1:

\texttt{\textbf{\small{}`-f FLUKA:/data/sdave\_numu07.dat{[}14{]},/data/sdave\_nue07.dat{[}12{]}'}}{\small \par}

This option will instruct GENIE to use the \textit{`/data/sdave\_numu07.dat'}
FLUKA flux simulation file for $\nu_{\mu}$ and the \textit{`/data/sdave\_nue07.dat'}
file for $\nu_{e}$. No other flux species will be considered in this
MC job.\\
Example 2:

\texttt{\textbf{\small{}`-f BGLRS:/data/flux10\_271003\_z.kam\_nue{[}12{]}'}}\texttt{\textbf{ }}

This option will instruct GENIE to use the \textit{`/data/flux10\_271003\_z.kam\_nue'}
BGLRS flux simulation file for $\nu_{e}$. No other flux species will
be considered in this MC job.\\
\\
\textbf{\textcolor{magenta}{-g }}\textbf{\small{}Specifies the input
detector geometry.} This option can be used to specify any of: 
\begin{itemize}
\item A ROOT file containing a ROOT/Geant4-based geometry description (\textit{TGeoManager}).
\\
\\
Example: \\
To use the ROOT detector geometry description stored in the `\textit{nd280-geom.root}'
file, type:\\
 \texttt{\textbf{\small{}`-g /some/path/nd280-geom.root}}'\\
\\
By default the entire input geometry will be used. Use the `\texttt{\textbf{\small{}-t}}'
option to allow event generation only on specific geometry volumes.
\item A mix of target materials, each with its corresponding weight. \\
This option should only be used when the beam and/or detector are
sufficiently uniform. The target mix is specified as a comma-separated
list of nuclear PDG codes (in the PDG2006 convention: 10LZZZAAAI)
followed by their corresponding weight fractions in brackets, as in:\\
`\texttt{\textbf{\small{}-t code1{[}fraction1{]},code2{[}fraction2{]},...}}'
\\
 \\
 Example 1: \\
 To use a target mix of $88.79\%$ (weight fraction) $O^{16}$ and
$11.21\%$ $H$ (i.e. `water') type: \\
 \texttt{\textbf{\small{}`-g 1000080160{[}0.8879{]},1000010010{[}0.1121{]}'}}
\\
 \\
 Example 2: \\
 To use a target which is $100\%$ $C^{12}$, type: \\
\texttt{\textbf{\small{} `-g 1000060120}}'
\end{itemize}
\textbf{\textcolor{magenta}{-R }}\textbf{\small{}Specifies a rotation
from the default topocentric horizontal coordinate system to a user-defined
frame.} The rotation is specified by the 3 Euler angles $\varphi$,
$\vartheta$, $\psi$. The Euler angles are used for creting a ROOT
TRotation object which gets applied to the flux neutrino position
and momentum 4-vectors before that flux neutrino is fired towards
the detector. The user has the option to select between the X and
Y conventions. By default, the X-convention is used. Additionally,
the user can request GENIE to invert the rotation matrix before applying
it to the flux neutrino vectors. Please note the following extract
from the ROOT TRotation documentation: ``Euler angles usually define
the rotation of the new coordinate system with respect to the original
system, however, the TRotation class specifies the rotation of the
object in the original system (an active rotation). To recover the
usual Euler rotations (ie. rotate the system not the object), you
must take the inverse of the rotation.\textquotedbl{}\\
\\
The Euler angles are input as a comma separated list. The general
syntax for specifying the rotation is:\texttt{\textbf{\small{} `-R
convention:phi,theta,psi}}' where `\texttt{\textbf{\small{}convention}}'
is either X (for X-convention), Y (for Y-convention), X\textasciicircum{}-1
or Y\textasciicircum{}-1 (as previously, but using the inverse rotation
matrix instead).\\
\\
Example 1: \\
To set the Euler angles $\varphi$=3.14, $\vartheta$=1.28, $\psi$=1.0
using the X-convention, type: \texttt{\textbf{\small{}`-R 3.14,1.28,1.0'}},
or \texttt{\textbf{\small{}`-R X:3.14,1.28,1.0'}}.\\
\\
Example 2: \\
To set the Euler angles $\varphi$=3.14, $\vartheta$=1.28, $\psi$=1.0
using the Y-convention, type: \texttt{\textbf{\small{}`-R Y:3.14,1.28,1.0'}}.\\
 \\
Example 3: \\
To set the Euler angles $\varphi$=3.14, $\vartheta$=1.28, $\psi$=1.0
using the Y-convention, and then use the inverse rotation matrix,
type: \texttt{\textbf{\small{}`-R Y\textasciicircum{}-1:3.14,1.28,1.0'}}.\textbf{}\\
\textbf{}\\
\textbf{\textcolor{magenta}{-t }}\textbf{\small{}Specifies the input
top volume for event generation.} This is an optional argument. By
default, it is set to be the `master volume' of the input geometry
resulting in neutrino events being generated over the entire geometry
volume. If the `\texttt{\textbf{\small{}-t}}' option is set, event
generation will be confined in the specified detector volume. The
option can be used to simulate events at specific sub-detectors.\\
 \\
\textbf{\textcolor{magenta}{-m }}\textbf{\small{}Specifies an XML
file with the maximum density-weighted path-lengths for each nuclear
target in the input geometry.} This is an optional argument. If the
option is not set then, at the MC job initialization, GENIE will scan
the input geometry to determine the maximum density-weighted path-lengths
for all nuclear targets. The computed information is used for calculating
the neutrino interaction probability scale to be used in the MC job
(the tiny neutrino interaction probabilities get normalized to a probability
scale which is defined as the maximum possible total interaction probability,
corresponding to a maximum energy neutrino in a worst-case trajectory
maximizing its density-weighted path-length, summed up over all possible
nuclear targets). That probability scale is also used to calculate
the absolute normalization of generated sample in terms of kton{*}yrs.

Feeding-in pre-computed maximum density-weighted path-lengths results
in faster MC job initialization and ensures that the same interaction
probability scale is used across all MC jobs in a physics production
job (the geometry is scanned by a MC ray-tracing method and the calculated
safe maximum density-weighted path-lengths may differ between MC jobs).

The maximum density-weighted path-lengths for a Geant4/ROOT-based
detector geometry can be pre-computed using GENIE's \textit{gmxpl}
utility.\\
 \\
\textbf{\textcolor{magenta}{-L}}\textbf{ }\textbf{\small{}Specifies
the input geometry length units.} This is an optional argument. By
default it is set to `m'. Possible options include: `m', `cm', `mm',
...\\
 \\
\textbf{\textcolor{magenta}{-D}}\textbf{ }\textbf{\small{}Specifies
the input geometry density units.} This is an optional argument. By
default it is set to `kg\_m3'. Possible options include: `kg\_m3',
`g\_cm3', `clhep\_def\_density\_unit' (= $\sim$1.6E-19 x g/cm3 !),...
\\
\\
\textbf{\textcolor{magenta}{-n}}\textbf{ }\textbf{\small{}Specifies
how many events to generate.}\\
\textbf{}\\
\textbf{\textcolor{magenta}{-e}}\textbf{ }\textbf{\small{}Specifies
the requested exposure in terms of kton{*}yrs.}{\small{}}\\
{\small \par}

\textcolor{red}{Not implemented yet.}\texttt{\textbf{\small{} }}\\
\\
\textbf{\textcolor{magenta}{-E}}\textbf{ }\textbf{\small{}Specifies
an energy range in GeV}. This is an optional argument. Must be a set
of comma-separated values. By default GENIE will generate atmospheric
neutrinos between 0.5 and 50 GeV.

Example: To generate events between 1 and 100 GeV type:\texttt{\textbf{\small{} `-E
1,100'}} \\
\textbf{}\\
\textbf{\textcolor{magenta}{-o}}\textbf{ }\textbf{\small{}Sets the
prefix of the output event file.} This is an optional argument. It
allows you to override the output event file prefix. In GENIE, the
output filename is built as:

\texttt{\small{}prefix.run\_number.event\_tree\_format.file\_format}
where, in \textit{gevgen\_atmo}, by default, \texttt{\small{}prefix}:
`gntp' and \texttt{\small{}event\_tree\_format}: `ghep' and \texttt{\small{}file\_format}:
`root'.\\
 \\
\textbf{\textcolor{magenta}{-r}}\textbf{ }\textbf{\small{}Specifies
the MC run number.} This is an optional argument. By default a run
number of `100000000' is used.\textbf{\textcolor{magenta}{}}\\
\textbf{\textcolor{magenta}{}}\\
\textbf{\textcolor{magenta}{--seed}}\textbf{ Specifies the random
number seed} for the current job.\textbf{}\\
\textbf{}\\
\textbf{\textcolor{magenta}{--cross-sections}} Specifies the name
(incl. full path) of an input XML file with pre-computed neutrino
cross-sections\textbf{}\\
\textbf{}\\
\textbf{\textcolor{magenta}{--event-generator-list}}\textbf{ Specifies
the list of event generators to use in the MC job.} By default, GENIE
is loading a list of of tuned and fully-validated generators which
allow comprehensive neutrino interaction modelling the medium-energy
range.\textbf{ }Valid settings are the XML block names appearing in
\textit{\$}\texttt{\textbf{\small{}GENIE}}\textit{/config/EventGeneratorListAssembler.xml}'.
Please, make sure you read Sec. \ref{sec:ObtainingSpecialSamples}
explaining why, almost invariantly, for physics studies you should
be using a comprehensive collection of event generators.\texttt{\textbf{\small{} }}\textbf{}\\
\textbf{}\\
\textbf{\textcolor{magenta}{--message-thresholds}}\textbf{ Specifies
the GENIE verbosity level}. The verbosity level is controlled with
an XML file allowing users to customize the threshold of each message
stream. The XML schema can be seen in \textit{`\$}\texttt{\textbf{\small{}GENIE}}\textit{/config/Messenger.xml}'.
The \textit{`Messenger.xml' }file contains the default thresholds
used by GENIE. The \textit{`Messenger\_laconic.xml' }and \textit{`Messenger\_rambling.xml'
}files define, correspondingly, less and more verbose configurations.\texttt{\textbf{\small{}}}~\\
\texttt{\textbf{\small{}}}~\\
\textbf{\textcolor{magenta}{--unphysical-event-mask }}Specify a 16-bit
mask to allow certain types of unphysical events to be written in
the output event file. By default, all unphysical events are rejected.\\
\texttt{\textbf{\small{}}}~\\
\textbf{\textcolor{magenta}{--event-record-print-level}}\texttt{\textbf{\small{}
}}Allows users to set the level of information shown when the event
94 record is printed in the screen. See GHepRecord::Print() for allowed
settings.\texttt{\textbf{\small{}}}~\\
\texttt{\textbf{\small{}}}~\\
\textbf{\textcolor{magenta}{--mc-job-status-refresh-rate}}\texttt{\textbf{\small{}
}}Allows users to customize the refresh rate of the status file.\texttt{\textbf{\small{}}}~\\
\texttt{\textbf{\small{}}}~\\
\textbf{\textcolor{magenta}{--cache-file}} Allows users to specify
a ROOT file so that results of calculation cached throughout a MC
job can be re-used in subsequent MC jobs. \textbf{}\\
\textbf{}\\
\textbf{\textcolor{magenta}{-h}}\textbf{ }Prints out the \textit{gevgen\_atmo}
syntax and exits.{\small{}}\\
{\small \par}

\subsubsection*{Examples}
\begin{enumerate}
\item Generate 100k events (run number `100000013') using the FLUKA 3-D
flux simulation output files \textit{`/data/flux/atmo/sdave\_numu07.dat',}
for $\nu_{\mu}$, and \textit{`/data/flux/atmo/sdave\_nue07.dat',
}for $\nu_{e}$. Do not consider any other flux neutrino species.
Generate events for water (weight fraction: 88.79$\%$ $O^{16}$ and
11.21\% H) and only in the 1-15 GeV energy range. Read pre-computed
cross-section splines from `\textit{/data/xsec/xsec.xml}'. Use seed
number 87218 and production mode verbosity level (all message thresholds
set to warning).\\
\texttt{\textbf{\small{}\$ gevgen\_atmo -n 100000 -r 100000013 -e
1,15}}~\\
\texttt{\textbf{\small{}-f FLUKA:/data/flux/atmo/sdave\_numu07.dat{[}14{]},/data/flux/atmo/sdave\_nue07.dat{[}12{]}
}}~\\
\texttt{\textbf{\small{}-g 1000080160{[}0.8879{]},1000010010{[}0.1121{]}}}~\\
\texttt{\textbf{\small{}-{}-cross-sections /data/xsec/xsec.xml}}~\\
\texttt{\textbf{\small{}-{}-seed 87218 -{}-message-thresholds Messenger\_laconic.xml
}}~\\
{\small \par}
\item Like above but, instead of generating events in water, generate events
using the detailed ROOT-based detector geometry description in file
\textit{`/data/geo/HyperKamionande.root'. }Let GENIE know that the
geometry file expresses length in `mm' and densities in `gr/cm$^{3}$'.
Don't generate events over the the entire volume but only within the
volume named `InnerDetector'.\\
\texttt{\textbf{\small{}\$ gevgen\_atmo -n 100000 -r 100000013 -e
1,15}}~\\
\texttt{\textbf{\small{}-f FLUKA:/data/flux/atmo/sdave\_numu07.dat{[}14{]},/data/flux/atmo/sdave\_nue07.dat{[}12{]}
}}~\\
\texttt{\textbf{\small{}-g /data/geo/HyperKamiokande.root -t InnerDetector
-L mm -D g\_cm3}}~\\
\texttt{\textbf{\small{}-{}-cross-sections /data/xsec/xsec.xml }}~\\
\texttt{\textbf{\small{}-{}-seed 87218 -{}-message-thresholds Messenger\_laconic.xml }}{\small \par}
\end{enumerate}

\chapter{Analyzing Output Event Samples \label{cha:AnalyzingOutputs}}

\section{Introduction}

{[}to be added in future revision{]}

\section{The GHEP event structure}

Events generated by GENIE are stored in a custom, \textit{STDHEP}-like,
event record called \textit{GHEP}. Each \textit{GHEP} event record,
an instance of the \textit{GHepRecord} class, is a ROOT \textit{TClonesArray}
container of \textit{GHepParticle} objects representing individual
particles. Other than being a container for the generated particles,
the event record holds additional information with event-, rather
than particle-, scope such as the cross sections for the selected
event and the differential cross section for the selected event kinematics,
the event weight, a series of customizable event flags and an interaction
summary. Additionally, the event record includes a host of methods
for querying / setting event properties including many methods that
allow querying for specific particles within the event (such as for
example methods to return the target nucleus, the final state primary
lepton or a list of all the stable descendants of any intermediate
particle).

The event record features a `spontaneous re-arrangement' feature which
maintains the compactness of the daughter lists at any given time.
This is necessary for the correct interpretation of the stored particle
associations as the daughter indices correspond to a contiguous range.
The particle mother and daughter indices for all particles in the
event record are automatically updated as a result of any such spontaneous
particle rearrangement.

The \textit{GHEP} structure is highly compatible with the event structures
used in most HEP generators. That allows us to call other generators
(such as for example PYTHIA / JETSET) as part of an event generation
chain and convert / append their output into the current \textit{GHEP}
event. Additionally the \textit{GHEP} events can be converted to many
other formats for facilitating the GENIE interface with experiment-specific
offline software systems and cross-generator comparisons.

\subsection{GHEP information with event-wide scope}

{[}to be added in future revision{]}

\subsection{Interaction summary}

All particles generated by GENIE for each simulated event are stored
into a \textit{GHEP} record which represents the most complete description
of a generated event. Certain external heavy-weight applications such
as specialized event-reweighing schemes or realistic, experiment-level
MC simulation chains using the generator as the physics front-end
require that detailed particle-level information.

However, many of the actual physics models employed by the generator,
such as cross section, form factor or structure function models, require
a much simpler event description. An event description based on simple
summary information, typically including a description of the initial
state, the process type and the scattering kinematics, is sufficient
for driving the algorithmic objects implementing these physics models.
In the interest of decoupling the physics models from event generation
and the particle-level event description, GENIE uses an \textit{Interaction}
object to store summary event information. Whenever possible, algorithmic
objects implementing physics models accept a single \textit{Interaction}
object as their sole source of information about an event. That enables
the use of these models both within the event generation framework
but also within a host of external applications such as model validation
frameworks, event re-weighting tools and user physics analysis code.

An \textit{Interaction} objects is an aggregate, hierarchical structure,
containing many specialised objects holding information for the initial
state (\textit{InitialState} object), the event kinematics (\textit{Kinematics}
object), the process type (\textit{ProcessInfo} object) and potential
additional information for tagging exclusive channels (\textit{XclsTag}
object).

Users can easily instantiate \textit{Interaction} objects and use
them to drive physics models. Creating this aggregate hierarchical
structure is streamlined using the `named constructor' C++ idiom.
For example, in order to define a 5 GeV QELCC $\nu_{\mu}+neutron$
interaction, where the neutron is bound in a $Fe^{56}$ nucleus, ($\nu_{\mu}$
PDG code = 14, $neutron$ PDG code = 2112, $Fe^{56}$ PDG code = 1000260560),
one needs to instantiate an \textit{Interaction} object as in: \textit{\emph{}}\\
\texttt{\textbf{\emph{\small{}Interaction}}}\texttt{\textbf{\small{}
{*} qelcc = Interactions::QELCC(1000260560, 2112, 14, 5.0);}} \\
That interaction definition can be used as is to drive a QELCC cross
section algorithm.

The \textit{Interaction} objects can serialize themselves as a unique
string codes which, within the GENIE framework, play the role of the
`reaction codes' of the old procedural systems. These string codes
are used extensively whenever there is a need to map information to
or from interaction types (as for example, mapping interaction types
to pre-computed cross section splines, or mapping interaction types
to specialized event generation code)

Each generated event has an \textit{Interaction} summary object attached
to it and written out in the output event trees. Despite the implications
of having a certain amount of redundancy (in the sense that this summary
information can be recreated entirely from the information at the
\textit{GHEP} record) this strategy presents many advantages during
both event generation and analysis of generated events.

\subsection{GHEP particles}

The basic output unit of the event generation process is a `particle'.
This is an overloaded term used to describe both particles and nuclei
appearing in the initial, intermediate or final state, or generator-specific
pseudo-particles used for facilitating book-keeping of the generator
actions.

Each such 'particle' generated by GENIE is an instance of the \textit{GHepParticle}
class. These objects contain information with particle-scope such
as its particle and status codes, its pdg mass, charge and name, the
indices of its mother and daughter particles marking possible associations
with other particles in the same event, its 4-momentum and 4-position
(in the target nucleus coordinate system), its polarization vector,
and other properties. The GHepParticle class includes methods for
setting and querying these properties.

GENIE has adopted the standard PDG particle codes. For ions it has
adopted a PDG extension, using the 10-digit code 10LZZZAAAI where
AAA is the total baryon number, ZZZ is the total charge, L is the
number of strange quarks and I is the isomer number (I=0 corresponds
to the ground state). GENIE-specific pseudo-particles have PDG code
>= 2000000000 and can convey important information about the underlying
physics model. Pseudo-particles generated by other specialized MCs
that may be called by GENIE (such as PYTHIA) are allowed to retain
the codes specified in that MC.

GENIE marks each particle with a status code. This signifies the position
of an particle in an event and helps navigating within the event record.
Most generated particles are typically marked as one of the following: 
\begin{itemize}
\item \textbf{\textit{`initial state}}' typically the first two particles
of the event record corresponding to the incoming neutrino and the
nuclear target. 
\item \textbf{\textit{`nucleon target'}}, corresponding to the hit nucleon
(if any) within the nuclear target. 
\item \textbf{\textit{`intermediate state'}}, typically referring to the
remnant nucleus, fragmentation intermediates such as quarks, diquarks,
some intermediate pseudo-particles etc. 
\item \textbf{\textit{`hadron in the nucleus'}}, referring to a particle
of the primary hadronic system, that is the particles emerging from
the primary interaction vertex before their possible re-interactions
during their intranuclear hadron transport. 
\item \textbf{\textit{`decayed state'}}, such as for example unstable particles
that have been decayed. 
\item \textbf{\textit{`stable final state'}} for the relatively long-lived
particles emerging from the nuclear targets 
\end{itemize}
All particles generated by GENIE during the simulation of a single
neutrino interaction are stored in a dynamic container representing
an `event'.

\subsection{Mother / daughter particle associations}

{[}to be added in future revision{]}\\

\begin{table*}[htb]
\begin{centering}
\begin{tabular}{|c|r|r|r|r|r|r|r|r|r|}
\hline 
\textbf{\footnotesize{}Idx } & \textbf{\footnotesize{}Name} & \textbf{\footnotesize{}ISt } & \textbf{\footnotesize{}PDG } & \textbf{\footnotesize{}Mom } & \textbf{\footnotesize{}Kids } & \textbf{\footnotesize{}E } & \textbf{\footnotesize{}px } & \textbf{\footnotesize{}py} & ...\tabularnewline
\hline 
\textbf{\footnotesize{}0} & {\footnotesize{}nu\_mu } & {\footnotesize{}0 } & {\footnotesize{}14 } & {\footnotesize{}-1 } & {\footnotesize{}4 \hspace{1pt} 4 } & {\footnotesize{}...} & {\footnotesize{}...} & {\footnotesize{}...} & {\footnotesize{}...}\tabularnewline
\textbf{\footnotesize{}1} & {\footnotesize{}Fe56 } & {\footnotesize{}0 } & {\footnotesize{}1000260560 } & {\footnotesize{}-1 } & {\footnotesize{}2 \hspace{1pt} 3 } & {\footnotesize{}... } & {\footnotesize{}...} & {\footnotesize{}...} & {\footnotesize{}...}\tabularnewline
\textbf{\footnotesize{}2} & {\footnotesize{}neutron } & {\footnotesize{}11 } & {\footnotesize{}2112 } & {\footnotesize{}1 } & {\footnotesize{}5 \hspace{1pt} 7 } & {\footnotesize{}... } & {\footnotesize{}...} & {\footnotesize{}...} & {\footnotesize{}...}\tabularnewline
\textbf{\footnotesize{}3} & {\footnotesize{}Fe55 } & {\footnotesize{}2 } & {\footnotesize{}1000260550 } & {\footnotesize{}1 } & {\footnotesize{}10 \hspace{1pt}10 } & {\footnotesize{}... } & {\footnotesize{}...} & {\footnotesize{}...} & {\footnotesize{}...}\tabularnewline
\textbf{\footnotesize{}4} & {\footnotesize{}mu- } & {\footnotesize{}1 } & {\footnotesize{}13 } & {\footnotesize{}0 } & {\footnotesize{}-1 \hspace{1pt}-1 } & {\footnotesize{} ...} & {\footnotesize{}...} & {\footnotesize{}...} & {\footnotesize{}...}\tabularnewline
\textbf{\footnotesize{}5} & {\footnotesize{}HadrSyst } & {\footnotesize{}12 } & {\footnotesize{}2000000001 } & {\footnotesize{}2 } & {\footnotesize{}-1 \hspace{1pt}-1 } & {\footnotesize{}... } & {\footnotesize{}...} & {\footnotesize{}...} & {\footnotesize{}...}\tabularnewline
\textbf{\footnotesize{}6} & {\footnotesize{}proton } & {\footnotesize{}14 } & {\footnotesize{}211 } & {\footnotesize{}2 } & {\footnotesize{}-1 \hspace{1pt}-1 } & {\footnotesize{}... } & {\footnotesize{}...} & {\footnotesize{}...} & {\footnotesize{}...}\tabularnewline
\textbf{\footnotesize{}7} & {\footnotesize{}pi0} & {\footnotesize{}14 } & {\footnotesize{}111 } & {\footnotesize{}2 } & {\footnotesize{}8 \hspace{1pt} 9 } & {\footnotesize{}... } & {\footnotesize{}...} & {\footnotesize{}...} & {\footnotesize{}...}\tabularnewline
\textbf{\footnotesize{}8} & {\footnotesize{}proton } & {\footnotesize{}1 } & {\footnotesize{}22 } & {\footnotesize{}7 } & {\footnotesize{}-1 \hspace{1pt}-1 } & {\footnotesize{}... } & {\footnotesize{}...} & {\footnotesize{}...} & {\footnotesize{}...}\tabularnewline
\textbf{\footnotesize{}9} & {\footnotesize{}pi- } & {\footnotesize{}1 } & {\footnotesize{}-211 } & {\footnotesize{}7 } & {\footnotesize{}-1 \hspace{1pt}-1 } & {\footnotesize{}...} & {\footnotesize{}...} & {\footnotesize{}...} & {\footnotesize{}...}\tabularnewline
10 & \textbf{\footnotesize{}HadrBlob} & {\footnotesize{}15} & {\footnotesize{}2000000002} & {\footnotesize{}3} & {\footnotesize{}-1 \hspace{1pt}-1} & {\footnotesize{}...} & {\footnotesize{}...} & {\footnotesize{}...} & {\footnotesize{}...}\tabularnewline
\hline 
\end{tabular}
\par\end{centering}

\caption{{[}{]}}

\label{table:ghep1}
\end{table*}

\begin{table*}[htb]
\begin{centering}
\begin{tabular}{|c|r|r|r|r|r|r|r|r|r|}
\hline 
\textbf{\footnotesize{}Idx } & \textbf{\footnotesize{}Name} & \textbf{\footnotesize{}ISt } & \textbf{\footnotesize{}PDG } & \textbf{\footnotesize{}Mom } & \textbf{\footnotesize{}Kids } & \textbf{\footnotesize{}E } & \textbf{\footnotesize{}px } & \textbf{\footnotesize{}py} & ...\tabularnewline
\hline 
\textbf{\footnotesize{}0} & {\footnotesize{}nu\_mu } & {\footnotesize{}0 } & {\footnotesize{}14 } & {\footnotesize{}-1 } & {\footnotesize{}5 \hspace{1pt} 5 } & {\footnotesize{}...} & {\footnotesize{}...} & {\footnotesize{}...} & {\footnotesize{}...}\tabularnewline
\textbf{\footnotesize{}1} & {\footnotesize{}Fe56 } & {\footnotesize{}0 } & {\footnotesize{}1000260560 } & {\footnotesize{}-1 } & {\footnotesize{}2 \hspace{1pt} 3 } & {\footnotesize{}... } & {\footnotesize{}...} & {\footnotesize{}...} & {\footnotesize{}...}\tabularnewline
\textbf{\footnotesize{}2} & {\footnotesize{}proton } & {\footnotesize{}11 } & {\footnotesize{}2212 } & {\footnotesize{}1 } & {\footnotesize{}4 \hspace{1pt} 4 } & {\footnotesize{}... } & {\footnotesize{}...} & {\footnotesize{}...} & {\footnotesize{}...}\tabularnewline
\textbf{\footnotesize{}3} & {\footnotesize{}Mn55 } & {\footnotesize{}2 } & {\footnotesize{}1000250550 } & {\footnotesize{}1 } & {\footnotesize{}12 \hspace{1pt}12 } & {\footnotesize{}... } & {\footnotesize{}...} & {\footnotesize{}...} & {\footnotesize{}...}\tabularnewline
\textbf{\footnotesize{}4} & {\footnotesize{}Delta++ } & {\footnotesize{}3 } & {\footnotesize{}2224 } & {\footnotesize{}2 } & {\footnotesize{}6 \hspace{1pt}7 } & {\footnotesize{}... } & {\footnotesize{}...} & {\footnotesize{}...} & {\footnotesize{}...}\tabularnewline
\textbf{\footnotesize{}5} & {\footnotesize{}mu- } & {\footnotesize{}1 } & {\footnotesize{}13 } & {\footnotesize{}0 } & {\footnotesize{}-1 \hspace{1pt}-1 } & {\footnotesize{}... } & {\footnotesize{}...} & {\footnotesize{}...} & {\footnotesize{}...}\tabularnewline
\textbf{\footnotesize{}6} & {\footnotesize{}proton } & {\footnotesize{}14 } & {\footnotesize{}2112 } & {\footnotesize{}4 } & {\footnotesize{}8 \hspace{1pt}8 } & {\footnotesize{}... } & {\footnotesize{}...} & {\footnotesize{}...} & {\footnotesize{}...}\tabularnewline
\textbf{\footnotesize{}7} & {\footnotesize{}pi+} & {\footnotesize{}14 } & {\footnotesize{}211 } & {\footnotesize{}4 } & {\footnotesize{}11 \hspace{1pt} 11 } & {\footnotesize{}... } & {\footnotesize{}...} & {\footnotesize{}...} & {\footnotesize{}...}\tabularnewline
\textbf{\footnotesize{}8} & {\footnotesize{}proton } & {\footnotesize{}3 } & {\footnotesize{}2212 } & {\footnotesize{}6 } & {\footnotesize{}9 \hspace{1pt}10 } & {\footnotesize{}... } & {\footnotesize{}...} & {\footnotesize{}...} & {\footnotesize{}...}\tabularnewline
\textbf{\footnotesize{}9} & {\footnotesize{}proton } & {\footnotesize{}1 } & {\footnotesize{}2212 } & {\footnotesize{}8 } & {\footnotesize{}-1 \hspace{1pt}-1 } & {\footnotesize{}...} & {\footnotesize{}...} & {\footnotesize{}...} & {\footnotesize{}...}\tabularnewline
10 & {\footnotesize{}proton} & {\footnotesize{}1} & {\footnotesize{}2212} & {\footnotesize{}8} & {\footnotesize{}-1 \hspace{1pt}-1} & {\footnotesize{}...} & {\footnotesize{}...} & {\footnotesize{}...} & {\footnotesize{}...}\tabularnewline
11 & {\footnotesize{}pi+} & {\footnotesize{}1} & {\footnotesize{}211} & {\footnotesize{}7} & {\footnotesize{}-1 \hspace{1pt}-1} & {\footnotesize{}...} & {\footnotesize{}...} & {\footnotesize{}...} & {\footnotesize{}...}\tabularnewline
12 & {\footnotesize{}HadrBlob} & {\footnotesize{}15} & {\footnotesize{}2000000002} & {\footnotesize{}3} & {\footnotesize{}-1 \hspace{1pt}-1} & {\footnotesize{}...} & {\footnotesize{}...} & {\footnotesize{}...} & {\footnotesize{}...}\tabularnewline
\hline 
\end{tabular}
\par\end{centering}

\caption{}

\label{table:ghep2}
\end{table*}

\begin{table*}[htb]
\begin{centering}
\begin{tabular}{|c|r|r|r|r|r|r|r|r|r|}
\hline 
\textbf{\footnotesize{}Idx } & \textbf{\footnotesize{}Name} & \textbf{\footnotesize{}ISt } & \textbf{\footnotesize{}PDG } & \textbf{\footnotesize{}Mom } & \textbf{\footnotesize{}Kids } & \textbf{\footnotesize{}E } & \textbf{\footnotesize{}px } & \textbf{\footnotesize{}py} & ...\tabularnewline
\hline 
\textbf{\footnotesize{}0} & {\footnotesize{}nu\_mu } & {\footnotesize{}0 } & {\footnotesize{}14 } & {\footnotesize{}-1} & {\footnotesize{}4 \hspace{1pt} 4 } & {\footnotesize{}...} & {\footnotesize{}...} & {\footnotesize{}...} & {\footnotesize{}...}\tabularnewline
\textbf{\footnotesize{}1} & {\footnotesize{}Fe56 } & {\footnotesize{}0 } & {\footnotesize{}1000260560 } & {\footnotesize{}-1} & {\footnotesize{}2 \hspace{1pt} 3 } & {\footnotesize{} ...} & {\footnotesize{}...} & {\footnotesize{}...} & {\footnotesize{}...}\tabularnewline
\textbf{\footnotesize{}2} & {\footnotesize{}neutron } & {\footnotesize{}11 } & {\footnotesize{}2112 } & {\footnotesize{}1} & {\footnotesize{}5 \hspace{1pt} 5 } & {\footnotesize{}... } & {\footnotesize{}...} & {\footnotesize{}...} & {\footnotesize{}...}\tabularnewline
\textbf{\footnotesize{}3} & {\footnotesize{}Fe55 } & {\footnotesize{}2 } & {\footnotesize{}1000260550 } & {\footnotesize{}1} & {\footnotesize{}22 \hspace{1pt}22 } & {\footnotesize{}... } & {\footnotesize{}...} & {\footnotesize{}...} & {\footnotesize{}...}\tabularnewline
4 & {\footnotesize{}mu} & {\footnotesize{}1} & {\footnotesize{}13} & {\footnotesize{}0} & {\footnotesize{}-1 \hspace{1pt}-1} & {\footnotesize{}...} & {\footnotesize{}...} & {\footnotesize{}...} & {\footnotesize{}...}\tabularnewline
5 & {\footnotesize{}HadrSyst } & {\footnotesize{}12 } & {\footnotesize{}2000000001} & {\footnotesize{}2} & {\footnotesize{}6 \hspace{1pt}7 } & {\footnotesize{}... } & {\footnotesize{}...} & {\footnotesize{}...} & {\footnotesize{}...}\tabularnewline
6 & {\footnotesize{}u } & {\footnotesize{}12 } & {\footnotesize{}2} & {\footnotesize{}5} & {\footnotesize{}8\hspace{1pt}8 } & {\footnotesize{}... } & {\footnotesize{}...} & {\footnotesize{}...} & {\footnotesize{}...}\tabularnewline
7 & {\footnotesize{}ud\_1 } & {\footnotesize{}12 } & {\footnotesize{}2103} & {\footnotesize{}5} & {\footnotesize{}-1\hspace{1pt}-1 } & {\footnotesize{}... } & {\footnotesize{}...} & {\footnotesize{}...} & {\footnotesize{}...}\tabularnewline
8 & {\footnotesize{}string} & {\footnotesize{}12 } & {\footnotesize{}92} & {\footnotesize{}6} & {\footnotesize{}9\hspace{1pt} 11 } & {\footnotesize{}... } & {\footnotesize{}...} & {\footnotesize{}...} & {\footnotesize{}...}\tabularnewline
9 & {\footnotesize{}pi0} & {\footnotesize{}14 } & {\footnotesize{}111} & {\footnotesize{}8} & {\footnotesize{}14\hspace{1pt}14 } & {\footnotesize{}... } & {\footnotesize{}...} & {\footnotesize{}...} & {\footnotesize{}...}\tabularnewline
10 & {\footnotesize{}proton} & {\footnotesize{}14 } & {\footnotesize{}2212} & {\footnotesize{}8} & {\footnotesize{}15 \hspace{1pt}15 } & {\footnotesize{}...} & {\footnotesize{}...} & {\footnotesize{}...} & {\footnotesize{}...}\tabularnewline
11 & {\footnotesize{}omega} & {\footnotesize{}12} & {\footnotesize{}223} & {\footnotesize{}8} & {\footnotesize{}12 \hspace{1pt}13} & {\footnotesize{}...} & {\footnotesize{}...} & {\footnotesize{}...} & {\footnotesize{}...}\tabularnewline
12 & {\footnotesize{}pi-} & {\footnotesize{}14} & {\footnotesize{}-211} & {\footnotesize{}11} & {\footnotesize{}16 \hspace{1pt}16} & {\footnotesize{}...} & {\footnotesize{}...} & {\footnotesize{}...} & {\footnotesize{}...}\tabularnewline
13 & {\footnotesize{}pi+} & {\footnotesize{}14} & {\footnotesize{}211} & {\footnotesize{}11} & {\footnotesize{}21 \hspace{1pt}21} & {\footnotesize{}...} & {\footnotesize{}...} & {\footnotesize{}...} & {\footnotesize{}...}\tabularnewline
14 & {\footnotesize{}pi0} & {\footnotesize{}1} & {\footnotesize{}111} & {\footnotesize{}9} & {\footnotesize{}-1 \hspace{1pt}-1} & {\footnotesize{}...} & {\footnotesize{}...} & {\footnotesize{}...} & {\footnotesize{}...}\tabularnewline
15 & {\footnotesize{}proton} & {\footnotesize{}1} & {\footnotesize{}2212} & {\footnotesize{}10} & {\footnotesize{}-1 \hspace{1pt}-1} & {\footnotesize{}...} & {\footnotesize{}...} & {\footnotesize{}...} & {\footnotesize{}...}\tabularnewline
16 & {\footnotesize{}pi-} & {\footnotesize{}3} & {\footnotesize{}-211} & {\footnotesize{}12} & {\footnotesize{}17 \hspace{1pt}20} & {\footnotesize{}...} & {\footnotesize{}...} & {\footnotesize{}...} & {\footnotesize{}...}\tabularnewline
17 & {\footnotesize{}neutron} & {\footnotesize{}1} & {\footnotesize{}2112} & {\footnotesize{}16} & {\footnotesize{}-1 \hspace{1pt}-1} & {\footnotesize{}...} & {\footnotesize{}...} & {\footnotesize{}...} & {\footnotesize{}...}\tabularnewline
18 & {\footnotesize{}neutron} & {\footnotesize{}1} & {\footnotesize{}2112} & {\footnotesize{}16} & {\footnotesize{}-1 \hspace{1pt}-1} & {\footnotesize{}...} & {\footnotesize{}...} & {\footnotesize{}...} & {\footnotesize{}...}\tabularnewline
19 & {\footnotesize{}proton} & {\footnotesize{}1} & {\footnotesize{}2212} & {\footnotesize{}16} & {\footnotesize{}-1 \hspace{1pt}-1} & {\footnotesize{}...} & {\footnotesize{}...} & {\footnotesize{}...} & {\footnotesize{}...}\tabularnewline
20 & {\footnotesize{}proton} & {\footnotesize{}1} & {\footnotesize{}2212} & {\footnotesize{}16} & {\footnotesize{}-1 \hspace{1pt}-1} & {\footnotesize{}...} & {\footnotesize{}...} & {\footnotesize{}...} & {\footnotesize{}...}\tabularnewline
21 & {\footnotesize{}pi+} & {\footnotesize{}1} & {\footnotesize{}211} & {\footnotesize{}13} & {\footnotesize{}-1 \hspace{1pt}-1} & {\footnotesize{}...} & {\footnotesize{}...} & {\footnotesize{}...} & {\footnotesize{}...}\tabularnewline
22 & {\footnotesize{}HadrBlob} & {\footnotesize{}15} & {\footnotesize{}2000000002} & {\footnotesize{}3} & {\footnotesize{}-1 \hspace{1pt}-1} & {\footnotesize{}...} & {\footnotesize{}...} & {\footnotesize{}...} & {\footnotesize{}...}\tabularnewline
\hline 
\end{tabular}
\par\end{centering}

\caption{}

\label{table:ghep3}
\end{table*}

\clearpage

\newpage

\section{Printing-out events}

\subsection{The \textit{gevdump} utility}

\subsubsection*{Name}

\textit{gevdump} - A GENIE utility printing-out GENIE GHEP event records.

\subsubsection*{Source}

The source code for this utility may be found in \textit{`\$}\texttt{\textbf{\small{}GENIE}}\textit{/src/stdapp/}gEvDump.cxx'.

\subsubsection*{\texttt{\small{}Synopsis}}

\texttt{\textbf{\small{}\$ gevdump -f filename {[}-n n1{[},n2{]}{]}}}\\
\\
where {[}{]} denotes an optional argument.

\subsubsection*{Description}

The following options are available:
\begin{itemize}
\item \textbf{-f} Specifies a GENIE GHEP event file.
\item \textbf{-n} Specifies an event number or a range of event numbers.
This is an optional argument. By default all events will be printed-out.
\end{itemize}

\subsubsection*{Notes}

You can fine-tune the amount of information that gets printed-out
by tweaking the \textit{`}\texttt{\textbf{\small{}GHEPPRINTLEVEL}}\textit{'}
environmental variable (see Appendix \ref{cha:AppendixEnvironment})

\subsubsection*{Examples}
\begin{enumerate}
\item To print-out all events from `\textit{/data/sample.ghep.root}', type:
\\
\\
\texttt{\textbf{\small{}\$ gevdump -f /data/sample.ghep.root}}\\

\item To print-out the first 500 events from `/\textit{data/sample.ghep.root}',
type: \\
\\
\texttt{\textbf{\small{}\$ gevdump -f /data/sample.ghep.root -n 0,499}}\\

\item To print-out event 178 from `\textit{/data/sample.ghep.root}', type:\\
\\
\texttt{\textbf{\small{}\$ gevdump -f /data/sample.ghep.root -n 178 }}{\small \par}
\end{enumerate}

\section{Event loop skeleton program}

An`event loop' skeleton is given below. You may insert your event
analysis code where is indicated below. Please look at the next section
for information on how to extract information from the `event' object.\\

\begin{verbatim}

{
  ...
  // Open the GHEP/ROOT file
  string filename = /data/sample.ghep.root;
  TFile file(filename.c_str(), READ);

  // Get the tree header & print it 
  NtpMCTreeHeader * header = 
    dynamic_cast<NtpMCTreeHeader*> (file.Get("header"));
  LOG(test, pINFO) << *header;

  // Get the GENIE GHEP tree and set its branch address
  TTree * tree = dynamic_cast<TTree*> (file.Get(gtree));
  NtpMCEventRecord * mcrec = 0;
  tree->SetBranchAddress(gmrec, &mcrec);

  // Event loop 
  for(Long64_t i=0; i<tree->GetEntries(); i++){
    tree->GetEntry(i);

    // print-out the event
    EventRecord & event = *(mcrec->event);
    LOG(test, pINFO) << event;

    // put your event analysis code here
    ...
    ...


    mcrec->Clear();
  }
  ...
}

\end{verbatim}

An`event loop' skeleton can be found in \textit{`\$GENIE/src/test/testEveltLoop.cxx'.
}Copy this file and use it as a starting point for your event loop. 

\clearpage

\section{Extracting event information}

The readers are instructed to spend some time browsing the GENIE doxygen
documentation, especially the classes defined in the \textit{Interaction}
and \textit{GHEP} packages, and familiarize themselves with the public
methods. Some examples on how to extract information from an `event'
objects are given below.

\subsubsection*{Examples}
\begin{enumerate}
\item Extract the interaction summary for the given event and check whether
it is a QEL CC event (excluding QEL CC charm production): \\
\begin{verbatim}
{
 ...

 Interaction * in = event.Summary();

 const ProcessInfo & proc = in->ProcInfo(); 
 const XclsTag & xclsv    = in->ExclTag();

 bool qelcc = proc.IsQuasiElastic() && proc.IsWeakCC();
 bool charm = xclsv.IsCharm();

 if (qelcc && !charm) 
 {
    ...
 }     
 ...
}
\end{verbatim}
\item Get the energy threshold for the given event:\\
\begin{verbatim}
{
 ...

 Interaction * in = event.Summary();

 double Ethr = in->PhaseSpace().Threshold();
 ...
}
\end{verbatim}
\item Get the momentum transfer $Q^{2}$ and hadronic invariant mass $W$,
as generated during kinematical selection, for RES CC event:\\
\begin{verbatim}
{
 ...
 const ProcessInfo & proc  = in->ProcInfo(); 
 const Kinematics  & kine = in->Kine();

 bool selected = true;

 if (proc.IsResonant() && proc.IsWeakCC())
 {
     double Q2s = kine.Q2(selected);
     double Ws  = kine.W (selected);
 }     
 ...
}
\end{verbatim}
\item Calculate the momentum transfer $Q^{2}$, the energy transfer $\nu$,
the Bjorken x variable, the inelasticity y and the hadronic invariant
mass $W$directly from the event record:\\
\begin{verbatim}
{
  ...
  // get the neutrino, f/s primary lepton and hit
  // nucleon event record entries
  //
  GHepParticle * neu = event.Probe();
  GHepParticle * fsl = event.FinalStatePrimaryLepton();
  GHepParticle * nuc = event.HitNucleon();

  // the hit nucleon may not be defined
  // (eg. for coherent, or ve- events)
  //
  if(!nuc) return;

  // get their corresponding 4-momenta (@ LAB)
  //
  const TLorentzVector & k1 = *(neu->P4());
  const TLorentzVector & k2 = *(fsl->P4());
  const TLorentzVector & p1 = *(nuc->P4());
 
  // calculate the kinematic variables
  // (eg see Part.Phys. booklet, page 191)
  //
  double M = kNucleonMass;

  TLorentzVector q = k1 - k2;

  double Q2 = -1 * q.M2();
  double v  = q.Energy();
  double x  = Q2 / (2*M*v);
  double y  = v / k1.Energy();
  double W2 = M*M - 2*M*v - Q2;
  double W  = TMath::Sqrt(TMath::Max(0., W2));

  ...
}
\end{verbatim}
\item Loop over particles and count the number of final state pions:\\
\begin{verbatim}
{
 ...
 int npi = 0;

 TObjArrayIter iter(&event);
 GHepParticle * p = 0;

 // loop over event particles
 for((p = dynamic_cast<GHepParticle *>(iter.Next()))) {

    int pdgc   = p->Pdg();
    int status = p->Status();
 
    if(status != kIStStableFinalState) continue;

    bool is_pi = (pdgc == kPdgPiP ||
                  pdgc == kPdgPi0 ||
                  pdgc == kPdgPiM);

    if(is_pi) npi++;
 }

 ...
}

\end{verbatim}
\item Get the corresponding NEUT reaction code for a GENIE event:\\
\begin{verbatim}

{
  ...

  int neut_code = utils::ghep::NeutReactionCode(&event);
  ...
}

\end{verbatim}
\end{enumerate}

\section{Event tree conversions}

You do not need to convert the GENIE \textit{GHEP} trees in order
to analyze the generated samples or pass them on to a detector-level
Monte Carlo. But you can do so if:
\begin{itemize}
\item you need to pass GENIE events to legacy systems using already standardized
formats,
\item you want to be able to read-in GENIE events without loading any GENIE
libraries (eg bare-ROOT, or XML formats), 
\item you want to extract just summary information and write it out in simpler
ntuples.
\end{itemize}

\subsection{The \textit{\normalsize{}gntpc} ntuple conversion utility}

\subsubsection*{Name}

\textit{gntpc} -- A utility tp converts the native GENIE \textit{GHEP}
event file to a host of plain text, XML or bare-ROOT formats. \\

\subsubsection*{Source}

The source code can be found in \textit{`\$}\texttt{\textbf{\small{}GENIE}}\textit{/src/stdapp/gNtpConv.cxx'
.}\\

\subsubsection*{Synopsis}

\texttt{\textbf{\small{}\$ gntpc }}{\small \par}

\texttt{\textbf{\small{}-i input\_file\_name }}{\small \par}

\texttt{\textbf{\small{}-f format\_of\_output\_file }}{\small \par}

\texttt{\textbf{\small{}{[}-v format\_version\_number{]} }}{\small \par}

\texttt{\textbf{\small{}{[}-c copy\_MC\_job\_metadata?{]}}}{\small \par}

\texttt{\textbf{\small{}{[}-o output\_file\_name{]}}}{\small \par}

\texttt{\textbf{\small{}{[}-n number\_of\_events\_to\_convert{]}}}\\
\\
where {[}{]} denotes an optional argument.\\

\subsubsection*{Description}

The following options are available: \\
\textbf{}\\
\textbf{-i} \textbf{Specifies the name of the GENIE GHEP file to convert}.
\\
\\
\textbf{-f} \textbf{Specifies the output file format}. \\

This can be any of the following: 
\begin{itemize}
\item `gst': The standard \textbf{G}ENIE \textbf{S}ummary \textbf{T}ree
(gst) format (see subsection \ref{sub:gst_format}).
\item `gxml': The GENIE XML event format (see subsection \ref{sub:gxml_format}).
\item `ghep\_mock\_data': Identical format as the input GHEP file but all
information other than final state particles is hidden.
\item `rootracker': A bare-ROOT STDHEP-like event tree. Very similar to
the native GHEP tree but with no dependency on GENIE classes (see
subsection \ref{sub:rootracker_formats}).
\item `rootracker\_mock\_data': Like `rootracker' but with all information
other than final state particles hidden.
\item `t2k\_rootracker': A variation of the `rootracker' format used by
some T2K detector MC chains (nd280). Includes, in addition, tree branches
storing JNUBEAM flux simulation `pass-through info'\footnote{This refers to parent meson information for every flux neutrino for
which GENIE generated an interaction.} (see subsection \ref{sub:rootracker_formats}). 
\item `numi\_rootracker': A variation of the `rootracker' format used by
some NuMI beamline experiments. Includes, in addition, tree branches
storing gNuMI flux simulation pass-through info (see subsection \ref{sub:rootracker_formats}).
\item `t2k\_tracker': A tracker-type format with tweaks required by the
SuperK MC (SKDETSIM) (see subsection \ref{sub:tracker_formats}).
\item `nuance\_tracker': {[}Depreciated{]} The original tracker format (see
subsection \ref{sub:tracker_formats}).
\item `ghad': {[}Depreciated{]} NEUGEN-style text-based format for hadronization
studies.
\item `ginuke': A summary ntuple for intranuclear-rescattering studies using
simulated hadron-nucleus samples.
\end{itemize}
\textbf{-v} \textbf{Specifies the output file format version number}.\\

This is an optional arument. It defaults to the latest version of
each specified format. The option exists to maintain ability to generate
old versions of certain formats.\\
\textbf{}\\
\textbf{-o} \textbf{Specifies the output file name.} \\

This is an optional argument. By default, the output file name is
constructed from the input \textit{GHEP} file name by removing the
`.ghep.root' (or just the `.root' one if `.ghep' is not present) extension
and by appending:
\begin{itemize}
\item `gst' format files: `\textit{.gst.root}'
\item `gxml' format files: `\textit{.gxml}' 
\item `ghep\_mock\_data' format files: `\textit{.mockd.ghep.root}' 
\item `rootracker' format files: `\textit{.gtrac.root}'
\item `rootracker\_mock\_data' format files: `\textit{.mockd.gtrac.root}'
\item `t2k\_rootracker' format files: `\textit{.gtrac.root}'
\item `numi\_rootracker' format files: `\textit{.gtrac.root}'
\item `t2k\_tracker' format files: `\textit{.gtrac.dat}'
\item `nuance\_tracker' format files: `\textit{.gtrac\_legacy.dat}'
\item `ghad' format files: `\textit{.ghad.dat}'
\item `ginuke' format files: `\textit{.ginuke.root}'
\end{itemize}
\textbf{-n} \textbf{Specifies the number of events to convert}. \\

This is an optional argument. By default, gntpc will convert all events
in the input file.

\subsubsection*{Examples:}
\begin{enumerate}
\item To convert all events in the \textit{input GHEP} file `\textit{myfile.ghep.root}'
into the \\
`t2k\_rootracker' format, type:\\
\\
\texttt{\textbf{\small{}\$ gntpc -i myfile.ghep.root -f t2k\_rootracker}}\\
\\
The output file is automatically named `\textit{myfile.gtrac.root}'\\

\item To convert the first 20,000 events in the GHEP file `\textit{myfile.ghep.root}'
into the `gst' format and name the output file `\textit{out.root}',
type:\\
\\
\texttt{\textbf{\small{}\$ gntpc -i myfile.ghep.root -f gst -n 20000
-o out.root}}{\small \par}
\end{enumerate}

\subsection{Formats supported by \textit{gntpc}}

\subsubsection{The `gst' format \label{sub:gst_format}}

The `gst' is a GENIE summary ntuple format. It is a simple, plain
ntuple that can be easily used for plotting in interactive ROOT sessions.
The stored ROOT \textit{TTree} contains the following branches:
\begin{itemize}
\item \textbf{iev} (\textit{int}): Event number.
\item \textbf{neu} (\textit{int}): Neutrino PDG code.
\item \textbf{tgt} (\textit{int}): Nuclear target PDG code (10LZZZAAAI).
\item \textbf{Z} (\textit{int}): Nuclear target Z. 
\item \textbf{A} (\textit{int}): Nuclear target A. 
\item \textbf{hitnuc} (\textit{int}): Hit nucleon PDG code (not set for
coherent, inverse muon decay and ve- elastic events).
\item \textbf{hitqrk} (\textit{int}): Hit quark PDG code (set for deep-inelastic
scattering events only).
\item \textbf{sea} (\textit{bool}): Hit quark is from sea (set for deep-inelastic
scattering events only).
\item \textbf{resid} (\textit{bool}): Produced baryon resonance id (set
for resonance events only).
\item \textbf{qel} (\textit{bool}): Is it a quasi-elastic scattering event?
\item \textbf{res} (\textit{bool}): Is it a resonanec neutrino-production
event?
\item \textbf{dis} (\textit{bool}): Is it a deep-inelastic scattering event? 
\item \textbf{coh} (\textit{bool}): Is it a coherent meson production event? 
\item \textbf{dfr} (\textit{bool}): Is it a diffractive meson production
event?
\item \textbf{imd} (\textit{bool}): Is it an invese muon decay event?
\item \textbf{nuel} (\textit{bool}): Is it a ve- elastic event? 
\item \textbf{cc} (\textit{bool}): Is it a CC event? 
\item \textbf{nc} (\textit{bool}): Is it a NC event?
\item \textbf{charm} (\textit{bool}): Produces charm? 
\item \textbf{neut\_code} (\textit{int}): The equivalent NEUT reaction code
(if any). 
\item \textbf{nuance\_code} (\textit{int}): The equivalent NUANCE reaction
code (if any). 
\item \textbf{wght} (\textit{double}): Event weight. 
\item \textbf{xs} (\textit{double}): Bjorken x (as was generated during
the kinematical selection / off-shell kinematics). 
\item \textbf{ys} (\textit{double}): Inelasticity y (as was generated during
the kinematical selection / off-shell kinematics). 
\item \textbf{ts }(\textit{double}): Energy transfer to nucleus (nucleon)
at coherent (diffractive) production events (as was generated during
the kinematical selection). 
\item \textbf{Q2s} (\textit{double}): Momentum transfer $Q^{2}$ (as was
generated during the kinematical selection / off-shell kinematics)
(in $GeV^{2}$).
\item \textbf{Ws} (\textit{double}): Hadronic invariant mass W (as was generated
during the kinematical selection / off-shell kinematics). 
\item \textbf{x} (\textit{double}): Bjorken x (as computed from the event
record).
\item \textbf{y} (\textit{double}): Inelasticity y (as computed from the
event record). 
\item \textbf{t} (\textit{double}): Energy transfer to nucleus (nucleon)
at coherent (diffractive) production events (as computed from the
event record). 
\item \textbf{Q2} (\textit{double}): Momentum transfer $Q^{2}$ (as computed
from the event record) (in $GeV^{2}$).
\item \textbf{W} (\textit{double}): Hadronic invariant mass W (as computed
from the event record).
\item \textbf{Ev} (\textit{double}): Incoming neutrino energy (in GeV).
\item \textbf{pxv} (\textit{double}): Incoming neutrino px (in GeV).
\item \textbf{pyv} (\textit{double}): Incoming neutrino py (in GeV).
\item \textbf{pzv} (\textit{double}): Incoming neutrino pz (in GeV).
\item \textbf{En} (\textit{double}): Initial state hit nucleon energy (in
GeV).
\item \textbf{pxn} (\textit{double}): Initial state hit nucleon px (in GeV).
\item \textbf{pyn} (\textit{double}): Initial state hit nucleon py (in GeV).
\item \textbf{pzn} (\textit{double}): Initial state hit nucleon pz (in GeV).
\item \textbf{El} (\textit{double}): Final state primary lepton energy (in
GeV).
\item \textbf{pxl} (\textit{double}): Final state primary lepton px (in
GeV).
\item \textbf{pyl} (\textit{double}): Final state primary lepton py (in
GeV).
\item \textbf{pzl} (\textit{double}): Final state primary lepton pz (in
GeV).
\item \textbf{nfp} (\textit{int}): Number of final state $p$ and \textbf{\large{}$\bar{p}$}
(after intranuclear rescattering).
\item \textbf{nfn} (\textit{int}): Number of final state $n$ and $\bar{n}$. 
\item \textbf{nfpip} (\textit{int}): Number of final state $\pi^{+}$. 
\item \textbf{nfpim} (\textit{in}t): Number of final state $\pi^{-}$.
\item \textbf{nfpi0} (\textit{int}): Number of final state $\pi^{0}$.
\item \textbf{nfkp} (\textit{int}): Number of final state $K^{+}$. 
\item \textbf{nfkm} (\textit{int}): Number of final state $K^{-}$.
\item \textbf{nfk0} (\textit{int}): Number of final state $K^{0}$ and $\bar{K^{0}}$. 
\item \textbf{nfem} (\textit{int}): Number of final state $\gamma$, $e^{-}$and
$e^{+}$. 
\item \textbf{nfother} (\textit{int}): Number of heavier final state hadrons
(D+/-,D0,Ds+/-,Lamda,Sigma,Lamda\_c,Sigma\_c,...).
\item \textbf{nip} (\textit{int}): Number of `primary' (`primary' : before
intranuclear rescattering) $p$ and \textbf{\large{}$\bar{p}$.}{\large \par}
\item \textbf{nin} (int): Number of `primary' $n$ and $\bar{n}$.
\item \textbf{nipip} (\textit{int}): Number of `primary' $\pi^{+}$. 
\item \textbf{nipim} (\textit{int}): Number of `primary' $\pi^{-}$. 
\item \textbf{nipi0} (\textit{int}): Number of `primary' $\pi^{0}$. 
\item \textbf{nikp} (\textit{int}): Number of `primary' $K^{+}$. 
\item \textbf{nikm} (\textit{int}): Number of `primary' $K^{-}$.
\item \textbf{nik0} \textit{(int)}: Number of `primary' $K^{0}$ and $\bar{K^{0}}$.
\item \textbf{niem} (\textit{int}): Number of `primary' $\gamma$, $e^{-}$and
$e^{+}$ (eg from nuclear de-excitations or from pre-intranuked resonance
decays). 
\item \textbf{niother} (\textit{int}): Number of other `primary' hadron
shower particles. 
\item \textbf{nf} (\textit{int}): Number of final state particles in hadronic
system.
\item \textbf{pdgf} (\textit{int{[}kNPmax{]}}): PDG code of $k^{th}$ final
state particle in hadronic system. 
\item \textbf{Ef} (\textit{double{[}kNPmax{]}}): Energy of $k^{th}$ final
state particle in hadronic system (in GeV).
\item \textbf{pxf} (\textit{double{[}kNPmax{]}}): Px of $k^{th}$ final
state particle in hadronic system (in GeV).
\item \textbf{pyf} (\textit{double{[}kNPmax{]}}): Py of $k^{th}$ final
state particle in hadronic system (in GeV).
\item \textbf{pzf} (\textit{double{[}kNPmax{]}}): Pz of $k^{th}$ final
state particle in hadronic system (in GeV).
\item \textbf{ni} (\textit{int}): Number of particles in the `primary' hadronic
system (`primary' : before intranuclear rescattering).
\item \textbf{pdgi} (\textit{int{[}kNPmax{]}}): PDG code of $k^{th}$ particle
in `primary' hadronic system.
\item \textbf{Ei} (\textit{double{[}kNPmax{]}}): Energy of $k^{th}$ particle
in `primary' hadronic system (in GeV).
\item \textbf{pxi} (\textit{double{[}kNPmax{]}}): Px of $k^{th}$ particle
in `primary' hadronic system (in GeV).
\item \textbf{pyi} (\textit{double{[}kNPmax{]}}): Py of $k^{th}$ particle
in `primary' hadronic system (in GeV).
\item \textbf{pzi} (\textit{double{[}kNPmax{]}}): Pz of $k^{th}$ particle
in `primary' hadronic system (in GeV).
\item \textbf{vtxx} (\textit{double}): Vertex x in detector coord system
(in SI units). 
\item \textbf{vtxy} (\textit{double}): Vertex y in detector coord system
(in SI units).
\item \textbf{vtxx} (\textit{double}): Vertex z in detector coord system
(in SI units). 
\item \textbf{vtxt} (\textit{double}): Vertex t in detector coord system
(in SI units). 
\item \textbf{calresp0} (\textit{double}): An approximate calorimetric response
to the generated hadronic vertex actibity, calculated by summing up:
the kinetic energy for generated \{$\pi^{+}$, $\pi^{-}$, $p$, \textbf{\large{}$n$}\},
the energy+mass for generated \{$\bar{p}$, $\bar{n}$\}, the (e/h){*}energy
for generated \{$\pi^{0}$, $\gamma$, $e^{-}$, $e^{+}$\} (with
an e/h = 1.3) and the kinetic energy for any other generated particle.
\end{itemize}

\paragraph*{Using ROOT to plot quantities stored in a `gst' ntuple }

The `gst' summary ntuples make it especially easy to plot GENIE information
in a ROOT/CINT session. Some examples are given below:
\begin{enumerate}
\item To draw a histogram of the final state primary lepton energy for all
$\nu_{\mu}$ CC DIS interactions with an invariant mass $W$ > 3 GeV,
then type:\\
\texttt{\textbf{\small{}root{[}0{]} gst->Draw(``El'',''dis\&\&cc\&\&neu==14\&\&Ws>3'');}}\\

\item To draw a histogram of all final state $\pi^{+}$ energies in CC RES
interactions, then type: \\
\texttt{\textbf{\small{}root{[}0{]} gst->Draw(``Ef'',''pdgf==211\&\&res\&\&cc'');}}\\

\end{enumerate}

\subsubsection{The `gxml' format \label{sub:gxml_format}}

The `gxml' format is a GENIE XML-based event format\footnote{In the format description that follows, the curly braces within tags
are to be `viewed' as a single value of the specified type with the
specified semantics. For example `\{number of particles; int\}' is
to be thought of as an integer value describing a number particles. }. \\
\\
Each event is included within <ghep> </ghep> tags as in: \\
\begin{verbatim}

<ghep np         = "{number of particles; int}" 
      unphysical = "{is it physical?; boolean (T/F)}">

</ghep>

\end{verbatim}

Both information with event-wide scope such as:

\begin{verbatim}
   <wght>      {event weight; double}        </wght>
   <xsec_evnt> {event cross section; double} </xsec_evnt>
   <xsec_kine> {cross section for event kinematics; double} </xsec_kine>

   <vx> {vertex x in detector coord system (SI); double} </vx> 
   <vy> {vertex y in detector coord system (SI); double} </vy> 
   <vz> {vertex z in detector coord system (SI); double} </vz> 
   <vt> {vertex t (SI); double} </vt>
\end{verbatim}

and a full list of the generated particles is included between the
<ghep> tags. The information for each generated particle is expressed
as: \\
\begin{verbatim}
<p idx  = "{particle index in event record; int}" 
   type = "{particle type; char (F[ake]/P[article]/N[uleus])}">

   <pdg> {pdg code; int}    </pdg>
   <ist> {status code; int} </ist>

   <mother>   
            <fst> {first mother index; int} </fst> 
            <lst> {last  mother index; int} </lst> 
   </mother> 
   <daughter> 
            <fst> {first daughter index; int} </fst> 
            <lst> {last  daughter index; int} </lst> 
   </daughter> 

   <px> {Px in GeV; double}  </px> 
   <py> {Py in GeV; double} </py> 
   <pz> {Pz in GeV; double}  </pz> 
   <E>  {E  in GeV; double}  </E>
   <x>  {x  in fm;  double}  </x> 
   <y>  {y  in fm;  double}  </y> 
   <z>  {z  in fm;  double}  </z> 
   <t>  {t; always set to 0} </t>

   <ppolar> {polarization, polar angle;     in rad} </ppolar> 
   <pazmth> {polarization, azimuthal angle; in rad} </pazmth>
</p>

\end{verbatim}

\subsubsection{The `rootracker' formats \label{sub:rootracker_formats}}

The `rootracker' format is a standardized bare-ROOT GENIE event tree
format evolved from work on integrating the GENIE simulations with
the nd280, INGRID and 2km detector-level simulations. In recent versions
of GENIE that format was renamed to `t2k\_rootracker', with `rootracker'
now being a more generic, stripped-down (excudes pass-through JPARC
flux info etc.) version of the T2K variance.\\
\\
The `rootracker' tree branch names, leaf types and a short description
is given below. For the JNUBEAM branches please consult the corresponding
documentation:
\begin{itemize}
\item \textbf{EvtNum} (\textit{int}): Event number
\item \textbf{EvtFlags} (\textit{TBits}{*}): {[}GENIE{]} Event flags.
\item \textbf{EvtCode} (\textit{TObjString}{*}): {[}GENIE{]} A string event
code. 
\item \textbf{EvtXSec} (\textit{double}): {[}GENIE{]} Event cross section
(in $10^{38}cm^{2}$). 
\item \textbf{EvtDXSec} (\textit{double}): {[}GENIE{]} Differential cross
section for the selected kinematics in the $K^{n}$ space (in $10^{38}cm^{2}$/
$[K^{n}]$). Typically, $K^{n}$ is: \{$Q^{2}\}$ for QEL, \{$Q^{2},W\}$
for RES, $\{x,y\}$ for DIS and COH, $\{y\}$ for $ve^{-}$ etc.
\item \textbf{EvtWght} (\textit{double}): {[}GENIE{]} Event weight. 
\item \textbf{EvtProb} (\textit{double}): {[}GENIE{]} Event probability
(given cross section, density-weighted path-length, etc).
\item \textbf{EvtVtx} (\textit{double{[}4{]}}): {[}GENIE{]} Event vertex
position (x, y, z, t) in the detector coordinate system (in SI).
\item \textbf{StdHepN} (\textit{int}): {[}GENIE{]} Number of entries in
the particle array. 
\item \textbf{StdHepPdg} (\textit{int}): {[}GENIE{]} $k^{th}$ particle
PDG code. 
\item \textbf{StdHepStatus} (\textit{int}): {[}GENIE{]} $k^{th}$ particle
status code (Generator-specific: For GENIE see \textit{GHepStatus\_t}).
\item \textbf{StdHepRescat} (\textit{int}): {[}GENIE{]} $k^{th}$ particle
intranuclear rescattering code (Hadron-transport model specific: For
INTRANUKE/hA see INukeFateHA\_t).
\item \textbf{StdHepX4} (\textit{double {[}kNPmax{]}{[}4{]}}): {[}GENIE{]}
$k^{th}$ particle 4-position (x, y, z, t) in the hit nucleus rest
frame (in fm) 
\item \textbf{StdHepP4} (\textit{double {[}kNPmax{]}{[}4{]}}): {[}GENIE{]}
$k^{th}$ particle 4-momentum (px, py, pz, E) in the LAB frame (in
GeV).
\item \textbf{StdHepPolz} (\textit{double {[}kNPmax{]}{[}3{]}}): {[}GENIE{]}
$k^{th}$ particle polarization vector. 
\item \textbf{StdHepFd} (\textit{int {[}kNPmax{]}}): {[}GENIE{]} $k^{th}$
particle first-daughter index.
\item \textbf{StdHepLd} (\textit{int {[}kNPmax{]}}): {[}GENIE{]} $k^{th}$
particle last-daughter index.
\item \textbf{StdHepFm} (\textit{int {[}kNPmax{]}}): {[}GENIE{]} $k^{th}$
particle first-mother index.
\item \textbf{StdHepLm} (\textit{int {[}kNPmax{]}}): {[}GENIE{]} $k^{th}$
particle last-mother index.\\
\\
\\
The following branches exist only in the `t2k\_rootracker' variance:\\

\item \textbf{NuParentPdg} (\textit{int}): {[}JNUBEAM{]} Parent PDG code.
\item \textbf{NuParentDecMode} (\textit{int}): {[}JNUBEAM{]} Parent decay
mode. 
\item \textbf{NuParentDecP4} (\textit{double {[}4{]}}): {[}JNUBEAM{]} Parent
4-momentum at decay.
\item \textbf{NuParentDecX4} (\textit{double {[}4{]}}): {[}JNUBEAM{]} Parent
4-position at decay. 
\item \textbf{NuParentProP4} (\textit{double {[}4{]}}): {[}JNUBEAM{]} Parent
4-momentum at production.
\item \textbf{NuParentProX4} (\textit{double {[}4{]}}): {[}JNUBEAM{]} Parent
4-position at production.
\item \textbf{NuParentProNVtx} (\textit{int}): {[}JNUBEAM{]} Parent vertex
id.
\item \textbf{G2NeutEvtCode} (\textit{int}): corresponding NEUT reaction
code for the GENIE event.\\
\\
The following branches exist only in the `numi\_rootracker' variance\footnote{More details on the GNuMI beam simulation outputs can be found at\\
\textit{http://www.hep.utexas.edu/\textasciitilde{}zarko/wwwgnumi/v19/}}:\\

\item \textbf{NumiFluxRun} (\textit{int}): {[}GNUMI{]} Run number.
\item \textbf{NumiFluxEvtno} (\textit{int}): {[}GNUMI{]} Event number (proton
on target).
\item \textbf{NumiFluxNdxdz} (\textit{double}): {[}GNUMI{]} Neutrino direction
slope (dx/dz) for a random decay.
\item \textbf{NumiFluxNdydz} (\textit{double}): {[}GNUMI{]} Neutrino direction
slope (dy/dz) for a random decay.
\item \textbf{NumiFluxNpz} (\textit{double}): {[}GNUMI{]} Neutrino momentum
(GeV/c) along z direction (beam axis).
\item \textbf{NumiFluxNenergy} (\textit{double}): {[}GNUMI{]} Neutrino energy
(GeV/c) for a random decay. 
\item \textbf{NumiFluxNdxdznea} (\textit{double}): {[}GNUMI{]} Neutrino
direction slope (dx/dz) for a decay forced at center of near detector. 
\item \textbf{NumiFluxNdydznea} (\textit{double}): {[}GNUMI{]} Neutrino
direction slope (dy/dz) for a decay forced at center of near detector. 
\item \textbf{NumiFluxNenergyn} (\textit{double}): {[}GNUMI{]} Neutrino
energy for a decay forced at center of near detector. 
\item \textbf{NumiFluxNwtnear} (\textit{double}): {[}GNUMI{]} Neutrino weight
for a decay forced at center of near detector. 
\item \textbf{NumiFluxNdxdzfar} (\textit{double}): {[}GNUMI{]} Neutrino
direction slope (dx/dz) for a decay forced at center of far detector. 
\item \textbf{NumiFluxNdydzfar} (\textit{double}): {[}GNUMI{]} Neutrino
direction slope (dy/dz) for a decay forced at center of far detector. 
\item \textbf{NumiFluxNenergyf} (\textit{double}): {[}GNUMI{]} Neutrino
energy for a decay forced at center of far detector. 
\item \textbf{NumiFluxNwtfar} (\textit{double}): {[}GNUMI{]} Neutrino weight
for a decay forced at center of far detector. 
\item \textbf{NumiFluxNorig} (\textit{int}): {[}GNUMI{]} Obsolete
\item \textbf{NumiFluxNdecay} (\textit{int}): {[}GNUMI{]} Decay mode that
produced neutrino\footnote{\begin{itemize}
\item 1: K0L -> nue pi- e+ 
\item 2: K0L -> nuebar pi+ e- 
\item 3: K0L -> numu pi- mu+
\item 4: K0L -> numubar pi+ mu- 
\item 5: K+ -> numu mu+ 
\item 6: K+ -> nue pi0 e+ 
\item 7: K+ -> numu pi0 mu+ 
\item 8: K- -> numubar mu- 
\item 9: K- -> nuebar pi0 e- 
\item 10: K- -> numubar pi0 mu- 
\item 11: mu+ -> numubar nue e+ 
\item 12: mu- -> numu nuebar e- 
\item 13: pi+ -> numu mu+ 
\item 14: pi- -> numubar mu- \end{itemize}
}
\item \textbf{NumiFluxNtype} (\textit{int}): {[}GNUMI{]} Neutrino flavor. 
\item \textbf{NumiFluxVx} (\textit{double}): {[}GNUMI{]} Position of hadron/muon
decay, X coordinate. 
\item \textbf{NumiFluxVy} (\textit{double}): {[}GNUMI{]} Position of hadron/muon
decay, Y coordinate.
\item \textbf{NumiFluxVz} (\textit{double}): {[}GNUMI{]} Position of hadron/muon
decay, Z coordinate.
\item \textbf{NumiFluxPdpx} (\textit{double}): {[}GNUMI{]} Parent momentum
at decay point, X - component. 
\item \textbf{NumiFluxPdpy} (\textit{double}): {[}GNUMI{]} Parent momentum
at decay point, Y - component. 
\item \textbf{NumiFluxPdpz} (\textit{double}): {[}GNUMI{]} Parent momentum
at decay point, Z - component.
\item \textbf{NumiFluxPpdxdz} (\textit{double}): {[}GNUMI{]} Parent dx/dz
direction at production. 
\item \textbf{NumiFluxPpdydz} (\textit{double}): {[}GNUMI{]} Parent dy/dz
direction at production. 
\item \textbf{NumiFluxPppz} (\textit{double}): {[}GNUMI{]} Parent Z momentum
at production. 
\item \textbf{NumiFluxPpenergy} (\textit{double}): {[}GNUMI{]} Parent energy
at production.
\item \textbf{NumiFluxPpmedium} (\textit{int}): {[}GNUMI{]} Tracking medium
number where parent was produced. 
\item \textbf{NumiFluxPtype} (\textit{int}): {[}GNUMI{]} Parent particle
ID (PDG) 
\item \textbf{NumiFluxPpvx} (\textit{double}): {[}GNUMI{]} Parent production
vertex, X coordinate (cm). 
\item \textbf{NumiFluxPpvy} (\textit{double}): {[}GNUMI{]} Parent production
vertex, Y coordinate (cm). 
\item \textbf{NumiFluxPpvz} (\textit{double}): {[}GNUMI{]} Parent production
vertex, Z coordinate (cm). 
\item \textbf{NumiFluxMuparpx} (\textit{double}): {[}GNUMI{]} Repeat of
information above, but for muon neutrino parents. 
\item \textbf{NumiFluxMuparpy} (\textit{double}): {[}GNUMI{]} -//-. 
\item \textbf{NumiFluxMuparpz} (\textit{double}): {[}GNUMI{]} -//-. 
\item \textbf{NumiFluxMupare} (\textit{double}): {[}GNUMI{]} -//-. 
\item \textbf{NumiFluxNecm} (\textit{double}): {[}GNUMI{]} Neutrino energy
in COM frame. 
\item \textbf{NumiFluxNimpwt} (\textit{double}): {[}GNUMI{]} Weight of neutrino
parent. 
\item \textbf{NumiFluxXpoint} (\textit{double}): {[}GNUMI{]} Unused. 
\item \textbf{NumiFluxYpoint} (\textit{double}): {[}GNUMI{]} Unused. 
\item \textbf{NumiFluxZpoint} (\textit{double}): {[}GNUMI{]} Unused. 
\item \textbf{NumiFluxTvx} (\textit{double}): {[}GNUMI{]} Exit point of
parent particle at the target, X coordinate. 
\item \textbf{NumiFluxTvy} (\textit{double}): {[}GNUMI{]} Exit point of
parent particle at the target, Y coordinate. 
\item \textbf{NumiFluxTvz} (\textit{double}): {[}GNUMI{]} Exit point of
parent particle at the target, Z coordinate. 
\item \textbf{NumiFluxTpx} (\textit{double}): {[}GNUMI{]} Parent momentum
exiting the target, X - component.
\item \textbf{NumiFluxTpy} (\textit{double}): {[}GNUMI{]} Parent momentum
exiting the target, Y- component. 
\item \textbf{NumiFluxTpz} (\textit{double}): {[}GNUMI{]} Parent momentum
exiting the target, Z - component.
\item \textbf{NumiFluxTptype} (\textit{double}): {[}GNUMI{]} Parent particle
ID exiting the target. 
\item \textbf{NumiFluxTgen} (\textit{double}): {[}GNUMI{]} Parent generation
in cascade\footnote{\begin{itemize}
\item 1: Primary proton, 
\item 2: Particles produced by proton interaction, 
\item 3: Particles produced by interactions of the 2's, 
\item ...\end{itemize}
}.
\item \textbf{NumiFluxTgptype} (\textit{double}): {[}GNUMI{]} Type of particle
that created a particle flying of the target.
\item \textbf{NumiFluxTgppx} (\textit{double}): {[}GNUMI{]} Momentum of
a particle, that created a particle that flies off the target (at
the interaction point), X - component.
\item \textbf{NumiFluxTgppy} (\textit{double}): {[}GNUMI{]} Momentum of
a particle, that created a particle that flies off the target (at
the interaction point), Y - component.
\item \textbf{NumiFluxTgppz} (\textit{double}): {[}GNUMI{]} Momentum of
a particle, that created a particle that flies off the target (at
the interaction point), Z - component.
\item \textbf{NumiFluxTprivx} (\textit{double}): {[}GNUMI{]} Primary particle
interaction vertex, X coordinate. 
\item \textbf{NumiFluxTprivy} (\textit{double}): {[}GNUMI{]} Primary particle
interaction vertex, Ycoordinate.
\item \textbf{NumiFluxTprivz} (\textit{double}): {[}GNUMI{]} Primary particle
interaction vertex, Z coordinate.
\item \textbf{NumiFluxBeamx} (\textit{double}): {[}GNUMI{]} Primary proton
origin, X coordinate.
\item \textbf{NumiFluxBeamy} (\textit{double}): {[}GNUMI{]} Primary proton
origin, Y coordinate.
\item \textbf{NumiFluxBeamz} (\textit{double}): {[}GNUMI{]} Primary proton
origin, Z coordinate.
\item \textbf{NumiFluxBeampx} (\textit{double}): {[}GNUMI{]} Primary proton
momentum, X - component. 
\item \textbf{NumiFluxBeampy} (\textit{double}): {[}GNUMI{]} Primary proton
momentum, Y - component. 
\item \textbf{NumiFluxBeampz} (\textit{double}): {[}GNUMI{]} Primary proton
momentum, Z - component. 
\end{itemize}

\subsubsection{The `tracker' formats \label{sub:tracker_formats}}

The `tracker'-type format is a legacy event format used by some fortran-based
event generators (eg. NUANCE) and detector-level simulations (eg.
SuperK's Geant3-based SKDETSIM). GENIE includes a number of `tracker'
format variations:

\subsubsection*{{*} `t2k\_tracker': }

This is tracker-type format with all the tweaks required for passing
GENIE events into the Geant3-based SuperK detector MC. In the `t2k\_tracker'
files:
\begin{itemize}
\item The begging of event file is marked with a \texttt{\textbf{\$begin}}
line, while the end of the file is marked by an\texttt{\textbf{ \$end
}}line.
\item Each new event is marked with a \texttt{\textbf{\$genie}} line. What
follows is a reaction code. Since GENIE doesn't use integer reaction
codes, it is writting-out the corresponding NEUT reaction code for
the generated GENIE event. This simplifies comparisons between the
GENIE and NEUT samples in SuperK physics analyses.
\item The \texttt{\textbf{\$vertex}} line is being used to pass the interaction
vertex position in the detector coordinate system in SI units
\item The \texttt{\textbf{\$track }}lines are being used to pass minimal
information on (some) initial / intermediate state particles (as expected
by SKDETSIM) and all final state particles to be tracked by the detector
simulation. Each \texttt{\textbf{\$track }}line includes the particle
PDG code, its energy, its direction cosines and a `status code' (Not
to be confused with GENIE's status code. The `tracker' file status
code expected by SKDETSIM is `-1' for initial state particles, `0'
for stable final states and `-2' for intermediate particles). 
\end{itemize}
Some further clarifications are in order here:
\begin{itemize}
\item $K^{0}$, $\bar{K^{0}}$ generated by GENIE are converted to $K_{L}^{0}$,
$K_{S}^{0}$ (as expected by SKDETSIM)
\item Since no mother / daughter associations are stored in \texttt{\textbf{\$track}}
lines only one level of intermediates can exist (the `primary' hadronic
system). Any intermediate particles corresponding to states evolved
from the `primary' hadronic state but before reaching the `final state'
are neglected.
\item The \texttt{\textbf{\$track}} line ordering is the one expected by
SKDETSIM with all the primaries, intermediates and final states grouped
together.
\end{itemize}
The `t2k\_tracker' format includes a set of \texttt{\textbf{\$info}}
lines. They include the exact same information as the one stored `t2k\_rootracker'
format files (complete event information generated by GENIE and JPARC
/ JNUBEAM flux pass-through information). This is partially redundant
information (some of it was included in the minimal \texttt{\textbf{\$track}}
lines) that is not intended for pushing particles through the detector
simulation. The \texttt{\textbf{\$info}} lines are read-in by SKDETSIM
and are passed-through to the DST stage so that the identical, full
MC information is available for events simulated on both SuperK and
the near detector complex (thus enabling global systematic studies).
\\
\\
A complete event in `t2k\_tracker' format looks-like:

\begin{verbatim}

$begin

$genie  {neut_like_event_type}
$vertex {vtxx} {vtxy} {vtxz} {vtxt}

$track  {pdg code} {E} {dcosx} {dcosy} {dcosz} {status}
$track  {pdg code} {E} {dcosx} {dcosy} {dcosz} {status}
...
$track  {pdg code} {E} {dcosx} {dcosy} {dcosz} {status}

$info {event_num} {error_code} {genie_event_type}
$info {event_xsec} {event_kinematics_xsec} {event_weight} {event_probability}
$info {vtxx} {vtxy} {vtxz} {vtxt}

$info {nparticles}
$info {idx}{pdg}{status}{fd}{ld}{fm}{lm}{px}{py}{pz}{E}{x}{y}{z}{t}{polx}{poly}{polz}
$info {idx}{pdg}{status}{fd}{ld}{fm}{lm}{px}{py}{pz}{E}{x}{y}{z}{t}{polx}{poly}{polz}
...
$info {idx}{pdg}{status}{fd}{ld}{fm}{lm}{px}{py}{pz}{E}{x}{y}{z}{t}{polx}{poly}{polz}

$info (jnubeam_parent_pdg) (jnubeam_parent_decay_mode)
$info (jnubeam_dec_px) (jnubeam_dec_py) (jnubeam_dec_pz) (jnubeam_dec_E)
$info (jnubeam_dec_x)  (jnubeam_dec_y)  (jnubeam_dec_z)  (jnubeam_dec_t)
$info (jnubeam_pro_px) (jnubeam_pro_py) (jnubeam_pro_pz) (jnubeam_pro_E)
$info (jnubeam_pro_x)  (jnubeam_pro_y)  (jnubeam_pro_z)  (jnubeam_pro_t)
$info (jnubeam_nvtx)

$end

\end{verbatim}

\section{Units}

GENIE is using the natural system of units ($\hbar=c=1$) so (almost)
everything is expressed in $[GeV]^{n}$. Notable exceptions are the
event vertex (in SI units, in the detector coordinate system) and
particle positions (in $fm$, in the hit nucleus coordinate system).
Additionally, although internally all cross sections are expressed
in the natural system units, values copied to certain files (eg `rootracker'-
or `tracker'-format files) are converted to $10^{38}$$cm^{2}$ (See
the corresponding documentation for these file formats).\\
\\
GENIE provides an easy way for converting back and forth between its
internal, natural system of units and other units. The conversion
factors are included in `\textit{\$GENIE/src/Conventions/Units.h}'.
\\
\\
For exampe, in order to convert a cross section value returned by
`\textit{a\_function()}' from the natural system of units to $10^{38}$$cm^{2}$,
type:

\begin{verbatim}

double xsec = a_function() / (1E-38 * units::cm2);

\end{verbatim}

\chapter{Non-Neutrino Event Generation Modes}

\section{Introduction}

{[}to be added in future revision{]}

\section{Hadron (and Photon) - Nucleus scattering }

{[}to be added in future revision{]}

\subsection{The \textit{gevgen\_hadron} event generation application}

\subsubsection*{Name}

\textit{gevgen\_hadron} - A GENIE hadron+nucleus event generation
application.

\subsubsection*{Source}

The source code for this utility may be found in \textit{`\$}\texttt{\textbf{\small{}GENIE}}\textit{/src/stdapp/}gEvGenHadronNucleus.cxx'.

\subsubsection*{\texttt{\small{}Synopsis}}

\texttt{\textbf{\small{}\$ gevgen\_hadron }}{\small \par}

\texttt{\textbf{\small{}{[}-n }}\texttt{number\_of\_events}\texttt{\textbf{\small{}{]}
-p }}\texttt{probe\_pdg\_code}\texttt{\textbf{\small{} -t }}\texttt{target\_pdg\_code}\texttt{\textbf{\small{} }}{\small \par}

\texttt{\textbf{\small{}-k }}\texttt{kinetic\_energy}\texttt{\textbf{\small{}
{[}-m }}\texttt{mode}\texttt{\textbf{\small{}{]} }}{\small \par}

\texttt{\textbf{\small{}{[}-f }}\texttt{flux}\texttt{\textbf{\small{}{]}
{[}-o }}\texttt{output\_file\_prefix}\texttt{\textbf{\small{}{]}{[}-r
}}\texttt{run\#}{]}

\texttt{\textbf{\small{}{[}-seed }}\texttt{\small{}random\_number\_seed}\texttt{\textbf{\small{}{]}
{[}-{}-message-thresholds }}\texttt{\small{}xml\_file}\texttt{\textbf{\small{}{]} }}{\small \par}

\texttt{\textbf{\small{}{[}-{}-event-record-print-level }}\texttt{\small{}level}\texttt{\textbf{\small{}{]}
{[}-{}-mc-job-status-refresh-rate }}\texttt{\small{}rate}\texttt{\textbf{\small{}{]}}}~\\
\texttt{\textbf{\small{} }}{\small \par}

\subsubsection*{Description}

The following options are available:\\
\\
\textbf{\textcolor{magenta}{-n}} \textbf{Specifies the number of events
to generate}. This is an optional argument. By default it is set to
`10000'.\\
\\
\textbf{\textcolor{magenta}{-p }}\textbf{Specifies the incoming hadron
PDG code. }The choice is limited to the hadrons that can be handled
by the intranuclear cascade code that is invoked by the application
(choice made via the -m option).\\
\\
\textbf{\textcolor{magenta}{-t}} \textbf{Specifies the nuclear target
PDG code. }As usual the PDG2006 convention is used (10LZZZAAAI). So,
for example, $O^{16}$ code = 1000080160, $Fe^{56}$ code = 1000260560.
For more details see Appendix \ref{cha:AppendixStatAndPdgCodes}.
\\
\\
\textbf{\textcolor{magenta}{-k}} \textbf{Specifies the incoming hadron's
kinetic energy (range).} This option can be use to specify either
a single kinetic energy value (eg \texttt{\textbf{\small{}`-k 0.5'}})
or a kinetic energy range as a comma-separated set of numbers (eg\texttt{\textbf{\small{}
`-k 0.1,1.2'}}). The input values are taken to be in GeV. If no flux
is specified then hadrons will be fired towards the nucleus with a
uniform kinetic energy distribution within the specified range. If
a kinetic energy spectrum is supplied then the hadron kinetic energies
will be generated using the input spectrum within the specified range.\\
 \\
\textbf{\textcolor{magenta}{-f}} \textbf{Specifies the incoming hadron's
kinetic energy spectrum.} This is an optional argument. It can be
either: a) a function, eg \texttt{\textbf{\small{}`x{*}x+4{*}exp(-x)'}},
or b) a text file containing 2 columns corresponding to (kinetic energy
\{GeV\}, 'flux'). If you do specify a flux then you need to specify
a kinetic energy range (not kust a single value). \\
\\
\textbf{\textcolor{magenta}{-o}} \textbf{Specifies the output filename
prefix.} This is an optional argument. It allows you to override the
output event file prefix. In GENIE, the output filename is built as:

\texttt{\small{}`prefix.run\_number.event\_tree\_format.file\_format'}
where, in \textit{gevgen\_hadro}, by default, \texttt{\small{}prefix}:
`gntp' and \texttt{\small{}event\_tree\_format}: `ghep' and \texttt{\small{}file\_format}:
`root'.\\
\\
\textbf{\textcolor{magenta}{-m}} \textbf{Specifies which intranuclear
cascade model to use.} This is an optional argument. Possible options
are `hA' (for the INTRANUKE hA model), `hN' (for the INTRANUKE hN
model). By default it is set to `hA'.\\
\\
\textbf{\textcolor{magenta}{-r}} \textbf{Specifies the run number.}
This is an optional argument. By default it is set to `0'.\textbf{\textcolor{magenta}{}}\\
\textbf{\textcolor{magenta}{}}\\
\textbf{\textcolor{magenta}{--seed}}\textbf{ Specifies the random
number seed} for the current job.\textbf{}\\
\textbf{}\\
\textbf{\textcolor{magenta}{--message-thresholds}}\textbf{ Specifies
the GENIE verbosity level}. The verbosity level is controlled with
an XML file allowing users to customize the threshold of each message
stream. The XML schema can be seen in \textit{`\$}\texttt{\textbf{\small{}GENIE}}\textit{/config/Messenger.xml}'.
The \textit{`Messenger.xml' }file contains the default thresholds
used by GENIE. The \textit{`Messenger\_laconic.xml' }and \textit{`Messenger\_rambling.xml'
}files define, correspondingly, less and more verbose configurations.\texttt{\textbf{\small{}}}~\\
\texttt{\textbf{\small{}}}~\\
\textbf{\textcolor{magenta}{--event-record-print-level}}\texttt{\textbf{\small{}
}}Allows users to set the level of information shown when the event
94 record is printed in the screen. See GHepRecord::Print() for allowed
settings.\texttt{\textbf{\small{}}}~\\
\texttt{\textbf{\small{}}}~\\
\textbf{\textcolor{magenta}{--mc-job-status-refresh-rate}}\texttt{\textbf{\small{}
}}Allows users to customize the refresh rate of the status file.\texttt{\textbf{\small{}}}~\\
{\small \par}

\subsubsection*{Examples}
\begin{enumerate}
\item Generate 100k $\pi^{+}$ + $Fe^{56}$ events with a $\pi^{+}$ kinetic
energy of 165 MeV. Use seed number 10010.\\
\\
\texttt{\textbf{\small{}\$ gevgen\_hadron -n 100000 -p 211 -t 1000260560
-k 0.165 -{}-seed 10010}}~\\
{\small \par}
\item Generate 100k $\pi^{+}$+$Fe^{56}$ events with the $\pi^{+}$ kinetic
energy distributed uniformly in the {[}165 MeV, 1200 MeV{]} range.
Use default seed number.\\
\\
\texttt{\textbf{\small{}\$ gevgen\_hadron -n 100000 -p 211 -t 1000260560
-k 0.165,1.200}}~\\
{\small \par}
\item Generate 100k $\pi^{+}$+$Fe^{56}$ events with the $\pi^{+}$ kinetic
energy distributed as f(KE) = 1/KE in the {[}165 MeV, 1200 MeV{]}
range. Use seed number 10010 and production-mode verbosity level (all
message thresholds set to warning).\\
\\
\texttt{\textbf{\small{}\$ gevgen\_hadron -n 100000 -p 211 -t 1000260560
-k 0.165,1.200 -f '1/x' }}~\\
\texttt{\textbf{\small{}-{}-seed 10010 -{}-message-thresholds Messenger\_laconic.xml}}{\small \par}
\end{enumerate}

\section{Charged Lepton - Nucleus scattering }

{[}to be added in future revision{]}

\section{Nucleon decay}

The simulated nucleon decay modes are given in Tab.\ref{table:NucleonDecayChannels}.
The primary decay is simulated using a phase-space-decay generator.
For bound nucleons, the nuclear environment is simulated as in neutrino
scattering. The nucleon is assigned a Fermi momentum and removal energy
and it is off the mass shell. The propagation of decay products is
simulated using an intranuclear cascade Monte Carlo.

{[}expand{]}

\textbf{}
\begin{table}
\center

\begin{tabular}{|c|c|c|}
\hline 
ID & Decay channel & Current limit ($\times$$10^{34}$ yrs)\tabularnewline
\hline 
\hline 
0 & $p\rightarrow e^{+}\pi^{0}$ & 1.3\tabularnewline
\hline 
1 & $p\rightarrow\mu^{+}\pi^{0}$ & 1.1\tabularnewline
\hline 
2 & $p\rightarrow e^{+}\eta^{0}$ & 0.42\tabularnewline
\hline 
3 & $p\rightarrow\mu^{+}\eta^{0}$ & 0.13\tabularnewline
\hline 
4 & $p\rightarrow e^{+}\rho^{0}$ & 0.07\tabularnewline
\hline 
5 & $p\rightarrow\mu^{+}\rho^{0}$ & 0.02\tabularnewline
\hline 
6 & $p\rightarrow e^{+}\omega^{0}$ & 0.03\tabularnewline
\hline 
7 & $p\rightarrow\mu^{+}\omega^{0}$ & 0.08\tabularnewline
\hline 
8 & $n\rightarrow e^{+}\pi^{-}$ & 0.2\tabularnewline
\hline 
9 & $n\rightarrow\mu^{+}\pi^{-}$ & 0.1\tabularnewline
\hline 
10 & $p\rightarrow\bar{\nu}K^{+}$ & 0.4\tabularnewline
\hline 
\end{tabular}

\textbf{\caption{\textbf{Nucleon decay modes simulated in GENIE.}}
\label{table:NucleonDecayChannels}}
\end{table}

\subsection{The \textit{gevgen\_ndcy} event generation application}

\subsubsection*{Name}

\textit{gevgen\_ndcy} - A GENIE-based nucleon decay event generation
application.

\subsubsection*{Source and build options}

The source code for \textit{gevgen\_ndcy} may be found in \textit{}\\
\textit{`\$}\texttt{\textbf{\small{}GENIE}}\textit{/src/support/ndcy/}EvGen/gNucleonDecayEvGen.cxx'.\\
To enable it add `\texttt{\textbf{\small{}-{}-enable-nucleon-decay}}'
during the GENIE build configuration.

\subsubsection*{\texttt{\small{}Synopsis}}

\texttt{\textbf{\small{}\$ gevgen\_ndcy}}{\small \par}

\texttt{\textbf{\small{}-n }}\texttt{number\_of\_events}\texttt{\textbf{\small{}
-m }}\texttt{nucleon\_decay\_mode}\texttt{\textbf{\small{} }}{\small \par}

\texttt{\textbf{\small{}-g }}\texttt{geometry}\texttt{\textbf{\small{}
{[}-t }}\texttt{geometry\_top\_volume\_name}\texttt{\textbf{\small{}{]} }}{\small \par}

\texttt{\textbf{\small{}{[}-L }}\texttt{geometry\_length\_units}\texttt{\textbf{\small{}{]}
{[}-D }}\texttt{geometry\_density\_units}\texttt{\textbf{\small{}{]} }}{\small \par}

\texttt{\textbf{\small{}{[}-o }}\texttt{output\_event\_file\_prefix}\texttt{\textbf{\small{}{]}
{[}-r }}\texttt{run\#}\texttt{\textbf{\small{}{]}}}{\small \par}

\texttt{\textbf{\small{}{[}-seed }}\texttt{\small{}random\_number\_seed}\texttt{\textbf{\small{}{]}
{[}-{}-message-thresholds }}\texttt{\small{}xml\_file}\texttt{\textbf{\small{}{]} }}{\small \par}

\texttt{\textbf{\small{}{[}-{}-event-record-print-level }}\texttt{\small{}level}\texttt{\textbf{\small{}{]}
{[}-{}-mc-job-status-refresh-rate }}\texttt{\small{}rate}\texttt{\textbf{\small{}{]}}}{\small \par}

\texttt{\textbf{\small{}{[}-h{]}}}~\\
\\
where {[}{]} denotes an optional argument.

\subsubsection*{Description}

The following options are available:\\
\\
\textbf{\textcolor{magenta}{-n}} \textbf{Specifies the number of events
to generate}.\\
\textbf{}\\
\textbf{\textcolor{magenta}{-m}} \textbf{Specifies the nucleon decay
channel ID}. The list of decay channels and the corresponding ID is
given in Tab. \ref{table:NucleonDecayChannels} \textbf{}\\
\textbf{}\\
\textbf{\textcolor{magenta}{-g}}\textbf{ }\textbf{\small{}Specifies
the input detector geometry.} This option can be used to specify any
of: 
\begin{itemize}
\item A ROOT file containing a ROOT / Geant4-based geometry description
(\textit{TGeoManager}). \\
\\
Example: \\
To use the ROOT detector geometry description stored in the `\textit{/data/geo/laguna.root}'
file, type:\\
 \texttt{\textbf{\small{}`-g /data/geo/laguna.root}}'\\
\\
By default the entire input geometry will be used. Use the `\texttt{\textbf{\small{}-t}}'
option to allow event generation only on specific geometry volumes.
\item A mix of target materials, each with its corresponding weight. \\
The target mix is specified as a comma-separated list of nuclear PDG
codes (in the PDG2006 convention: 10LZZZAAAI) followed by their corresponding
weight fractions in brackets, as in:\\
`\texttt{\textbf{\small{}-t code1{[}fraction1{]},code2{[}fraction2{]},...}}'
\\
 \\
 Example 1: \\
 To use a target mix of $88.79\%$ (weight fraction) $O^{16}$ and
$11.21\%$ $H$ (aka `water') type: \\
 \texttt{\textbf{\small{}`-g 1000080160{[}0.8879{]},1000010010{[}0.1121{]}'}}
\\
 \\
 Example 2: \\
 To use a target which is $100\%$ $C^{12}$, type: \\
\texttt{\textbf{\small{} `-g 1000060120}}'
\end{itemize}
\textbf{\textcolor{magenta}{-t }}\textbf{\small{}Specifies the input
top volume for event generation.} This is an optional argument, relevant
only for ROOT-based detector geometry descriptions. By default, it
is set to be the `master volume' of the input geometry resulting in
neutrino events being generated over the entire geometry volume. If
the `\texttt{\textbf{\small{}-t}}' option is set, event generation
will be confined in the specified detector volume. The option can
be used to simulate events at specific sub-detectors.\\
You can use the `\texttt{\textbf{\small{}-t}}' option to switch generation
on / off at multiple volumes. For further details, see similar discussion
in the description of other event generation applications (eg. \textit{gevgen\_t2k}).
\\
\\
\textbf{\textcolor{magenta}{-L}}\textbf{ }\textbf{\small{}Specifies
the input geometry length units.} This is an optional argument, relevant
only for ROOT-based detector geometry descriptions. By default, that
option is set to `mm'. Possible options include: `m', `cm', `mm',
...\\
 \\
\textbf{\textcolor{magenta}{-D}}\textbf{ }\textbf{\small{}Specifies
the input geometry density units.} This is an optional argument, relevant
only for ROOT-based detector geometry descriptions. By default, that
option is set to `g\_cm3'. Possible options include: `kg\_m3', `g\_cm3',
`clhep\_def\_density\_unit',... \\
\\
\textbf{\textcolor{magenta}{-o}} \textbf{Specifies the output filename
prefix.} This is an optional argument. It allows you to override the
output event file prefix. In GENIE, the output filename is built as:

\texttt{\small{}`prefix.run\_number.event\_tree\_format.file\_format'}
where, in \textit{gevgen\_hadro}, by default, \texttt{\small{}prefix}:
`gntp' and \texttt{\small{}event\_tree\_format}: `ghep' and \texttt{\small{}file\_format}:
`root'.\\
\\
\textbf{\textcolor{magenta}{-r}} \textbf{Specifies the run number.}
This is an optional argument. By default it is set to `0'.\textbf{\textcolor{magenta}{}}\\
\\
\textbf{\textcolor{magenta}{--seed}}\textbf{ Specifies the random
number seed} for the current job.\textbf{}\\
\textbf{}\\
\textbf{\textcolor{magenta}{--message-thresholds}}\textbf{ Specifies
the GENIE verbosity level}. The verbosity level is controlled with
an XML file allowing users to customize the threshold of each message
stream. The XML schema can be seen in \textit{`\$}\texttt{\textbf{\small{}GENIE}}\textit{/config/Messenger.xml}'.
The \textit{`Messenger.xml' }file contains the default thresholds
used by GENIE. The \textit{`Messenger\_laconic.xml' }and \textit{`Messenger\_rambling.xml'
}files define, correspondingly, less and more verbose configurations.\texttt{\textbf{\small{}}}~\\
\texttt{\textbf{\small{}}}~\\
\textbf{\textcolor{magenta}{--event-record-print-level}}\texttt{\textbf{\small{}
}}Allows users to set the level of information shown when the event
94 record is printed in the screen. See GHepRecord::Print() for allowed
settings.\texttt{\textbf{\small{}}}~\\
\texttt{\textbf{\small{}}}~\\
\textbf{\textcolor{magenta}{--mc-job-status-refresh-rate}}\texttt{\textbf{\small{}
}}Allows users to customize the refresh rate of the status file.\textbf{}\\

\chapter{Event Reweighting }

\section{Introduction}

This chapter describes strategies for propagating neutrino interaction
uncertainties. The reweighting schemes described here are tied to
the default physics choices made in GENIE and they have been implemented
in the GENIE ReWeight package. As GENIE evolves, by including better-motivated
theoretical models and integrating new data in its effective models\footnote{ The GENIE development roadmap is outlined at: http://releases.genie-mc.org },
the reweighting schemes need to be updated. This evolution can not
always be transparent but, to aid the user, we strive to keep this
part of the user and physics manual up to date. Also, as our understanding
and the systematic analysis of the GENIE model improves, new reweighting
schemes are added.

For each neutrino-generator input physics quantity $P$, whose uncertainty
is taken into account in this work, we introduce a systematic parameter\footnote{ The terms `systematic parameter', `nuisance parameter', `tweaking
dial' may be used interchangeably in this paper and our presentations/discussions
of this work. } $x_{P}$. Tweaking this systematic parameter modifies the corresponding
physics parameter $P$ as follows: 
\begin{equation}
{\displaystyle P\rightarrow P^{\prime}=P(1+x_{P}*\frac{\delta{P}}{P})}
\end{equation}
where $\delta{P}$ is the estimated standard deviation of $P$. Setting
the systematic parameter to zero corresponds to using the nominal
value of the physics parameter. Tweaking the systematic parameter
by $\pm$1 modifies the corresponding physics quantity $P$ by $\pm\delta{P}$
. In this section, we provide a summary of all the systematic parameters
included in this work, and a brief description of the corresponding
tweaked physics quantities $P$ and, wherever possible, the assumed
fractional errors $\delta{P}/P$. The quantity $P$ may be a single
configurable parameter (eg. $CCQE$ axial mass), or it may be a simple
function of a kinematical parameter (eg. a hadron-nucleus cross-section
as a function of the hadron energy), or, more generally, it may be
any nominal MC prediction, which can not be easily expressed analytically
or tabulated. For that reason, it is always preferable to formulate
the problem (eg. oscillation fits in presence of neutrino-interaction
nuisance parameters) in terms of $x_{P}$.

A number of neutrino cross section systematics are considered in this
chapter, and a complete list of these is given in Tab. \ref{table:NuXSecKnobs}.
The dominant systematics, for neutrino interactions in the few-GeV
energy range, include the axial mass for charged-current quasi-elastic
scattering and the axial and vector masses for both charged-current
and neutral-current resonance neutrino production. Uncertainties in
nuclear effects (Pauli supression) in charged-current quasi-elastic
reactions are taken into account by modifying the Fermi momentum level
$k_{F}$. Uncertainties in the choice of vector form factors (dipole
vs BBA2005) for charged-current quasi-elastic reactions are also taken
into account. Charged-current and neutral-current coherent pion production
uncertainties are taken into account by modifying the corresponding
axial mass and the nuclear size parameter $R_{0}$, which controls
the pion absorption factor in the Rein-Sehgal (RS) model. Uncertainties
in the level of the non-resonance background are considered for all
neutrino charged-current and neutral-current $1\pi$- and $2\pi$-production
channels. Finally, in order to consider uncertainties in charged-current
and neutral-current deep inelastic scattering, the most important
parameters of the Bodek-Yang (BY) model are taken into account. These
BY uncertainties are considered only for events in the `safe' deep-inelastic
kinematic regime ($Q^{2}>$ 1 $GeV^{2}/c^{2}$ and $W>$ 2 $GeV/c^{2}$)
to avoid double counting uncertainties in the resonance / transition
region that have already been taken into account.

\begin{table*}[htb]
\center

\global\long\def\arraystretch{1.75}


\begin{tabular}{llll}
\hline 
$x_{P}$  & Description of $P$  & $\delta{P}/{P}$  & \tabularnewline
\hline 
$x_{M_{A}^{NCEL}}$  & Axial mass for NC elastic  & $\pm$25\%  & \tabularnewline
$x_{\eta^{NCEL}}$  & Strange axial form factor $\eta$ for NC elastic  & $\pm$30\%  & \tabularnewline
$x_{M_{A}^{CCQE}}$  & Axial mass for CC quasi-elastic  & -15\% +25\%  & \tabularnewline
$x_{CCQE-Norm}$ & Normalization factor for CCQE  &  & \tabularnewline
$x_{CCQE-PauliSup}$  & CCQE Pauli suppression (via changes in Fermi level $k_{F}$)  & $\pm$35\%  & \tabularnewline
$x_{CCQE-VecFF}$  & Choice of CCQE vector form factors (BBA05 $\leftrightarrow$ Dipole)  & -  & \tabularnewline
$x_{CCRES-Norm}$ & Normalization factor for CC resonance neutrino production &  & \tabularnewline
$x_{NCRES-Norm}$ & Normalization factor for NC resonance neutrino production &  & \tabularnewline
$x_{M_{A}^{CCRES}}$  & Axial mass for CC resonance neutrino production  & $\pm$20\%  & \tabularnewline
$x_{M_{V}^{CCRES}}$  & Vector mass for CC resonance neutrino production  & $\pm$10\%  & \tabularnewline
$x_{M_{A}^{NCRES}}$  & Axial mass for NC resonance neutrino production  & $\pm$20\%  & \tabularnewline
$x_{M_{V}^{NCRES}}$  & Vector mass for NC resonance neutrino production  & $\pm$10\%  & \tabularnewline
$x_{M_{A}^{COHpi}}$  & Axial mass for CC and NC coherent pion production  & $\pm$50\%  & \tabularnewline
$x_{R_{0}^{COHpi}}$  & Nuclear size param. controlling $\pi$ absorption in RS model  & $\pm$10\%  & \tabularnewline
$x_{R_{bkg}^{\nu p,CC1\pi}}$  & Non-resonance bkg in $\nu p$ $CC1\pi$ reactions  & $\pm$50\%  & \tabularnewline
$x_{R_{bkg}^{\nu p,CC2\pi}}$  & Non-resonance bkg in $\nu p$ $CC2\pi$ reactions  & $\pm$50\%  & \tabularnewline
$x_{R_{bkg}^{\nu n,CC1\pi}}$  & Non-resonance bkg in $\nu n$ $CC1\pi$ reactions  & $\pm$50\%  & \tabularnewline
$x_{R_{bkg}^{\nu n,CC2\pi}}$  & Non-resonance bkg in $\nu n$ $CC2\pi$ reactions  & $\pm$50\%  & \tabularnewline
$x_{R_{bkg}^{\nu p,NC1\pi}}$  & Non-resonance bkg in $\nu p$ $NC1\pi$ reactions  & $\pm$50\%  & \tabularnewline
$x_{R_{bkg}^{\nu p,NC2\pi}}$  & Non-resonance bkg in $\nu p$ $NC2\pi$ reactions  & $\pm$50\%  & \tabularnewline
$x_{R_{bkg}^{\nu n,NC1\pi}}$  & Non-resonance bkg in $\nu n$ $NC1\pi$ reactions  & $\pm$50\%  & \tabularnewline
$x_{R_{bkg}^{\nu n,NC2\pi}}$  & Non-resonance bkg in $\nu n$ $NC2\pi$ reactions  & $\pm$50\%  & \tabularnewline
$x_{A_{HT}^{BY}}$  & $A_{HT}$ higher-twist param in BY model scaling variable $\xi_{w}$  & $\pm$25\%  & \tabularnewline
$x_{B_{HT}^{BY}}$  & $B_{HT}$ higher-twist param in BY model scaling variable $\xi_{w}$  & $\pm$25\%  & \tabularnewline
$x_{C_{V1u}^{BY}}$  & $C_{V1u}$ u valence GRV98 PDF correction param in BY model  & $\pm$30\%  & \tabularnewline
$x_{C_{V2u}^{BY}}$  & $C_{V2u}$ u valence GRV98 PDF correction param in BY model  & $\pm$40\%  & \tabularnewline
$x_{CCDIS}$ & Inclusive CC cross-section normalization factor &  & \tabularnewline
$x_{CC\bar{\nu}/\nu}$ & $\bar{\nu}/\mbox{\ensuremath{\nu}}$ CC ratio &  & \tabularnewline
$x_{DIS-NuclMod}$ & DIS nuclear modification (shadowing, anti-shadowing, EMC) &  & \tabularnewline
\hline 
\end{tabular}\\[2pt] \caption{Neutrino interaction cross-section systematic parameters considered
in GENIE. For some of the above parameters there are two reweighting
implementations: One which includes the full effect of the systematic
(shape + normalization) and one which includes only its effect on
the shape of observable distributions (maintains normalization). Note
that some systematics have overlapping effects so care is needed to
avoid double counting.}

\label{table:NuXSecKnobs} 
\end{table*}

We consider a number of uncertainties in neutrino-induced hadronization
and resonance decays. We include uncertainties in the assignment of
pion kinematics in $N\pi$ hadronic states generated by the Andreopoulos-Gallagher-Kehayias-Yang
(AGKY) GENIE hadronization model, as well as uncertainties in the
in-medium modifications of the hadronization process. Uncertainties
in the pion angular distribution in $\Delta\rightarrow\pi N$ decays
and uncertainties in certain resonance-decay branching ratios are
also taken into account. The complete list is given in Tab. \ref{table:HadronzKnobs}.

Finally, we consider two kinds of uncertainties affecting the INTRANUKE
(hA) intranuclear hadron transport model: Uncertainties in the total
rescattering probability (mean free path) for hadrons within the target
nucleus and uncertainties in the conditional probability of each hadron
rescattering mode (elastic, inelastic, charge exchange, pion production
and absorption / multi-nucleon knock-out), given that a rescattering
did occur. These physics uncertainties are considered separately for
nucleons and pions. The complete list of systematic parameters is
given in Tab. \ref{table:HadTranspKnobs}.

\begin{table*}[htb]
\center

\global\long\def\arraystretch{1.75}


\begin{tabular}{llll}
\hline 
$x_{P}$  & Description of $P$  & $\delta{P}/{P}$  & \tabularnewline
\hline 
$x_{AGKY}^{pT1\pi}$  & Pion transverse momentum ($p_{T}$) for $N\pi$ states in AGKY  & -  & \tabularnewline
$x_{AGKY}^{xF1\pi}$  & Pion Feynman x ($x_{F}$) for $N\pi$ states in AGKY  & -  & \tabularnewline
$x_{fz}$  & Hadron formation zone  & $\pm$50\%  & \tabularnewline
$x_{\theta_{\pi}}^{\Delta\rightarrow\pi N}$  & Pion angular distribution in $\Delta\rightarrow\pi N$ (isotropic
$\leftrightarrow$ RS)  & -  & \tabularnewline
$x_{BR}^{R\rightarrow X+1\gamma}$  & Branching ratio for radiative resonance decays  & $\pm$50\%  & \tabularnewline
$x_{BR}^{R\rightarrow X+1\eta}$  & Branching ratio for single-$\eta$ resonance decays  & $\pm$50\%  & \tabularnewline
\hline 
\end{tabular}\\[2pt] \caption{ Neutrino-induced hadronization and resonance-decay systematic parameters
considered in this work. }

\label{table:HadronzKnobs} 
\end{table*}

\begin{table*}[htb]
\center

\global\long\def\arraystretch{1.75}


\begin{tabular}{llll}
\hline 
$x_{P}$  & Description of $P$  & $\delta{P}/{P}$  & \tabularnewline
\hline 
$x_{mfp}^{N}$  & Nucleon mean free path (total rescattering probability)  & $\pm$20\%  & \tabularnewline
$x_{cex}^{N}$  & Nucleon charge exchange probability  & $\pm$50\%  & \tabularnewline
$x_{el}^{N}$  & Nucleon elastic reaction probability  & $\pm$30\%  & \tabularnewline
$x_{inel}^{N}$  & Nucleon inelastic reaction probability  & $\pm$40\%  & \tabularnewline
$x_{abs}^{N}$  & Nucleon absorption probability  & $\pm$20\%  & \tabularnewline
$x_{\pi}^{N}$  & Nucleon $\pi$-production probability  & $\pm$20\%  & \tabularnewline
$x_{mfp}^{\pi}$  & $\pi$ mean free path (total rescattering probability)  & $\pm$20\%  & \tabularnewline
$x_{cex}^{\pi}$  & $\pi$ charge exchange probability  & $\pm$50\%  & \tabularnewline
$x_{el}^{\pi}$  & $\pi$ elastic reaction probability  & $\pm$10\%  & \tabularnewline
$x_{inel}^{\pi}$  & $\pi$ inelastic reaction probability  & $\pm$40\%  & \tabularnewline
$x_{abs}^{\pi}$  & $\pi$ absorption probability  & $\pm$20\%  & \tabularnewline
$x_{\pi}^{\pi}$  & $\pi$ $\pi$-production probability  & $\pm$20\%  & \tabularnewline
\hline 
\end{tabular}\\[2pt] \caption{ Intranuclear hadron transport systematic parameters considered in
this work. }

\label{table:HadTranspKnobs} 
\end{table*}

\clearpage

\section{Propagating neutrino-cross section uncertainties}

\label{sec:xsec_syst}

Unlike the propagation of hadronic simulation uncertainties (to be
discussed later), which is challenging as the probability for a generated
multi-particle configuration is difficult to calculate analytically,
the propagation of neutrino interaction cross-section modelling uncertainties
is relatively straightforward using a generic reweighing scheme less
strongly tied to the details of the physics modeling. Cross section
reweighing is modifying the neutrino interaction probability directly
and, therefore the considerations on unitarity conservation developed
in the hadron transport reweighing section are not relevant here.

The neutrino event weight, $w_{\sigma}^{evt}$, to account for changes
in physics parameters controlling neutrino cross sections is calculated
as 
\begin{equation}
{\displaystyle w_{\sigma}^{evt}=(d^{n}{\sigma}_{\nu}^{\prime}/dK^{n})/(d^{n}{\sigma}_{\nu}/dK^{n})}
\end{equation}
 where $d^{n}\sigma/dK^{n}$ is the nominal differential cross section
for the process at hand, $d^{n}\sigma^{\prime}/dK^{n}$ is the differential
cross section computed using the modified input physics parameters.
The differential cross section is evaluated at the kinematical phase
space $\{K^{n}\}$\footnote{ In GENIE, typically, the $K^{n}$ kinematical phase space is $\{Q^{2}\}$
for CC quasi-elastic and NC elastic, $\{Q^{2},W\}$ for resonance
neutrino production, $\{x,y\}$ for deep inelastic scattering and
coherent or diffractive meson production, $\{y\}$ for ${\nu}e^{-}$
elastic scattering or inverse muon decay where $Q^{2}$ is the momentum
transfer, $W$ the hadronic invariant mass, $x$ is Bjorken scaling
variable and $y$ the inelasticity. The choice is not significant.
The differential cross section calculation can be mapped from the
$K^{n}$ to the $K^{n\prime}$ kinematic phase space through the Jacobian
for the $K^{n}\rightarrow K^{n\prime}$ transformation. }. A critical point in implementing the cross section reweighing scheme
for scattering off nuclear targets, is that the correct off-shell
kinematics, as used in the original simulation, must be recreated
before evaluating the differential cross sections. This is trivial
as long as detailed information for the bound nucleon target has been
maintained by the simulation.

\begin{figure}[htb]
\center \includegraphics[width=19pc]{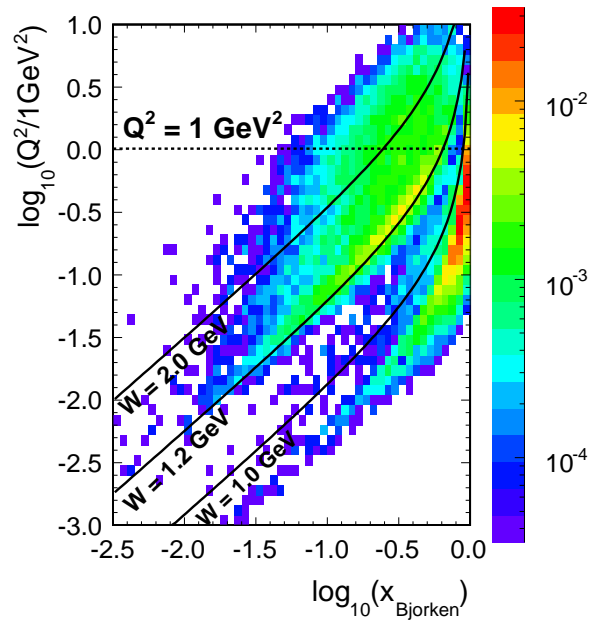} 

\caption{ JPARC neutrino beam kinematic coverage at the nd280. Cross section
uncertainties of different magnitude are appropriate for different
parts of the kinematic phase space. }

\label{fig:JPARCKineCoverage} 
\end{figure}

\begin{figure}[htb]
\center \includegraphics[width=19pc]{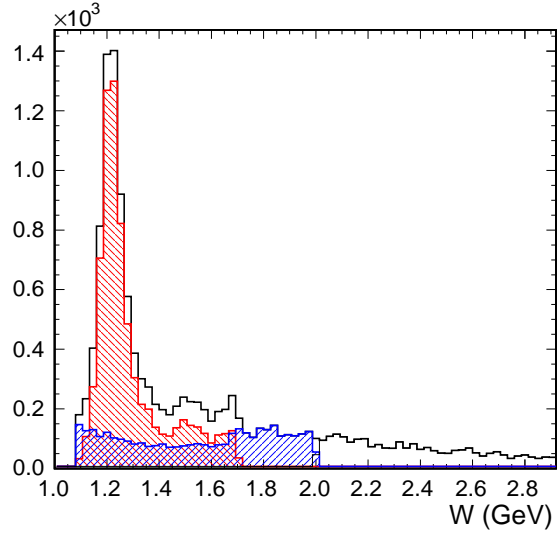} 

\caption{ True hadronic invariant mass, W, distribution for inelastic events
in nd280 (shown with the black solid line). The red hatched area shows
the resonance contributions while the blue hatched area shows the
contributions from the type of inelastic events dubbed in GENIE as
`transition-DIS'. The remaining contributions are coming from the
`safe-DIS' and `low $Q^{2}$ DIS' components. Different uncertainties
are associated with each component: The resonance uncertainty is of
the order of 20\%, while the `transition-DIS' uncertainty is of the
order of 50\%. The uncertainty associated with the remaining DIS component
at higher invariant masses is significantly lower (of the order of
5\% at an energy of 5 GeV and lower at higher energies) and have not
been included at this first iteration of deploying the reweighing
tools. }

\label{fig:Wdep} 
\end{figure}

\section{Propagating hadronization and resonance decay uncertainties}

\begin{figure}[ht]
\begin{minipage}[t]{0.45\linewidth}%
\centering \includegraphics[width=15pc]{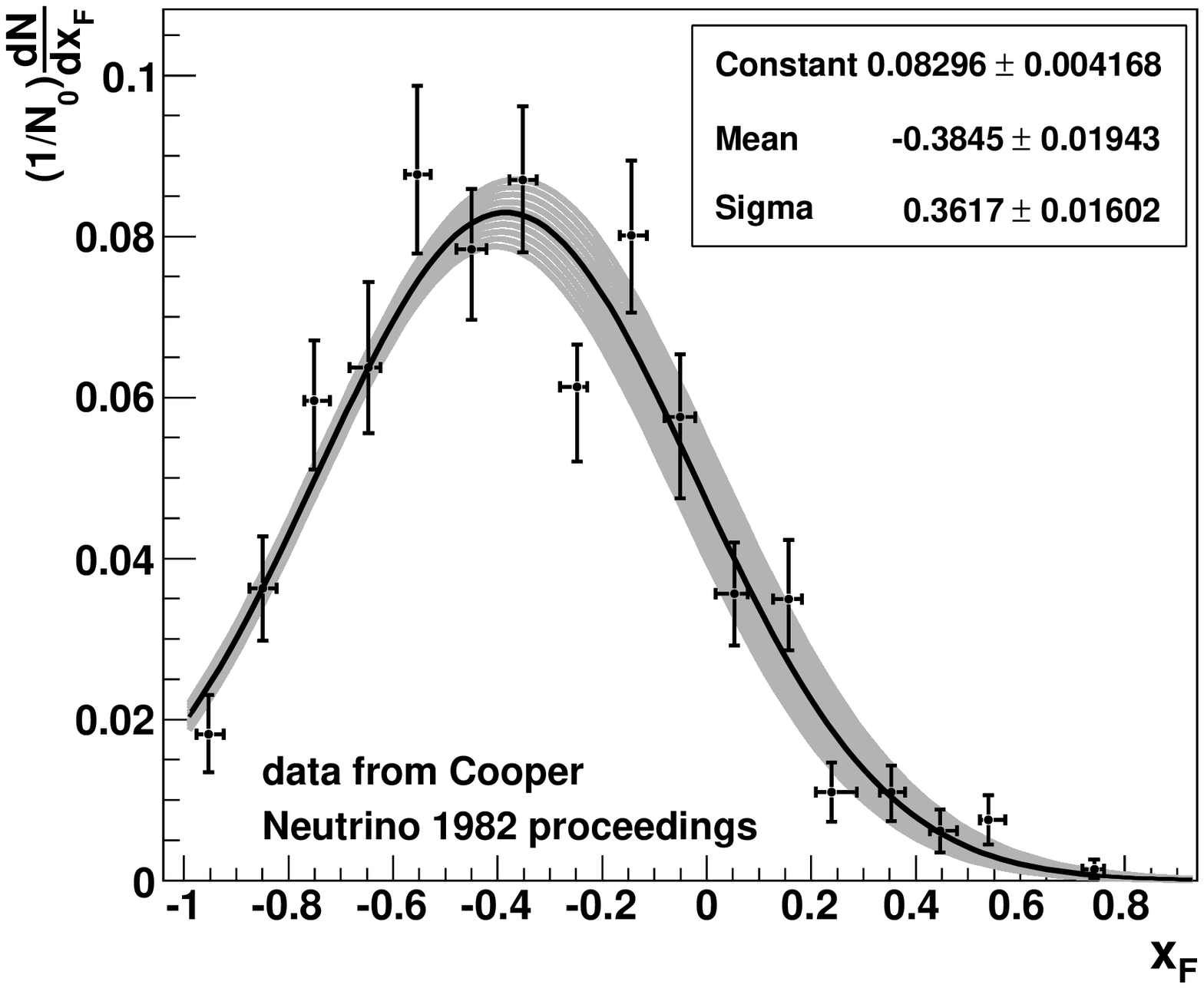}
\caption{ Nucleon Feynman x ($x_{F}$) pdf used in the GENIE AGKY model for
generating the kinematics of 2-body $N+\pi$ primary hadronic systems. }

\label{fig:AgkyNucXF} %
\end{minipage}\hspace{1.5cm} %
\begin{minipage}[t]{0.45\linewidth}%
\centering \includegraphics[width=15pc]{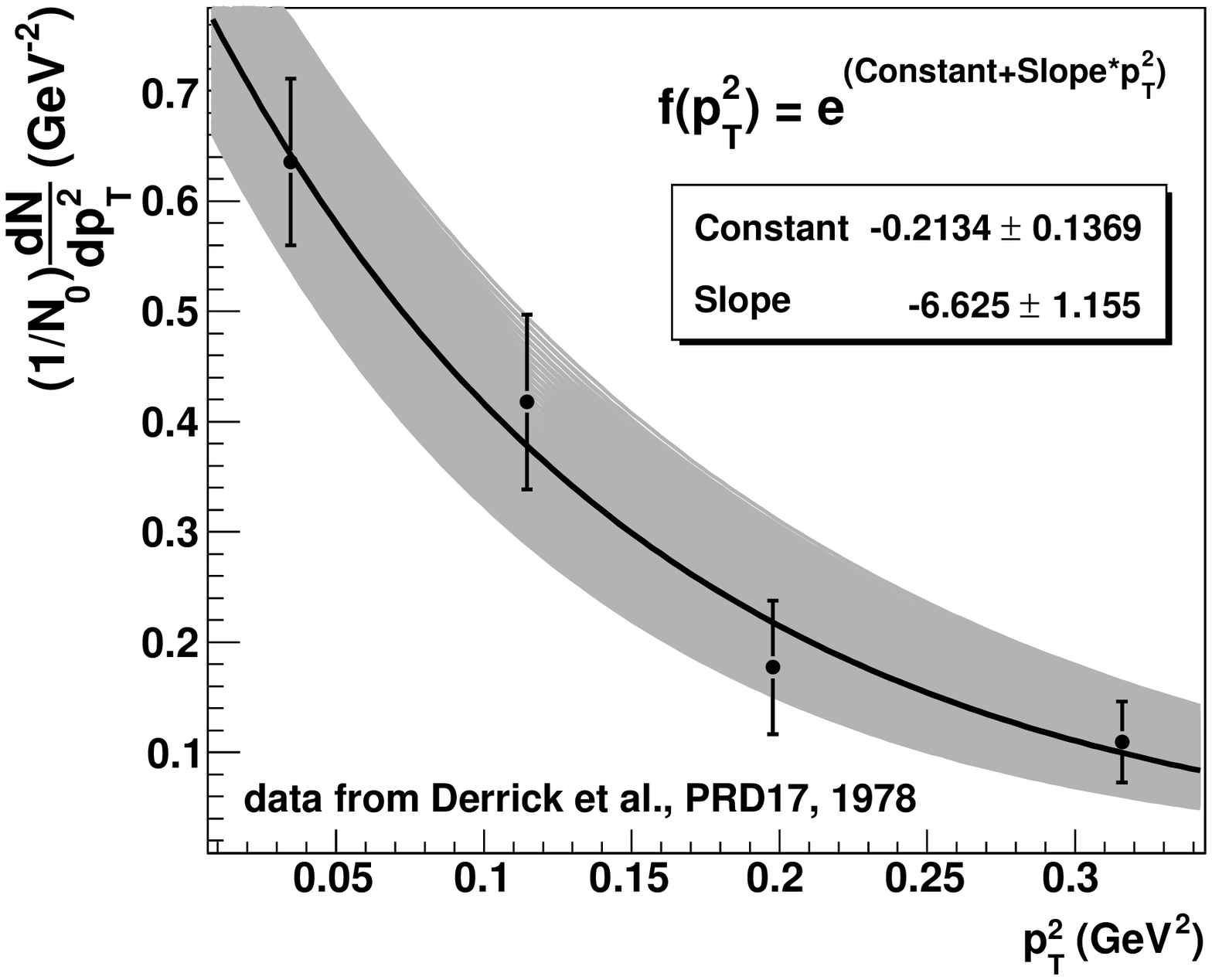}
\caption{ Nucleon transverse momentum ($p_{T}^{2}$) pdf used in the GENIE
AGKY model for generating the kinematics of 2-body $N+\pi$ primary
hadronic systems. }

\label{fig:AgkyNucPT} %
\end{minipage}
\end{figure}

Significant uncertainties exist in the modelling of neutrino-induced
hadronization for neutrinos in the few-GeV energy range. In the energy
range of T2K, possibly the most important hadronization uncertainty
is that in the assignment of pion kinematics for $N\pi$ hadronic
states. In GENIE, low invariant-mass hadronization is handled exclusively
by the KNO-based model included in AGKY. This model uses target-fragment
Feynman x ($x_{F}$) and transverse momentum ($p_{T}^{2}$) pdfs extracted
from bubble chamber data. The pdf used for $x_{F}$ has a particularly
large effect on the characteristics of the generated hadronic system
since a preferentially backward-going (in the hadronic CM frame) heavy
target-fragment (nucleon) leads to a preferentially forward-going
fast current-fragment (pion). This allows GENIE to reproduce the experimental
data on the backward/forward $x_{F}$ asymmetry. There is, however,
experimental ambiguity on whether this backward/forward asymmetry
also exists for lower-multiplicity events. The $x_{F}$ and $p_{T}^{2}$
pdfs used in GENIE (v2.6.0) are shown in Figs. \ref{fig:AgkyNucXF}
and \ref{fig:AgkyNucPT} respectively. They are parametrized as 
\begin{equation}
{\displaystyle f(x_{F})=Ae^{0.5(x_{F}-<x_{F}>)^{2}/\sigma_{x_{F}}^{2}}}\label{eq:AgkyNucXFpdfdef}
\end{equation}
and 
\begin{equation}
{\displaystyle f(p_{T}^{2})=Be^{-p_{T}^{2}/<p_{T}^{2}>}}\label{eq:AgkyNucPT2pdfdef}
\end{equation}

\begin{figure}[ht]
\begin{minipage}[t]{0.45\linewidth}%
\centering \includegraphics[width=15pc]{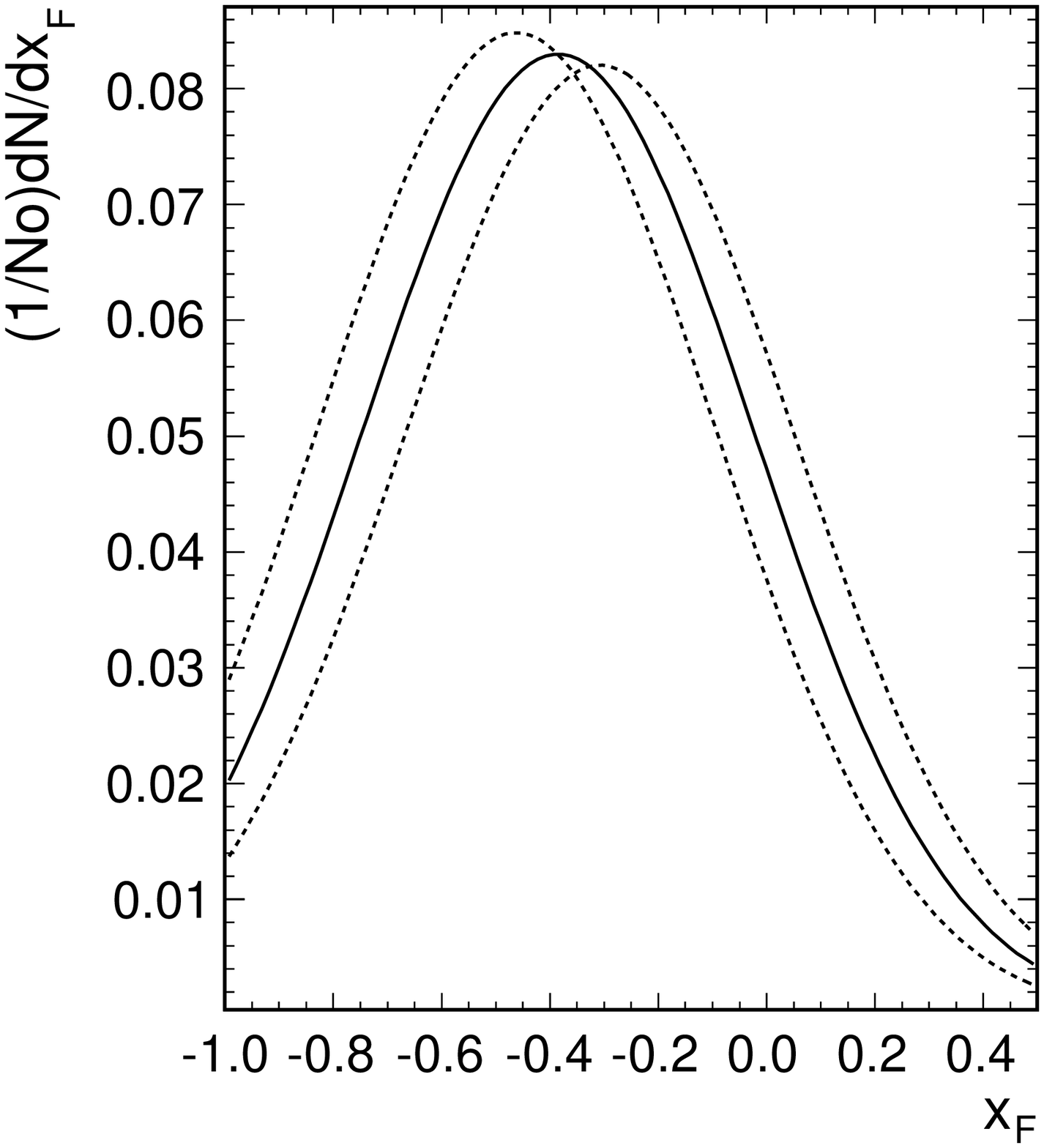}
\caption{ Default $x_{F}$ pdf (solid line) and tweaked pdfs (dotted lines)
resulting from modifying the $x_{AGKY}^{xF1\pi}$ systematic parameter
by $\pm$1. }

\label{fig:AgkyNucXFtweak} %
\end{minipage}\hspace{1.5cm} %
\begin{minipage}[t]{0.45\linewidth}%
\centering \includegraphics[width=15pc]{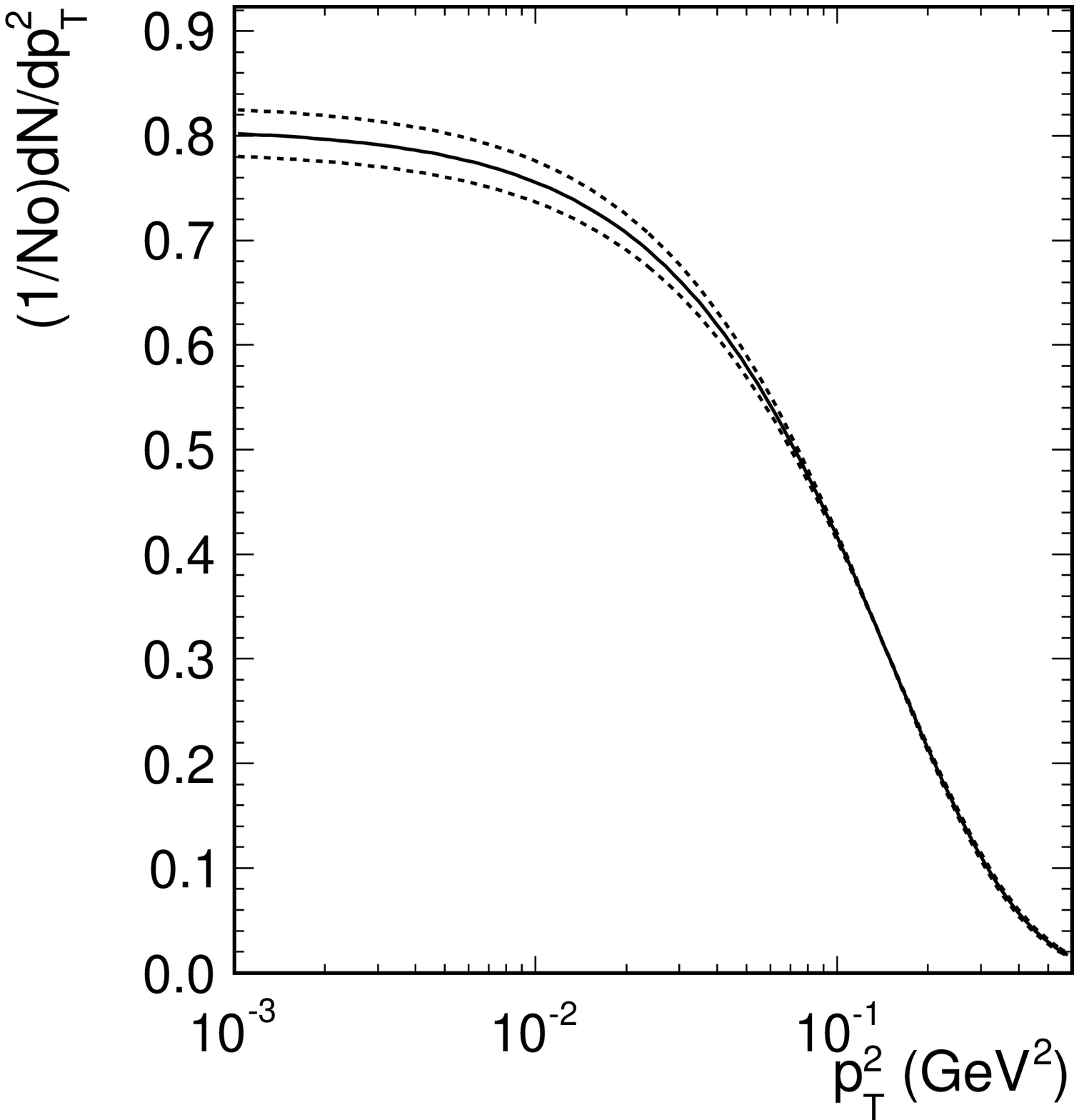}
\caption{ Default $p_{T}^{2}$ pdf (solid line) and tweaked pdfs (dotted lines)
resulting from modifying the $x_{AGKY}^{pT1\pi}$ systematic parameter
by $\pm$1. }

\label{fig:AgkyNucPTtweak} %
\end{minipage}
\end{figure}

In the reweighting scheme employed in this work, the systematic parameter
$x_{AGKY}^{xF1\pi}$ ($x_{AGKY}^{pT1\pi}$) is used to tweak $<x_{F}>$
($<p_{T}^{2}>$) in Eqs. \ref{eq:AgkyNucXFpdfdef} (\ref{eq:AgkyNucPT2pdfdef}).
This modifies the $x_{F}$ and $p_{T}^{2}$ pdfs as shown in Figs.
\ref{fig:AgkyNucXFtweak} and \ref{fig:AgkyNucPTtweak}. Our reweighting
code identifies events with a $N\pi$ hadronic state produced by the
AGKY model and extracts the pion $x_{F}$, $p_{T}^{2}$ and the hadronic
invariant mass $W$. For each such event, 2 $\times$ 10$^{4}$ $N\pi$
hadronic decays, with invariant mass $W$, are performed for both
the default and tweaked values of the $x_{AGKY}^{xF1\pi}$ and $x_{AGKY}^{pT1\pi}$
systematic parameters. The generated decays are analysed to obtain
the default and tweaked 2-dimensional pion-kinematics pdfs $f_{\pi}^{def}(x_{F},p_{T}^{2};W)$
and $f_{\pi}^{twk}(x_{F},p_{T}^{2};W)$. The event weight is computed
as 
\begin{equation}
{\displaystyle w=f_{\pi}^{twk}(x_{F},p_{T}^{2};W)/f_{\pi}^{def}(x_{F},p_{T}^{2};W)}\label{eq:AgkyWght}
\end{equation}

\subsubsection{Formation-zone uncertainties}

\label{sec:FZone}

It is well established that hadrons produced in the nuclear environment
do not immediately reinteract with their full cross section. Initially
quarks propagate through the nucleus with a dramatically reduced probability
of interaction as they have not yet materialized as hadrons. This
is implemented in GENIE as a `free step' for all hadrons produced
in deep-inelastic reactions. The `free step', $f_{z}$, which comes
from a formation time of $\tau_{0}=$ 0.342 fm/c, is calculated as
\begin{equation}
{\displaystyle f_{z}=pc\tau_{0}/m}\label{eq:FormZone}
\end{equation}
where $p$ is the hadron momentum, $m$ is the hadron mass and $c$
is the speed of light.

In the reweighting scheme employed in this work, the original formation
zone assigned to each hadron during event generation is recovered
from the distance between the intranuclear event vertex and the hadron
position as recorded at the beginning of the intranuclear cascade
step. As usual, the systematic parameter $x_{fz}$ modifies the formation
zone: 
\begin{equation}
{\displaystyle f_{z}\rightarrow f_{z}^{\prime}=f_{z}(1+x_{fz}*\delta{f_{z}}/f_{z})}
\end{equation}
Weights are calculated in a way similar to that used when modifying
the hadron mean free path (see section \ref{sec:RewHadronTransport}).
When the formation zone is tweaked, it alters the amount of nuclear
matter through which the hadron must propagate before it exits the
target nucleus. The nominal and tweaked survival probabilities are
calculated as in Eq. \ref{Eq:SurvProb} and a hadron weight is assigned
as in Eq. \ref{eq:ParticleWeightMFP}. An event weight is calculated
as the product of particle weights for all particles in the primary
hadronic system.

\subsubsection{Pion angular distribution uncertainties in $\Delta\rightarrow N\pi$
decay}

\label{sec:RDecThetaPi}

In general, the pion angular distribution $W_{\pi}(cos\theta)$ in
$\Delta\rightarrow N\pi$ decay can be expressed as 
\begin{equation}
{\displaystyle W_{\pi}(cos\theta)=1-p(\frac{3}{2})P2(cos\theta)+p(\frac{1}{2})P2(cos\theta)}\label{eq:WDeltaTheta}
\end{equation}
where $\theta$ is the pion production angle in the $\Delta$ center
of mass frame with respect to the $\Delta$ angular momentum quantization
axis, $P2$ is the 2nd order Legendre polynomial and $p(\frac{3}{2})$,
$p(\frac{1}{2})$ are coefficients for each state of $\Delta$ angular
momentum projection ($\frac{3}{2}$, $\frac{1}{2}$).

For simplicity, GENIE decays baryon resonances isotropically during
event generation. Isotropy requires $p(\frac{3}{2})=p(\frac{1}{2})=0.5$
but the Rein-Sehgal (RS) model predicts $p(\frac{3}{2})=0.75$ and
$p(\frac{1}{2})=0.25$. In this work, we employ a reweighting scheme
to quantify the uncertainty over the $\pi$ angular momentum distribution.
A measure of this uncertainty is taken to be the difference between
the isotropic and RS predictions. The reweighting code identifies
events with a $\Delta++(1232)$ decaying to a $N\pi$ state. In a
nuclear environment, where hadronic rescattering is possible, the
$N\pi$ state produced may not necessarily be the final hadronic state.
Once the event is identified, the daughter $\pi$ and parent $\Delta$
4-momenta in the LAB frame are used to calculate the $\pi$ 4-momentum
in the $\Delta$ center-of-mass frame. Then the $\pi$ production
angle $\theta$ is calculated with respect to an arbitrarily-defined
angular momentum quantization axis ($+z$). Let $W_{\pi}^{iso}(cos\theta)$
and $W_{\pi}^{RS}(cos\theta)$ be, respectively, the $\pi$ production-angle
probability density for the isotropic and RS cases, computed from
Eq. \ref{eq:WDeltaTheta}. An event weight is constructed as follows:
\begin{equation}
{\displaystyle w=\Big(x_{\theta_{\pi}}^{\Delta\rightarrow\pi N}W_{\pi}^{RS}(cos\theta)+(1-x_{\theta_{\pi}}^{\Delta\rightarrow\pi N})W_{\pi}^{iso}(cos\theta)\Big)/W_{\pi}^{iso}(cos\theta)}
\end{equation}
where $x_{\theta_{\pi}}^{\Delta\rightarrow\pi N}$ is the corresponding
nuisance parameter. For $x_{\theta_{\pi}}^{\Delta\rightarrow\pi N}=0$,
all weights are equal to 1, i.e.\ this setting corresponds to the
default case of isotropic $\Delta$ decays. For $x_{\theta_{\pi}}^{\Delta\rightarrow\pi N}=1$,
the calculated weight is equal to $W_{\pi}^{RS}(cos\theta)/W_{\pi}^{iso}(cos\theta)$;
this reweights the isotropic pion angular distributions to those predicted
by RS. For values of $x_{\theta_{\pi}}^{\Delta\rightarrow\pi N}$
between 0 and 1, there is a linear transition between the isotropic
and RS angular distributions.

\begin{figure}[ht]
\centering \includegraphics[width=32pc]{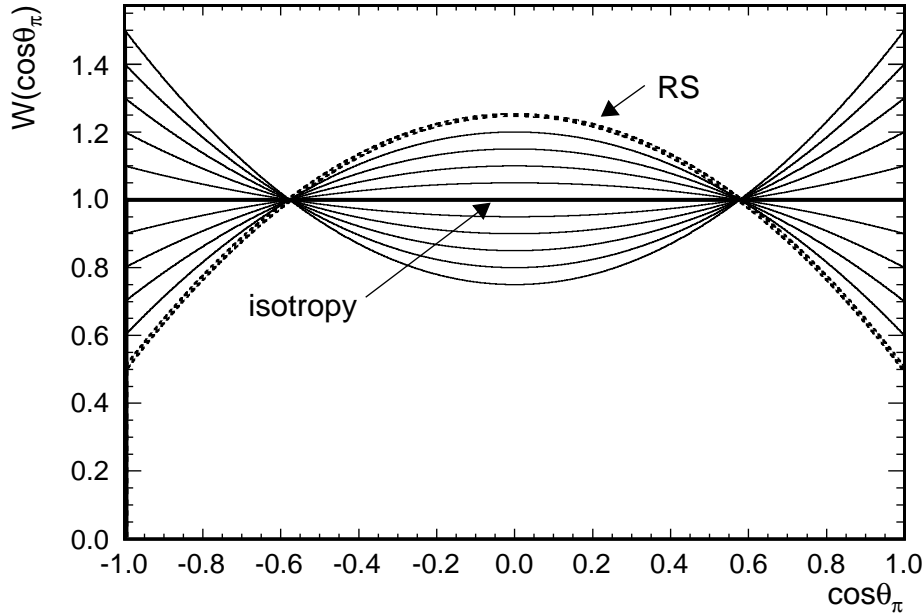}
\caption{ Angular distributions for pions from $\Delta\rightarrow\pi N$ decays
for various values of the $x_{\theta_{\pi}}^{\Delta\rightarrow\pi N}$
nuisance parameter between -1 and 1. The isotropic distribution (GENIE
simulation default) is obtained for $x_{\theta_{\pi}}^{\Delta\rightarrow\pi N}=0$.
The RS model prediction is obtained for $x_{\theta_{\pi}}^{\Delta\rightarrow\pi N}=1$. }

\label{fig:Wcostheta_DeltaNpi} 
\end{figure}

\subsubsection{Branching ratio uncertainties}

\label{sec:RDecBR}

Reweighting events to account for changes in decay branching ratios
is straightforward. It is important to ensure that the sum of all
branching ratios for each unstable particle remains unchanged.

Let $x_{d}^{p}$ be a nuisance parameter which affects the branching
ratio $f_{d}^{p}$ for the decay channel $d$ which is available to
particle $p$. As usual in this work, the nuisance parameter modifies
the corresponding physics parameter (branching ratio) as $f_{d}^{p}\rightarrow f_{d}^{\prime p}=f_{d}^{p}(1+x_{d}^{p}*\sigma_{f_{d}^{p}}/f_{d}^{p})$,
where $\sigma_{f_{d}^{p}}$ is the uncertainty in the branching ratio.
In the reweighting scheme employed in this work, if {\em any} branching
ratio of a given particle is tweaked, then {\em all} decays of
that particle are reweighted so that the sum of all branching ratios
remains unchanged. If $x_{d}^{p}$ is tweaked, the weight for decay
$d$ is computed as follows: 
\begin{equation}
{\displaystyle w_{d}^{p}=\frac{f_{d}^{\prime p}}{f_{d}^{p}}}
\end{equation}
For every other decay $d\prime\ne d$ of that particle a weight is
computed as: 
\begin{equation}
{\displaystyle w_{d\prime}^{p}=\frac{1-f_{d}^{\prime p}}{1-f_{d}^{p}}}
\end{equation}
The above weight is assigned to a single unstable particle for which
the branching ratio of any decay channel has been altered. The event
weight is the product of weights for all such particles.

\clearpage

\section{Propagating intranuclear hadron transport uncertainties}

\label{sec:RewHadronTransport}

\begin{figure}[htb]
\center \includegraphics[width=19pc]{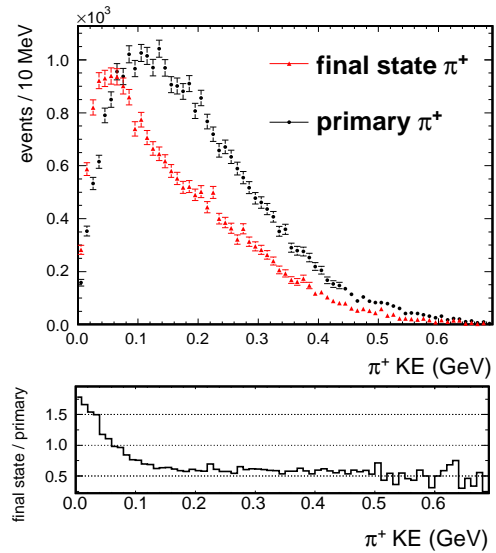} 

\caption{Kinetic energy spectrum of final state and primary (before rescattering)
$\pi^{+}$ produced in $\nu_{\mu}Fe^{56}$ interactions at 1 GeV.}

\label{fig:INukeSpectrum} 
\end{figure}

Hadrons produced in the nuclear environment may rescatter on their
way out of the nucleus, and these reinteractions significantly modify
the observable distributions. The simulated effect of hadronic reiniteractions
is illustrated in Tab. \ref{tab:INukeTopo} and Fig. \ref{fig:INukeSpectrum}.
The sensitivity of a particular experiment to intranuclear rescattering
depends strongly on the detector technology, the energy range of the
neutrinos, and the physics measurement being made.

\begin{sidewaystable*}
\global\long\def\arraystretch{1.75}
\begin{tabular}{c||cccccccccc}
\hline 
Final-  & \multicolumn{10}{c}{ Primary Hadronic System }\tabularnewline
State  & $0\pi X$  & $1\pi^{0}X$  & $1\pi^{+}X$  & $1\pi^{-}X$  & $2\pi^{0}X$  & $2\pi^{+}X$  & $2\pi^{-}X$  & $\pi^{0}\pi^{+}X$  & $\pi^{0}\pi^{-}X$  & $\pi^{+}\pi^{-}X$ \tabularnewline
\hline 
$0\pi X$  & \textbf{293446}  & 12710  & 22033  & 3038  & 113  & 51  & 5  & 350  & 57  & 193 \tabularnewline
$1\pi^{0}X$  & 1744  & \textbf{44643}  & 3836  & 491  & 1002  & 25  & 1  & 1622  & 307  & 59 \tabularnewline
$1\pi^{+}X$  & 2590  & 1065  & \textbf{82459}  & 23  & 14  & 660  & 0  & 1746  & 5  & 997 \tabularnewline
$1\pi^{-}X$  & 298  & 1127  & 1  & \textbf{12090}  & 16  & 0  & 46  & 34  & 318  & 1001 \tabularnewline
$2\pi^{0}X$  & 0  & 0  & 0  & 0  & \textbf{2761}  & 2  & 0  & 260  & 40  & 7 \tabularnewline
$2\pi^{+}X$  & 57  & 5  & 411  & 0  & 1  & \textbf{1999}  & 0  & 136  & 0  & 12 \tabularnewline
$2\pi^{-}X$  & 0  & 0  & 0  & 1  & 0  & 0  & \textbf{134}  & 0  & 31  & 0 \tabularnewline
$\pi^{0}\pi^{+}X$  & 412  & 869  & 1128  & 232  & 109  & 106  & 0  & \textbf{9837}  & 15  & 183 \tabularnewline
$\pi^{0}\pi^{-}X$  & 0  & 0  & 1  & 0  & 73  & 0  & 8  & 5  & \textbf{1808}  & 154 \tabularnewline
$\pi^{+}\pi^{-}X$  & 799  & 7  & 10  & 65  & 0  & 0  & 0  & 139  & 20  & \textbf{5643} \tabularnewline
\hline 
\end{tabular}\\[2pt] 

\caption{ Occupancy of primary and final state hadronic systems for interactions
off $O^{16}$ computed with GENIE v2.4.0. The off-diagonal elements
illustrate and quantify the topology changing effect of intranuclear
rescattering.}

\label{tab:INukeTopo} 
\end{sidewaystable*}

Neutrino generators typically use intranuclear cascade simulations
to handle the propagation of hadronic multi-particle states. At each
simulation step a large number of outcomes is accessible with the
probabilities of those outcomes being conditional upon the hadron
transport history up to that point. The complexity of intranuclear
hadron transport makes it difficult to evaluate the probability for
a generated multi-particle final state, given a primary hadronic multi-particle
system, without resorting to a Monte Carlo method. Subsequently, is
not possible to evaluate how that probability ought to be modified
in response to changes in the fundamental physics inputs. As a result
it is generally not possible to build comprehensive reweighing schemes
for intranuclear hadron-transport simulations.

In this regard GENIE's INTRANUKE/hA model is unique by virtue of the
simplicity of the simulation while, at the same time, it exhibiting
very reliable aspects by being anchored to key hadron-nucleon and
hadron-nucleus data. Its simplicity allows a rather straightforward
probability estimate for the generated final state making it amenable
to reweighing. A full systematic analysis of the model is therefore
possible making it a unique tool in the analysis of neutrino data.
The event reweighing strategy to be presented here is \em specific
\em to GENIE's INTRANUKE/hA model. The current reweighing implementation
has been tied to the physics choices made in the GENIE v2.4.0\footnote{ The validity of the current reweighing implementation in future versions
of GENIE is dependent upon the INTRANUKE/hA changes that may be installed.
The T2KReWeight package will always be updated and kept in sync with
GENIE. In case of important updates a follow-up internal note will
be posted.}.

\begin{figure}[htb]
\center \includegraphics[width=23pc]{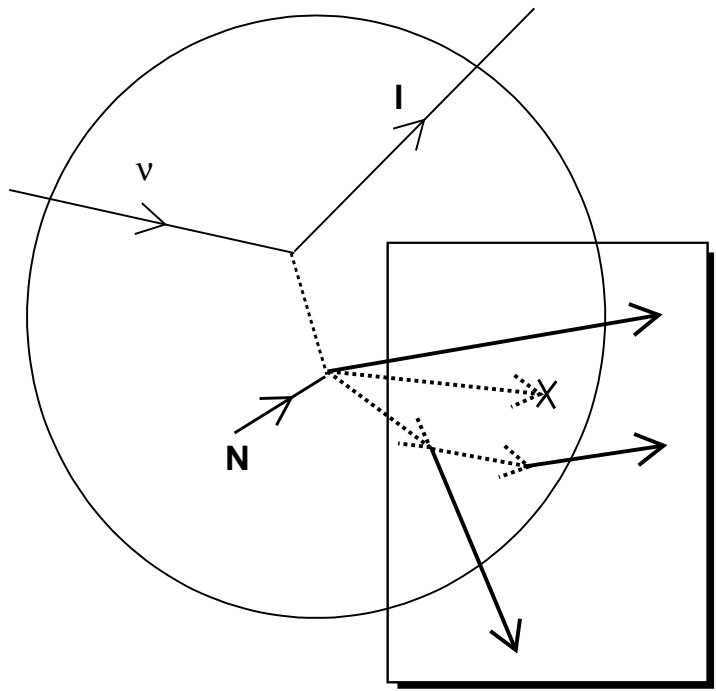} 

\caption{ Consider the effect of modifying the intranuclear hadron-transport
physics (affecting the particles within the \em box\em) from the
perspective of an observer who is blind to the hadronic system emerging
from the nucleus and measures only the primary lepton. One can easily
assert that, from the perspective of that observer, the hadron-transport
reweighing scheme should have no effect on the leptonic system characteristics
of samples that have not been selected for hadronic system characteristics.
The event weights must cancel each other so as the sum of weights
is conserved, therefore maintaining the sample normalization. We will
be referring to that condition as the \em `unitarity constraint'\em.
As we will see in the reweighing validation section, the scheme discussed
in this note satisfies the unitarity constraint, by \em virtue of
construction\em, to better than 1 part in 5000. }

\label{fig:Unitarity} 
\end{figure}

Any intranuclear hadron-transport reweighing strategy should, by \em
virtue of construction\em, have no effect on the inclusive leptonic
distributions of the reweighted sample, as illustrated in Fig. \ref{fig:Unitarity}.
In this paper we will be referring to that probability conservation
condition as the \em `unitarity constraint'\em. We emphasize the
fact that the constraint needs to hold only for unselected samples.
It does not need to hold for selected samples, where the normalization
is expected to vary due to the effect of the cut acceptance.

The unitarity constraint is obviously very difficult to satisfy by
virtue of construction and has had a significant role in determining
the reweighing strategy. Additionally, the constraint played an important
role in validating the reweighing scheme and in matching exactly all
physics assumptions of the original simulation. The most profound
effect of weighting artifacts is to cause the unitarity constraint
to be violated. We will revisit the issue of the unitarity constraint
in later sections and, particularly, on the discussion of the reweighing
validation.

In the reweighing strategy developed here we consider 2 kinds of physics
uncertainties: 
\begin{itemize}
\item Uncertainties in the total rescattering probability for hadrons within
the target nucleus. 
\item Uncertainties in the relative probability of rescattering modes available
to each hadron once it interacts. 
\end{itemize}
These physics uncertainties are considered separately for nucleons
and pions. The determination of simulation parameters linked with
these physics uncertainties and the prescription for calculating event
weights to account for variations in these parameters is discussed
next.

\subsubsection{Reweighting the rescattering rate}

During event generation, for each hadron being propagated within the
nuclear environment its rescattering probability, $P_{rescat}^{h}$
(or, equivalently the survival probability, $P_{surv}^{h}$) is calculated
as 
\begin{equation}
{\displaystyle P_{rescat}^{h}=1-P_{surv}^{h}=1-\int{e^{-r/\lambda^{h}(\vec{r},h,E_{h})}dr}}\label{Eq:SurvProb}
\end{equation}
 where $\lambda^{h}$ is the mean free path and the integral is evaluated
along the hadron trajectory. The mean free path is a function of the
hadron type, $h$, the hadron energy, $E_{h}$, and its position,
$\vec{r}$, within the target nucleus and is computed as 
\begin{equation}
{\displaystyle \lambda^{h}=1/(\rho_{nucl}(r)*\sigma^{hN}(E_{h}))}
\end{equation}
 where $\rho_{nucl}(r)$ is the nuclear density profile and $\sigma^{hN}(E_{h})$
the corresponding hadron-nucleon total cross section.

\begin{figure}[htb]
\center \includegraphics[width=19pc]{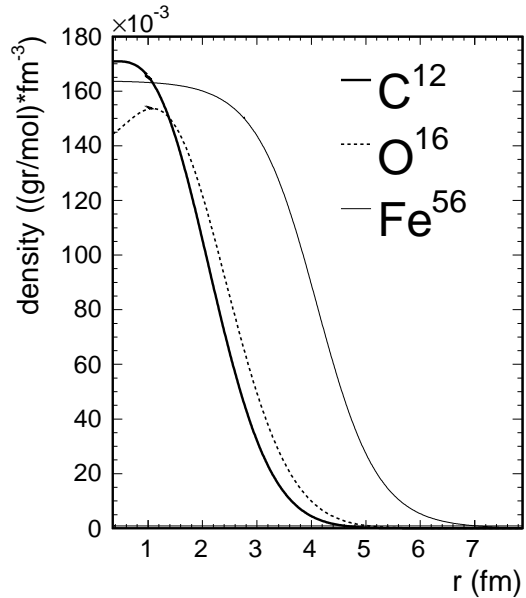} 

\caption{Nuclear density profiles for $C^{12}$, $O^{16}$ and $Fe^{56}$.}

\label{fig:NuclDensity} 
\end{figure}

\begin{figure}[htb]
\center \includegraphics[width=19pc]{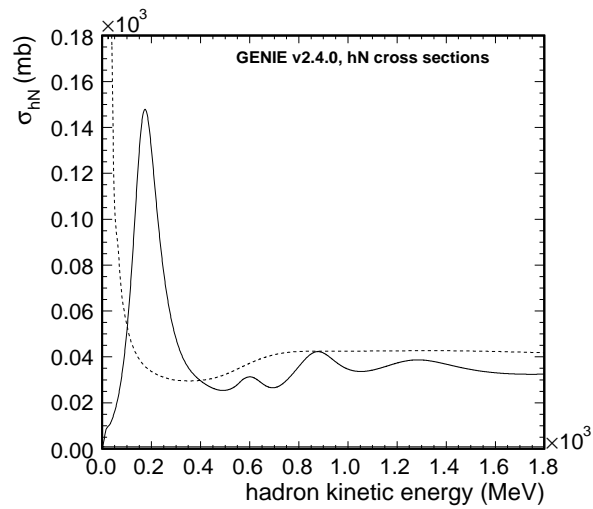} 

\caption{The nucleon-nucleon (dashed line) and pion-nucleon (solid line) cross
sections used in INTRANUKE/hA (GENIE v2.4.0) for determining the hadron
mean free path.}

\label{fig:MfpXSec} 
\end{figure}

\begin{figure}[htb]
\center \includegraphics[width=19pc]{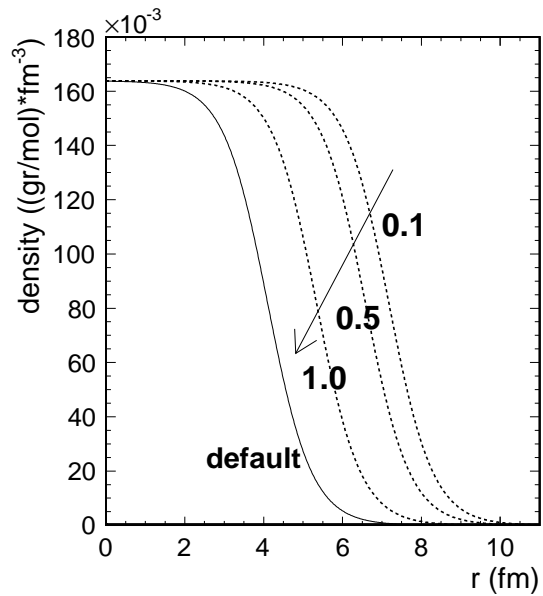} 

\caption{Nuclear density profiles for $Fe^{56}$ \em`stretched' \em by the
de-Broglie wave-length corresponding to hadrons with a momentum of
0.1 GeV, 0.5 GeV and 1.0 GeV. The default nuclear density distribution
is also shown. }

\label{fig:NuclDensityRing} 
\end{figure}

\begin{figure}[htb]
\center \includegraphics[width=19pc]{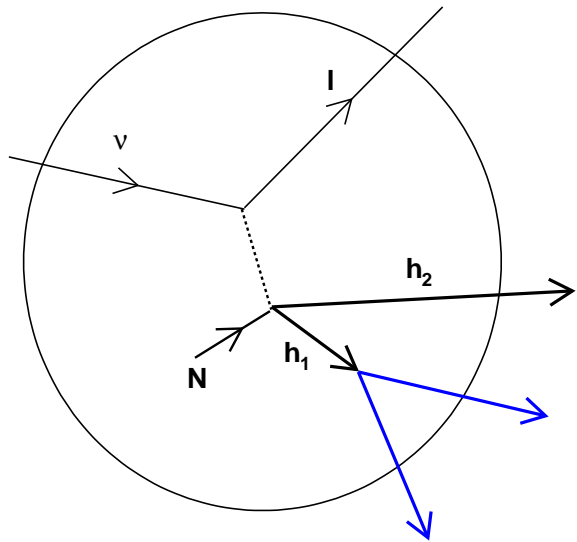} 

\caption{An example event with two primary hadrons, $h_{1}$ and $h_{2}$,
one of which ($h_{1}$) re-interacts within the target nucleus while
the other escapes ($h_{2}$). See text for a description of the weights
to be assigned to each hadron if the mean free path has been tweaked.}

\label{fig:RewMfp} 
\end{figure}

During the reweighing procedure, using the positions and 4-momenta
of the simulated primary hadronic system particles (that is the hadrons
emerging from the primary interaction vertex before any intranuclear
rescattering ever took place) we calculate the exact same hadron survival
probabilities as in the original simulation. In doing so we match
exactly the physics choices of the hadron transport simulation code
so as to avoid weighting artifacts. More importantly: 
\begin{itemize}
\item The reweighing code accesses the same hadron-nucleon cross section
and nuclear density profile functions as the simulation code. The
nuclear density profiles for $^{12}$C, $^{16}$O and $^{56}$Fe and
the nucleon-nucleon and pion-nucleon cross sections used by INTRANUKE/hA
in GENIE v2.4.0 are shown in Figs. \ref{fig:NuclDensity} and \ref{fig:MfpXSec}
respectively. 
\item The hadrons are being transported in steps of 0.05 fm as in the original
simulation. 
\item Each hadron is traced till it reaches a distance of $r=N*R_{nucl}=N*R_{0}*A^{1/3}$,
where $R_{0}$ = 1.4 fm and $N$ = 3.0. This allows taking into account
the effect the nuclear density distribution tail has on the hadron
survival probability. (For example, the nuclear radius, $R_{nucl}$
for $C^{12}$, $O^{16}$ and $Fe^{56}$ is 3.20 fm, 3.53 fm and 5.36
fm respectively. The reweighing, as the actual simulation code, integrates
Eq. \ref{Eq:SurvProb} for distances up to 9.62 fm, 10.58 fm and 16.07
fm respectively. Compare these values with the nuclear density profiles
shown in Fig. \ref{fig:NuclDensity}.) 
\item The nuclear density distribution through which the hadron is tracked
is increased by $n*\lambda_{B}$, where $\lambda_{B}$ is the de Broglie
wave-length of the hadron and $n$ is a tunable parameter (in GENIE
v2.4.0, INTRANUKE/hA uses $n=1$ for nucleons and $n=0.5$ for pions).
As explained earlier, this empirical approach is an important feature
of the INTRANUKE/hA mean free path tuning, accounting for the effects
of wave-like processes to the hadron survival probability which are
typically not well described within the context of an INC model. The
reweighing code matches that feature so as to emulate the hadron survival
probabilities calculated during event generation. The effect on the
nuclear density profile is shown in Fig. \ref{fig:NuclDensityRing}. 
\end{itemize}
The reweighing scheme allows the mean free path, $\lambda^{h}$, for
a hadron type $h$ to be modified in terms of its corresponding error,
$\delta{\lambda^{h}}$: 
\begin{equation}
{\displaystyle \lambda^{h}\rightarrow\lambda^{h\prime}=\lambda^{h}(1+x_{mfp}^{h}*\delta\lambda^{h}/\lambda^{h})}
\end{equation}
 where $\lambda^{h\prime}$ is the modified mean free path and $x_{mfp}$
is a tweaking knob. Then, by re-evaluating the integral in Eq. \ref{Eq:SurvProb},
we are able to compute the hadron survival probabilities that the
simulation code would have computed, had it been using the modified
mean free path. The nominal, $P_{surv}^{h}$, and tweaked, $P_{surv}^{h\prime}$,
survival probabilities can be used to calculate the weight that accounts
for that change in the hadron mean free path. The choice of how to
weight each hadron depends critically on its intranuclear transport
history. Consider the case illustrated in Fig. \ref{fig:RewMfp} where
a neutrino event has 2 primary hadrons, $h_{1}$ and $h_{2}$, one
of which ($h_{1}$) re-interacts while the other ($h_{2}$) escapes.
Had the mean free path been larger than the one used in the simulation
(and therefore, had the the interaction probability been lower) then
$h_{1}$'s history would have been more unlikely while, on the other
hand, $h_{2}$'s history would have been more likely. Therefore, in
order to account for an increase in mean free path, $h_{1}$ has to
be weighted down while $h_{2}$ has to be weighted up (and vice versa
for a mean free path decrease). The desired qualitative behavior of
single-hadron weights in response to mean free path changes is summarized
in Tab. \ref{table:RewMfpArr}. The following weighting function exhibits
the desired qualitative characteristics: 
\begin{equation}
{\displaystyle w_{mfp}^{h}=\left\{ \begin{array}{ll}
\frac{1-P_{surv}^{h\prime}}{1-P_{surv}^{h}} & \mbox{if \ensuremath{h}re-interacts}\\
\\
\frac{P_{surv}^{h\prime}}{P_{surv}^{h}} & \mbox{if \ensuremath{h}escapes }
\end{array}\right.}\label{eq:ParticleWeightMFP}
\end{equation}
 where $P_{surv}^{h}$ is the hadron survival probability corresponding
to mean free path $\lambda^{h}$ and $P_{surv}^{h\prime}$ is the
hadron survival probability corresponding to the tweaked mean free
path $\lambda^{h\prime}$.

\clearpage{}

\begin{table*}[htb]
\global\long\def\arraystretch{1.75}


\begin{tabular}{cccc}
\hline 
$\lambda^{h}$ change & $P_{rescat}^{h}$change  & Weight  & Weight \tabularnewline
 &  & (hadrons interacting)  & (hadrons escaping) \tabularnewline
\hline 
$\Uparrow$  & $\Downarrow$  & $\Downarrow$  & $\Uparrow$ \tabularnewline
$\Downarrow$  & $\Uparrow$  & $\Uparrow$  & $\Downarrow$ \tabularnewline
\hline 
\end{tabular}\\[2pt] 

\caption{ The intended qualitative behavior of hadron weights in response to
mean free path, $\lambda^{h}$, changes depending on whether the simulated
hadron had been rescattered or escaped. Had the mean free path been
larger in reality than the one used in the simulation (and therefore,
had the the interaction probability, $P_{rescat}^{h}$, been lower)
then rescattered hadrons would have been over-represented in the generated
sample and they would need to be weighted-down to match reality, while
escaping hadrons would have been under-represented and they would
need to be weighted-up. Vice versa for a mean free path decrease.
See text for description of the hadron weighting functions. }

\label{table:RewMfpArr} 
\end{table*}

\subsubsection{Reweighting the rescattering fates}

Once INTRANUKE/hA determines that a particular hadron is to be rescattered,
then a host of scattering modes are available to it. We will be referring
to these scattering modes as the \em hadron fates\em. Many fates
are considered for both pions and nucleons. The fates considered here
are: elastic, inelastic, charge exchange\footnote{Only single charge exchange is considered},
absorption\footnote{ Followed by emission of 2 or more nucleons with no pions in the final
state. The term \em `absorption' \em is usually used for pions while
the term \em `multi-nucleon knock-out' \em is used for nucleons.
Here, for simplicity and in the interest of having common fate names
for both pions and nucleons we will be using the term \em `absorption'
\em for both. }, and pion production. Each such fate may include many actual rescattering
channels \footnote{ For example, the \em `pion absorption' \em fate includes rescattering
modes with any of the np, pp, npp, nnp, nnpp final states }.

\begin{figure}[htb]
\center \includegraphics[width=17pc]{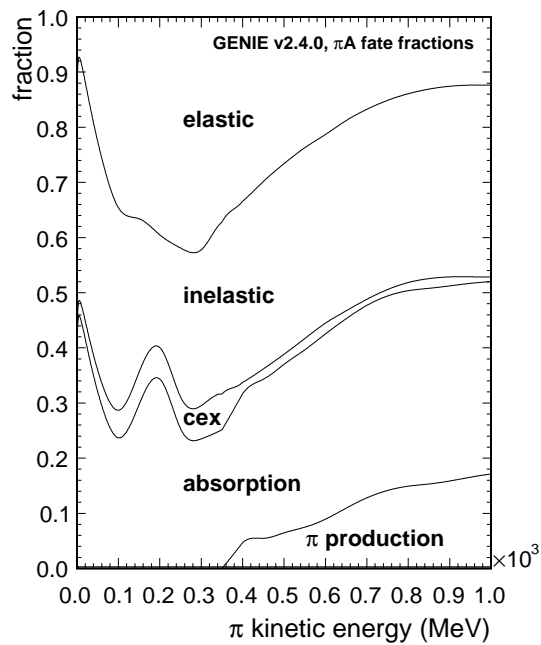} 

\caption{ The default fate fractions for rescattered pions in INTRANUKE/hA
(GENIE v2.4.0). The area that corresponds to each pion fate represents
the probability for that fate as a function of the pion kinetic energy.
The probabilities shown here conditional upon the pion interacting
so they always add up to 1. }

\label{fig:FatePiDef} 
\end{figure}

\begin{figure}[htb]
\center \includegraphics[width=17pc]{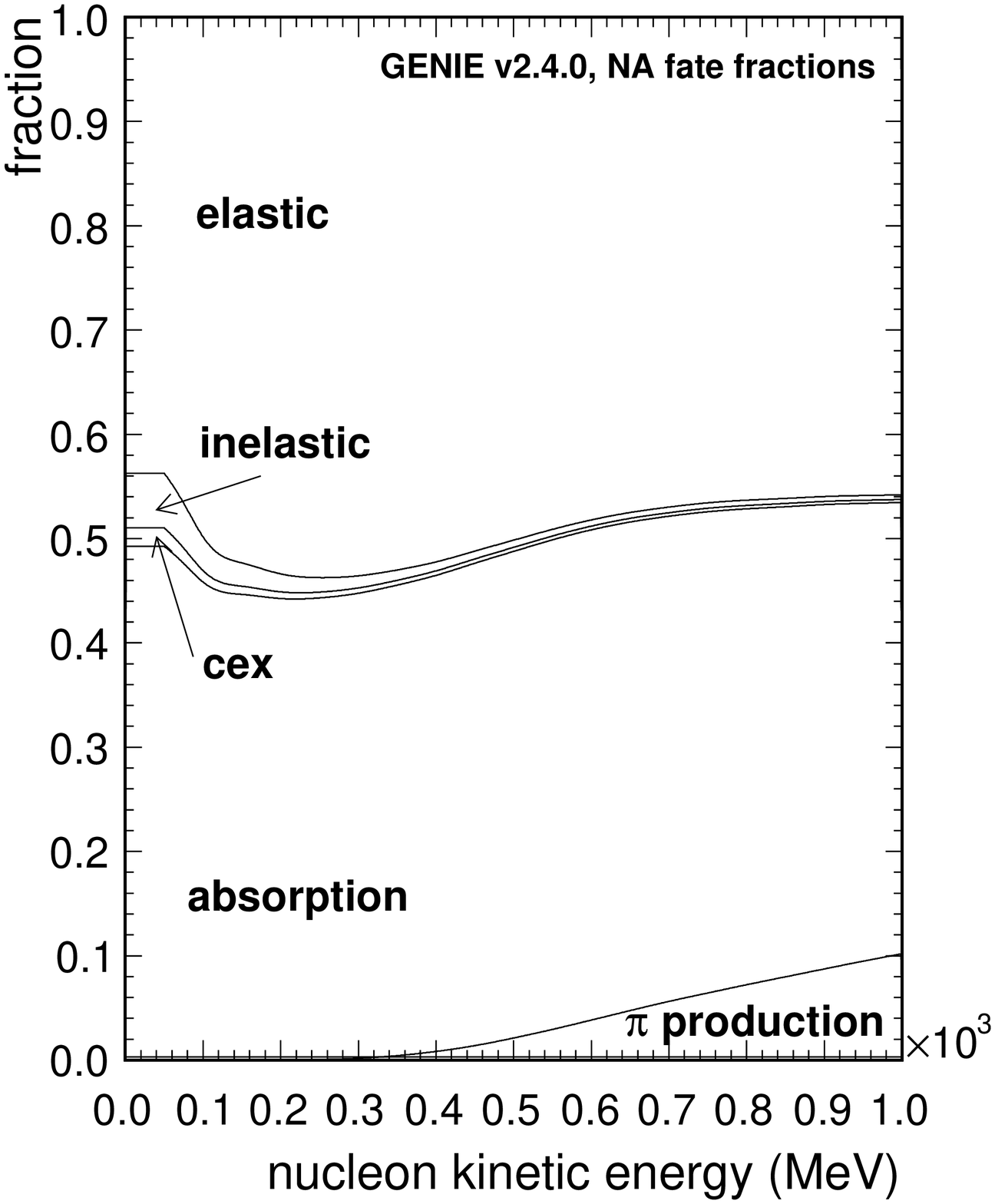} 

\caption{ The default fate fractions for rescattered nucleons in INTRANUKE/hA
(GENIE v2.4.0). The area that corresponds to each nucleon fate represents
the probability for that fate as a function of the nucleon kinetic
energy. The probabilities shown here conditional upon the nucleon
interacting so they always add up to 1. }

\label{fig:FateNDef} 
\end{figure}

In order to calculate the probability of each fate INTRANUKE/hA, being
an effective data-driven hadron transport MC, switches to a more macroscopic
description of hadron rescattering: Rather than building everything
up from hadron-nucleon cross sections, at this point in event simulation,
INTRANUKE/hA determines the probability for each fate using built-in
hadron-nucleus cross sections coming primarily from data. The probability
for a hadron fate $f$ is 
\begin{equation}
{\displaystyle P_{f}^{h}=\sigma_{f}^{hA}/\sigma_{total}^{hA}}
\end{equation}
 where $\sigma_{f}^{hA}$ is the hadron-nucleus cross section for
that particular fate and $\sigma_{total}^{hA}$ is the total hadron-nucleus
cross section. The calculated probabilities are conditional upon a
hadron being rescattered and the sum of these probabilities over all
possible fates should always add up to 1. The default probability
fractions for pions and nucleons in INTRANUKE/hA (GENIE v2.4.0) are
shown in Fig. \ref{fig:FatePiDef} and \ref{fig:FateNDef}.

The generation strategy leads to a conceptually simple and technically
straight-forward fate reweighing strategy: The hadron-nucleus cross
section for a particular fate may be modified in terms of its corresponding
error, $\delta\sigma_{f}^{hA}$ as in: 
\begin{equation}
{\displaystyle \sigma_{f}^{hA}\rightarrow\sigma_{f}^{\prime hA}=\sigma_{f}^{hA}(1+x_{f}^{h}*\delta\sigma_{f}^{hA}/\sigma_{f}^{hA})}
\end{equation}
 where $x_{f}$ is a fate tweaking knob.

It follows that the single-hadron fate weight is 
\begin{equation}
{\displaystyle w_{fate}^{h}=\sum_{f}\delta_{f;f^{\prime}}*x_{f}^{h}*\delta\sigma_{f}^{hA}/\sigma_{f}^{hA}}
\end{equation}
 where $f$ runs over all possible fates \{elastic, inelastic, charge
exchange, absorption, pion production\}, $f\prime$ is the actual
fate for that hadron as it was determined during the simulation and
$\delta_{f;f^{\prime}}$ is a factor which is 1 if $f=f^{\prime}$
and 0 otherwise.

Not all 5 hadron fates may be tweaked simultaneously. Since the sum
of all fractions should add up to 1 then, at most, at most 4 out of
the 5 fates may be tweaked directly. The fates not tweaked directly
(\em cushion \em terms) are adjusted automatically to conserve the
sum. The choice of which fates act as a cushion terms is configurable. 

\begin{figure}[htb]
\center \includegraphics[width=17pc]{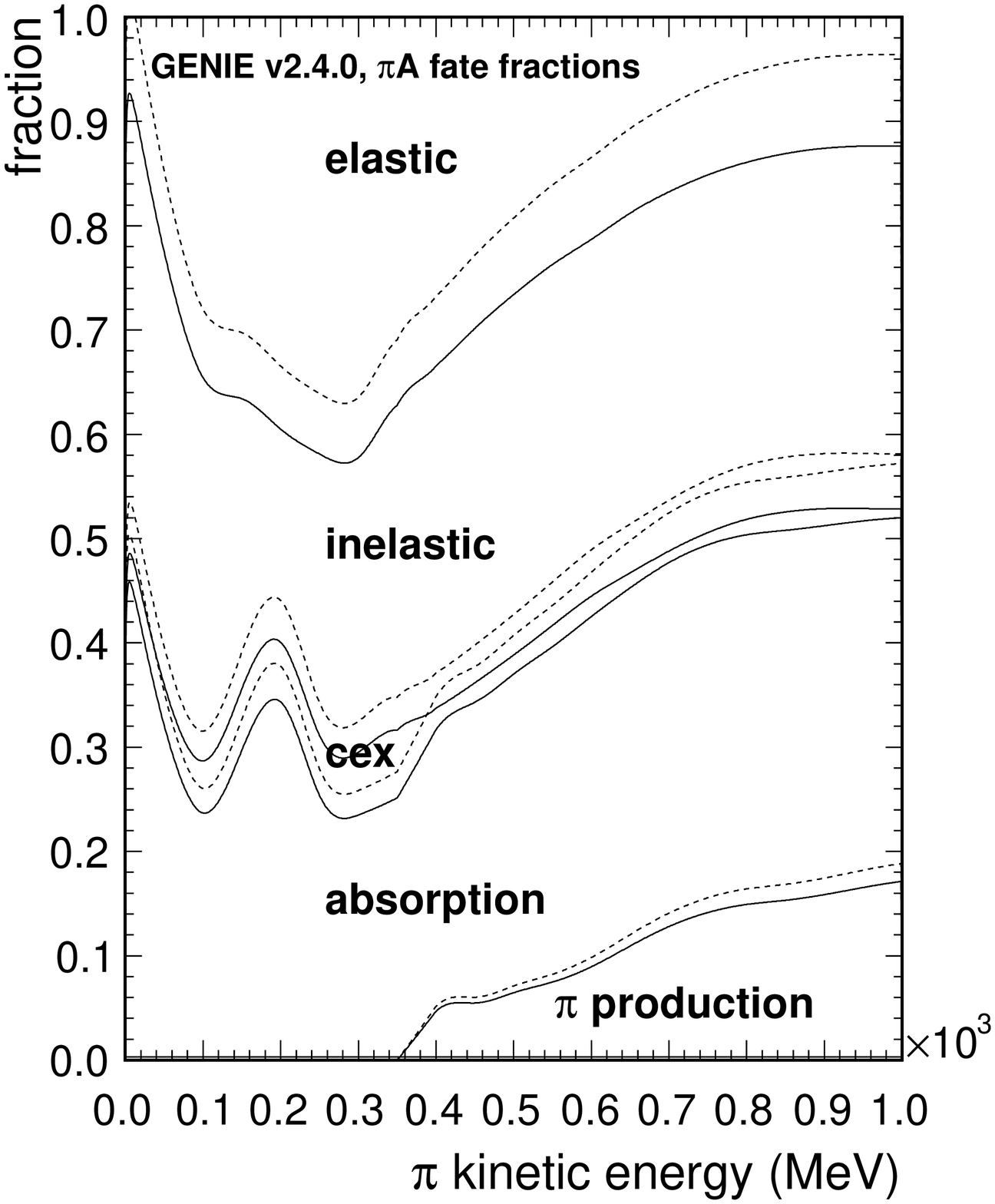} 

\caption{The default (solid lines) and tweaked (dashed lines) pion fate fractions.
The tweaked pion fate fractions are shown for a case where the pion
production, absorption, charge exchange and inelastic cross sections
have been increased by 10\%. Here it is the elastic cross section
term that is being used as the cushion term. See text for details.}

\label{fig:FatePiTwk} 
\end{figure}

In Fig. \ref{fig:FatePiTwk} we show the tweaked pion fate fraction
(dashed lines) obtained by simultaneously increasing the pion production,
absorption, charge exchange and inelastic cross sections by 10\%.
In this example the elastic component is being used as a cushion term
absorbing the changes in all other terms so as to maintain the total
probability. The default pion fate fractions (solid lines) are superimposed
for reference.

\subsubsection{Computing event weights}

The scheme outlined above, provides a detailed prescription for calculating
single-hadron weights so as to take into account the effect that modified
hadron-nucleon and hadron-nucleus cross sections would have had on
that hadron ($w_{mfp}^{h}$ and $w_{fate}^{h}$ respectively). The
total single-hadron weight is 
\begin{equation}
{\displaystyle w^{h}=w_{mfp}^{h}*w_{fate}^{h}}
\end{equation}

The corresponding hadron transport (HT) related weight for a neutrino
interaction event, $w_{HT}^{evt}$, is, obviously, the product of
single-hadron weights 
\begin{equation}
{\displaystyle w_{HT}^{evt}=\prod_{j}w_{j}^{h}}
\end{equation}
 where the index $j$ runs over all the primary hadronic system particles
in the event.

\subsubsection{Computing penalty terms}

A penalty term can easily be calculated from the physics tweaking
knobs which can be included as nuisance parameters in physics fits.
The penalty has components, penalizing deviations from the default
total rescattering rate and from the default fractions of rescattering
modes. It can be written as

\begin{equation}
{\displaystyle \chi_{penalty}^{2}=\sum_{h=\pi,N}\{(x_{mfp}^{h})^{2}+\sum_{f\ne fc}(x_{f}^{h})^{2}+(\widehat{x_{fc}^{h}})^{2}\}}
\end{equation}
 where the $x's$ correspond to mean free path and fate tweaking knobs
for pions and nucleons The sum over fates, $f$, excludes the cushion
term, $fc$, which is added separately. The reason is technical: All
directly tweaked hadron-nucleus cross sections are tweaked in units
of their own (typically hadron energy-dependent) uncertainty, therefore
having a corresponding contribution to penalty term which is energy
independent. The change in the cushion term, being forced to absorb
the other changes, is not well defined in terms of its own uncertainty.
Therefore, its contribution in the penalty term, $\widehat{x_{fc}^{h}}^{2}$,
is averaged over the hadron energy range.

\subsubsection{Unitarity expectations}

This section demonstrates why both the intranuclear reweighting schemes
presented earlier are expected to maintain unitarity. In general,
when reweighting an event, we multiply by a weight $w$ 
\begin{equation}
{\displaystyle w=\frac{P^{\prime}}{P}.\label{eq:0}}
\end{equation}

where $P$ and $P^{\prime}$ are the probabilities for getting that
event\footnote{In this section an event is defined as the transport of a single hadron.},
for the nominal and tweaked cases respectively, and they depend on
the particular event being reweighted.

When describing processes where multiple discrete outcomes are possible
then the analytical form of the above probabilities will change depending
on the outcome. An example of this is the case of mean free path (rescattering
rate) reweighting where the fate of an event can be divided into two
categories: Those that rescattered and those that escaped the nucleus.
The two forms of $P$ in this case are,

\begin{equation}
{\displaystyle P_{rescat}=1-e^{\frac{-x}{\lambda}}}
\end{equation}

and

\begin{equation}
{\displaystyle P_{surv}=e^{\frac{-x}{\lambda}}.}
\end{equation}

Thus a hadron that rescattered will receive a weight, reflecting a
change in mean free path of $\lambda\rightarrow\lambda^{\prime}$,
of 
\begin{equation}
{\displaystyle w_{rescat}=\frac{1-e^{\frac{-x}{\lambda^{\prime}}}}{1-e^{\frac{-x}{\lambda}}}}
\end{equation}
 whereas one that escaped the nucleus will get a weight 
\begin{equation}
{\displaystyle w_{surv}=\frac{e^{\frac{-x}{\lambda^{\prime}}}}{e^{\frac{-x}{\lambda}}}}
\end{equation}

Take the general case where there are $n$ possible outcomes and where
the $i$'th outcome occurs with a probability $P_{i}$. For a set
of $N_{tot}$ events one expects 
\begin{equation}
N_{i}=N_{tot}\times\frac{P_{i}}{\sum_{j=1}^{n}P_{j}}\label{eq:1}
\end{equation}
 events corresponding to the $i$'th outcome.

Now consider reweighting all $N_{tot}$ events. Events corresponding
to the $i$'th outcome get weighted by $w_{i}$ so that the after
reweighting the number of events for the $i$'th outcome is given
by 
\begin{eqnarray}
N_{i}^{\prime}=w_{i}\times N_{i}.\label{eq:2.1}
\end{eqnarray}
 Note that Eq. \ref{eq:2.1} holds only if we consider just the functional
dependance of the weights on the weighting parameters\footnote{We neglect any functional dependance on kinematical quantities. This
is a valid assumption if the density of events, defined as the number
in a given volume of kinematical phase space, is high enough such
that a statistically significant number of neighboring events cover
a small enough volume in the kinematical phase space over which the
effect of the variation in kinematical quantities is negligible.}. The number of events in the new reweighted sample is given by 
\begin{eqnarray*}
N_{tot}^{\prime} & = & \sum_{j=1}^{N_{tot}}w_{j}^{evt}\\
 & = & \sum_{i=1}^{n}w_{i}^{outcome}\times N_{i}\\
 & = & \sum_{i=1}^{n}\frac{P_{i}^{\prime}}{P_{i}}\times N_{i}.
\end{eqnarray*}

Substituting Eq. \ref{eq:1} we get, 
\[
N_{tot}^{\prime}=N_{tot}\times\frac{\sum_{i=1}^{n}{P_{i}^{\prime}}}{\sum_{j=1}^{n}P_{j}}.
\]
 So if 
\begin{equation}
\sum_{i=1}^{n}P_{i}=\sum_{i=1}^{n}P_{i}^{\prime}\label{eq:3}
\end{equation}
 then $N_{tot}^{\prime}=N_{tot}$ and unitarity is conserved.

In the case of rescattering,

\begin{eqnarray*}
\sum_{i=1}^{n}P_{i} & = & P_{rescat}+P_{surv}\\
 & = & 1-e^{\frac{-x}{\lambda}}+e^{\frac{-x}{\lambda}}\\
 & = & 1-e^{\frac{-x}{\lambda^{\prime}}}+e^{\frac{-x}{\lambda^{\prime}}}\\
 & = & P_{rescat}^{\prime}+P_{surv}^{\prime}\\
 & = & \sum_{i=1}^{n}P_{i}^{\prime}
\end{eqnarray*}

So for the rescattering scheme we expect unitarity to be a built in
feature. This is also true for the fate reweighting where the cushion
term ensures Eq. \ref{eq:3} is satisfied. It is worth highlighting
that the unitarity constraint is sensitive to any differences between
the generator and the reweighting scheme. This is also why a particular
implementation of a reweighting scheme is not generator agnostic.

\section{Event reweighting applications}

\subsection{Built-in applications}

\subsubsection{The \textit{grwght1scan} utility }

\subsubsection*{Name}

\textit{grwght1scan} - Generates weights given an input GHEP event
file and for a given systematic parameter (supported by the ReWeight
package). It outputs a ROOT file containing a tree with an entry for
every input event. Each such tree entry contains a \textit{TArrayF}
of all computed weights and a \textit{TArrayF} of all used tweak dial
values.

\subsubsection*{Source and build options}

The source code for this application is in `\texttt{\textbf{\small{}\$GENIE}}\textit{/src/support/rwght/gRwght1Scan.cxx}'.\\
To enable this application (and, also, to build the ReWeight package
library) add `\texttt{\textbf{\small{}-{}-enable-rwght}}' during the
GENIE build configuration step.

\subsubsection*{Synopsis}

\texttt{\textbf{\small{}grwght1scan }}{\small \par}

\texttt{\textbf{\small{}-f input\_filename }}{\small \par}

\texttt{\textbf{\small{}{[}-n number\_of\_events{]} }}{\small \par}

\texttt{\textbf{\small{}-s systematic\_name }}{\small \par}

\texttt{\textbf{\small{}-t number\_of\_tweaking\_diall\_values }}{\small \par}

\texttt{\textbf{\small{}{[}-p neutrino\_codes{]} }}~\\
\texttt{\textbf{\small{}}}~\\
where {[}{]} is an optional argument.

\subsubsection*{Description}

The following options are available:\\
\\
\textbf{-f Specifies an input GHEP event file.}\\
\\
 \textbf{-n Specifies the number of events to process.}\\

This is an optional argument. By default GENIE will process all events.
\\
\\
\textbf{-s Specifies the name of the systematic param to tweak. }\\
\\
\textbf{-t Specifies the number of the systematic parameter tweaking
dial values between -1 and 1.}\\

Note: This must be an odd number so as to include al; -1, 0 and 1.
If it is an even number then it will be incremented by 1. \\
\\
\textbf{-p If set, specifies which neutrino species to reweight.}\\

This is an optional argument. By default GENIE will reweight all neutrino
species. The expected input is a comma separated list of PDG codes.

\subsubsection*{Examples}

\subsection{Writing a new reweighting application}

Writing a new reweighting application is relatively trivial. The built-in
applications described above can be used as a template and be modified
accordingly. A \textit{GReWeight} object provides an interface between
the user and the GENIE event reweighting objects (weight calculators).\textit{
GReWeight} holds both a list of weight calculators (\textit{GReWeightI
}subclasses), each one referred-to by a user-specified name, and a
set of tweaked systematic parameters (\textit{GSystSet} object). 

Typically, in an event reweighting application one would have to include
at least the following steps:
\begin{itemize}
\item Instantiate a \textit{GReWeight} object and add to it a set of concrete
weight calculators. For example (modify accordingly by adding / removing
weight calculators from this list): \\
\begin{verbatim}

GReWeight rw;

rw.AdoptWghtCalc( "xsec_ccqe",       new GReWeightNuXSecCCQE      );   
rw.AdoptWghtCalc( "xsec_ccqe_vec",   new GReWeightNuXSecCCQEvec   );   
rw.AdoptWghtCalc( "xsec_ccres",      new GReWeightNuXSecCCRES     );   
rw.AdoptWghtCalc( "xsec_ncres",      new GReWeightNuXSecNCRES     );   
rw.AdoptWghtCalc( "xsec_nonresbkg",  new GReWeightNonResonanceBkg );   
rw.AdoptWghtCalc( "xsec_dis",        new GReWeightNuXSecDIS       );   
rw.AdoptWghtCalc( "xsec_coh",        new GReWeightNuXSecCOH       );   
rw.AdoptWghtCalc( "nuclear_qe",      new GReWeightFGM             );   
rw.AdoptWghtCalc( "nuclear_dis",     new GReWeightDISNuclMod      );   
rw.AdoptWghtCalc( "hadro_res_decay", new GReWeightResonanceDecay  );   
rw.AdoptWghtCalc( "hadro_fzone",     new GReWeightFZone           );   
rw.AdoptWghtCalc( "hadro_intranuke", new GReWeightINuke           );   
rw.AdoptWghtCalc( "hadro_agky",      new GReWeightAGKY            );

\end{verbatim}\\

\item Retrieve and fine-tune weight calculators. This is an optional step.
Each calculator is retrieved from \textit{GReWeight} using the user-defined
name specified in the previous step. Fine-tuning methods are specific
to each weight calculator, so please refer to the documentation for
each individual calculator. For example, to disable $\nu_{e}$, $\bar{\nu_{e}}$
and $\bar{\nu_{\mu}}$ reweighting in \textit{GReWeightNuXSecCCQE}
stored with the ``xsec\_calc'' name, type: \\
\begin{verbatim}

GReWeightNuXSecCCQE * rwccqe =      
    dynamic_cast<GReWeightNuXSecCCQE *> (
      rw.WghtCalc("xsec_ccqe"));   
rwccqe -> RewNue    (false);    
rwccqe -> RewNuebar (false);    
rwccqe -> RewNumubar(false); 

\end{verbatim}\\

\item Get the \textit{GSystSet} object held by \textit{GReWeight} and tweak
all systematic params you wish to consider (complete list to be found
in `\texttt{\textbf{\small{}\$GENIE}}\textit{/src/ReWeight/GSyst.h}').
What you are actually setting is the value $d$ of a tweaking dial
(default value: 0) which modifies a corresponding physics parameter
$p$ as $p\rightarrow p^{\prime}=p\times(1+d\times(dp/p))$. Setting
a tweaking dial to $+/-$1 modifies a physics quantity by $+/-$ 1
$\sigma$ respectivelly. The default fractional errors $dp/p$ are
defined in \textit{GSystUncertainty} and can be overriden. The following
example sets non-default values to a series of systematics parameters
handled by the weight calculators included in the previous step. After
all parameters have been tweaked, invoke \textit{GReWeight::Reconfigure()}
so that tweaked parameters can be propagated across GENIE. You probably
need to be setting these parameters and reconfiguring GENIE inside
a `parameter loop' or a `minimization function'. \\
\\
\begin{verbatim}

GSystSet & syst = rw.Systematics();

syst.Set(kXSecTwkDial_NormCCQE,        +1.0);   
syst.Set(kXSecTwkDial_MaCCQEshape,     +1.0); 
syst.Set(kXSecTwkDial_NormCCRES,       -1.0);   
syst.Set(kXSecTwkDial_VecFFCCQEshape,  -1.0);
syst.Set(kXSecTwkDial_MaCCRESshape,    -1.0);   
syst.Set(kXSecTwkDial_MvCCRESshape,    +0.5); 
syst.Set(kXSecTwkDial_NormNCRES,       +1.0);   
syst.Set(kXSecTwkDial_MaNCRESshape,    -0.7); 
syst.Set(kXSecTwkDial_MvNCRESshape,    +0.3);   
syst.Set(kXSecTwkDial_RvpCC1pi,        +0.5); 
syst.Set(kXSecTwkDial_RvnCC1pi,        +0.5);   
syst.Set(kXSecTwkDial_MaCOHpi,         -0.5);
syst.Set(kINukeTwkDial_MFP_pi,         +1.0);   
syst.Set(kINukeTwkDial_MFP_N,          -1.0); 
syst.Set(kINukeTwkDial_FrPiProd_pi,    -0.7);   
syst.Set(kHadrAGKYTwkDial_xF1pi,       -1.0); 
syst.Set(kHadrAGKYTwkDial_pT1pi,       +1.0);   
syst.Set(kHadrNuclTwkDial_FormZone,    +1.0); 
syst.Set(kRDcyTwkDial_Theta_Delta2Npi, +1.0);

rw.Reconfigure();

\end{verbatim}\\

\item Calculate an event weight by invoking \textit{GReWeight::CalcWeight().
}The function expects an \textit{EventRecord} object as input. The
return value is the calculated weight and is computed as the product
of the weights computed by all included weight calculators for the
current set of systematics / tweaking dial values stored in \textit{GSystSet.
}You can also calculate a penalty factor, $\chi_{penalty}^{2}$, for
the current set of systematic tweaking dial values by invoking \textit{GReWeight::CalcChisq().}
\end{itemize}

\paragraph*{Important notes}

The reweighting package includes a large number of weight calculators
handling a large numbers of systematic parameters. Alternative reweighting
schemes may exist for the same systematic parameter. It is the user's
responsibilty to make sure that all parameters tweaked in \textit{GSystSet}
are handled by exactly one weight calculator added via \textit{GReWeight::AdoptWeightCalc()}.
Additionally, certain systematic parameters should not be combined
together. For example, you should tweak either \texttt{\textbf{\small{}kXSecTwkDial\_MaCCQE}}
(tweakes the axial mass used in the $CCQE$ cross section model and
allows it to change both the shape and the normalization of the output
$d\sigma/dQ^{2}$ distribution at fixed energy), OR \texttt{\textbf{\small{}kXSecTwkDial\_NormCCQE}}
and \texttt{\textbf{\small{}kXSecTwkDial\_MaCCQEshape}} (where the
normalization and shape-effects have been separated) and you should
never mix them all together. All in all, a good understanding of the
effect of each included systematic parameter and weight calculator
(see this Chapter) is imperative in order to get meaningfull results.

\section{Adding a new event reweighting class}

A large number of event reweighting classes (weight calculators) exist
within GENIE and can serve as examples. One can easily add a new concrete
weight calculator which can be integrated with the existing reweighting
framework. This new calculator should subclass \textit{GReWeightI}
and implement, at least, the following methods:
\begin{itemize}
\item `\textit{bool IsHandled(genie::GSyst\_t syst)}' :\\
Declare whether the weight calculator handles the input systematic
parameter.
\item `\textit{void SetSystematic(genie::GSyst\_t syst, double val)}' :\\
Update the current value for the specified systematic parameter.
\item `\textit{void Reset(void)}' :\\
Set all handled systematic parameters to default values.
\item `\textit{void Reconfigure(void)}' :\\
Propagate updated systematic parameter values to actual GENIE MC code,
if needed. 
\item `\textit{double CalcWeight(const genie::EventRecord \& event)}' :
\\
Calculate a weight for the input event using the current values of
all handled systematic parameters. 
\item `\textit{double CalcChisq(void)}' :\\
Calculate a penalty factor for the current deviation of all handled
systematic params from their default values.
\end{itemize}
This is the minimum set of methods required by GENIE itself. More
methods, specific to each weight calculator, can be added and used
in the user's event reweighting application so as to fine-tune the
behaviour of each calculator.

Note that if you are adding a weight calculator to quantify the effect
of a new systematic parameter, one which is not already included in
\textit{`}\texttt{\textbf{\small{}\$GENIE}}\textit{/src/ReWeight/GSyst.h}',
then also you need to:
\begin{itemize}
\item add the new parameter in \textit{`}\texttt{\textbf{\small{}\$GENIE}}\textit{/src/ReWeight/GSyst.h}',
and 
\item define a default 1 $\sigma$ error in \textit{`}\texttt{\textbf{\small{}\$GENIE}}\textit{/src/ReWeight/GSystUncertainty.cxx}'.
\end{itemize}
\appendix

\chapter{Copyright Notice and Citation Guidelines \label{cha:AppendixCitingGENIE}}

\textbf{\textsc{\Large{}(c) 2003-2015, GENIE Collaboration}}\\
\\
\\
For all communications:\\
\textbf{Dr. Constantinos Andreopoulos} < costas.andreopoulos@stfc.ac.uk
>\\
\\
\begin{tabular}{|l|l|}
\hline 
University of Liverpool & STFC Rutherford Appleton Laboratory\tabularnewline
\hline 
Physics Department & Department of Particle Physics\tabularnewline
\hline 
\textit{Liverpool L69 7ZE, UK} & \textit{Harwell Oxford Campus, Oxfordshire OX11 0QX, UK}\tabularnewline
\hline 
TEL: +44-(0)1517-943201 & TEL: +44-(0)1235-445091\tabularnewline
\hline 
 & FAX: +44-(0)1235-446733\tabularnewline
\hline 
\end{tabular}\\
\\
The license conditions may be found in \url{http://copyright.genie-mc.org}

\section{Guidelines for Fair Academic Use}

The authors of GENIE endorse the MCNET guidelines\footnote{Full text may be found at \textit{http://www.montecarlonet.org/GUIDELINES}}
for fair academic use. In particular, users are invited to consider
which GENIE components are important for a particular analysis and
cire them, in addition to the main references.

\section{Main references}

All derivative works should cite:\\
\\
C.Andreopoulos et al., `The GENIE Neutrino Monte Carlo Generator',
Nucl.Instrum.Meth. A614:87-104,2010.\\
\\
Corresponding Bib\TeX{} entry:\\
\begin{verbatim}
  @Article{Andreopoulos:2009rq,
     author    = "Andreopoulos, C. and others",
     title     = "{The GENIE Neutrino Monte Carlo Generator}",
     journal   = "Nucl. Instrum. Meth.",
     volume    = "A614",
     year      = "2010",
     pages     = "87-104",
     eprint    = "0905.2517",
     archivePrefix = "arXiv",
     primaryClass  =  "hep-ph",
     doi       = "10.1016/j.nima.2009.12.009",
     SLACcitation  = "%%CITATION = 0905.2517;%%" 
  } 

\end{verbatim}

\newpage

\chapter{Special Topics, FAQs and Troubleshooting}

\section{Installation / Versioning }

\subsection{Making user-code conditional on the GENIE version}

User-code can be made conditional upon the GENIE version number, in
similar way as with ROOT, by including `\texttt{\textbf{\small{}\$GENIE}}\textit{/src/Conventions/GVersion.h}'.
This header file is automatically generated during the GENIE installation.
If, for example, one wishes to do something different before / after
version 2.16.22, then simply type: \\
\\
\texttt{\textbf{\small{}\#if \_\_GENIE\_RELEASE\_CODE\_\_ >= GRELCODE(2,16,22)
}}~\\
\texttt{\textbf{\small{}... }}~\\
\textit{<your code here>}\texttt{\textbf{\small{} }}~\\
\texttt{\textbf{\small{}... }}~\\
\texttt{\textbf{\small{}\#else }}~\\
\texttt{\textbf{\small{}... }}~\\
\textit{<your code here>}\texttt{\textbf{\small{} }}~\\
\texttt{\textbf{\small{}... }}~\\
\texttt{\textbf{\small{}\#endif}}{\small \par}

\section{Software framework}

\subsection{Calling GENIE algorithms directly}

GENIE provides a host of event generation applications and utilities
and most users will only ever interact with these. It is only for
the most advanced GENIE uses-cases that one may need to access and
run algorithms directly. This is typically a 4-step process, as outlined
below:
\begin{enumerate}
\item Get an algorithm factory (\textit{AlgFactory}) instance. The algorithm
factory provides access to configured instances of all GENIE algorithms.
\\
\\
\texttt{\textbf{\small{}AlgFactory {*} algf = AlgFactory::Instance();}}~\\
{\small \par}
\item Request a concrete algorithm from the factory. Each algorithm is uniquely
specified by its name and the name of its configuration parameter
set.\\
\\
\texttt{\textbf{\small{}const Algorithm {*} alg\_base = algf->GetAlgorithm(``name'',
``config'');}}~\\
{\small \par}
\item Type-cast \textit{Algorithm }to the specific algorithmic interface
(\textit{XyzI}) being implemented. For example, for cross section
algorithms type-cast to \textit{XSecAlgorithmI}, for hadronization
models to \textit{HadronizationModelI}, for strucrure function models
to \textit{DISStructureFuncModelI}, for event generation modules to
\textit{EventRecordVisitorI} etc (please consult the GENIE doxygen
code reference for a full list of possibilities).\\
\\
\texttt{\textbf{\small{}const XzyI {*} alg = dynamic\_cast<const XyzI
{*}>(alg\_base); }}~\\
{\small \par}
\item Prepare the algorithm inputs and run it (please consult GENIE doxygen
code reference for documentation on each algorithmic interface).
\end{enumerate}

\subsubsection*{Example 1}

The following example shows how to get the Rein-Sehgal resonance neutrino-production
model, calculate the differential cross section $d^{2}\sigma/dWdQ^{2}$
for $\nu_{\mu}+n$ (bound in $Fe^{56}$) $\rightarrow\mu^{-}+\mbox{P11(1440)}$
at $E_{\nu}=$2.4 $GeV$, $W$=1.35 $GeV$, $Q^{2}$=1.1 $GeV^{2}$
and then calculate the integrated cross section at the same energy:\\
\begin{verbatim}

{
  ...

  // get the algorithm factory
  AlgFactory * algf = AlgFactory::Instance();                   

  // get the cross section algorithm
  const Algorithm * algbase = 
     algf->GetAlgorithm("genie::ReinSeghalRESPXSec", "Default"));
  const XSecAlgorithmI * xsec_model =        
     dynamic_cast<const XSecAlgorithmI *> (algbase);

  // prepare the cross section algorithm inputs
  Interaction * interaction 
          = Interaction::RESCC(kPdgTgtFe56,kPdgNeutron,kPdgNuMu);
  interaction->InitStatePtr()->SetProbeE(2.4);  
  interaction->KinePtr()->SetW(1.35);   
  interaction->KinePtr()->SetQ2(1.1);   
  interaction->ExclTagPtr()->SetResonance(kP11_1440);

  // calculate d2sigma/dWdQ2 differential cross section
  // (in 1E-38 cm^2 / GeV^3)
  double diff_xsec = xsec_model->XSec(
          interaction, kPSWQ2fE) / (1E-38 * units::cm2);

  // get the integrated cross section
  // (in 1E-38 cm^2)
  double intg_xsec = xsec_model->Integral(
          interaction) / (1E-38 * units::cm2);
  
  ...
}

\end{verbatim}

\subsection{Plugging-in to the message logging system}

The message logging system is based on the \textit{log4cpp} library.
GENIE provides the \textit{Messenger} class which enforces common
formatting for messages emitted by GENIE classes and provides an easier
interface to the log4cpp library. Messages are sent using one of the 
\begin{itemize}
\item LOG(stream, priority), 
\item LOG\_FATAL(stream), 
\item LOG\_ALERT(stream), 
\item LOG\_CRIT(stream), 
\item LOG\_ERROR(stream), 
\item LOG\_WARN(stream), 
\item LOG\_NOTICE(stream), 
\item LOG\_INFO(stream) 
\item LOG\_DEBUG(stream) 
\end{itemize}
Messenger macros as shown in \ref{algUsingMessenger}. Each message
is assigned a priority level (see Table \ref{tablePriorityLevels})
that can be used for message filtering using the\\

\textsf{\footnotesize{}void genie::Messenger::SetPriorityLevel(const
char {*} stream log4cpp::Priority::Value priority)}\\
\textsf{\footnotesize{}}\\
method as shown in \ref{algUsingMessenger}. Each message is 'decorated'
with its time stamp, its priority level, its stream name and the name
space / class name / method name / line of code from where it was
emitted\\

\textsf{\footnotesize{}time priority stream name : <method signature
(line of code)> : actual message}\\
\\
For example:\\

\textsf{\scriptsize{}10891167}\textsf{\footnotesize{} }\textsf{\scriptsize{}ERROR}\textsf{\footnotesize{}
Config:<bool genie::ConfigPool::LoadXMLConfig() (100)>: Parsing failed}\\
{\footnotesize \par}

\begin{table}
\center

\begin{tabular}{|c|}
\hline 
Message Priority Levels\tabularnewline
\hline 
\hline 
pFATAL\tabularnewline
\hline 
pALERT\tabularnewline
\hline 
pCRIT\tabularnewline
\hline 
pERROR\tabularnewline
\hline 
pWARN\tabularnewline
\hline 
pNOTICE\tabularnewline
\hline 
pINFO\tabularnewline
\hline 
pDEBUG\tabularnewline
\hline 
\end{tabular}

\caption{Priority levels in GENIE / log4cpp shown in decreasing importance.
\label{tablePriorityLevels}}
\end{table}

\begin{algorithm}
\{

$\qquad$...

$\qquad$\textsf{\footnotesize{}LOG(``stream-name'', pFATAL) <\textcompwordmark{}<
`` a fatal message'';}{\footnotesize \par}

$\qquad$\textsf{\footnotesize{}LOG(``stream-name'', pERROR) <\textcompwordmark{}<
`` an error message'';}{\footnotesize \par}

$\qquad$\textsf{\footnotesize{}LOG(``stream-name'', pWARN) <\textcompwordmark{}<
`` a warning'';}{\footnotesize \par}

$\qquad$

$\qquad$\textsf{\footnotesize{}// alternative ways to send messages}{\footnotesize \par}

$\qquad$\textsf{\footnotesize{}LOG\_ERROR(``stream-name'') <\textcompwordmark{}<
`` another error message'';}{\footnotesize \par}

$\qquad$\textsf{\footnotesize{}LOG\_WARN(``stream-name'') <\textcompwordmark{}<
`` another warning'';}{\footnotesize \par}

$\qquad$...

$\qquad$\textsf{\footnotesize{}Messenger {*} msg = Messenger::Instance();
// get a messenger instance}{\footnotesize \par}

$\qquad$...

$\qquad$\textsf{\footnotesize{}msg->SetPriorityLevel(``stream-name'',pERROR);
// set message threshold to 'ERROR'}{\footnotesize \par}

$\qquad$...

$\qquad$\textsf{\footnotesize{}LOG(``stream-name'', pALERT) <\textcompwordmark{}<
`` an alert -- passes the message thershold'';}{\footnotesize \par}

$\qquad$\textsf{\footnotesize{}LOG(``stream-name'', pDEBUG) <\textcompwordmark{}<
`` a debug message -- filtered / not shown'';}{\footnotesize \par}

$\qquad$...

\}

\caption{Example use of the GENIE / log4cpp message logging. \label{algUsingMessenger}}
\end{algorithm}

\section{Particle decays}

\subsection{Deciding which particles to decay}

GENIE attempts to simulate the complex physics within the nuclear
environment and, by default, it considers that every particle which
escapes the target nucleus has left its realm. It is the responsibility
of the detector simulation to handle particles that propagate more
than a few fermis before decaying. GENIE, for example, in its default
mode, will not decay charmed hadrons. If, like many others, you think
that these are ``short-lived'' particles GENIE ought to decay then
consider this: If a $C^{12}$ nucleus was as big as the Earth, then
these particles would decay more than a light year away ($c\tau_{0}(\Lambda_{c}^{+})/(C^{12}radius)\sim$
2 $\times$ $10^{10}$, $c\tau_{0}(D_{s})/(C^{12}radius)\sim$ 5 $\times$
$10^{10}$, etc). Similarly, GENIE won't decay $\tau$ leptons. The
default GENIE settings are appropriate as we do not want to be making
any assumption regarding the user's detector technology and its ability
to detect these short tracks. (Decaying $\tau$ leptons is obviously
not desirable for an emulsion detector.) By default, GENIE does not
inhibit any kinematically allowed channel. Users can modify these
options (see next chapter).

\subsection{Setting particle decay flags}

The default particle decay flag choices were described in the previous
chapter. One can easily override the default GENIE choices by setting
a series of \texttt{\textbf{\small{}``DecayParticleWithCode=i}}''
flags at the \textit{`}\texttt{\textbf{\small{}\$GENIE}}\textit{/config/UserPhysicsOptions.xml'}
configuration file, where $i$ is the particle's PDG code. \\

For example, to enable decays of $\tau^{-}$ leptons (PDG code = 15),
one needs to change: \\
\texttt{\textbf{\small{}<param type=''bool'' name=''DecayParticleWithCode=15''>
false </param>}}\\
\\
to:\\
\texttt{\textbf{\small{}<param type=''bool'' name=''DecayParticleWithCode=15''>
true </param>}}{\small \par}

\subsection{Inhibiting decay channels}

By default, GENIE does not inhibit any kinematically allowed channel.
However, for certain studies, a user may wish to inhibit certain uninteresting
decay channels in order to speed up event generation. This can be
done by setting a series of \texttt{\textbf{\small{}``InhibitDecay/Particle=i,Channel=j}}''
configuration options at the \textit{`}\texttt{\textbf{\small{}\$GENIE}}\textit{/config/UserPhysicsOptions.xml'}
file, where $i$ is the particle's PDG code and $j$ the decay channel
ID. To figure out the decay channel code numbers use the \textit{print\_decay\_channels.C}
script in `\texttt{\textbf{\small{}\$GENIE}}\textit{/src/contrib/misc/'
}(GENIE uses the ROOT `\textit{TDecayChannel}' IDs).\\

For example, to inbibit the $\tau^{-}$ lepton (PDG code = 15) $\tau^{-}\rightarrow\nu_{\tau}e^{-}\bar{\nu_{e}}$
decay channel (decay channel ID = 0), one needs to type:\\
\\
\texttt{\textbf{\small{}<param type=''bool'' name=InhibitDecay/Particle=15,Channel=0''>
true </param>}}~\\
{\small \par}

\section{Numerical algorithms}

\subsection{Random number periodicity}

GENIE is using ROOT's Mersenne Twistor random number generator with
periodicity of $10^{6000}$. See the ROOT \textit{TRandom3} class
for details. In addition GENIE is structured to use several random
number generator objects each with its own \textquotedbl{}independent\textquotedbl{}
random number sequence (see discussion in ROOT \textit{TRandom} class
description). GENIE provides different random number generators for
different types of GENIE modules: As an example, \textit{RandomGen::RndHadro()
}returns the generator to by used in hadronization models, \textit{RandomGen::RndDec()}
returns the generator to be used by decayers, \textit{RandomGen::RndKine()}
returns the generator to be used by kinematics generators, \textit{RandomGen::RndFsi()}
returns the generator to be used by intranuclear rescattering MCs
and so on... (see \textit{RandomGen} for the list of all generators).
This is an option reserved for the future as currently all modules
are passed the same random number generator (no problems with the
generator periodicity have been found or reported so far).

\subsection{Setting required numerical accurancy}

...

\chapter{Summary of Important Physics Parameters \label{cha:AppendixPhysParams}}

\begin{longtable}{|l|c|l|}
\hline
\hline 
Physics Parameter & Default value & GENIE parameter name\tabularnewline
\hline 
\hline
\endhead
\hline 
CKM element $V_{ud}$  & 0.97377  & \texttt{\small{}CKM-Vud}\tabularnewline
\hline 
CKM element $V_{us}$  & 0.2257  & \texttt{\small{}CKM-Vus}\tabularnewline
\hline 
CKM element $V_{cd}$  & 0.230  & \texttt{\small{}CKM-Vcd}\tabularnewline
\hline 
CKM element $V_{cs}$  & 0.957  & \texttt{\small{}CKM-Vcs}\tabularnewline
\hline 
Cabbibo angle, $\theta_{c}$  & 0.22853207  & \texttt{\small{}CabbiboAngle}\tabularnewline
\hline 
Fermi coupling constant, $G_{F}$  & 1.16639E-5 $GeV^{-2}$ & \tabularnewline
\hline 
Fine structure constant, $\alpha_{em}$  &  & \tabularnewline
\hline 
Weinberg angle, $\theta_{w}$  & 0.49744211  & \texttt{\small{}WeinbergAngle}\tabularnewline
\hline 
Charm mass, $m_{charm}$  & 1.430 $GeV$  & \tabularnewline
\hline 
Anomalous magnetic moment of the proton, $\mu_{p}$  & 2.7930  & \texttt{\small{}AnomMagnMoment-P}\tabularnewline
\hline 
Anomalous magnetic moment of the neutron, $\mu_{n}$  & -1.913042  & \texttt{\small{}AnomMagnMoment-N}\tabularnewline
\hline 
Nucleon e/m f/f, BBA2005 - $G_{ep}(a0)$  & 1.  & \tabularnewline
\hline 
Nucleon e/m f/f, BBA2005 - $G_{ep}(a1)$  & -0.0578  & \tabularnewline
\hline 
Nucleon e/m f/f, BBA2005 - $G_{ep}(a2)$  & 0.  & \tabularnewline
\hline 
Nucleon e/m f/f, BBA2005 - $G_{ep}(b1)$  & 11.100  & \tabularnewline
\hline 
Nucleon e/m f/f, BBA2005 - $G_{ep}(b2)$  & 13.60  & \tabularnewline
\hline 
Nucleon e/m f/f, BBA2005 - $G_{ep}(b3)$  & 33.00  & \tabularnewline
\hline 
Nucleon e/m f/f, BBA2005 - $G_{ep}(b4)$  & 0.  & \tabularnewline
\hline 
Nucleon e/m f/f, BBA2005 - $G_{{\mu}p}(a0)$  & 1.  & \tabularnewline
\hline 
Nucleon e/m f/f, BBA2005 - $G_{{\mu}p}(a1)$  & 0.1500  & \tabularnewline
\hline 
Nucleon e/m f/f, BBA2005 - $G_{{\mu}p}(a2)$  & 0.  & \tabularnewline
\hline 
Nucleon e/m f/f, BBA2005 - $G_{{\mu}p}(b1)$  & 11.100  & \tabularnewline
\hline 
Nucleon e/m f/f, BBA2005 - $G_{{\mu}p}(b2)$  & 19.600  & \tabularnewline
\hline 
Nucleon e/m f/f, BBA2005 - $G_{{\mu}p}(b3)$  & 7.540  & \tabularnewline
\hline 
Nucleon e/m f/f, BBA2005 - $G_{{\mu}p}(b4)$  & 0.  & \tabularnewline
\hline 
Nucleon e/m f/f, BBA2005 - $G_{en}(a0)$  & 0.  & \tabularnewline
\hline 
Nucleon e/m f/f, BBA2005 - $G_{en}(a1)$  & 1.250  & \tabularnewline
\hline 
Nucleon e/m f/f, BBA2005 - $G_{en}(a2)$  & 1.30  & \tabularnewline
\hline 
Nucleon e/m f/f, BBA2005 - $G_{en}(b1)$  & -9.86  & \tabularnewline
\hline 
Nucleon e/m f/f, BBA2005 - $G_{en}(b2)$  & 305.0  & \tabularnewline
\hline 
Nucleon e/m f/f, BBA2005 - $G_{en}(b3)$  & -758.0  & \tabularnewline
\hline 
Nucleon e/m f/f, BBA2005 - $G_{en}(b4)$  & 802.0  & \tabularnewline
\hline 
Nucleon e/m f/f, BBA2005 - $G_{{\mu}n}(a0)$  & 1.  & \tabularnewline
\hline 
Nucleon e/m f/f, BBA2005 - $G_{{\mu}n}(a1)$  & 1.810  & \tabularnewline
\hline 
Nucleon e/m f/f, BBA2005 - $G_{{\mu}n}(a2)$  & 0.  & \tabularnewline
\hline 
Nucleon e/m f/f, BBA2005 - $G_{{\mu}n}(b1)$  & 14.100  & \tabularnewline
\hline 
Nucleon e/m f/f, BBA2005 - $G_{{\mu}n}(b2)$  & 20.70  & \tabularnewline
\hline 
Nucleon e/m f/f, BBA2005 - $G_{{\mu}n}(b3)$  & 68.7  & \tabularnewline
\hline 
Nucleon e/m f/f, BBA2005 - $G_{{\mu}n}(b4)$  & 0.  & \tabularnewline
\hline 
$P{}_{33}(1232)$ resonance mass & 1.232 GeV & \tabularnewline
\hline 
$S_{11}(1535)$ resonance mass  & 1.535 GeV  & \tabularnewline
\hline 
$D_{13}(1520)$ resonance mass & 1.520 GeV  & \tabularnewline
\hline 
$S_{11}(1650)$ resonance mass & 1.650 GeV  & \tabularnewline
\hline 
$D_{13}(1700)$ resonance mass  & 1.700 GeV  & \tabularnewline
\hline 
$D_{15}(1675)$ resonance mass & 1.675 GeV & \tabularnewline
\hline 
$S_{31}(1620)$ resonance mass & 1.620 GeV  & \tabularnewline
\hline 
$D_{33}(1700)$ resonance mass  & 1.700 GeV  & \tabularnewline
\hline 
$P_{11}(1440)$ resonance mass & 1.440 GeV  & \tabularnewline
\hline 
$P_{13}(1720)$ resonance mass & 1.720 GeV  & \tabularnewline
\hline 
$F_{15}(1680)$ resonance mass & 1.680 GeV  & \tabularnewline
\hline 
$P_{31}(1910)$ resonance mass & 1.910 GeV  & \tabularnewline
\hline 
$P_{33}(1920)$ resonance mass & 1.920 GeV  & \tabularnewline
\hline 
$F_{35}(1905)$ resonance mass & 1.905 GeV  & \tabularnewline
\hline 
$F_{37}(1950)$ resonance mass & 1.950 GeV  & \tabularnewline
\hline 
$P_{11}(1710)$ resonance mass & 1.710 GeV  & \tabularnewline
\hline 
$P{}_{33}(1232)$ resonance width & 0.120 GeV & \tabularnewline
\hline 
$S_{11}(1535)$ resonance width & 0.150 GeV & \tabularnewline
\hline 
$D_{13}(1520)$ resonance width & 0.120 GeV & \tabularnewline
\hline 
$S_{11}(1650)$ resonance width & 0.150 GeV & \tabularnewline
\hline 
$D_{13}(1700)$ resonance width & 0.100 GeV & \tabularnewline
\hline 
$D_{15}(1675)$ resonance width & 0.150 GeV & \tabularnewline
\hline 
$S_{31}(1620)$ resonance width & 0.150 GeV  & \tabularnewline
\hline 
$D_{33}(1700)$ resonance width & 0.300 GeV  & \tabularnewline
\hline 
$P_{11}(1440)$ resonance width & 0.350 GeV  & \tabularnewline
\hline 
$P_{13}(1720)$ resonance width & 0.150 GeV  & \tabularnewline
\hline 
$F_{15}(1680)$ resonance width & 0.130 GeV  & \tabularnewline
\hline 
$P_{31}(1910)$ resonance width & 0.250 GeV  & \tabularnewline
\hline 
$P_{33}(1920)$ resonance width & 0.200 GeV  & \tabularnewline
\hline 
$F_{35}(1905)$ resonance width & 0.350 GeV  & \tabularnewline
\hline 
$F_{37}(1950)$ resonance width & 0.300 GeV  & \tabularnewline
\hline 
$P_{11}(1710)$ resonance width & 0.100 GeV  & \tabularnewline
\hline 
NCEL axial mass,$M_{A}$  & 0.990 $GeV$  & \texttt{\small{}EL-Ma}\tabularnewline
\hline 
NCEL vector mass,$M_{V}$  & 0.840 $GeV$  & \texttt{\small{}EL-Mv}\tabularnewline
\hline 
NCEL strange axial form factor,$\eta_{axial}$  & 0.12  & \texttt{\small{}EL-Axial-Eta}\tabularnewline
\hline 
CCQE axial mass,$M_{A}$  & 0.990 $GeV$  & \texttt{\small{}QEL-Ma}\tabularnewline
\hline 
CCQE vector mass,$M_{V}$  & 0.840 $GeV$  & \texttt{\small{}QEL-Mv}\tabularnewline
\hline 
$F_{A}(Q^{2}=0)$  & -1.2670  & \texttt{\small{}QEL-FA0}\tabularnewline
\hline 
CC/NC resonance (Rein-Sehgal), $\Omega$  & 1.05  & \tabularnewline
\hline 
CC/NC resonance (Rein-Sehgal), Z  & 0.762  & \tabularnewline
\hline 
CC/NC resonance axial mass (Rein-Sehgal), $M_{A}$  & 1.120 $GeV$  & \tabularnewline
\hline 
CC/NC resonance vector mass (Rein-Sehgal),$M_{V}$  & 0.840 $GeV$  & \tabularnewline
\hline 
CC/NC DIS (Bodek-Yang), $A$  & 0.538  & \tabularnewline
\hline 
CC/NC DIS (Bodek-Yang), $B$  & 0.305  & \tabularnewline
\hline 
CC/NC DIS (Bodek-Yang), $C_{s}^{u}$  & 0.363  & \tabularnewline
\hline 
CC/NC DIS (Bodek-Yang), $C_{s}^{d}$  & 0.621  & \tabularnewline
\hline 
CC/NC DIS (Bodek-Yang), $C_{v1}^{u}$  & 0.291  & \tabularnewline
\hline 
CC/NC DIS (Bodek-Yang), $C_{v2}^{u}$  & 0.189  & \tabularnewline
\hline 
CC/NC DIS (Bodek-Yang), $C_{v1}^{d}$  & 0.202  & \tabularnewline
\hline 
CC/NC DIS (Bodek-Yang), $C_{v2}^{d}$  & 0.255  & \tabularnewline
\hline 
CC/NC DIS (Bodek-Yang), $X_{0}$  & -0.00817  & \tabularnewline
\hline 
CC/NC DIS (Bodek-Yang), $X_{1}$  & 0.0506  & \tabularnewline
\hline 
CC/NC DIS (Bodek-Yang), $X_{2}$  & 0.0798  & \tabularnewline
\hline 
CC/NC DIS (Bodek-Yang), PDF $Q_{min}^{2}$  & 0.800 $GeV^{2}$  & \tabularnewline
\hline 
CC/NC DIS (Bodek-Yang), Uncorr. PDF set  & GRV98LO  & \tabularnewline
\hline 
CC/NC DIS (Bodek-Yang), Nuclear mod.?  & true  & \tabularnewline
\hline 
CC/NC DIS (Bodek-Yang), Whitlow R ($F_{L}$)?  & true  & \tabularnewline
\hline 
CC/NC coherent $\pi$ (Rein-Sehgal), $M_{A}$  & 1.000 $GeV$  & \tabularnewline
\hline 
CC/NC coherent $\pi$ (Rein-Sehgal), $R_{0}$  & 1.000 $fm$  & \tabularnewline
\hline 
CC/NC coherent $\pi$ (Rein-Sehgal), $Re/Im{Ampl}$  & 0.300  & \tabularnewline
\hline 
CC/NC coherent $\pi$ (Rein-Sehgal), Mod. PCAC?  & true  & \tabularnewline
\hline 
Transition region modeling (neugen3), $R_{bkg}^{\nu pCC1\pi}$  & 0.100  & \tabularnewline
\hline 
Transition region modeling (neugen3), $R_{bkg}^{\nu pCC2\pi}$  & 1.000  & \tabularnewline
\hline 
Transition region modeling (neugen3), $R_{bkg}^{\nu pNC1\pi}$  & 0.100  & \tabularnewline
\hline 
Transition region modeling (neugen3), $R_{bkg}^{\nu pNC2\pi}$  & 1.000  & \tabularnewline
\hline 
Transition region modeling (neugen3), $R_{bkg}^{\nu nCC1\pi}$  & 0.300  & \tabularnewline
\hline 
Transition region modeling (neugen3), $R_{bkg}^{\nu nCC2\pi}$  & 1.000  & \tabularnewline
\hline 
Transition region modeling (neugen3), $R_{bkg}^{\nu nNC1\pi}$  & 0.300  & \tabularnewline
\hline 
Transition region modeling (neugen3), $R_{bkg}^{\nu nNC2\pi}$  & 1.000  & \tabularnewline
\hline 
Transition region modeling (neugen3), $R_{bkg}^{\bar{\nu}pCC1\pi}$  & 0.300  & \tabularnewline
\hline 
Transition region modeling (neugen3), $R_{bkg}^{\bar{\nu}pCC2\pi}$  & 1.000  & \tabularnewline
\hline 
Transition region modeling (neugen3), $R_{bkg}^{\bar{\nu}pNC1\pi}$  & 0.300  & \tabularnewline
\hline 
Transition region modeling (neugen3), $R_{bkg}^{\bar{\nu}pNC2\pi}$  & 1.000  & \tabularnewline
\hline 
Transition region modeling (neugen3), $R_{bkg}^{\bar{\nu}nCC1\pi}$  & 0.100  & \tabularnewline
\hline 
Transition region modeling (neugen3), $R_{bkg}^{\bar{\nu}nCC2\pi}$  & 1.000  & \tabularnewline
\hline 
Transition region modeling (neugen3), $R_{bkg}^{\bar{\nu}nNC1\pi}$  & 0.100  & \tabularnewline
\hline 
Transition region modeling (neugen3), $R_{bkg}^{\bar{\nu}nNC2\pi}$  & 1.000  & \tabularnewline
\hline 
Transition region modeling (neugen3), $W_{cut}$  & 1.7 GeV  & \tabularnewline
\hline 
Average chg. hadron multiplicity factor (KNO), $a_{\nu p}$ & 0.4 & \tabularnewline
\hline 
Average chg. hadron multiplicity factor (KNO), $a_{\nu n}$ & -0.2 & \tabularnewline
\hline 
Average chg. hadron multiplicity factor (KNO, $a_{\bar{\nu p}}$ & 0.02 & \tabularnewline
\hline 
Average chg. hadron multiplicity factor (KNO), $a_{\bar{\nu n}}$ & 0.80 & \tabularnewline
\hline 
Average chg. hadron multiplicity factor (KNO), $b_{\nu p}$ & 1.42 & \tabularnewline
\hline 
Average chg. hadron multiplicity factor (KNO), $b_{\nu n}$ & 1.42 & \tabularnewline
\hline 
Average chg. hadron multiplicity factor (KNO), $b_{\bar{\nu p}}$ & 1.28 & \tabularnewline
\hline 
Average chg. hadron multiplicity factor (KNO), $b_{\bar{\nu n}}$ & 0.95 & \tabularnewline
\hline 
Average strange baryon multiplicity factor (KNO), $a_{hyp}$ & 0.021951447 & \tabularnewline
\hline 
Average strange baryon multiplicity factor (KNO), $b_{hyp}$ & 0.041969985 & \tabularnewline
\hline 
KNO parameterization (Levy), $c_{\nu p}$ & 7.93 & \tabularnewline
\hline 
KNO parameterization (Levy), $c_{\nu n}$ & 5.22 & \tabularnewline
\hline 
KNO parameterization (Levy), $c_{\bar{\nu p}}$ & 5.22 & \tabularnewline
\hline 
KNO parameterization (Levy), $c_{\bar{\nu n}}$ & 7.93 & \tabularnewline
\hline 
$\pi^{0}\pi^{0}$ probability (AGKY/KNO) & 0.3133 & \tabularnewline
\hline 
$\pi^{+}\pi^{-}$ probability (AGKY/KNO) & 0.6267 & \tabularnewline
\hline 
$K^{0}\bar{K^{0}}$ probability (AGKY/KNO) & 0.03 & \tabularnewline
\hline 
$K^{+}K^{-}$ probability (AGKY/KNO) & 0.03 & \tabularnewline
\hline 
Phase-space $p_{T}$reweighting factor (AGKY/KNO) & 3.5 & \tabularnewline
\hline 
$s\bar{s}$ suppression factor (PYTHIA6) & 0.30 & \tabularnewline
\hline 
Gaussian $p{}_{T}^{2}$ (PYTHIA6) & 0.44 & \tabularnewline
\hline 
Non-gaussian $p{}_{T}^{2}$ tail (PYTHIA6) & 0.01 & \tabularnewline
\hline 
Remaining energy cut-off (PYTHIA6) & 0.20 & \tabularnewline
\hline 
Formation time, $c\tau_{0}$ & 0.342 fm & \tabularnewline
\hline 
Binding energy (RFG), $E_{b}$(Li$^{6}$)  & 0.017 GeV & \tabularnewline
\hline 
Binding energy (RFG), $E_{b}$(C$^{12}$)  & 0.025 GeV & \tabularnewline
\hline 
Binding energy (RFG), $E_{b}$(O$^{16}$) & 0.027 GeV & \tabularnewline
\hline 
Binding energy (RFG), $E_{b}$(Mg$^{24}$)  & 0.032 GeV & \tabularnewline
\hline 
Binding energy (RFG), $E_{b}$(Ca$^{40}$)  & 0.028 GeV & \tabularnewline
\hline 
Binding energy (RFG), $E_{b}$(Fe$^{56}$)  & 0.036 GeV & \tabularnewline
\hline 
Binding energy (RFG), $E_{b}$(Ni$^{58}$)  & 0.036 GeV & \tabularnewline
\hline 
Binding energy (RFG), $E_{b}$(Pb$^{208}$)  & 0.044 GeV & \tabularnewline
\hline 
Fermi momentum (RFG), $k_{F}$(Li$^{6}$, p)  & 0.169 GeV & \tabularnewline
\hline 
Fermi momentum (RFG), $k_{F}$(Li$^{6}$, n)  & 0.169 GeV  & \tabularnewline
\hline 
Fermi momentum (RFG), $k_{F}$(C$^{12}$, p)  & 0.221 GeV & \tabularnewline
\hline 
Fermi momentum (RFG), $k_{F}$(C$^{12}$, n) & 0.221 GeV & \tabularnewline
\hline 
Fermi momentum (RFG), $k_{F}$(O$^{16}$, p)  & 0.225 GeV & \tabularnewline
\hline 
Fermi momentum (RFG), $k_{F}$(O$^{16}$, n)  & 0.225 GeV & \tabularnewline
\hline 
Fermi momentum (RFG), $k_{F}$(Mg$^{24}$, p)  & 0.235 GeV & \tabularnewline
\hline 
Fermi momentum (RFG), $k_{F}$(Mg$^{24}$, n) & 0.235 GeV & \tabularnewline
\hline 
Fermi momentum (RFG), $k_{F}$(Si$^{28}$, p)  & 0.239 GeV & \tabularnewline
\hline 
Fermi momentum (RFG), $k_{F}$(Si$^{28}$, n)  & 0.239 GeV & \tabularnewline
\hline 
Fermi momentum (RFG), $k_{F}$(Ar$^{40}$, p)  & 0.242 GeV & \tabularnewline
\hline 
\end{longtable}

\chapter{Common Status and Particle Codes \label{cha:AppendixStatAndPdgCodes}}

\section{Status codes}

\begin{tabular}{|l|l|l|}
\hline 
\textbf{Description} & \textbf{\textit{GHepStatus\_t}} & \textbf{As }\textbf{\textit{int}}\tabularnewline
\hline 
Undefined & \textit{kIStUndefined} & -1\tabularnewline
\hline 
Initial state & \textit{kIStInitialState} & 0\tabularnewline
\hline 
Stable final state & \textit{kIstStableFinalState} & 1\tabularnewline
\hline 
Intermediate state & \textit{kIStIntermediateState} & 2\tabularnewline
\hline 
Decayed state & \textit{kIStDecayedState} & 3\tabularnewline
\hline 
Nucleon target & \textit{kIStNucleonTarget} & 11\tabularnewline
\hline 
DIS pre-fragm. hadronic state & \textit{kIStDISPreFragmHadronicState} & 12\tabularnewline
\hline 
Resonant pre-decayed state & \textit{kIStPreDecayResonantState} & 13\tabularnewline
\hline 
Hadron in the nucleus & \textit{kIStHadronInTheNucleus} & 14\tabularnewline
\hline 
Final state nuclear remnant & \textit{kIStFinalStateNuclearRemnant} & 15\tabularnewline
\hline 
Nucleon cluster target & \textit{kIStNucleonClusterTarget} & 16\tabularnewline
\hline 
\end{tabular}

\section{Particle codes}

See PDG `Monte Carlo Particle Numbering Scheme' for a complete list.\url{http://pdg.lbl.gov/2008/mcdata/mc_particle_id_contents.shtml}\\

{\small{}}%
\begin{tabular}{|l|l||l|l||l|l||l|l||l|l|}
\hline 
{\small{}$\nu_{e}$ ($\bar{\nu_{e}}$) } & {\small{}12 (-12)} & {\small{}$p$} & {\small{}2212} & {\small{}$\pi^{0}$} & {\small{}111} & {\small{}$uu$ $(s=1)$} & {\small{}2203} & {\small{}$g$} & {\small{}21}\tabularnewline
\hline 
{\small{}$\nu_{\mu}$ ($\bar{\nu_{\mu}}$)} & {\small{}14 (-14)} & {\small{}$n$} & {\small{}2112} & {\small{}$\pi^{+}$ ($\pi^{-}$)} & {\small{}211 (-211)} & {\small{}$ud$ $(s=0)$} & {\small{}2101} & {\small{}$\gamma$} & {\small{}22}\tabularnewline
\hline 
{\small{}$\nu_{\tau}$ ($\bar{\nu_{\tau}}$)} & {\small{}16 (-16)} & {\small{}$\Lambda^{0}$} & {\small{}3122} & {\small{}$\rho^{0}$} & {\small{}113} & {\small{}$ud$ $(s=1)$} & {\small{}2103} & {\small{}$Z^{0}$} & {\small{}23}\tabularnewline
\hline 
{\small{}$e^{-}$ ($e^{+}$)} & {\small{}11 (-11)} & {\small{}$\Sigma^{+}$} & {\small{}3222} & {\small{}$\rho^{+}$ ($\rho^{-}$)} & {\small{}213 (-213)} & {\small{}$su$ $(s=0)$} & {\small{}3201} & {\small{}$W^{+}$ ($W^{-})$} & {\small{}24 (-24)}\tabularnewline
\hline 
{\small{}$\mu^{-}$ ($\mu^{+}$)} & {\small{}13 (-13)} & {\small{}$\Sigma^{0}$} & {\small{}3212} & {\small{}$\eta$} & {\small{}221} & {\small{}$su$ $(s=1)$} & {\small{}3203} &  & \tabularnewline
\hline 
{\small{}$\tau^{-}$ ($\tau^{+}$)} & {\small{}15 (-15)} & {\small{}$\Sigma^{-}$} & {\small{}3112} & {\small{}$\eta^{\prime}$} & {\small{}331} & {\small{}$sd$ $(s=0)$} & {\small{}3101} &  & \tabularnewline
\hline 
{\small{}$d$ ($\bar{d}$)} & {\small{}1 (-1)} & {\small{}$\Xi^{0}$ } & {\small{}3322 } & {\small{}$\omega$} & {\small{}223} & {\small{}$sd$ $(s=1)$} & {\small{}3103} &  & \tabularnewline
\hline 
{\small{}$u$ ($\bar{u}$)} & {\small{}2 (-2)} & {\small{}$\Xi^{-}$ } & {\small{}3312 } & {\small{}$\phi$} & {\small{}333} & {\small{}$ss$ $(s=1)$} & {\small{}3303} &  & \tabularnewline
\hline 
{\small{}$s$ ($\bar{s}$)} & {\small{}3 (-3)} & {\small{}$\Omega^{-}$} & {\small{}3332} & {\small{}$\eta_{c}$} & {\small{}441} &  &  &  & \tabularnewline
\hline 
{\small{}$c$ ($\bar{c}$)} & {\small{}4 (-4)} & {\small{}$\Lambda_{c}^{+}$} & {\small{}4122} & {\small{}$J/\psi$} & {\small{}443} &  &  &  & \tabularnewline
\hline 
{\small{}$b$ ($\bar{b}$)} & {\small{}5 (-5)} & {\small{}$\Sigma_{c}^{0}$} & {\small{}4112} & {\small{}$K^{0}$ ($\bar{K^{0}}$)} & {\small{}311 (-311)} &  &  &  & \tabularnewline
\hline 
{\small{}$t$ ($\bar{t}$)} & {\small{}6 (-6)} & {\small{}$\Sigma_{c}^{+}$} & {\small{}4212} & {\small{}$K^{+}$ ($K^{-}$)} & {\small{}321 (-321)} &  &  &  & \tabularnewline
\hline 
 &  & {\small{}$\Sigma_{c}^{++}$} & {\small{}4222} & {\small{}$K_{L}^{0}$} & {\small{}130} &  &  &  & \tabularnewline
\hline 
 &  & {\small{}$\Xi_{c}^{0}$} & {\small{}4132} & {\small{}$K_{S}^{0}$} & {\small{}310} &  &  &  & \tabularnewline
\hline 
 &  & {\small{}$\Xi_{c}^{+}$} & {\small{}4232} & {\small{}$D^{0}$ ($\bar{D^{0}}$)} & {\small{}421 (-421)} &  &  &  & \tabularnewline
\hline 
 &  & {\small{}$\Omega_{c}^{0}$} & {\small{}4332} & {\small{}$D^{+}$ ($D^{-}$)} & {\small{}411 (-411)} &  &  &  & \tabularnewline
\hline 
 &  &  &  & {\small{}$D_{s}^{+}$ ($D_{s}^{-}$)} & {\small{}431 (-431)} &  &  &  & \tabularnewline
\hline 
\end{tabular}{\small \par}

\section{Baryon resonance codes}

{\small{}}%
\begin{tabular}{|l|l||l|l||l|l||l|l|}
\hline 
{\small{}$P_{33}(1232)$; $\Delta^{-}$} & {\small{}1114} & {\small{}$S_{11}(1650)$; $N^{0}$} & {\small{}32112} & {\small{}$D_{13}(1700)$; $N^{0}$} & {\small{}21214} & {\small{}$P_{31}(1910)$; $\Delta^{-}$} & {\small{}21112}\tabularnewline
\hline 
{\small{}$P_{33}(1232)$; $\Delta^{0}$} & {\small{}2114} & {\small{}$S_{11}(1650)$; $N^{+}$} & {\small{}32212} & {\small{}$D_{13}(1700)$; $N^{+}$ } & {\small{}22124} & {\small{}$P_{31}(1910)$; $\Delta^{0}$} & {\small{}21212}\tabularnewline
\hline 
{\small{}$P_{33}(1232)$; $\Delta^{+}$} & {\small{}2214} & {\small{}$D_{15}(1675)$; $N^{0}$} & {\small{}2116} & {\small{}$P_{11}(1710)$; $N^{0}$} & {\small{}42112} & {\small{}$P_{31}(1910)$; $\Delta^{+}$} & {\small{}22122}\tabularnewline
\hline 
{\small{}$P_{33}(1232)$; $\Delta^{++}$} & {\small{}2224} & {\small{}$D_{15}(1675)$; $N^{+}$} & {\small{}2216} & {\small{}$P_{11}(1710)$; $N^{+}$} & {\small{}42212} & {\small{}$P_{31}(1910)$; $\Delta^{++}$} & {\small{}22222}\tabularnewline
\hline 
{\small{}$P_{11}(1440)$; $N^{0}$} & {\small{}12112} & {\small{}$F_{15}(1680)$; $N^{0}$} & {\small{}12116} & {\small{}$P_{13}(1720)$; $N^{0}$} & {\small{}31214} & {\small{}$P_{33}(1920)$; $\Delta^{-}$} & {\small{}21114}\tabularnewline
\hline 
{\small{}$P_{11}(1440)$; $N^{+}$} & {\small{}12212} & {\small{}$F_{15}(1680)$; $N^{+}$} & {\small{}12216} & {\small{}$P_{13}(1720)$; $N^{+}$} & {\small{}32124} & {\small{}$P_{33}(1920)$; $\Delta^{0}$} & {\small{}22114}\tabularnewline
\hline 
{\small{}$D_{13}(1520)$; $N^{0}$ } & {\small{}1214} & {\small{}$D_{33}(1700)$; $\Delta^{-}$} & {\small{}11114} & {\small{}$F_{35}(1905)$; $\Delta^{-}$} & {\small{}1116} & {\small{}$P_{33}(1920)$; $\Delta^{+}$} & {\small{}22214}\tabularnewline
\hline 
{\small{}$D_{13}(1520)$; $N^{+}$} & {\small{}2124} & {\small{}$D_{33}(1700)$; $\Delta^{0}$} & {\small{}12114} & {\small{}$F_{35}(1905)$; $\Delta^{0}$} & {\small{}1216} & {\small{}$P_{33}(1920)$; $\Delta^{++}$} & {\small{}22224}\tabularnewline
\hline 
{\small{}$S_{11}(1535)$; $N^{0}$} & {\small{}22112} & {\small{}$D_{33}(1700)$; $\Delta^{+}$} & {\small{}12214} & {\small{}$F_{35}(1905)$; $\Delta^{+}$} & {\small{}2126} & {\small{}$F_{37}(1950)$; $\Delta^{-}$} & {\small{}1118}\tabularnewline
\hline 
{\small{}$S_{11}(1535)$; $N^{+}$} & {\small{}22212} & {\small{}$D_{33}(1700)$; $\Delta^{++}$} & {\small{}12224} & {\small{}$F_{35}(1905)$; $\Delta^{++}$} & {\small{}2226} & {\small{}$F_{37}(1950)$; $\Delta^{0}$} & {\small{}2118}\tabularnewline
\hline 
{\small{}$S_{31}(1620)$; $\Delta^{-}$} & {\small{}11112} &  &  &  &  & {\small{}$F_{37}(1950)$; $\Delta^{+}$} & {\small{}2218}\tabularnewline
\hline 
{\small{}$S_{31}(1620)$; $\Delta^{0}$} & {\small{}1212} &  &  &  &  & {\small{}$F_{37}(1950)$; $\Delta^{++}$} & {\small{}2228}\tabularnewline
\hline 
{\small{}$S_{31}(1620)$; $\Delta^{+}$} & {\small{}2122} &  &  &  &  &  & \tabularnewline
\hline 
{\small{}$S_{31}(1620)$; $\Delta^{++}$} & {\small{}2222} &  &  &  &  &  & \tabularnewline
\hline 
\end{tabular}{\small \par}

\section{Ion codes}

GENIE has adopted the standard PDG (2006) particle codes. For ions
it has adopted a PDG extension, using the 10-digit code 10LZZZAAAI
where AAA is the total baryon number, ZZZ is the total charge, L is
the number of strange quarks and I is the isomer number (I=0 corresponds
to the ground state). \\
\\
So, for example:\\
\\
1000010010 $\rightarrow$ $H^{1}$\\
1000060120 $\rightarrow$ $C12:$ \\
1000080160 $\rightarrow$ $O^{16}$ :\\
1000260560 $\rightarrow$ $Fe^{56}$ :\\
\\
and so on.

\section{GENIE pseudo-particle codes}

GENIE-specific pseudo-particles have PDG codes >= 2000000000. 

\newpage

\chapter{3rd Party Softw. Installation Instructions \label{app:ThirdPartySoftw}}

The following dependencies need to be installed, in the following
order.

\section*{LOG4CPP}

\subsubsection{Before installing log4cpp}

Check whether log4cpp is already installed at your system. The library
filename contains liblog4cpp, so if you cannot find a file with a
filename containing liblog4cpp, then you probably do not have the
software installed.

\subsubsection{Getting the source code}

Download the source code from the sourceforge anonymous CVS repository
(when prompted for a password, simply hit enter): \\
\texttt{\textbf{\small{}\$ cd /dir/for/external/src/code}}\texttt{\textbf{\footnotesize{}}}~\\
\texttt{\textbf{\footnotesize{}\$ cvs -d :pserver:anonymous@log4cpp.cvs.sourceforge.net:/cvsroot/log4cpp
login}}~\\
\texttt{\textbf{\footnotesize{}\$ cvs -d :pserver:anonymous@log4cpp.cvs.sourceforge.net:/cvsroot/log4cpp
-z3 co log4cpp}}{\footnotesize \par}

\subsubsection{Configuring and building}

Enter the log4cpp directory and run `\texttt{\textbf{\small{}autogen}}'
and `\texttt{\textbf{\small{}configure}}'. Replace {[}location{]}
with the installation directory of your choice; you cannot install
it in the same directory as the source (where you are now). You can
choose not to use the \texttt{\textbf{\small{}`-{}-prefix'}} tag,
in which case the default install directory is `\textit{/usr/local}'.\\
\texttt{\textbf{\small{}\$ cd log4cpp}}~\\
\texttt{\textbf{\small{}\$ ./autogen.sh}}~\\
\texttt{\textbf{\small{}\$ ./configure -{}-prefix={[}location{]}}}~\\
\texttt{\textbf{\footnotesize{}}}~\\
What's left is to run `\texttt{\textbf{\small{}make}}' and `\texttt{\textbf{\small{}make
install}}'. If make install gives you an error while copying or moving
files stating that the files are identical, then you probably choose
the source folder as your install folder in the above configure step.
Rerun configure with a different location (or simply leave the \texttt{\textbf{\small{}`-{}-prefix'}}
option out for the default).\\
\texttt{\textbf{\small{}\$ make}}~\\
\texttt{\textbf{\small{}\$ make install}}{\small \par}

\subsubsection{Notes:}
\begin{itemize}
\item Alternatively, you may install pre-compiled binaries. For example,
if you are using `yum' on LINUX then just type:\\
\texttt{\textbf{\small{}\$ yum install log4cpp}}~\\
On MAC OS X you can do the same using `DarwinPorts':\\
\texttt{\textbf{\small{}\$ sudo port install log4cpp}}~\\
{\small \par}
\end{itemize}

\section*{LIBXML2}

\subsubsection{Before installing libxml2}

Check whether libxml2 is already installed at your system - most likely
it is. Look for a libxml2.{*} library (typically in `\textit{/usr/lib}')
and for a libxml2 include folder (typically in `\textit{/usr/include}').

\subsubsection{Getting the source code}

Download the source code from the GNOME subversion repository:\\
\texttt{\textbf{\small{}\$ cd /dir/for/external/src/code}}~\\
\texttt{\textbf{\small{}\$ svn co https://svn.gnome.org/svn/libxml2/trunk
libxml2}}~\\
\texttt{\textbf{\small{}}}~\\
Alternatively, you download the code as a gzipped tarball from:\\
\textit{http://xmlsoft.org/downloads.html}.

\subsubsection{Configuring and building }

\texttt{\textbf{\small{}\$ cd libxml2}}~\\
\texttt{\textbf{\small{}\$ ./autogen.sh -{}-prefix={[}location{]}}}~\\
\texttt{\textbf{\small{}\$ make}}~\\
\texttt{\textbf{\small{}\$ make install}}{\small \par}

\subsubsection{Notes:}
\begin{itemize}
\item Alternatively, you may install pre-compiled binaries. For example,
if you are using `yum' on LINUX then just type:\\
\texttt{\textbf{\small{}\$ yum install libxml2}}~\\
On MAC OS X you can do the same using `DarwinPorts':\\
\texttt{\textbf{\small{}\$ sudo port install libxml2}}~\\
{\small \par}
\end{itemize}

\section*{LHAPDF}

\subsubsection{Getting the source code}

Get the LHAPDF code (and PDF data files) from \textit{http://projects.hepforge.org/lhapdf/}.
The tarball corrsponding to version \textit{`x.y.z'} is named \textit{`lhapdf-x.y.z.tar.gz}'.
\\
\texttt{\textbf{\small{}\$ mv lhapdf-x.y.z.tar.gz /dir/for/external/src/code}}~\\
\texttt{\textbf{\small{}\$ cd /directory/to/download/external/code}}~\\
\texttt{\textbf{\small{}\$ tar xzvf lhapdf-x.y.z.tar.gz}}~\\
{\small \par}

\subsubsection{Configuring and building }

\texttt{\textbf{\small{}\$ cd lhapdf-x.y.z/}}~\\
\texttt{\textbf{\small{}\$ ./configure -{}-prefix={[}location{]}}}~\\
\texttt{\textbf{\small{}\$ make}}~\\
\texttt{\textbf{\small{}\$ make install}}\\

\section*{PYTHIA6}

Installation of PYTHIA6 is simplified by using a script provided by
Robert Hatcher (`\textit{build\_pythia6.sh}'). The file is included
in the GENIE source tree (see `\texttt{\textbf{\small{}\$GENIE}}\textit{/src/scripts/build/ext/build\_pythia6.sh}').
You can also get a copy from the web\footnote{Visit:{\scriptsize{} }\textit{\footnotesize{}http://projects.hepforge.org/genie/trac/browser/trunk/src/scripts/build/ext/}\textit{}\\
Click on the file and then download it by clicking on `Download in
other formats / Original format' towards the end of the page.}:\\
\\
You can run the script (please, also read its documentation) as:\\
\texttt{\textbf{\small{}\$ source build\_pythia6.sh {[}version{]}}}\\
\\
For example, in order to download and install version 6.4.12, type:\\
\texttt{\textbf{\small{}\$ source build\_pythia6.sh 6412}}~\\
{\small \par}

\section*{ROOT}

\subsubsection{Getting the source code}

Get the source code from the ROOT subversion repository. To get the
development version, type:\\
\texttt{\textbf{\small{}\$ cvs co http://root.cern.ch/svn/root/trunk
root}}\textbf{}\\
To get a specific version `\textit{x.y.z}', type:\\
\texttt{\textbf{\small{}\$ cvs co http://root.cern.ch/svn/root/tags/vx-y-z
root}}~\\
\texttt{\textbf{\small{}\$ cvs co http://root.cern.ch/svn/root/tags/v5-22-00
root}}~\\
\texttt{\textbf{\small{}}}~\\
See \textit{http://root.cern.ch/drupal/content/downloading-root/}

\subsubsection{Configuring and building }

\texttt{\textbf{\small{}\$ export ROOTSYS=/path/to/install\_root}}~\\
\texttt{\textbf{\small{}\$ cd \$ROOTSYS}}~\\
\texttt{\textbf{\small{}\$ ./configure {[}arch{]} {[}other options{]}
-{}-enable-pythia6 -{}-with-pythia6-libdir=\$PYTHIA6 -{}-enable-mathmore}}~\\
\texttt{\textbf{\small{}\$ make}}{\small \par}

\subsubsection{Testing}

Accessing root is an easy test to see if it has installed correctly.
If you are not familiar with root, use ``.q'' in root prompt to
quit.\\
\\
\texttt{\textbf{\small{}\$ root -l}}~\\
\\
\texttt{\textbf{\small{}root {[}0{]} .q}}\\
\\
See \textit{http://root.cern.ch/root/Install.html} for more information
on installing ROOT from source.

\chapter{Finding More Information \label{cha:AppendixGetMoreInfo}}

\section{The GENIE web page}

The GENIE web page, hosted at HepForge is the exclusive official source
of information on GENIE. The page can be reached at \texttt{\textbf{\small{}http://www.genie-mc.org}}{\small \par}

\section{Subscribing at the GENIE mailing lists}

The GENIE mailing lists are hosted at JISCmail, UK's National Academic
Mailing List Service. We currently maintain two mailing lists
\begin{itemize}
\item \texttt{\textbf{\small{}neutrino-mc-support@jiscmail.ac.uk}} : This
is the GENIE support mailing list and is open to all users.
\item \texttt{\textbf{\small{}neutrino-mc-core@jiscmail.ac.uk}} : This is
the GENIE developers mailing list and is open only to members of the
GENIE collaboration.
\end{itemize}
To register at the GENIE support mailing list go to \\
\textit{https://www.jiscmail.ac.uk/cgi-bin/webadmin?A0=NEUTRINO-MC-SUPPORT}\\
(or follow the link the GENIE web page) and click on `Join or Leave
NEUTRINO-MC-SUPPORT'. In the registration page specify your name,
preferred e-mail address and subscription type and click on `Join
NEUTRINO-MC-SUPPORT'. This will generate a request that has to be
approved by a member of the GENIE collaboration. Upon approval a notification
and the JISCmail Data Protection policy will be forwarded at your
nominated e-mail address.

\section{The GENIE document database (DocDB)}

The GENIE internal note repository is hosted at Fermilab at the Projects
Document Database. Most documents are internal to the GENIE collaboration.
However certain documents are made publicly available. The Fermilab
Projects Documents Database can be reached at: \texttt{\textbf{\small{}}}~\\
\textit{http://projects-docdb.fnal.gov:8080/cgi-bin/ListBy?groupid=30}

\section{The GENIE issue tracker}

The issue tracker hosted at HepForge is a useful tool for monitoring
tasks and milestones, for submitting bug reports and getting information
about their resolution. It is available at:\\
\textit{http://projects.hepforge.org/genie/trac/report/}\\
\textit{}\\
Non-developers can also submit tickets. A general `guest' account
has been setup (the password is available upon request).

\section{The GENIE Generator repository browser}

\textit{http://projects.hepforge.org/genie/trac/browser/generator}

\section{The GENIE doxygen documentation}

\textit{http://doxygen.genie-mc.org/}

\chapter{Glossary \label{cha:AppendixGlossary}}
\begin{itemize}
\item \textbf{\textit{\large{}A}}{\large \par}

\begin{itemize}
\item AGKY: A home-grown neutrino-induced hadronic multiplarticle production
model developed by C.Andreopoulos, H.Gallagher, P.Kehayias and T.Yang.
\end{itemize}
\item \textbf{\textit{\large{}B}}{\large \par}

\begin{itemize}
\item BGLRS: An atmospheric neutrino simulation developed by G. Barr, T.K.
Gaisser, P. Lipari, S. Robbins and T. Stanev.
\item BY: Bodek-Yang.
\end{itemize}
\item \textbf{\textit{\large{}C}}{\large \par}

\begin{itemize}
\item COH: 
\item CVS:
\end{itemize}
\item \textbf{\textit{\large{}D}}{\large \par}

\begin{itemize}
\item DIS: Deep Inelastic Scattering.
\end{itemize}
\item \textbf{\textit{\large{}E}}{\large \par}
\item \textbf{\textit{\large{}F}}{\large \par}

\begin{itemize}
\item FGM: Fermi Gas Model.
\item FLUKA:
\end{itemize}
\item \textbf{\textit{\large{}G}}{\large \par}

\begin{itemize}
\item GEF: Geocentric Earth-Fixed Coordinate System (+z: Points to North
Pole / xy: Equatorial plane / +x: Points to the Prime Meridian / +y:
As needed to make a right-handed coordinate system).
\item Geant4:
\item GDML:
\item GENEVE: A legacy, fortran77-based neutrino generator by F.Cavanna
et al.
\item GENIE: \textbf{\textcolor{red}{G}}enerates \textbf{\textcolor{red}{E}}vents
for \textbf{\textcolor{red}{N}}eutrino \textbf{\textcolor{red}{I}}nteraction
\textbf{\textcolor{red}{E}}xperiments.
\item GiBUU: A fortran2003-based state-of-the-art particle transport simulation
using the Boltzmann-Uehling-Uhlenbeck (BUU) framework. Developed primarily
by the theory group at Giessen University (U.Mosel et al.)
\item GNuMI: Geant3- and Geant4-based NuMI beamline simulation software.
\item GSL: GNU Scientific Library
\item gevdump: A GENIE application for printing-out event records.
\item gevpick: A GENIE event topology cherry-picking application.
\item gevgen: A simple, generic GENIE event generation application.
\item gevgen\_hadron: A GENIE hadron+nucleus event generation application.
\item gevgen\_atmo: A GENIE event generation application for atmospheric
neutrinos.
\item gevgen\_ndcy: A GENIE nucleon decay event generation application.
\item gevgen\_t2k: A GENIE event generation application customized for T2K.
\item gevgen\_fnal: A GENIE event generation application customized for
the NuMI beamline experiments.
\item gmkspl: A GENIE application for generating cross section spline files
(evet generation inputs).
\item gntpc: A GENIE ntuple conversion application.
\item gspladd: A GENIE XML cross section spline file merging application. 
\item gspl2root: A GENIE XML to ROOT cross section spline file conversion
utility. 
\item gevgen\_numi: Alias for gevgen\_fnal maintained for historical reasons.
\end{itemize}
\item \textbf{\textit{\large{}H}}{\large \par}

\begin{itemize}
\item hA: See INTRANUKE.
\item hN: See INTRANUKE.
\end{itemize}
\item \textbf{\textit{\large{}I}}{\large \par}

\begin{itemize}
\item IMD: Inverse Muon Decay
\item INTRANUKE: A home-grown intranuclear hadron transport MC. Intranuke
was initially developed within NEUGEN for the Soudan-2 experiment
by W.A.Mann, R.Merenyi, R.Edgecock, H.Gallagher, G.F.Pearce and others.
Since then it was significantly improved and is now extensively used
by MINOS and other experiments. Current INTRANUKE development is led
by S.Dytman. INTRANUKE, in fact, contains two independent models (called
`hN' and `hA').
\end{itemize}
\item \textbf{\textit{\large{}J}}{\large \par}

\begin{itemize}
\item JNUBEAM: Geant3-based JPARC neutrino beamline simulation software.
\item JPARC: Japan Proton Accelerator Research Complex. Home of T2K neutrino
beamline.
\end{itemize}
\item \textbf{\textit{\large{}K}}{\large \par}

\begin{itemize}
\item KNO: Koba, Nielsen and Olesen scaling law.
\end{itemize}
\item \textbf{\textit{\large{}L}}{\large \par}

\begin{itemize}
\item LHAPDF: Les Houches Accord PDF Interface.
\item libxml2: The XML C parser and toolkit of Gnome (see http://xmlsoft.org).
\item log4cpp: A library of C++ classes for fleible loggingto files, syslog,
IDSA and other destinations (see http://log4cpp.sourceforge.net).
\end{itemize}
\item \textbf{\textit{\large{}M}}{\large \par}

\begin{itemize}
\item MacPorts: An open-source community initiative to design an easy-to-use
system for compiling, installing, and upgrading either command-line,
X11 or Aqua based open-source software on the Mac OS X operating system.
\item Mersenne Twistor: The default random number generator in GENIE (via
ROOT TRandom3 whose implementation is based on M. Matsumoto and T.
Nishimura, Mersenne Twistor: A 623-diminsionally equidistributed uniform
pseudorandom number generator ACM Transactions on Modeling and Computer
Simulation, Vol. 8, No. 1, January 1998, pp 3--30.)
\end{itemize}
\item \textbf{\textit{\large{}N}}{\large \par}

\begin{itemize}
\item NeuGEN: A legacy, fortran77-based neutrino generator by H.Gallagher
et al.
\item NEUT: A legacy, fortran77-based neutrino generator by Y.Hayato et
al.
\item NUANCE: A legacy, fortran77-based neutrino generator by D.Casper et
al.
\item NUX: A legacy, fortran77-based neutrino generator by A.Rubbia et al.
\item NuMI: Neutrinos at the Main Injector. A neutrino beamline at Fermilab.
\end{itemize}
\item \textbf{\textit{\large{}O}}{\large \par}
\item \textbf{\textit{\large{}P}}{\large \par}

\begin{itemize}
\item PREM: Preliminary Earth Model, The Encyclopedia of Solid Earth Geophysics,
David E. James, ed., Van Nostrand Reinhold, New York, 1989, p.331
\item PYTHIA:
\end{itemize}
\item \textbf{\textit{\large{}Q}}{\large \par}

\begin{itemize}
\item QEL: Quasi-Elastic.
\end{itemize}
\item \textbf{\textit{\large{}R}}{\large \par}

\begin{itemize}
\item RES: Resonance.
\item Registry:
\item ROOT:
\item RooTracker: A ROOT-only STDHEP-like event format (very similar to
GHEP event format but with no GENIE class dependencies) developed
in GENIE as an evolution of the Tracker format. See also Tracker.
\item RS: Rein-Sehgal
\item RSB: Rein-Sehgal-Berger
\item RSD: Remote Software Deployment Tools. A system for automated software
installation developed by Nick West (Oxford).
\end{itemize}
\item \textbf{\textit{\large{}S}}{\large \par}

\begin{itemize}
\item SF:
\item SVN: See Subversion.
\item SKDETSIM: The fortran77-based Super-Kamiokande detector simulation.
\item Subversion: 
\end{itemize}
\item \textbf{\textit{\large{}T}}{\large \par}

\begin{itemize}
\item THZ: Topocentric Horizontal Coordinate System (+z: Points towards
the Local Zenith / +x: On same plane as Local Meridian, pointing South
/ +y: as needed to make a right-handed coordinate system / Origin:
Detector centre). 
\item Tracker:
\end{itemize}
\item \textbf{\textit{\large{}U}}{\large \par}
\item \textbf{\textit{\large{}V}}{\large \par}
\item \textbf{\textit{\large{}W}}{\large \par}
\item \textbf{\textit{\large{}X}}{\large \par}

\begin{itemize}
\item XML: Extensible Markup Language.
\end{itemize}
\item \textbf{\textit{\large{}Y}}:

\begin{itemize}
\item Yum: Yellowdog Updater, Modified (YUM). An open-source command-line
package-management utility for RPM-compatible Linux operating systems.
\end{itemize}
\item \textbf{\textit{\large{}Z}}{\large \par}
\end{itemize}
\newpage

\clearpage{}

\bibliographystyle{ieeetr}
\phantomsection\addcontentsline{toc}{chapter}{\bibname}\bibliography{refs}

\end{document}